\documentclass{ws-ijmpa}
\usepackage{amsmath}
\usepackage{amssymb}
\usepackage{xspace}
\newcommand {\epen} {\ensuremath{\mathrm{e^+e^-}}\xspace}

\newcommand {\koralw} {K{\sc oral}W\xspace}
\newcommand {\grace} {G{\sc race4f}\xspace}
\newcommand {\yfsww} {Y{\sc fs}WW\xspace}
\newcommand {\yfszz} {Y{\sc fs}ZZ\xspace}
\newcommand {\raconww} {R{\sc acoon}WW\xspace}
\newcommand {\zzto} {ZZTO\xspace}

\newcommand {\pythia} {P{\sc ythia}\xspace}
\newcommand {\jetset} {J{\sc etset}\xspace}
\newcommand {\herwig} {H{\sc erwig}\xspace}
\newcommand {\geant} {G{\sc eant}\xspace}
\newcommand {\opal} {{OPAL}\xspace}
\newcommand {\nutev} {{NuTeV}\xspace}
\newcommand {\tevatron} {{Te\kern -0.1em Vatron}\xspace}
\newcommand {\delphi} {{DELPHI}\xspace}
\newcommand {\alephe} {{ALEPH}\xspace}
\newcommand {\ldrei} {{L3}\xspace}
\newcommand {\excalibur} {E{\sc xcalibur}\xspace}
\newcommand {\ariadne} {A{\sc riadne}\xspace}

\newcommand {\jade} {J{\sc ade}\xspace}
\newcommand {\durham} {D{\sc urham}\xspace}
\newcommand {\camjet} {C{\sc amjet}\xspace}
\newcommand {\diclus} {D{\sc clus}\xspace}
\newcommand {\MWC} {\ensuremath{\mathrm{M_W }}\xspace}
\newcommand {\W} {\ensuremath{\mathrm{W}}\xspace}
\newcommand {\Z} {\ensuremath{\mathrm{Z}^0}\xspace}
\newcommand {\ZX} {\ensuremath{\mathrm{Z}}\xspace}
\newcommand {\bW} {\ensuremath{\mathrm{\mathbf W }}\xspace}
\newcommand {\dmwc}    {\ensuremath{\mathrm{\Delta M_W^{4C} }}\xspace}

\newcommand {\prob} {{ P}}
\newcommand {\GeV}    {\ensuremath{\mathrm{Ge\kern -0.1em V}}\xspace}
\newcommand {\MeV}    {\ensuremath{\mathrm{Me\kern -0.1em V}}\xspace}
\newcommand {\fb}    {\ensuremath{\mathrm{fb}}\xspace}
\newcommand {\rms}    {RMS\xspace}
\newcommand {\vx}     {\ensuremath{\mathbf x}\xspace}
\newcommand {\vp}     {\ensuremath{\mathbf p}\xspace}
\newcommand {\ip}     {\ensuremath{\mathit p}\xspace}
\def\ifmath#1{\relax\ifmmode #1\else $#1$\fi}%
\def\TeV{\ifmmode {\mathrm{ Te\kern -0.1em V}\xspace}\else
                   \textrm{Te\kern -0.1em V}\xspace\fi}%
\def\keV{\ifmmode {\mathrm{ ke\kern -0.1em V}\xspace}\else
                   \textrm{ke\kern -0.1em V\xspace}\fi}%
\def\eV{\ifmmode  {\mathrm{ e\kern -0.1em V}\xspace}\else
                   \textrm{e\kern -0.1em V\xspace}\fi}%

\newcommand{\TEVATRON}{\ensuremath{\mathrm{\hbox{Te\kern -0.1em Vatron}}}\xspace}
\newcommand{\LEP}{\ensuremath{\mathrm{\hbox{LEP}}}\xspace}
\newcommand{\LEPI}{\ensuremath{\mathrm{\hbox{LEP-I}}}\xspace}
\newcommand{\LEPII}{\ensuremath{\mathrm{\hbox{LEP-II}}}\xspace}
\newcommand{\pb}{\ensuremath{\mathrm{pb^{-1}}}}

\def\mca#1#2 {\multicolumn{#1}{|c|}{#2}}
\newcommand{\antibar}[1]{\ensuremath{\mathrm{#1\overline{#1}}}}

\newcommand{\ff}{\antibar{f}}
\newcommand{\leptlept}{\ensuremath{\ell^+\ell^-}}
\newcommand{\ee}{\ensuremath{\mathrm{e}^+\mathrm{e}^-}}
\newcommand{\mumu}{\ensuremath{\mu^+\mu^-}}
\newcommand{\tautau}{\ensuremath{\tau^+\tau^-}}
\newcommand{\bb}{\antibar{b}}
\newcommand{\cc}{\antibar{c}}
\renewcommand{\ss}{\antibar{s}}
\newcommand{\dd}{\antibar{d}}
\newcommand{\uu}{\antibar{u}}
\newcommand{\qq}{\antibar{q}}

\newcommand{\WW}{\ensuremath{\mathrm{W^+W^-}}}
\newcommand{\Abb}{\ensuremath{A_{\mathrm{FB}}^{\bb}}}

\newcommand{\Acc}{\ensuremath{A_{\mathrm{FB}}^{\cc}}}

\newcommand{\alfas}{\alpha_{\mathrm{S}}}

\newcommand{\MZ}{m_{\mathrm{Z}}}

\newcommand{\MW}{\mbox{$m_{\mathrm{W}}$}}
\newcommand{\MH}{m_{\mathrm{H}}}

\newcommand{\Mt}{m_{\mathrm{t}}}

\newcommand{\Gnn}{\Gamma_{\nu\nu}}

\newcommand{\ord}{{\cal O}}

\newcommand{\Zzero}{\ensuremath{\mathrm{Z}}}

\newcommand{\thw}{\theta_{\mathrm{W}}}
\newcommand{\thweff}{\theta_{\mathrm{eff}}}
\newcommand{\thwefff}{\theta_{\mathrm{eff}}^{{\rm f}}}
\newcommand{\thweffq}{\theta_{\mathrm{eff}}^{{\rm q}}}
\newcommand{\thweffb}{\theta_{\mathrm{eff}}^{{\rm b}}}
\newcommand{\thweffc}{\theta_{\mathrm{eff}}^{{\rm c}}}

\newcommand{\swsqb}{\sin^2\overline{\theta}_{\mathrm{W}}}

\newcommand{\thz}{\theta_0}

\newcommand{\roots}{\ensuremath{\sqrt{s}}}

\newcommand{\ra}{\rightarrow}

\newcommand{\ptau}{\ensuremath{{\cal P}_{\tau}}}

\newcommand{\blX}{\mbox{${\mathrm b \rightarrow \ell^- X}$}}

\newcommand{\Dstar}{{\rm D}^{* }}

\newcommand{\Dstarpm}{\mbox{${\mathrm D}^{*\pm}$}}

\newcommand{\Dzero}{{\rm D}^0}
\newcommand{\Dplus}{{\rm D}^+}

\newcommand{\clX}{\rm c \rightarrow \ell X}

\newcommand{\Ds}{{\rm{D_s}}}
\newcommand{\Lc}{{\rm{\Lambda_c}}}
\def\Ups4s{\mbox{$\Upsilon(4S)$}}
\newcommand{\etal}{\mbox{{\it et al.}}}
\newcommand{\RSI}[3]  {Review of Scientific Instruments {\bf A#1} (#2) #3}
\newcommand{\NIM}[3]  {Nucl.\ Instr.\ and Meth.\ {\bf A#1} (#2) #3}
\newcommand{\PLB}[3]  {Phys.\ Lett.\ {\bf B#1} (#2) #3}
\newcommand{\EPJ}[3]  {Eur.\ Phys.\ J.\ {\bf C#1} (#2) #3}
\newcommand{\ZPC}[3]  {Zeit.\ Phys.\ {\bf C#1} (#2) #3}

\newcommand{\PRB}[3] {Phys.\ Rev.\ {\bf B#1} (#2) #3}
\newcommand{\PR}[3] {Phys.\ Rev.\ {\bf #1} (#2) #3}

\newcommand{\PRL}[3]  {Phys.\ Rev.\ Lett.\ \textbf{#1} (#2) #3}
\newcommand{\PRD}[3]  {Phys.\ Rev.\ {\bf D#1} (#2) #3}
\newcommand{\NPB}[3]  {Nucl.\ Phys.\ {\bf B#1} (#2) #3}

\newcommand{\CPC}[3]  {Comp.\ Phys.\ Comm.\ {\bf #1} (#2) #3}
\newcommand{\QW}{\ensuremath{Q_{\mathrm{W}}}}

\newcommand{\epem}{\mbox{$\mathrm{e^+e^-}$}}

\newcommand{\WWg}{\mbox{WW$\gamma$}\xspace}

\newcommand{\WWZ}{\mbox{WW$\Zzero$}\xspace}
\newcommand{\ZZ}{\mbox{\Zzero\Zzero}\xspace}

\newcommand{\lnu}{\mbox{$\ell\overline{\nu}_{\ell}$}}
\newcommand{\lnux}{\mbox{$\ell{\nu}_{\ell}$}}

\newcommand{\enu}{\mbox{$\mathrm{e\overline{\nu}_{e}}$}}
\newcommand{\mnu}{\mbox{$\mu\overline{\nu}_{\mu}$}}
\newcommand{\tnu}{\mbox{$\tau\overline{\nu}_{\tau}$}}
\newcommand{\WWqqqq}{\mbox{\WW$\rightarrow$\qq\qq}\xspace}

\newcommand{\qqqq}{\mbox{\qq\qq}}

\newcommand{\qqln}{\mbox{\qq\lnu}}
\newcommand{\lnln}{\mbox{\lnux\lnu}}
\newcommand{\WWqqln}{\mbox{\WW$\rightarrow$\qq\lnu}\xspace}

\newcommand{\WWqqen}{\mbox{\WW$\rightarrow$\qq\enu}\xspace}

\newcommand{\WWlnln}{\mbox{\WW$\rightarrow$ \lnux\lnu}}

\def\etal{\mbox{{\it et al.}}}

\begin{document}
\markboth{Raimund Str\"ohmer}
{Review of the Properties of the W Boson at LEP,
and the Precision Determination  of its Mass}

\catchline{}{}{}

\title
{Review of the Properties of the W Boson at LEP,
and the Precision Determination  of its Mass}

\author{\footnotesize Raimund Str\"ohmer}

\address{
Lehrstuhl Schaile, Ludwig-Maximilans Universit\"at M\"unchen,
Am Coulombwall 1, D-85748 Garching, Germany.}

\maketitle

\pub{Received (Day Month Year)}{Revised (Day Month Year)}

%
%
\begin{abstract}
We review the precision measurement of the mass and couplings 
of the W Boson at LEP. The total and differential $\W^+\W^-$ 
 cross section is  used to extract the \WWZ and \WWg couplings.
We discuss the techniques used by the four LEP experiments
to determine the \W mass in different decay
channels, and present the details of methods used to evaluate 
the sources of systematic uncertainty.
\end{abstract}
%
%

\section{Introduction }
Our current understanding of the fundamental forces 
is contained in the description of 
 the gravitational, the strong, the weak and the
electromagnetic interactions among elementary particles. 
In the 1960s, Glashow,
Salam and Weinberg unified the electromagnetic and weak interaction into the
 electroweak theory,\cite{bib:ew_theor} which, together with  QCD (the
theory of the strong interaction), forms the Standard Model of particle physics.  
The electroweak interaction among particles takes place through the exchange of
four bosons, namely, the massless photon, and the massive \Z, $\W^+$ and $\W^-$
bosons. In 1983, the UA1 and UA2 experiments at the CERN proton-antiproton  collider 
($Sp\bar{p}S$) were  first to   directly
observe the \Z and \W bosons.\cite{bib:ua1}

The  electron-positron collider \LEP at CERN provided the possibility 
of precise studies of the properties of the  \Z and \W.  
Unlike  proton-proton  or proton-antiproton colliders, where quarks or gluons 
carry an a priori unknown fraction of the hadron momentum,
the well defined initial state in
electron-positron collisions is  well suited for such precision 
measurements
In its first phase (\LEPI ), \LEP operated at
center-of-mass energies near the \Z resonance  of about 91 \GeV. 
In the second phase (\LEPII ),  the center-of-mass energy was increased  above the
threshold of twice the \W  mass ($M_\W$), which made it possible 
to  produce pairs of \W bosons in 
$e^+e^-$ collisions,
and thus offered the opportunity for a precision determination of the
\W mass, its couplings, and decay  branching ratios.
The precision measurements of the properties of the \Z and \W boson  provided
stringent tests  of  quantum corrections to  the Standard Model. 

In this section, we give a general overview of the \LEP collider and
its four major experiments. We discuss the pair production of 
\W bosons   and the event
topologies arising from different decay modes of the \W.
Event selection, the total cross sections, and 
branching fractions will be presented in Section~\ref{sec:cross}.
The extraction of triple and quartic gauge couplings from the total and
differential cross sections will be described in Section~\ref{sec:tgc}.
In Section~\ref{sec:mw_evt}, we describe
event reconstruction and the event-by-event estimation of the
mass of the W and its uncertainty.
Section~\ref{sec:mw_global} 
discusses different methods used to determine 
the Standard-Model parameter \MWC from these estimates.
Possible biases in determining the 
mass, its statistic stability, and the estimate in its uncertainty are 
examined using  Monte Carlo ensemble tests. 
Section~\ref{sec:syst} describes how the systematic uncertainties of the measurements 
are estimated and how the measurements can be optimized in order to
reduce the sum of the  statistical and systematic error. 
Section~\ref{sec:mw_lvlv} will  present the determination of the \W  mass
for events where both \W bosons decay leptonically.
Section ~\ref{sec:mw_results} will
summarize  measurements  of the mass of the \W boson mass and its width and
their implications for the Standard Model.

\subsection{Motivation}
\begin{figure}
\begin{center}
\epsfig{file=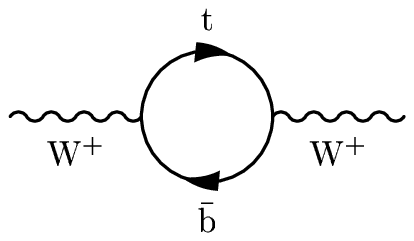,width=5.5cm} 
\epsfig{file=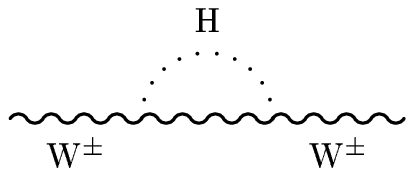,width=5.5cm}
\vspace*{.2cm}
\end{center}
\caption{Feynman diagram for 1-loop corrections to the \W propagator.
 \label{fig:mw_loop}
}
\end{figure}
\begin{figure}
\begin{center}
\epsfig{file=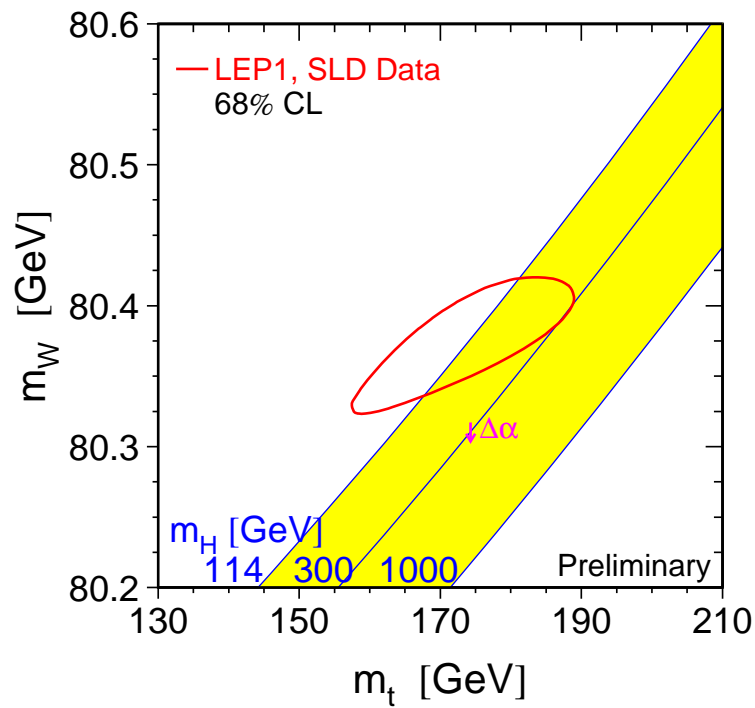,width=10.cm}
\end{center}
\caption{Result of a  fit of the electroweak data to the Standard Model,
and the expected correlation between  the \W-boson mass  
and the top mass\protect\cite{bib:elw_fits}.
\label{fig:mw_mt_od}}
\end{figure}
In the electroweak Standard Model,
the properties of the \Z and \W boson depend 
only  on a few fundamental parameters.
A comparison of the directly measured \W  mass with 
predictions based on precision measurements of properties of the  \Z boson provides 
therefore an important test of the Standard Model.
In the context of \LEPI analyses, $M_\W$ can be determined from
the relation:\cite{bib:w_br}
\[ G_\mu = \frac{\alpha \pi}{\sqrt{2}M_\W^2(1-M_\W^2/M_\ZX^2)}
              \frac{1}{1-\Delta r} .  \]
At lowest order (``tree'' level),
$M_\W$ depends  only on the Fermi constant $G_\mu$, which is  known  accurately from
muon decay, the fine structure constant $\alpha$, and the mass of the \Z
boson ($M_\Z$).
In  the  above  equation, loop corrections are parametrized by $\Delta r$, and lead
to a quadratic dependence of $M_\W^2$ on the top mass, and a logarithmic dependence
on the Higgs mass. At lowest order,   $\Delta r =0$.
As example, Fig.  \ref{fig:mw_loop} shows 1-loop
contributions to the \W propagator, including the top quark and  Higgs boson.
A fit of the precision electroweak observables to the Standard
Model (excluding the direct measurements of  $M_{top}$) yields:\cite{bib:elw_fits}
\[ M_\W = 80.373 \pm 0.033 \ \GeV. \]
Figure \ref{fig:mw_mt_od} shows the prediction for the \W boson and top quark
mass using a fit to all data, excluding the direct measurements of \MWC and $M_{top}$.
The figure also displays the  prediction of the Standard Model for \MWC as function of $M_{top}$
for three values of Higgs mass. The little arrow indicates the uncertainty due to
 the running of $\alpha$ at the scale of $M_\Z$.
The comparison of these predictions with direct measurements of $M_{top}$
and \MWC at the \LEP and \TEVATRON colliders are important tests of
the electroweak sector of the Standard Model, which to be fully exploited
require reduction of 
 the error on the observed \W mass.

Beyond just the mass, the measurement of the \W pair and other  four-fermion cross sections,
 and their angular distributions can also be used to study triple gauge-boson couplings.
Any deviations from predictions of the Standard Model can then be interpreted
as evidence  for physics beyond the Standard Model. This issue is addressed in the next section.

\subsection[\W-Pair Production and Decay]{\bW Pair Production and Decay} \label{sec:ww_prod}
\begin{figure}
\begin{center}
\epsfig{file=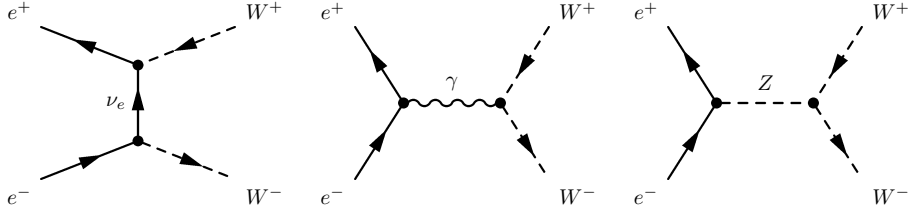,width=12.cm}
\vspace*{.2cm}
\end{center}
\caption{Feynman diagrams for \W pair production.
 \label{fig:fey_ww}}
\end{figure}
\begin{figure}[t]
\begin{center}
\parbox[b]{.6cm}{a) \vspace*{3cm}} \parbox[t]{5.5cm}{\epsfig{file=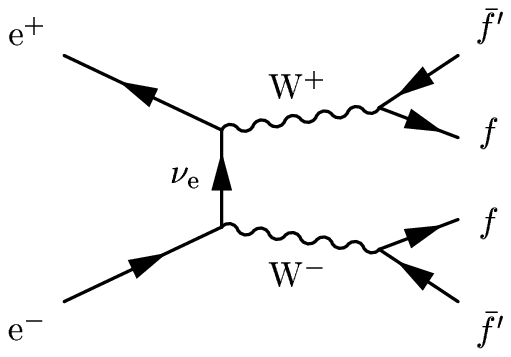,width=5.3cm}}
\parbox[b]{.6cm}{b) \vspace*{3cm}} \parbox[t]{5.5cm}{\epsfig{file=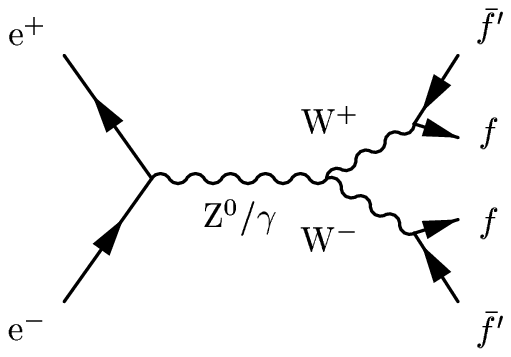,width=5.3cm}} 
\parbox[b]{.6cm}{c) \vspace*{3cm}} \parbox[t]{5.5cm}{\epsfig{file=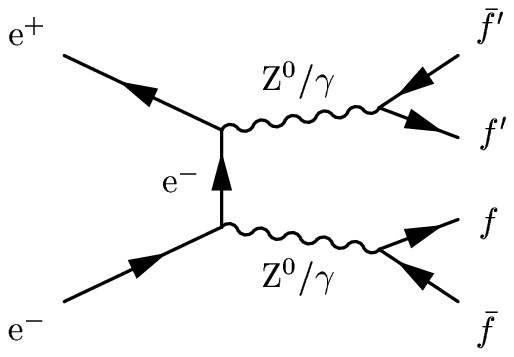,width=5.3cm}} 
\parbox[b]{.6cm}{d) \vspace*{3cm}} \parbox[t]{5.5cm}{\epsfig{file=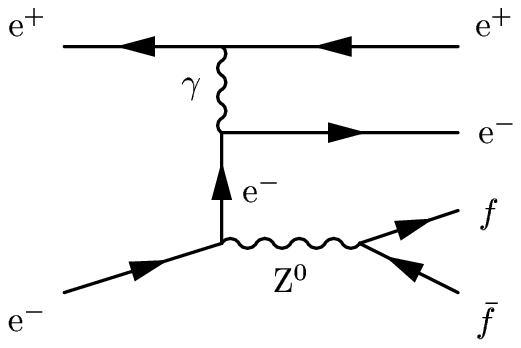,width=5.3cm}} 
\parbox[b]{.6cm}{e) \vspace*{3cm}} \parbox[t]{5.5cm}{\epsfig{file=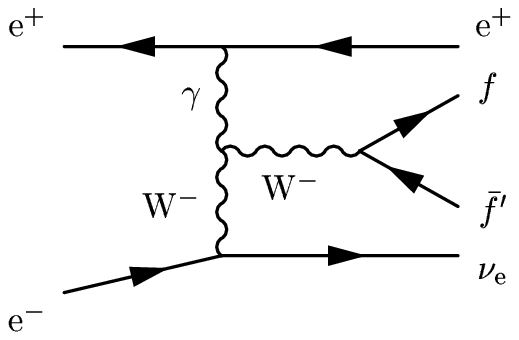,width=5.3cm}}
\parbox[b]{.6cm}{f) \vspace*{3cm}} \parbox[t]{5.5cm}{\epsfig{file=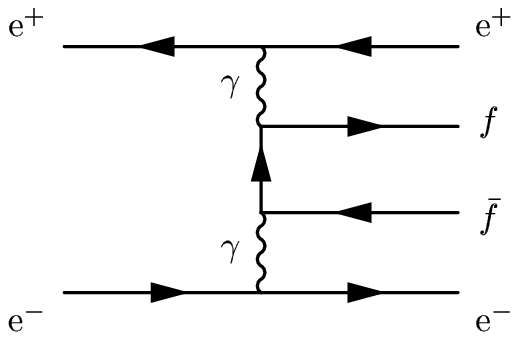,width=5.3cm}} 
\vspace*{.2cm}
\end{center}
\caption{Feynman diagrams leading to four-fermion final states. 
a) and b) display the CC03 diagrams, c) the NC02 diagrams involving
\Z or $\gamma$-boson pairs, d),e) and f) are additional diagrams, all of which lead to a four-fermion
final state.  \label{fig:fey_4f}}
\end{figure}

\W-boson pairs can 
be produced in $e^+e^-$ annihilations above the center-of-mass energy
threshold of $ \sqrt{s} = 2 \cdot \MWC$.
Figure \ref{fig:fey_ww} shows the tree-level Feynman diagrams contributing
to \W pair productions. These diagrams are called CC03 diagrams.
(CC refers to charged current, because \W bosons are involved, and 03 to the number
of diagrams.) The Feynman diagrams for \W pair production via a 
virtual \Z or virtual photon, contain a triple gauge-boson coupling, which can
be extracted by measuring the pair-production cross section, or by
studying the angular distribution of the \W bosons and their decay products.
The \W bosons are not stable but decay within $3 \cdot 10^{-25}$s, corresponding
to a decay width of about 2 \GeV.
Each \W decays into a pair of fermions,
as  indicated in Fig.\ref{fig:fey_4f}~a) and b). The
final state in  \W-boson pair production processes  therefore contain  four
fermions. Figures \ref{fig:fey_4f} c)-f) show additional electroweak processes
that also lead to four-fermion final states. Since these provide the same final state,
their interference with \W-pair production  
must be taken into proper account.
The analysis of the \W-boson cross section and mass  is based on events in which the four fermions
can be grouped into two pairs with  invariant masses  close to the \W mass.  
In this region of phase space, the CC03 diagrams dominate the four-fermion  
cross section, which is of the order of $15$ pb for most of the luminosity collected
at \LEPII. 
 
The \W boson decays 68.5 \%  into a quark-antiquark pair,
which is observed in the detector as two jets, and to 31.5 \% 
into a charged lepton and its neutrino.
Consequently, the  decays of the two \W-bosons provide three
distinct signatures. About 47 \% of the time, both \W bosons 
decay hadronically, which yields at least four  jets in the final state.
About  43 \% of the times, one \W decays hadronically and the other leptonically
(``semileptonic'' decay channel), leading to events with two jets, a high energetic
lepton, and missing energy due to the unobserved neutrino. 
The remaining 10\% of events, where  both \W bosons decay leptonically, 
contain two  energetic leptons and a large amount of missing 
energy.
For all these event topologies,  
it is important to be able to have detectors that can
 measure precisely the momenta and directions of the leptons and jets, and this
is the subject of the next section.

\subsection{The \LEP Collider and its Detectors}
\LEP  is an  electron-positron collider with a circumference of 27 km, located  at CERN near Geneva, Switzerland.\cite{bib:lep} 
The electron and positron beams collide at four interaction points
inside the \LEP tunnel at the positions of the   \alephe,\cite{bib:aleph_g} \delphi,\cite{bib:delphi_g}
 \ldrei,\cite{bib:l3_g} and \opal\cite{bib:opal_g} detectors. 
From 1989 until 1995,
\LEP was operated at center-of-mass energies near the \Z resonance.
At the end of 1995, the center-of-energy was increased in  steps, 
crossing the \W-pair production threshold in 1996, and  
reaching 209 \GeV for some of the data taking in the year 2000.
 Each \LEP experiment has collected more than $700 \  \pb$ 
at center-of-mass energies above the \W-pair production threshold,
each corresponding to the production of about $10,000$ \W-pair events.

\begin{figure}[t]
\begin{center}
\epsfig{file=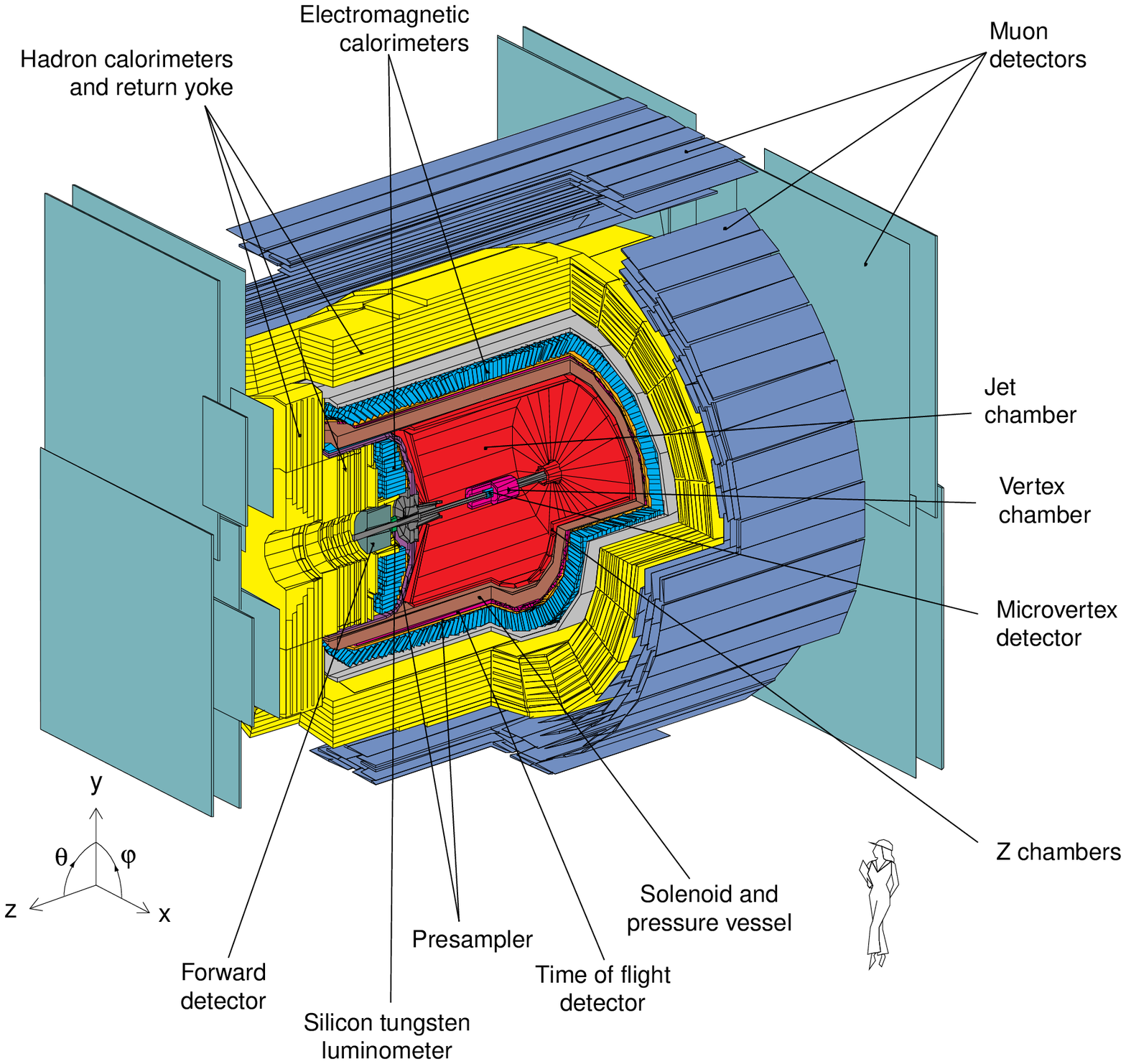,width=12cm}
\vspace*{.2cm}
\end{center}
\caption{The \opal detector.
 \label{fig:opal}}
\end{figure}
All \LEP detectors are cylindrically symmetric around the beam axis.
The detector components most important for the study of  \W bosons are
the inner tracking detectors, the calorimeters and the muon system.
As an example of a \LEP detector, Fig. \ref{fig:opal} shows the \opal detector.

In the inner gas-tracking detectors, 
the trajectory of charged particles  can be observed through the ionization
produced along their paths. When an electric field is applied in the gaseous volume, 
the  created-charge 
 drifts towards sensors, which,  in the case of \opal correspond to wires parallel to the
beam axis (drift chamber), and, in the case of the other three experiments, to sensors
located in a plane perpendicular to the beam, at the ends of the detectors (time-projection or
time-expansion chambers).
Points along the trajectory can  be determined 
from the time it takes the charge to reach the sensor. 
In order to measure a particle's momentum,
the track detectors are placed inside a solenoidal magnetic
field, and the momentum is determined from the curvature of the track 
in the magnetic field. The curvature is  proportional to $1/p_t$, where 
$p_t$ is the component of the momentum transverse 
to the magnetic field. The component parallel to the magnetic field can be
determined from $p_t$ and the initial direction of the track. 
The sign of a particle's charge is determined by the sense
of the curvature.

Electrons  and photons are detected in the electromagnetic calorimeter.
In matter, electrons radiate  photons through  bremsstrahlung, and photons can convert
into electron-positron pairs. The interplay of these two  processes leads
 to a ``shower'' of electrons, positrons and photons within the 
calorimeter.
The showering stops  when the energy loss of the particles  in the shower
from  bremsstrahlung is of the same order
as the energy loss from ionization. 
The energy of the primary particle 
can be obtained from the energy deposited  by the shower, which reflects
the total length of the trajectories of  all electrons 
and positrons in the shower. The effective   length of the trajectory can, in turn, be   
determined from  the amount of
\c{C}erenkov radiation produced in the detector  materials (lead-glass for the
\opal detector), or the amount of light produced in scintillating material (BGO Crystal in \ldrei), 
or  the amount of ionization
produced in  gas filled detectors sandwiched between sheets of lead  (sampling calorimeters of \alephe and \delphi).   

The electromagnetic calorimeter is followed by a hadronic calorimeter, in which energies of both
charged and neutral hadrons are measured by showers  they produce through strong interactions
within the detector material.

The only particles that  do not produce electromagnetic or hadronic showers, and
are therefore not stopped within the calorimeters,  are muons and neutrinos.
Muons do not interact strongly and, compared to electrons, their bremsstrahlung is 
greatly suppressed because of  their much larger mass.  
Muons are identified beyond the hadron calorimeter using  ionization
detectors.  Only  the muon detectors  of  \ldrei reside within 
a magnetic field, and can be used to measure the muon momentum. In the
other three \LEP experiments, the signals in the muon detectors  
have to be matched with a reconstructed track in
the inner tracking detectors, in order to determine the muon momentum.
Neutrinos  interact only weakly. Because of their extremely small scattering cross section,
they leave the detector without interacting, and can 
therefore  not  be detected directly in any \LEP detector.

\subsection{Monte Carlo Simulation}\label{sec:mc}
An important tool in interpreting \LEP data involves the comparison of
measurements with predictions based on Monte Carlo simulations.
Precision measurements are used typically to determine  parameters of theory (e.g., the mass of the \W boson) 
that agree best with the observed data.
The theory, however, only predicts the properties of fundamental particles
(e.g., the quarks and leptons from \W-boson decay). Monte Carlo simulations
must therefore be  used to estimate the effects of the fragmentation of quarks into
hadrons, and the  acceptance and resolution of the detector. Well-understood  processes are
used to test and to calibrate such Monte Carlo simulations, which 
can  be regarded as tools for extrapolating  the effects
of fragmentation and detector resolution from known test samples to 
events used in the precision measurement.
In addition to this extrapolation, Monte Carlo studies provide  an important tool for 
 the optimization of any analysis. The statistical precision and  effects of systematic
uncertainties can be checked using  large samples of Monte Carlo events, for which
the results are already known.
 
The first step in the generation of a Monte Carlo  event 
involves the simulation of the primary interaction. Several programs are usually used
to simulate the process
$e^+e^- \rightarrow 4f$. The \koralw  program\cite{bib:koralw}
provides either the CC03 diagrams, or, by interfacing with 
the \grace program,\cite{bib:grace}
the full leading-order four-fermion matrix elements. 
The program \excalibur\cite{bib:excalibur} can also be used to 
calculate  the four-fermion matrix
elements.
 Both programs contain a detailed implementation of 
the radiation of initial-state photons, and can be used to simulate the momenta of the 
four final-state fermions and such photons.
The programs \yfsww~\cite{bib:yfs} and \raconww~\cite{bib:raconww} can simulate the full 
${\cal O}(\alpha)$ QED correction
to the CC03 contributions in the double-pole approximation.\cite{bib:double_pole}

When the final state contains quarks, the fragmentation of these colored objects into
hadrons is simulated with the programs \pythia,\cite{bib:phytia} \herwig\cite{bib:herwig} 
or \ariadne.\cite{bib:ariadne}
In the first phase of the simulation, a QCD cascade is generated by simulating the radiation of
gluons and the splitting of gluons into quark anti-quark pairs. 
In \ariadne, the QCD cascade is simulated by a dipole
model, where quarks and gluons form color dipoles which then radiate gluons.
The QCD cascade is terminated when the virtuality
of the quarks and gluons falls below a cut off, of typically 1-2 \GeV.

The fragmentation of  quarks and gluons at the end of the QCD cascade can be
simulated  only  phenomenologically. 
\pythia and \ariadne  use string models.
A string connects  partons, which form a color singlet.
The string begins at a quark and connects it with a gluon. This gluon is connected by the
string to other gluons, until the string ends on an antiquark. The gluons can therefore 
be considered as kinks in a string that connects a quark with an antiquark.
The string is broken into pieces by the insertion of quark-antiquark and diquark-antidiquark pairs
into the string. These pieces, which consist of short strings connecting a quark and a
antiquark close in phase space, form hadrons. The fraction of  string energy carried by a 
quark or antiquark, and the fraction of  momentum perpendicular to the string axis,
are generated using functions that contain free parameters of the model.

In \herwig, all gluons  split into quarks and antiquarks. Hadrons are
then produced from clusters that are formed of color-neutral quark-antiquark pairs.
All fragmentation models have  parameters that can be adjusted 
to maximize  agreement with data,
but the fragmentation of \W bosons is well described using the same  parameters as
determined for the \Z at resonance.

The decay of unstable particles is simulated with the help of decay tables. For most 
 particles, the tables are based on their measured branching ratios in different
decay channels. When branching ratios are unknown,  estimates are used for which the
simulated inclusive-particle-rates agree  best  with available measurements. 
For particles with sufficiently  long  lifetimes  that 
interact within the detector, the decay is convoluted with the detector response.
A $K^0_S$, for example, decays typically within the  volume of the tracking detectors, creating a pair of
tracks that start in the middle of the tracking chambers, while most  $K^0_L$ do not decay
before they reach the calorimeters.
The propagation of any particle through the detector, its interaction with the 
detector material, and the response of the sensors, are simulated with the \geant
program package.\cite{bib:geant}
 
\subsection{Measurement of Jets}\label{sec:jet}
The quarks  produced in the primary weak interaction
cannot be observed directly in the detector. As described in
Section \ref{sec:mc}, they fragment into hadrons that can be  observed.
To estimate 
the energy and momentum of the primary partons
(quarks or any radiated gluons),
the observed detector signals are clustered into so-called jets. 
In general, jet algorithms 
 make use of the fact that hadrons from any primary particle
are close to each other.
Most analyses of \epen collider data  use   the \jade type
of  jet algorithms.\cite{bib:jade} 
In this algorithm, pseudo-particles are added together iteratively 
to form jets.
The algorithm starts with a list of pseudo-particles that consist of
all considered particles or detector hit information.
Based on an algorithm-dependent measure, the two pseudo-particles
that are closest to each other  are merged
into  a new pseudo-particle. 
This is iterated until a prescribed cutoff criterion is
reached. The remaining pseudo-particles are then called jets.
Different algorithms of the \jade variety use different definitions of
closeness between pseudo-particles, and  calculate the four-momenta of
the new pseudo-particles from the four-momenta of the
two parents (pseudo-particles from which they are formed)  in different ways.
The \durham algorithm,\cite{bib:durham} which is used by all four \LEP
experiments in the determination of the \W  mass, defines the closeness
between to pseudo-particles $i$ and $j$ as: 
\[ y(i,j) =  \frac{2\min (E_i^2,E_j^2) (1 -\cos \theta_{ij})}{E^2_{vis}} \]
where  $E_i$ and
$E_j$ are the energies of the two pseudo-particles, and $\theta_{ij}$ is
the angle between them, and $E_{vis}$ is the sum of the energies of all
initial pseudo-particles.
In the \durham scheme, two pseudo-particles are combined into a new
pseudo-particle by adding their four-momenta. The four-momentum
of a jet is therefore given by the sum of the four-momenta of all particles belonging 
to it. The jets reconstructed using the Durham algorithm  have 
significant masses. 
When the  algorithm is used to  reconstruct \W boson pairs, 
it is stopped when the number
of pseudo-particles equals the number of expected jets, i.e., at least four in
the fully hadronic channel (or five if  gluon
bremsstrahlung is considered), and two in the semileptonic channel (in
this channel, detector signals from the  identified lepton are not considered
in the initial list of pseudo-particles).
The smallest value of $y(i,j)$ of $n$ remaining pseudo-particles is called 
$y_{n-1,n}$. It can be used to distinguish  between  events with  
 an $n$-jet structure and events with an $(n-1)$-jet structure.  $y_{34}$ is one of the quantities 
used to distinguish  between signal events that have  both \W bosons decaying hadronically,
and background from 2-quark events that contain extra gluon bremsstrahlung. In certain
\LEP analyses, $y_{45}$ is used  to decide whether an event
should be reconstructed as a four-jet or a five-jet event.

When jets are reconstructed from data or from Monte Carlo events that include  full detector
simulation, the initial list of pseudo-particles consists of the charged tracks measured in the
tracking devices and any clusters of energy deposition in the calorimeters. 
Since most charged tracks are caused by pions, the pion mass is assumed in the
determination of the four-momentum of a pseudo-particle obtained from  tracking.
As most electromagnetic showers are due to photons (from $\pi^0$ decays), 
the  pseudo-particles formed from calorimeter clusters are assumed massless.
Charged particles produce tracks and also deposit energy 
in the calorimeter,  which can lead to a double counting of their energy.
This double counting is  corrected  either by estimating the average effect on  jet energy,
or by estimating for each cluster how much of its energy
is due to charged particles already measured in the tracking detector.
When jets are formed from Monte Carlo events generated  without
detector simulation, the four-momenta of the simulated particles are 
used directly as the initial pseudo-particles.

\section{Measurement of the Four-Fermion Production Cross Section}\label{sec:cross}
As described in Section~\ref{sec:ww_prod}, the production cross section of four fermions in $e^+e^-$
annihilation has contributions from 
Feynman diagrams containing triple gauge-boson couplings. 
A precise measurement of the cross section can therefore be used to
extract these couplings, and test whether they are consistent with prediction. 
\begin{figure}[t]
\begin{center}
\epsfig{file=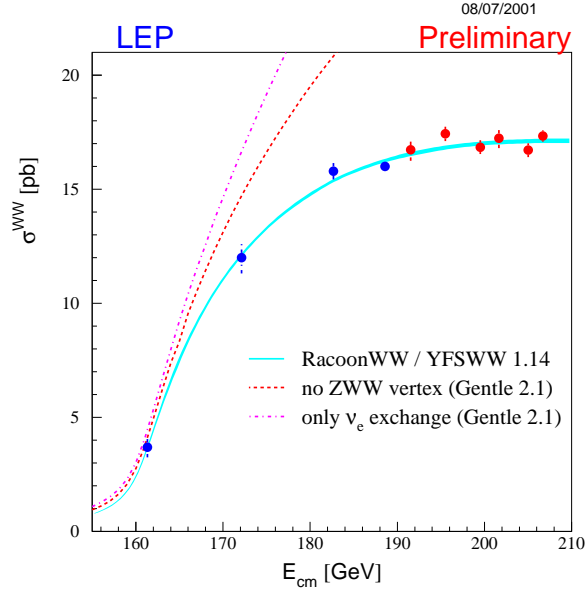,width=8.5cm}
\end{center}
\caption{Comparison of the \W-pair cross section with the Standard Model,
and with predictions that  exclude  either the \WWZ coupling or use only  the
diagram with  t-channel $\nu$ exchange. The points show preliminary results
for the combination of  \LEP measurements.\protect\cite{bib:cross_lep} 
\label{fig:ww_m_cross}}
\end{figure}
Figure~\ref{fig:ww_m_cross} shows a comparison between the measured \W-pair
cross section and the predictions of the Standard Model,\cite{bib:cross_lep} as well as two other models 
that  exclude either the \WWZ coupling or use only   the diagram with  t-channel neutrino exchange 
(see Fig.~\ref{fig:fey_4f} a).
The two variants of the  Standard Model lead to cross sections 
that  increase with rising center-of-mass energies  and would eventually  violate the unitarity bound. 
The contributions to the \W pair cross section from  diagrams containing gauge-boson couplings  
provide  cancellations  essential to avoid this  increase.
Clearly, models without  contributions from the omitted  gauge-boson couplings are
already ruled out by measurements at center-of-mass energies of about 10 to 20 \GeV beyond the
\W-pair production threshold. As can be expected,
the  cross section for  \W-pair production near threshold
is also very sensitive to the mass 
of the \W boson and to the $e^+e^-$  center-of-mass energy.

The Feynman diagrams discussed in Section~\ref{sec:ww_prod}, and shown 
in Figure~\ref{fig:fey_4f}, contribute in  different  regions of phase space 
of the four-fermion cross section.
The \LEP experiments have therefore measured the gauge-boson  cross sections in different ways.
In particular, they have extracted separately the \W-pair, the  \Z-pair and the inclusive single-\W 
cross section. 
The \W-pair cross section is defined as the contribution from the leading-order Feynman 
diagrams with two \W bosons
(the CC03 diagrams shown in Figs~\ref{fig:fey_4f} a) and b)).
The \Z-pair  cross section is  defined as the contribution from  the leading-order Feynman 
diagrams with two \Z bosons (the  NC02 diagram  are shown in Fig~\ref{fig:fey_4f} c)).
The contributions from other diagrams, or their  interference terms, are small in the
region of phase space  used to determine  the mass and cross section of the W,
and are estimated by comparing Monte  Carlo predictions
containing all four-fermion diagrams with those containing only the CC03 or NC02 diagrams.
The single-\W cross section is defined 
as the subset of t-channel Feynman diagrams contributing to the $e\nu ff^\prime$ final states, usually
with additional selections  on kinematic variables that exclude the regions of phase space 
dominated by ``two-photon'', or multi-peripheral, processes
(see Fig.~\ref{fig:fey_4f}e ).

\subsection[ Selection of \W Pair Events]{ Selection of \bW Pair Events}
As mentioned before, the \W boson can decay  either into a lepton-neutrino pair or into a quark-antiquark pair.
Consequently,
different selections must be used to define the different final states. The ten experimentally 
distinguishable final states are: $q\bar{q}q\bar{q}$, $q \bar{q} \mu^+\nu$, $q \bar{q} e^+\nu$, 
$q \bar{q} \tau^+\nu$, $ \mu^+\nu \tau^-\nu$, $ e^+\nu \tau^-\nu$, $ \tau^+\nu \tau^-\nu$,
$ \mu^+\nu e^-\nu$,  $ \mu^+\nu \mu^-\nu$ and $ e^+\nu e^-\nu$. (Charge conjugation invariance 
is assumed throughout.)
Event selections generally consist of  a  pre-selection, followed by the use of  neural-network 
or likelihood discriminant  to separate signal from the major backgrounds, 
which are dominated by $q\bar{q}\gamma$
events for the fully hadronic states, and both   $q\bar{q}\gamma$ and the four-fermion backgrounds
for  the other channels. 
 
The selection of fully leptonic \W  decays requires  the event to
contain exactly two jets
that can be identified   either as electrons or muons, or 
low multiplicity jets from
$\tau$-lepton decays.
The two jets (which may consist of a single particles) must not be coplanar 
with the collision axis, and 
the event must contain significant missing momentum.
One potential background for this class of events corresponds to  leptonic \Z decays that contain a high momentum 
 photon radiation in the initial state.
The leptonic \W-pair decays can be distinguished from such events because their missing momentum
does not  point preferentially along  the beam direction, which means that their missing momentum has a 
substantial component
perpendicular to the beam axis. Also, \W-pairs do not contain energetic isolated photons.
Neutrinos from $\tau$ decays  in $e^+e^- \rightarrow \tau^+ \tau^-$ events can have large missing
momentum.
But because of the large momentum of the $\tau^{\pm}$,  the missing momentum from the neutrinos points 
along  the respective $\tau$ directions.
Consequently, these  types of events  can be distinguished from  signal, for which
 the missing momentum is in general 
not parallel to the jet axis. The contribution from multi-peripheral events
(Fig.~\ref{fig:fey_4f} e) can be  reduced   further
by requiring a minimum total visible energy, because the visible energy in 
these events tends to be lower than for signal.
\begin{figure}[t]
\begin{center}
\epsfig{file=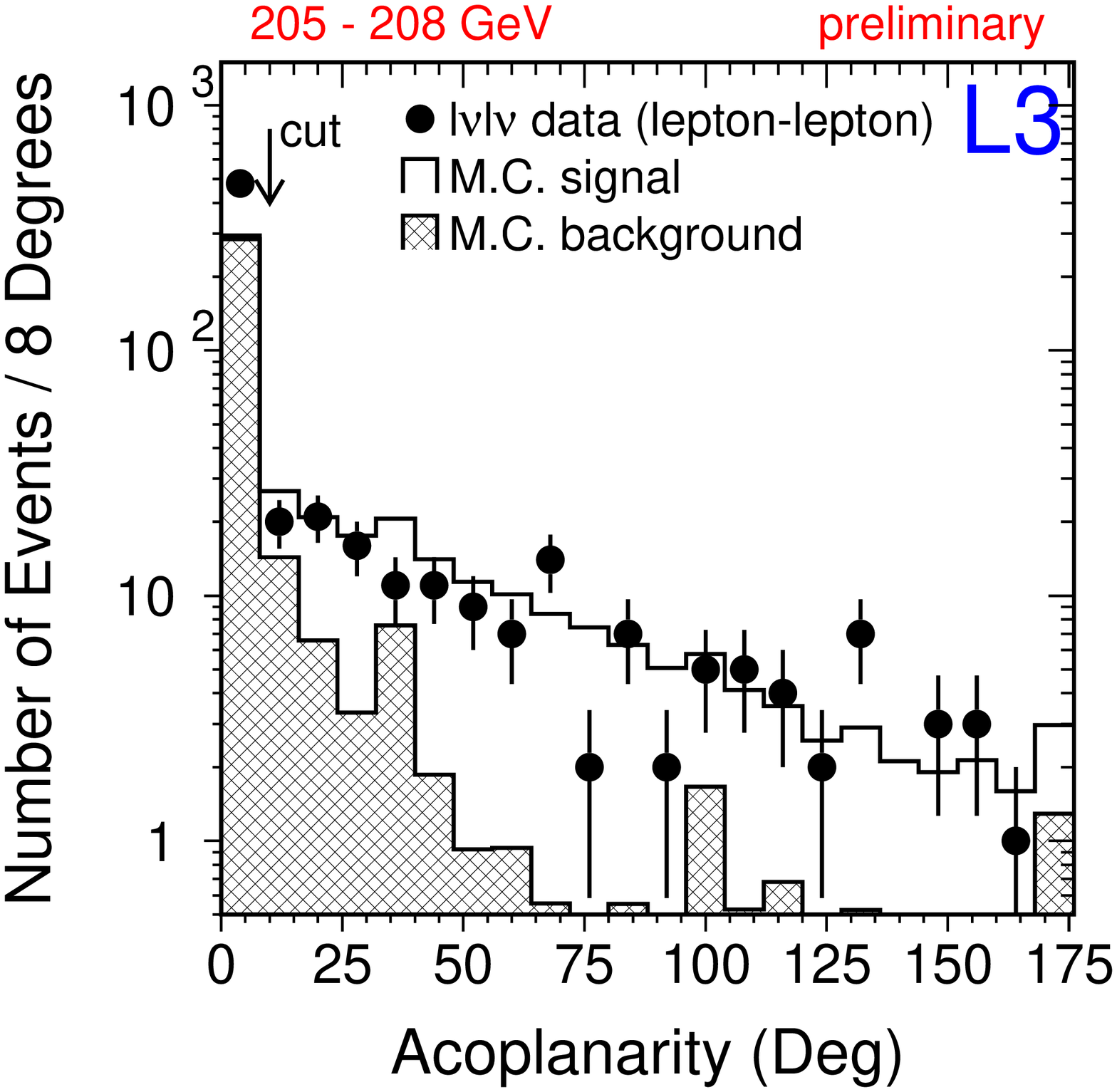,width=6.2cm}
\epsfig{file=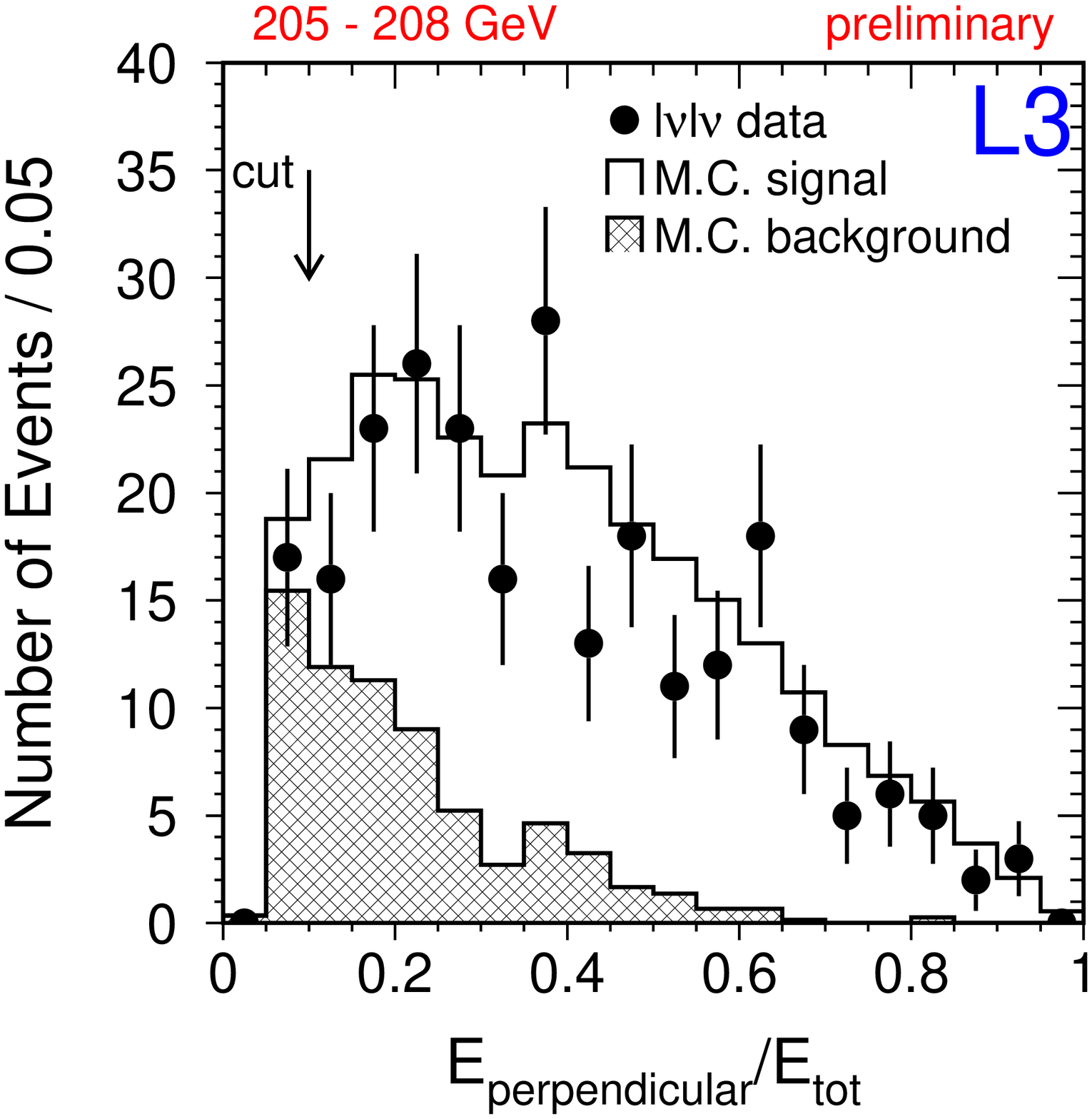,width=6.2cm}
\end{center}
\caption{Distributions in variables used to select $e^+e^- \rightarrow l\nu l\nu$ events,
shown after all selections, expect on the variable in the displayed plot.
Left: Acoplanarity between the two leptons. Right: Energy imbalance in the plane
transverse to the beam direction (normalized to the visible energy). 
The results are from  a preliminary analysis of the  205-208 \GeV  \ldrei 
data.\protect\cite{bib:l3_cross207}
\label{fig:lvlv_sel}}
\end{figure}
From a preliminary  analysis of 
the 205-208 \GeV \ldrei data,\cite{bib:l3_cross207} Fig.~\ref{fig:lvlv_sel} shows  
the acoplanarity between the two leptons (the acoplanarity is the difference between $180^\circ$ 
and  the angle between the
projection of the two leptons into the plane transverse to the beam direction), and the  
energy imbalance in the plane transverse to the beam direction ($E_{perpendicular}$), normalized by the total 
visible energy ($E_{tot}$)  for signal and background, after applying selections on all variables expect the one
displayed.
The selection efficiencies for different channels depend on the details of the individual analyses 
and the efficiency of  lepton identification. For the \LEP experiments, the latter  ranges from 80\% (for $\mu-\mu$ ) to about 30 \%
(for $\tau-\tau$) events. The selections provide  typical signal purities of about 80\%-90\% in all channels.

Semileptonic \W-pair channels must contain two jets from one of the \W bosons, and  
 one  energetic lepton
and missing energy from the second \W boson. Event selection is based typically  on a
loose pre-selection to remove events with low track multiplicity or low visible energy. 
For the accepted events, the most likely lepton candidate, which can contain more then one
track in the case of $\tau$-leptons,  is then identified, and the remaining
event information is forced into two jets.
The selection likelihood or neural network output is then based on kinematic quantities such as the size and direction
of the lepton momentum, the missing momentum, the angle between the two jets, the visible energy,
 and event-shape variables like thrust and sphericity.\cite{bib:event_shape} In addition, the invariant mass of the
jet pair or the lepton and the neutrino (defined by the missing momentum) can be used to improve signal/background. 
Alternatively, the masses can be determined in a kinematic fit, as will be described in Section~\ref{sec:mw_evt}.
\begin{figure}[t]
\begin{center}
\epsfig{file=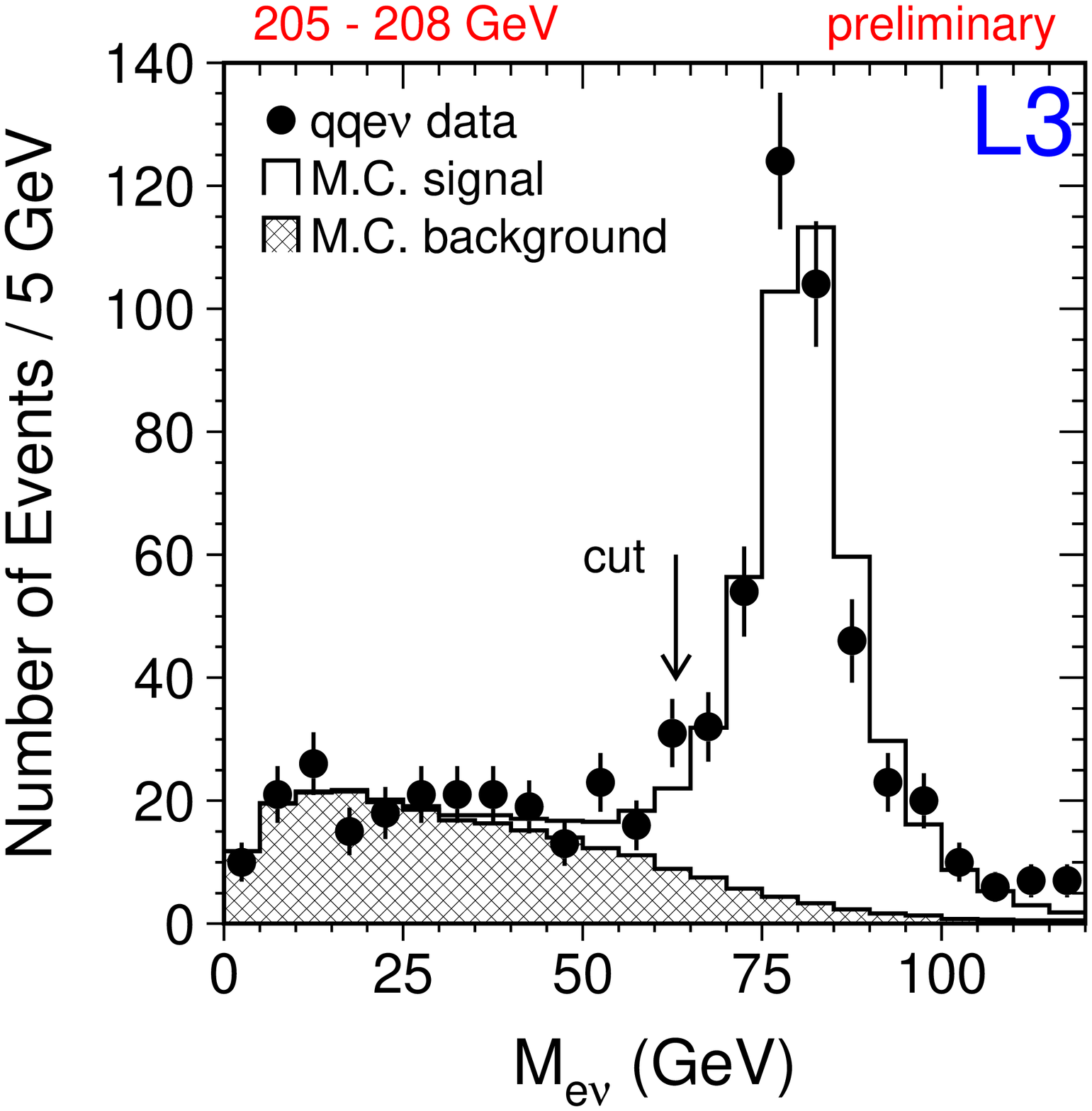,width=6.2cm}
\epsfig{file=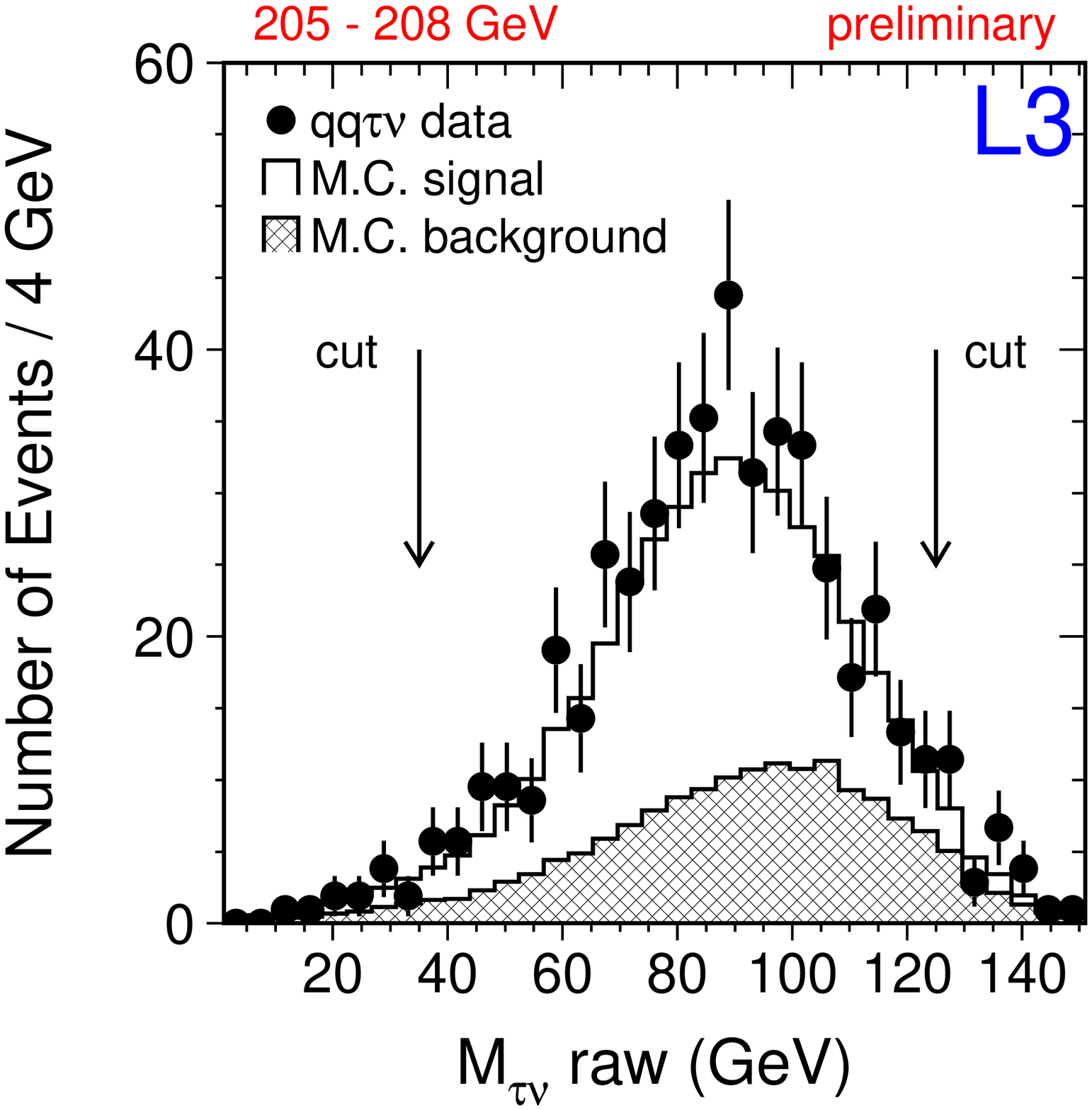,width=6.2cm}
\end{center}
\caption{Invariant mass of the lepton and neutrino  in the $e^+e^- \rightarrow qql\nu$ channel,
 from the  analysis of  205-208 \GeV \ldrei data,\protect\cite{bib:l3_cross207}
after all selections, expect for the $l\nu$ mass. Results for electrons are shown on the left,
  and  for the $\tau$ channel 
on the right.  \label{fig:lvqq_sel}}
\end{figure}
The  preliminary    analysis of the 205-208 \GeV \ldrei data,\cite{bib:l3_cross207}
in Fig.~\ref{fig:lvqq_sel} shows 
the invariant mass of the charged lepton and neutrino  in the electron and $\tau$ channels.
Typical efficiencies range from 70\%-90\% for  different  channels, and the purities for the electron and
muon channels  are about 95\%, and for the $\tau$ channel about 80\%.  

The pre-selection for \W-pair events where the \W bosons yield four-jet events., includes
loose requirements  on the multiplicity, and a veto on events already selected as semileptonic or leptonic \W-pair decays.
The most important  background is from  quark-pair events (including events with initial state radiation) 
$e^+e^- \rightarrow q\bar{q}(\gamma)$, with accompanying hard-gluon radiation. 
In quark-pair events that contain an  energetic initial-state photon, the  photon  is usually 
emitted close to the beam direction,
and is therefore not observed in the detector. These types of events 
can be rejected by requiring a minimal visible energy,  typically 70 \% of the center-of-mass energy. 
The difference in  event topology can be used to distinguish  hadronic  pairs from such  ``QCD'' background.  
QCD events that  are reconstructed as four-jet events consist of two jets from
the primary quark and two jets from hard gluon radiation (in some cases these are two quark jets evolved from
the splitting of a gluon into two quarks). 
The secondary jets are  typically of
lower energy, and their angles relative  to the  originating jets tend to be  small. 
These features 
 can be exploited by looking at event-shape distributions, such as the sphericity or the $y_{cut}$ value 
($y_{34}$ where  a three-jet event turns into
a four-jet event  or $y_{45}$ where  a four-jet event turns into a five-jet event, as mentioned Section~\ref{sec:jet}).
Another way to exploit these differences is to use the jet algorithm to reconstruct  the event as a four-jet event, 
and then use the four-momenta of the four jets to  calculate the QCD matrix  element ``$W_{420}$'' for four jets 
(defined as the sum of the ${\cal O} (\alpha_s^2)$ matrix elements for
$e^+e^- \rightarrow q\bar{q} \rightarrow  q\bar{q}gg$ and
 $e^+e^- \rightarrow q\bar{q} \rightarrow  q\bar{q}q\bar{q}$ processes),\cite{bib:w420}
and the matrix element for the CC03 diagrams. 
For true hadronic \W-pair events, the value of the  CC03 matrix element should be
larger than for QCD background and the value of the QCD matrix element
should be smaller for  hadronic \W-pair event decays than for the QCD background.
\begin{figure}[t]
\begin{center}
\epsfig{file=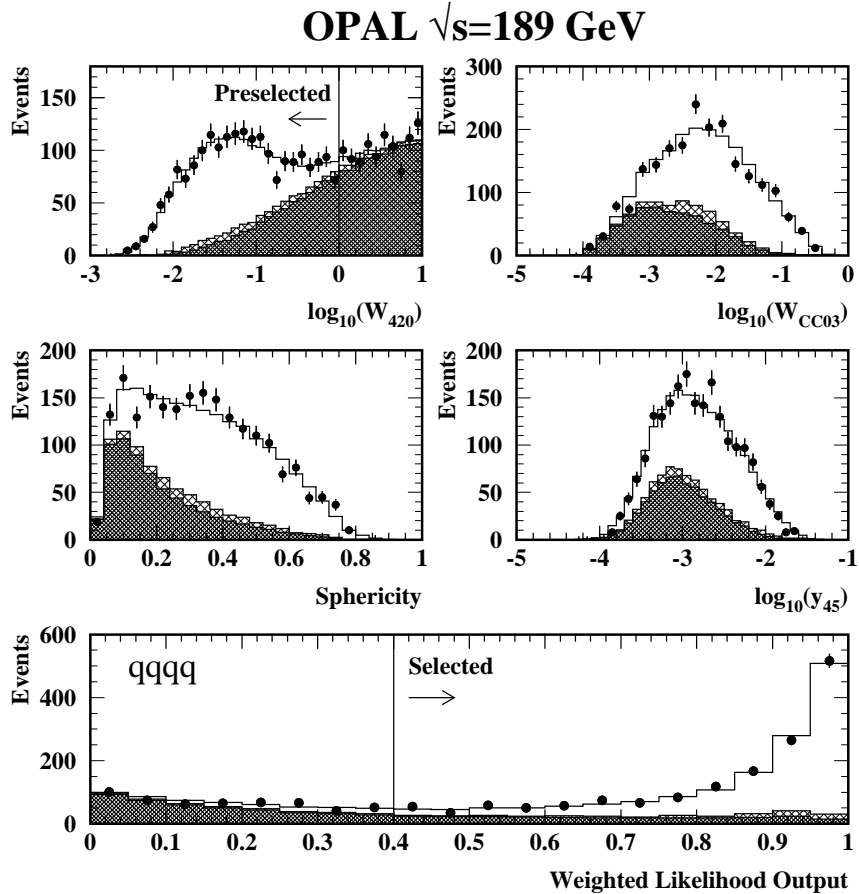,width=12.cm}
\end{center}
\caption{Distributions used in the selection of hadronic \W-pair events in the \opal
analysis of data taken at $\sqrt{s}= 189$ \GeV. Also shown is the likelihood output  based
on these four distributions, with background peaking at $\approx 0$, and signal at
$\approx 1$   \label{fig:qqqq_sel}}
\end{figure}
Figure~\ref{fig:qqqq_sel} shows  distribution in the four variables used to define the \opal likelihood
for hadronic \W-pairs at a center-of-mass energy of 189 \GeV.\cite{bib:opal_cross189}
The points correspond to data, the cross-hatched regions to background and the remainder to signal.
The final selection efficiencies in the all-hadronic channel
range from 80\% to 90 \%, and the purities from 70\% to~80\%, for all the major \LEP detectors.     

\subsection{Results}
The cross sections for individual \W-pair channels can
be determined from   neural-network or relative log-likelihood discriminants, 
 either  by counting the number of events above some cutoff, 
or by fitting the data to
signal and background templates obtained from Monte Carlo simulations.
The measurements yield the products of
production cross sections and the decay branching fractions of the two \W bosons
in any given final state.
The ratios of such results 
can be used to extract  decay fractions for \W bosons. 
\begin{figure}[t]
\begin{center}
\epsfig{file=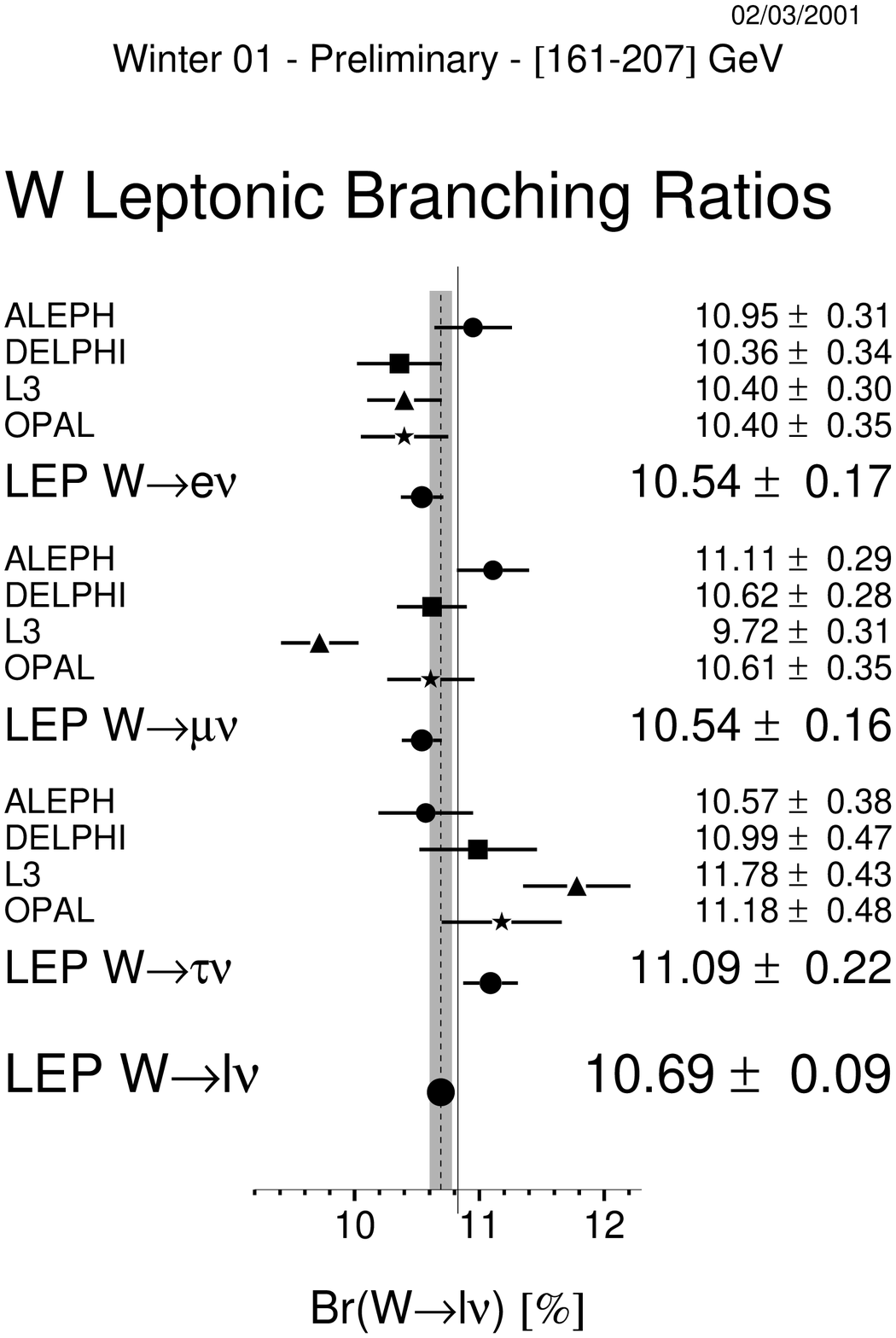,width=6.cm}
\epsfig{file=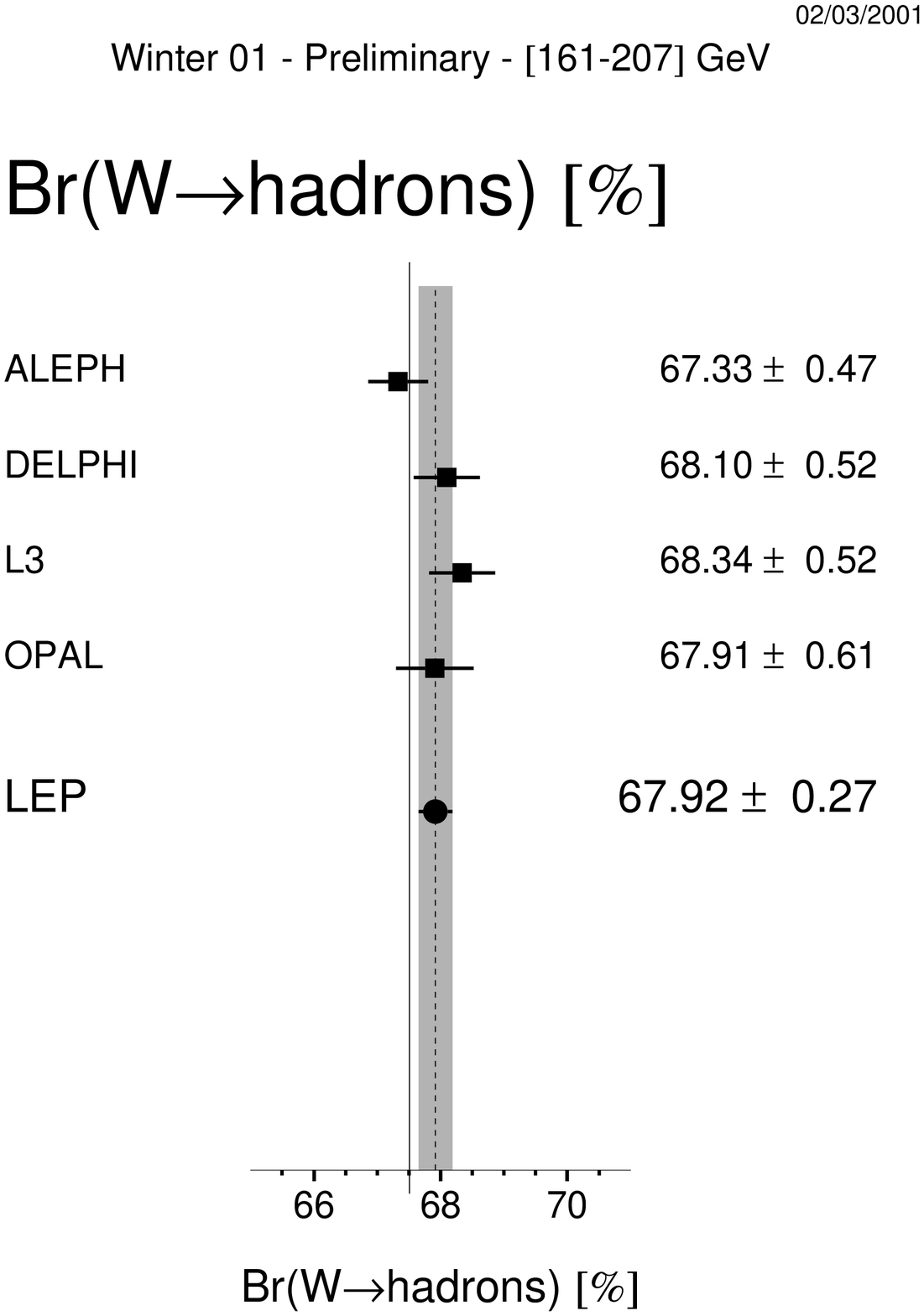,width=6.cm}
\end{center}
\caption{Summary of the leptonic and hadronic \W-decay fractions. The solid line indicates the
Standard Model prediction of 10.81\% and 67.51\%, respectivetly.\protect\cite{bib:cross_lep}. 
The dashed line
and the gray band show the  combined \LEP result and its uncertainty \label{fig:w_br}.}
\end{figure}
A summary of the  measurements of leptonic and hadronic \W decay fractions is shown in Fig.~\ref{fig:w_br}.
In the Standard Model, the branching fractions depend only on the six matrix elements $|V_{q\bar{q}^\prime}|$
of the Cabibbo-Kobayashi-Maskawa (CKM) quark mixing matrix. 
(The CKM matrix is a unitary matrix describing
the relative strength of the \W  coupling to different combinations of up and down types of quarks. 
The  coupling of the $\W^+$ to an up-quark and an anti-strange quark, for example, is  proportional to $V_{us}$. 
The off-diagonal CKM matrix elements do not vanish because the eigenstates of the weak interaction
are not the same as the mass eigenstates.)
Leaving out the top quark because it is too heavy to be produced in \W boson decays,
the branching fraction of the W into any lepton flavor ${\cal B} (W \rightarrow l \bar{\nu})$
can be related to the CKM matrix elements as follows:\cite{bib:cross_lep}
\[ \frac{1}{{\cal B} (W \rightarrow l \bar{\nu})} = 3 \left\{ 1+ \left[ 1+ \frac{\alpha_s(M^2_\W)}{\pi} \right]
   \sum_{\tiny \begin{array}{c} i=(u,c), \\ j=(d,s,b) \end{array}} |V_{ij}|^2 \right\} \]
where $\alpha_s(M_\W^2)$ is the strong coupling constant at  the scale of the \W-boson mass. 
Taking $\alpha_s(M_\W^2)=0.121 \pm 0.002$, and using 
the measured leptonic branching fractions of the \W, yields:\cite{bib:cross_lep}
\[  \sum_{\tiny \begin{array}{c} i=(u,c), \\ j=(d,s,b) \end{array}} |V_{ij}|^2 = 2.039 \pm
    0.025( {\cal B}_{\W \rightarrow l \bar{\nu}}) \pm 0.001 (\alpha_s). \]
Using the experimental values\cite{bib:pdg2000} for 
$|V_{ud}|^2+|V_{us}|^2+|V_{ub}|^2+|V_{cd}|^2+|V_{cb}|^2= 1.0477 \pm 0.0074 $, the result can be used to extract 
a measurement of $|V_{cs}|$, which is the least well know of these matrix elements:  $|V_{cs}| = 0.996 \pm 0.013$.

\begin{figure}[t]
\parbox{12.5cm}{
\begin{center}
\epsfig{file=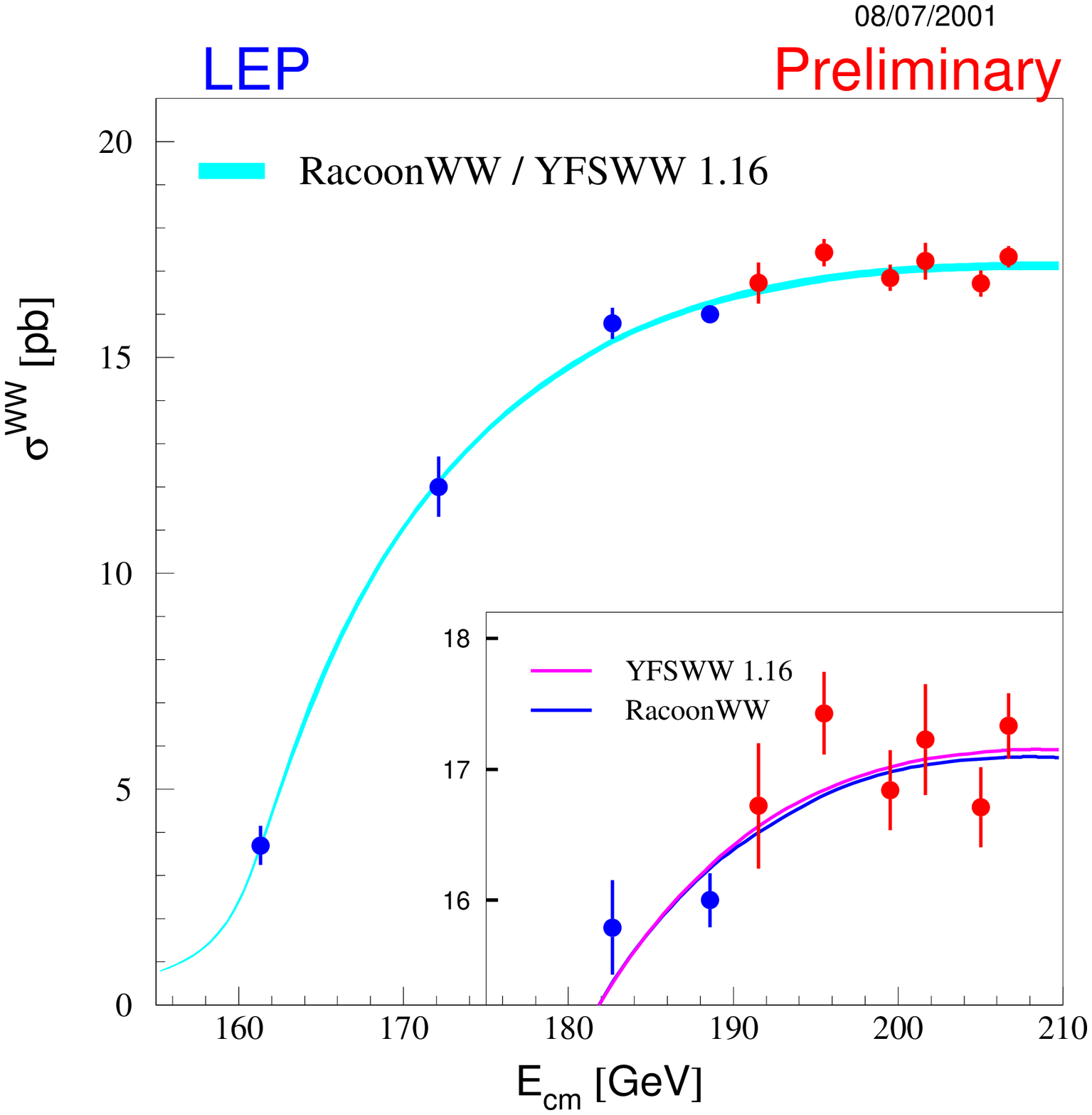,width=5.9cm}
\epsfig{file=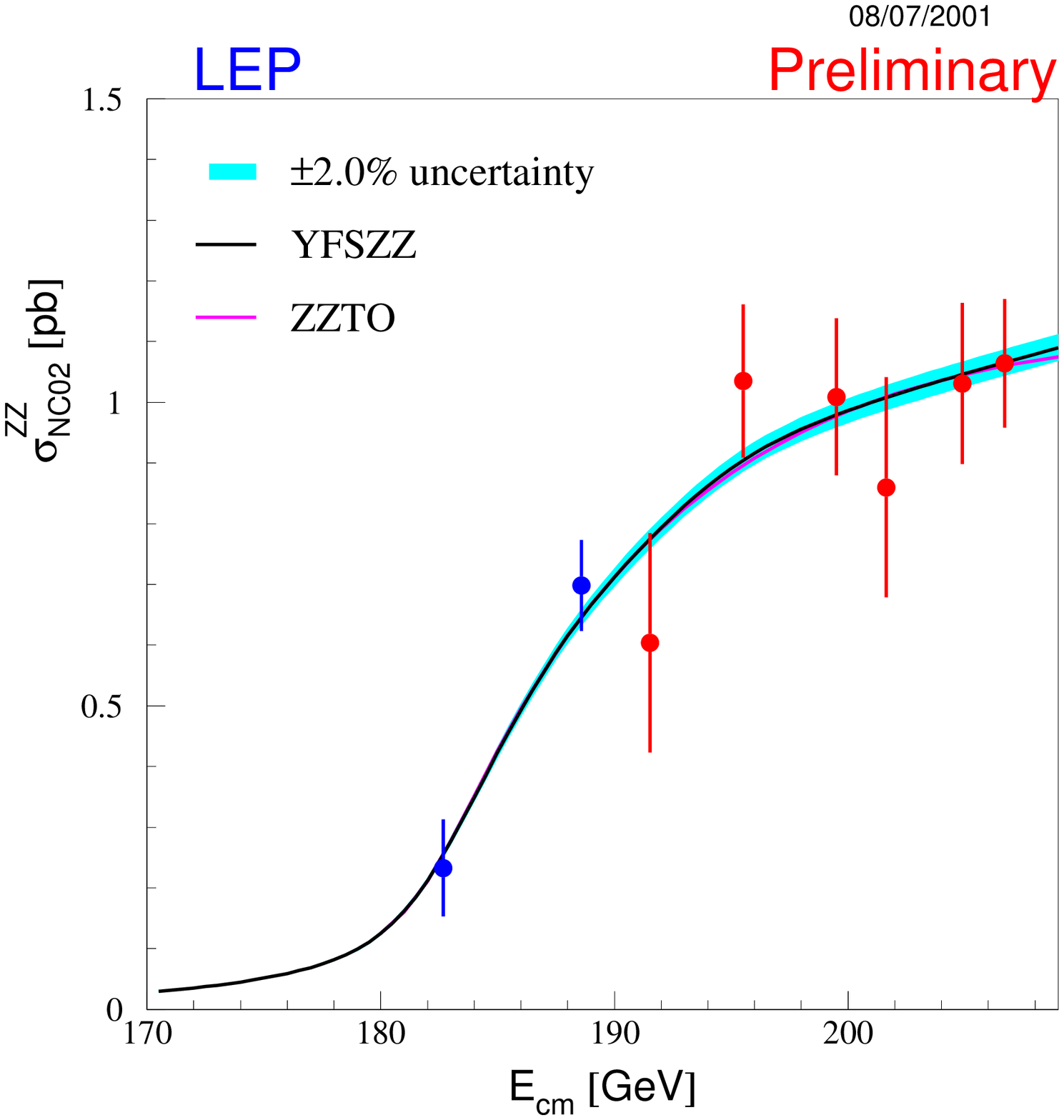,width=5.9cm}
\end{center}
\vspace*{-.35cm}
\caption{The \W-pair  and \Z-pair  cross sections, left and right, respectively \label{fig:cross}}
} \newline
\parbox{12.5cm}{
\begin{center}
\epsfig{file=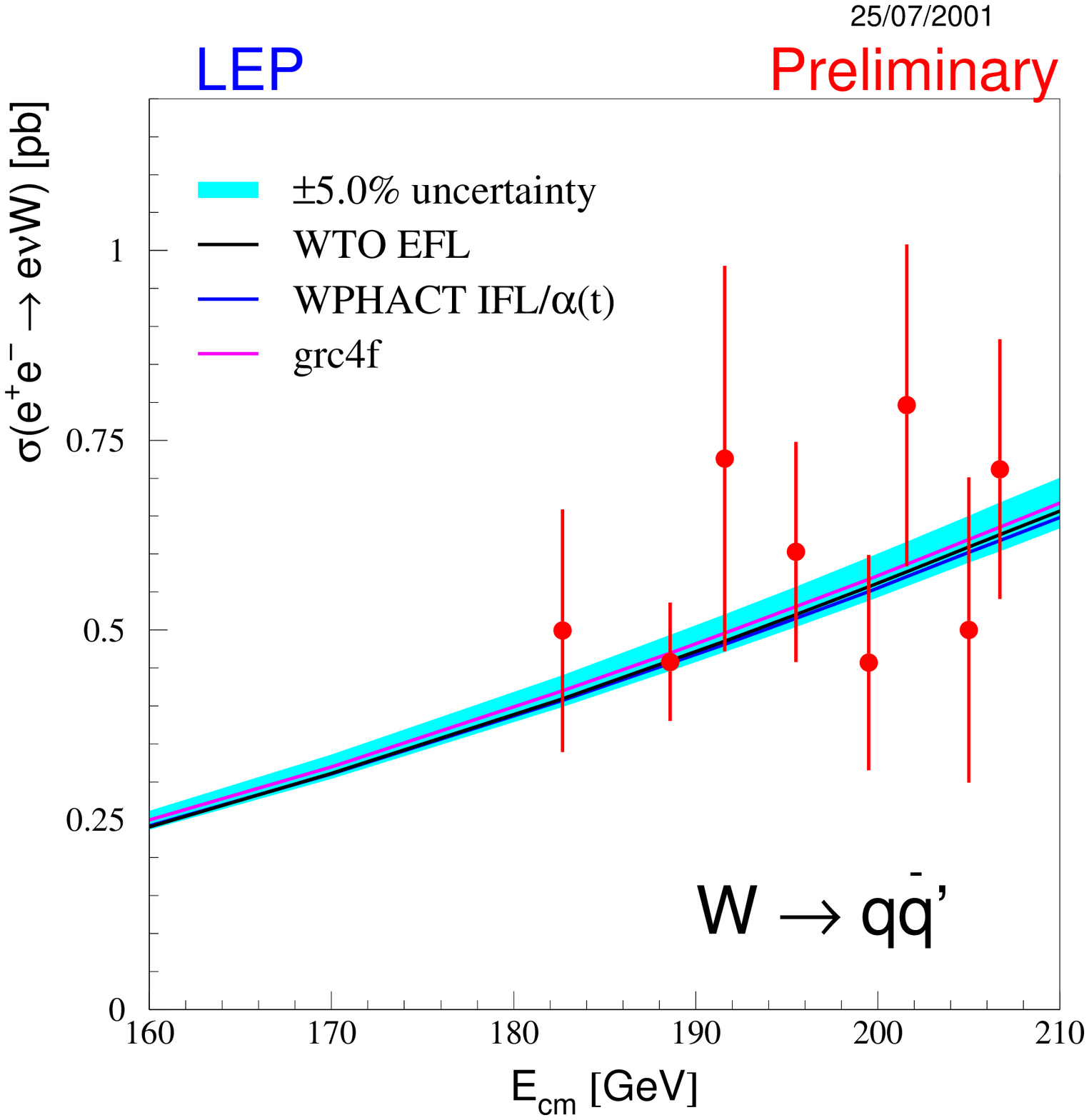,width=5.9cm}
\epsfig{file=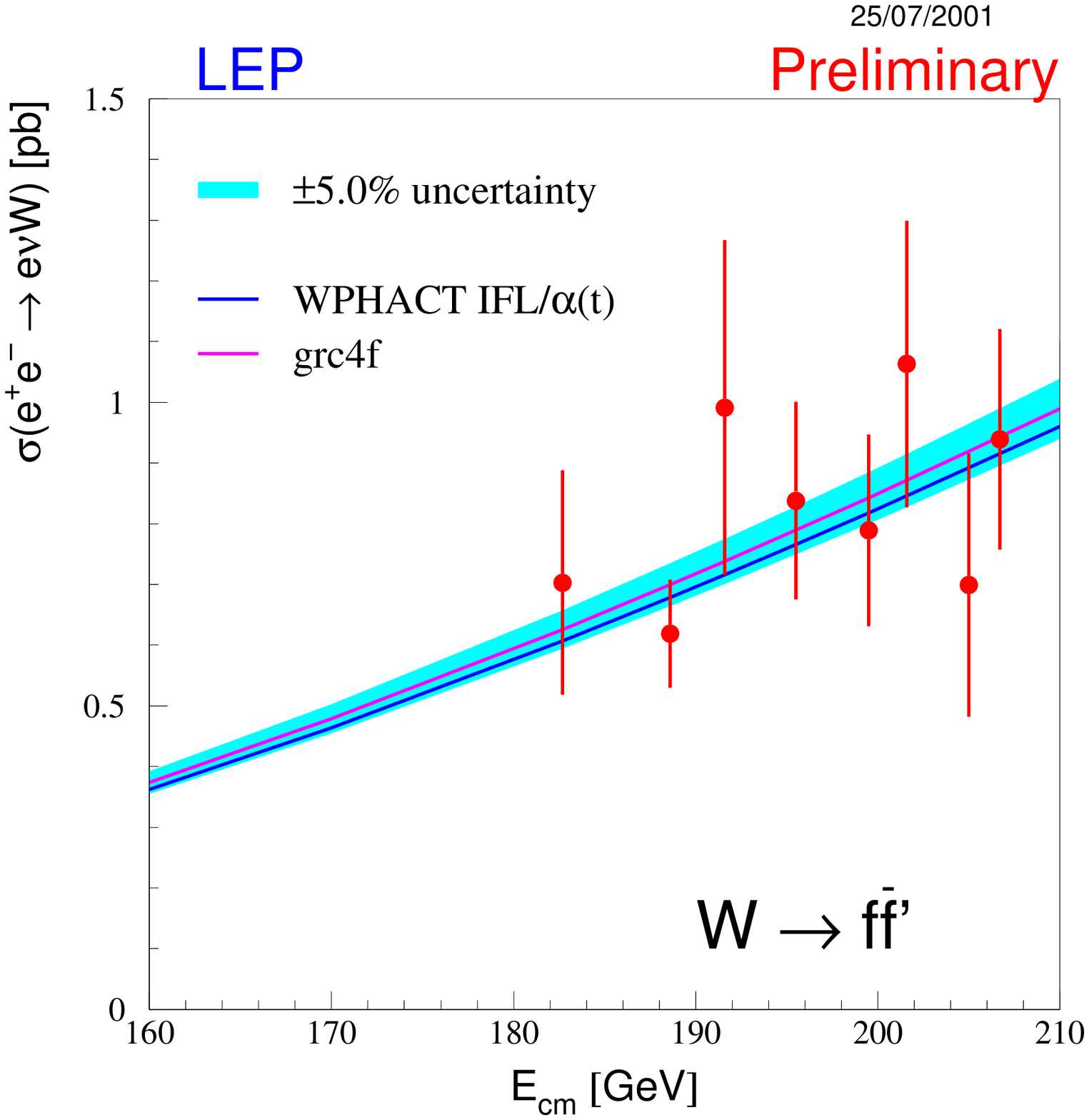,width=5.9cm}
\end{center}
\vspace*{-.35cm}
\caption{The single-\W cross section as function of $E_{cm}=\sqrt{s}$. \label{fig:cross_w}}
}
\end{figure}
Figure~\ref{fig:cross} shows the combined \LEP results for the measurement of
the cross section for \W-pairs (CC03 diagrams) and \Z-pairs (NC02 diagram).\cite{bib:cross_lep} 
The \W-pair cross section is compared to  predictions of  the Standard Model
based on the programs  \yfsww~\cite{bib:yfs} and \raconww.\cite{bib:raconww} The two programs
have been  compared extensively, and agree at a level of $<0.5$ \% at 
\LEPII energies.\cite{bib:ww_4f} The calculations above 170 \GeV have  uncertainties
decreasing from 0.7~\% at 170~\GeV to about 0.4~\% at center-of-mass energies larger than 200 \GeV.
An uncertainty of 50 \MeV on the \W mass  translates into an additional error of 0.1 \% ( 3.0 \%)
on the predicted cross section at 200 \GeV (161 \GeV).
The \Z pair cross section in Fig.~\ref{fig:cross} is compared with  predictions
based on the programs \yfszz\cite{bib:yfszz} and \zzto.\cite{bib:zzto} The uncertainty of the
theory  is estimated as 2\%.\cite{bib:ww_4f}
All results up to the highest
center-of-mass energies are in good agreement with the predictions.

As indicated previously, the single-\W  cross section is defined by the sum of all leading-order four-fermion t-channel 
processes  contributing to the $e\bar{\nu_e}f\bar{f}$ final state, with additional 
selections on kinematic variables used to exclude regions of phase space dominated by multi-peripheral diagrams,
where the calculation has large uncertainties. The criteria used to define signal
are: $m_{q\bar{q}}> 45$ \GeV/c$^2$ for $e\nu_eq\bar{q}$ final states, $E_l> 20$ \GeV for
$e\nu_e l \nu_l$ final states (with $l=\mu$ or $\tau$); and  $|cos \theta_{e^-}| > 0.95$,
$|cos \theta_{e^+}| < 0.95$, and $E_{e^+}>20$ \GeV (and similar charge conjugate selections) for the 
$e\nu_e e \nu_e$ final states.
The momentum transfer to the electron in these processes is small, and the electron 
is therefore often not detected (in the beam pipe). The signal therefore consists either of
a single high-energy lepton, or a jet-pair with a large invariant mass and missing momentum.
Figure~\ref{fig:cross_w} shows the combined \LEP results\cite{bib:cross_lep} for the measurement of
the single-\W cross section, both for the hadronic channel alone, and for the combination of the hadronic
and leptonic channels. Although the uncertainties are large, the data agree with  predictions.

\section{Gauge Boson Couplings }\label{sec:tgc}
An essential feature of the Standard Model is the non-abelian structure of the 
weak interaction.  The  gauge-boson couplings are a consequence of this structure, 
and the  direct observation of these  couplings is therefore an 
important test of the  non-abelian structure of the  weak interaction.
Any deviations from the  predictions of the Standard Model  would
imply the presence of new physics. 
This could arise from  
a violation of the gauge invariance of the theory, or from
additional particles or interactions that have  yet to be observed. 

Gauge-boson couplings also provide an essential contribution to 
the strong interaction, because
QCD is a non-abelian theory, with self couplings between its gauge bosons, the gluons.
These couplings are regarded as the reason for the confinement of colored particles within
objects that are neutral in color charge. Clear evidence for the triple-gluon vertex
has been found  in four-jet events  at \LEP.\cite{bib:tgc_lep_r}

\subsection{Triple-Gauge Couplings of  \W Boson}
\begin{figure}[t]
\begin{center}
\epsfig{file=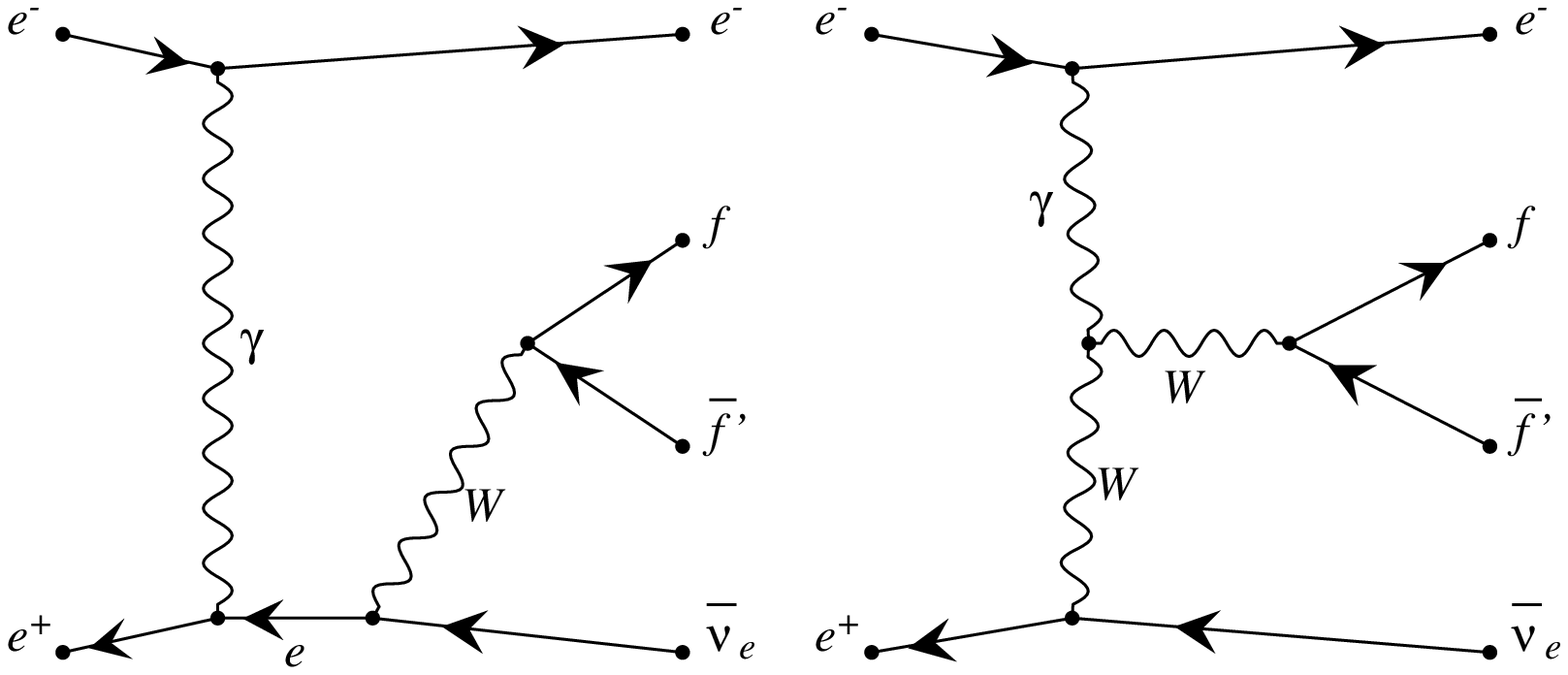,width=8.5cm} \hspace*{.4cm}
\epsfig{file=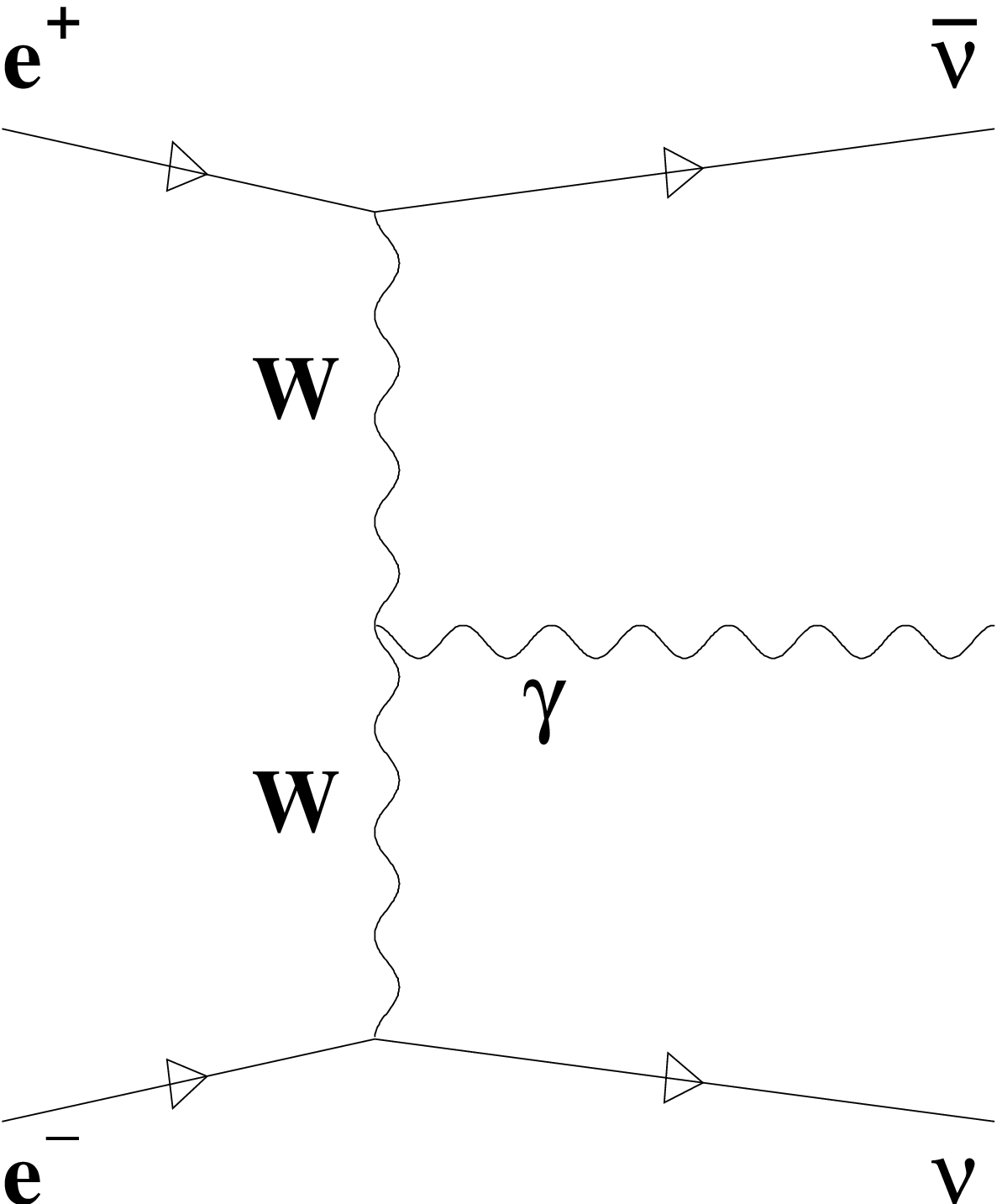,width=3.cm}
\end{center}
\caption{Feynman diagrams for single-\W  and single-photon production. \label{fig:tgc_fey}}
\end{figure}

The most general Lorentz-invariant Lagrangian has  seven independent
couplings describing each of the \WWg and \WWZ vertices.\cite{bib:tgc_lep2}  
By requiring electromagnetic gauge invariance, and both charge-conjugation (C) as well as parity (P) 
invariance, the number of parameters can be reduced to five. Thus, the parameters $g_1^Z$,
$\kappa_Z$, $\kappa_\gamma$, $\lambda_Z$ and  $\lambda_\gamma$
are often used to describe the \WWg and \WWZ couplings.\cite{bib:tgc_lep2}
The C and P conserving terms for the \WWg 
coupling
correspond to the lowest-order multi-pole expansion of the \W-photon interaction.
The charge $Q_\W$, the magnetic dipole moment $\mu_\W$, and the electric quadrupole moment $q_\W$ of
the $\W^+$, are given by:\cite{bib:tgc_exp}    
\[ Q_\W = e, \ \ \ \quad \mu_\W=\frac{e}{2 m_\W}(g_1^Z+\kappa_Z+\lambda_\gamma), \ \ \ \quad 
   q_\W = \frac{e}{m_\W^2}(\kappa_\gamma-\lambda_\gamma). \]
The expectation for the five parameters is $g_1^Z=\kappa_Z=\kappa_\gamma=1$, and
\mbox{$\lambda_Z=\lambda_\gamma=0$.}
Precision measurements at the \Z resonance are consistent with 
the following $SU(2) \times U(1)$ relations among the five couplings:\cite{bib:tgc_lep2}
\[ \begin{array}{r @{\quad=\quad} l} 
(1-\kappa_Z) & -(1-\kappa_\gamma) tan^2 \theta_w+(1-g_1^Z) \\
\lambda_Z & \lambda_\gamma \ , 
\end{array} \]
where $\theta_w$ is the weak mixing angle.
These relations leave only three independent couplings not  restricted significantly\cite{bib:tgc_su2}
by existing \Z data from \LEP and SLD.\cite{bib:sld_c}
In the following,  $\kappa_\gamma$, $g_1^Z$, and $\lambda = \lambda_Z = \lambda_\gamma$ will be used
to check triple gauge boson couplings of the \W bosons.  
The results will also be expressed in terms of  deviations from the Standard Model, e.g.,
$\Delta \kappa_\gamma= \kappa_\gamma-1$, and  $\Delta g_1^Z=g_1^Z-1$.  
Anomalous contributions to the couplings can yield additional contributions for  different 
helicity states of the  outgoing \W bosons in \W-pair events. This affects the  production angular distribution of the 
\W bosons, and, because of  correlation with the \W spin, it also affects the 
angular distribution of the decay products. Cross sections provide  additional information
about gauge couplings. The \W pair cross section and its angular dependence, are sensitive to 
all three couplings. In contrast, the cross sections for single-\W and single-photon production 
(through contributions from $e^+e^- \rightarrow \gamma \nu \nu $~)  are
 sensitive mainly to $\kappa_\gamma$ for t-channel processes at low momentum transfers
involving the \WWg vertex (see Fig.~\ref{fig:tgc_fey}).
Contributions from $\lambda_\gamma$ are  suppressed in this case, because,  they are  important only 
for processes involving large  \W momentum.

\subsubsection[Angular Distributions in Decays of W Pairs]{Angular Distributions in  Decays of \bW Pairs}
Ignoring small  effects due to the finite width of the \W, and the impact of initial state radiation, 
the production and decay of 
W bosons is described  completely by five angles. Conventionally, these are taken to be:\cite{bib:tgc_lep2}
the  polar production angle   $\theta_\W$ defined as the angle between the 
incident $e^-$ and the produced $\W^-$, and the polar and azimuthal angles $\theta_1^*$ and
$\phi^*_1$ of the decay fermion  from the $\W^-$, calculated in the $\W^-$ rest frame (``helicity'' frame), 
and the analogous angles $\theta_2^*$ and $\phi^*_2$ for the anti-fermion for the $\W^+$. 
(The axes of the right-handed coordinate system in the \W rest frame are defined
such that the z-axis is along the parent-\W line of flight ($\vec{W}$), and the y axis is along the 
direction $\vec{e^-} \times \vec{W}$, where $\vec{e^-}$ is the direction of the $e^-$ beam.)
\begin{figure}[h]
\begin{center}
\epsfig{file=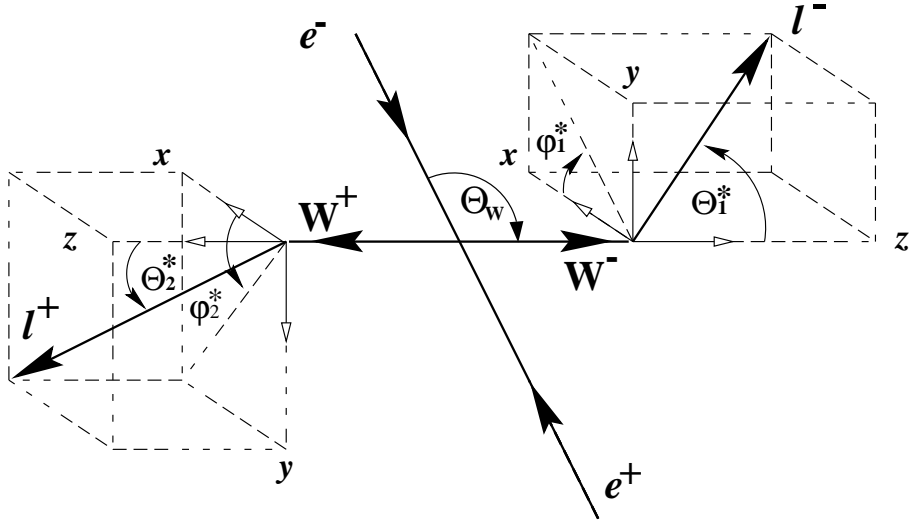,width=12cm}
\end{center}
\caption{Definition of the angles in  leptonic \W pair decay. \label{fig:tgc_adef}}
\end{figure}
Figure~\ref{fig:tgc_adef} illustrates these definitions. 

In most analyses using hadronic   
W decays, it is not possible to distinguish the quark jet from the antiquark jet. Hence, 
the angles in the \W rest frame are defined arbitraily using  the jet with the smaller azimuthal
angle $\phi^*$. 
\begin{figure}[t]
\begin{center}
\epsfig{file=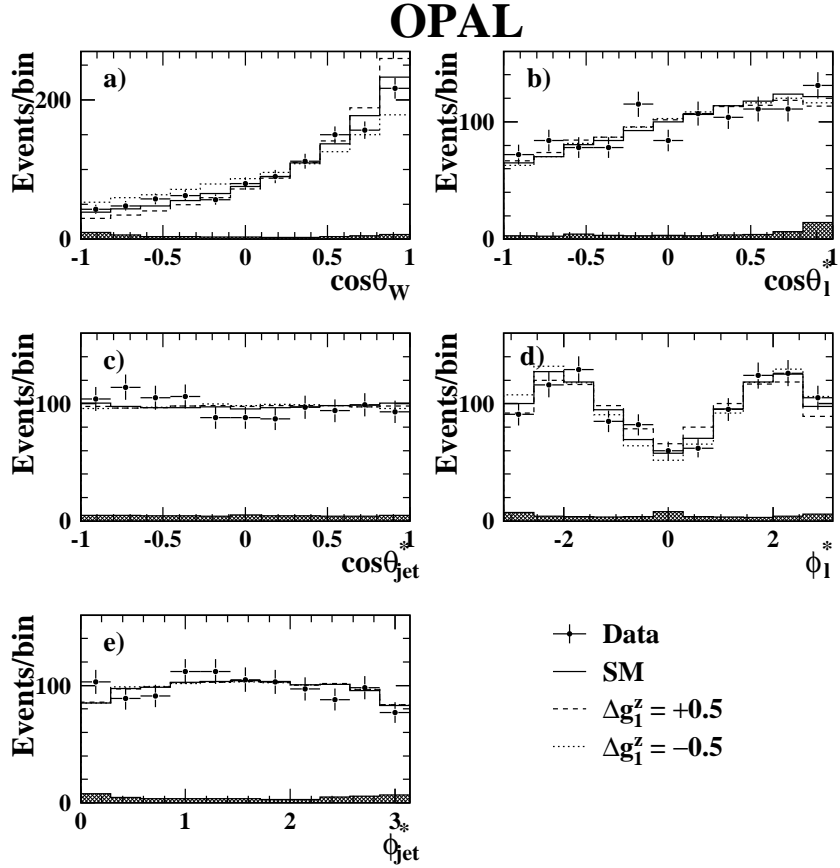,width=12.5cm}
\end{center}
\caption{ Comparison of  \opal data at $\sqrt{s}=189$ \GeV with the Standard Model and with predictions for 
anomalous gauge couplings $\Delta g_1^Z= \pm 0.5$. See the text for the definitions of the
angles. The shaded histograms show the  background from sources other than  $q\bar{q}l\bar{v}$. \label{fig:tgc_angv}}
\end{figure}
Figure~\ref{fig:tgc_angv} shows a comparison of  \opal data\cite{bib:tgc_opal189} 
for semileptonic \W-pairs at $\sqrt{s}=189$ \GeV~- to the 
Standard Model, and to  predictions assuming the  anomalous gauge couplings $\Delta g_1^Z = \pm 0.5$.
Comparing the distributions  from  leptonic
W decay (Figs.~\ref{fig:tgc_angv}b and d) with those  from  hadronic decay (Figs.~\ref{fig:tgc_angv}c and e),
indicates the clear loss of sensitivity 
when the fermion  cannot be distinguished from the antifermion.

The method of the ``Optimal Observable'' is often used 
to extract information about  gauge couplings from the $\le 5$-dimensional 
angular distributions (for semileptonic events, some analyses use only  the $\W^-$ production and lepton 
decay angles).\cite{bib:tgc_oo,bib:tgc_oo2} 
Since the Lagrangian is linear in the triple-gauge  couplings, the differential cross section can be expanded in the
couplings without using terms beyond second order:
\[ \frac{d\sigma}{d\Omega} = S_0(\Omega)(1+\sum_i {\cal O}_{i}^{(1)}(\Omega) g_i +
 \sum_{ij} {\cal O}_{ij}^{(2)}(\Omega) g_i g_j), \]
where $g_i$ denotes any type of additional coupling not included in the calculation of the  differential cross section 
denoted by $S_0(\Omega)$,  
and $\Omega$ denotes 
the phase space variables, taking
into account reconstruction ambiguities (e.g., the ambiguity between jets from quarks and antiquarks) 
for the individual $\W^+\W^-$ channels.
The functional dependence of the first-order coefficients ${\cal O}_{i}^{(1)}(\Omega)$ and of
the second-order coefficients ${\cal O}_{ij}^{(2)}(\Omega)$
on $\Omega$ are described in terms of a phenomenological  model of anomalous couplings.

Instead of the multidimensional phase space density $\Omega$,  the one dimensional projections
${\cal O}_{i}^{(1)}(\Omega)$ can be used to  determine the gauge couplings.
The information about the couplings $g_i$ can be extracted from the mean values of the ${\cal O}_{i}$. 
This can be seen as follows:
Neglecting the quadratic $g$ terms, the 
value for $d\sigma/d\Omega$ increases  faster with increasing  $g_i$  for
events with large ${\cal O}_{i}^{(1)}(\Omega)$ 
than for events with smaller  ${\cal O}_{i}^{(1)}(\Omega)$.
 Consequently, the probability for events with large  
${\cal O}_{i}^{(1)}(\Omega)$ and, therefore the mean value of ${\cal O}_{i}$,
increases with $g_i$.

Detector and hadronization effects can be included in the determination of the couplings by comparing
the measured mean value  $\left<{\cal O}_{i}(\Omega) \right>$ to the mean value extracted from a full Monte Carlo
simulation, where samples with  different anomalous couplings are generated using a reweighting 
technique (see Sect.~\ref{sec:fit_rew}).
For large values of $g_i$, the contribution from the quadratic terms  reduces the
sensitivity of an analysis based  on only the mean values $\left<{\cal O}_{i}^{(1)}\right>$ of the linear terms.
This can be mitigated  either by using an iterative approach, were $S_0(\Omega)$ is recalculated in each step, and the $g_i$ 
reflect the  additional contributions to
the gauge couplings compared to the previous iteration,  or by including of the mean values 
$\left<{\cal O}_{ij}^{(2)}\right>$ of the quadratic terms.
The second approach has the additional advantage that the second-order terms can be used
to resolve ambiguities of an analysis based on only  the first-order terms.

The \opal\cite{bib:tgc_opal189} and \alephe collaborations\cite{bib:tgc_a206,bib:tgc_a189} use the mean values of
the Optimal Observables to derive limits on anomalous gauge couplings. The \delphi collaboration uses the distribution of the Optimal 
Observables in a likelihood approach to determine limits on the couplings.\cite{bib:tgc_d189}
The \ldrei collaboration determines limits on anomalous gauge couplings using a binned maximum-likelihood fit to the
production and decay angles in \W-pair events,\cite{bib:tgc_l183} and  a likelihood fit to the distribution of
the Optimal Observables as a cross check.\cite{bib:tgc_l206}

\subsubsection{Results}
\begin{figure}[t]
\begin{center}
\epsfig{file=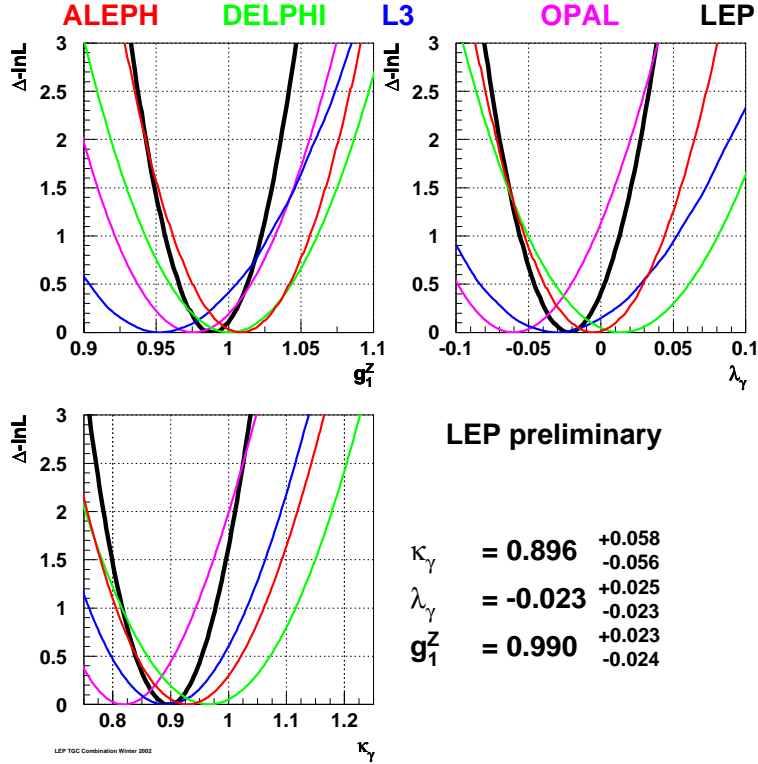,width=11.cm}
\end{center}
\vspace*{-.6cm}
\caption{The combined  one-parameter fits to  triple-gauge couplings of the $\W$ boson. \label{fig:tgc_lep}}
\end{figure}

The gauge couplings can be extracted  either by fitting one of the three couplings, 
while  keeping the other two at their Standard-Model predictions, or by simultaneously fitting two or three
couplings (in the fit of two couplings,  the third coupling is fixed  to the lowest-order prediction of the Standard Model).
These fits include information from the total cross section, and from the angular distributions in
\W-pair, single-\W, and single-photon events.
\begin{figure}[h]
\begin{center}
\epsfig{file=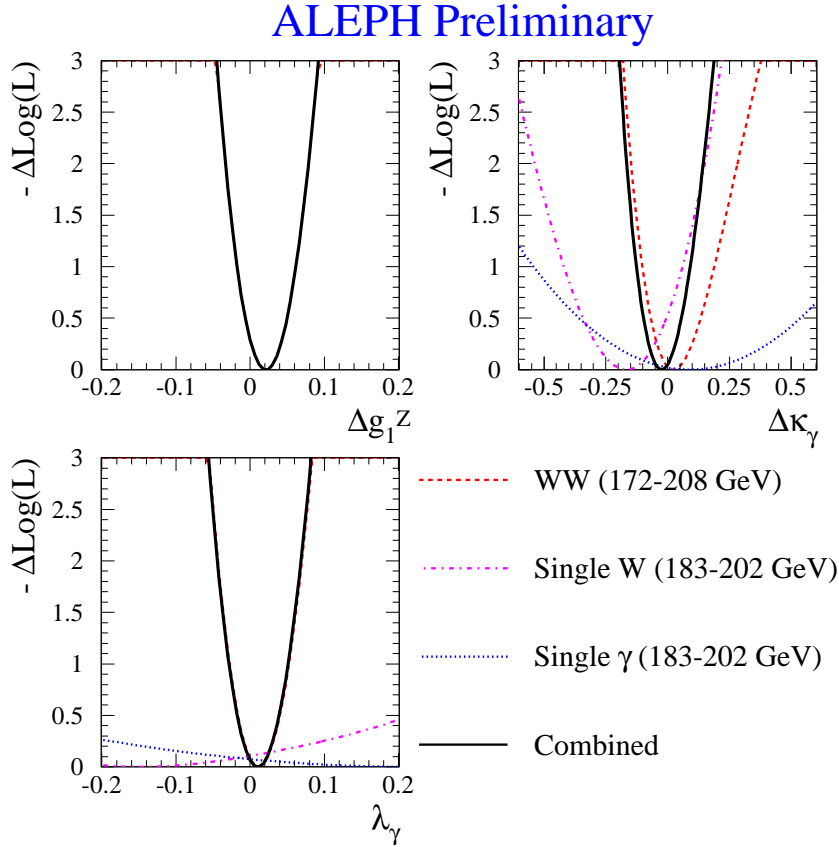,width=11cm}
\end{center}
\caption{Dependence of the likelihood  on triple-gauge couplings of \W bosons.  \label{fig:tgc_a1d}}
\end{figure}
A preliminary analysis from
 the \alephe collaboration,\cite{bib:tgc_a206} fitting
only one coupling at a time, is given in  Fig.~\ref{fig:tgc_a1d} in terms of the dependence of the log-likelihood on the couplings. 
The plot also indicates the separate contributions of the \W-pair, single-\W,
and  single-photon components to the likelihood. The  results are clearly dominated
 by the contribution from  from \W-pair events, except in the determination of $\Delta \kappa_\gamma$, where the contribution
from single-\W events is also important. The contribution of single-\W events is especially important in 
the two- and three-parameter fits, where they help
to reduce the correlation between parameters.
\begin{figure}
\begin{center}
\epsfig{file=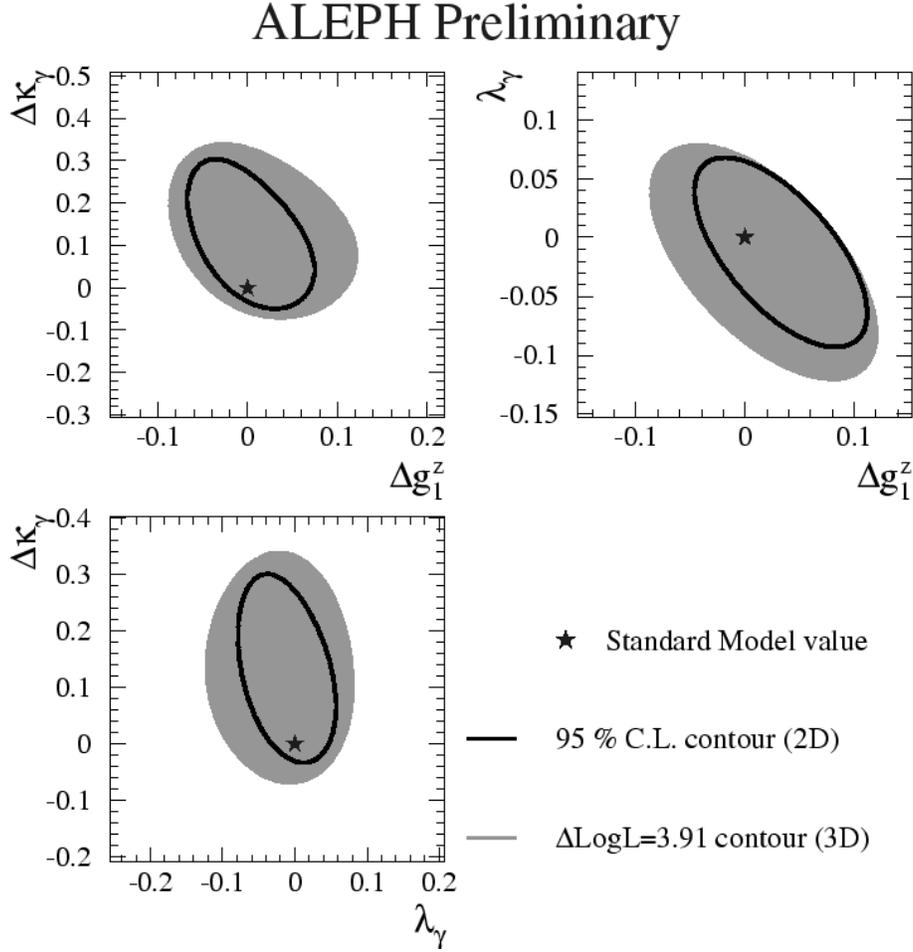,width=12.cm}
\end{center}
\caption{Contours of  95\% confidence  for  simultaneous fits of two gauge
couplings, and projections of the 95\% confidence volume for a simultaneous fit to all
three parameters (shaded region).  \label{fig:tgc_a2d}}
\end{figure}
Figure~\ref{fig:tgc_a2d} shows the results of the two- and three-parameter fits
to the same data.\cite{bib:tgc_a206}
The preliminary results from all 4 \LEP collaborations,~\cite{bib:tgc_lep_r} and their combined 
one-parameter fits,
are shown in Fig.~\ref{fig:tgc_lep} and in Table~\ref{tab:tgc_lep}.
\begin{table}
\tbl{Preliminary results of the combined \LEP fits to triple gauge couplings of the \W boson.
Given are the  68 \% and  95 \% C.L. intervals. In each case, only the listed parameter 
is varied, while the others are kept at the Standard-Model predictions.   
\label{tab:tgc_lep}}{ 
\begin{tabular}{|l||r|c|} \hline
Parameter & 68\%  C.L. &   95 \% C.L \\ \hline \hline
$g_1^Z$           &  $0.990 \begin{array}{c} +0.023 \\ -0.024 \end{array}$ & $[0.944,1.035]$ \\  \hline
$\kappa_\gamma$ &  $0.896 \begin{array}{c} +0.058 \\ -0.056 \end{array}$ & $[0.786,1.009]$ \\  \hline
$\lambda$& $-0.023 \begin{array}{c} +0.025 \\ -0.023 \end{array}$ & $[-0.069,0.026]$ \\  \hline
\end{tabular}
}
\end{table}

The experimental results can be compared to the Standard-Model loop corrections,
to the gauge couplings, and to  expectation for extensions of the Standard Model.\cite{bib:tgc_lep2}
The one-loop corrections are of the order 
$\Delta \kappa_\gamma \simeq (4.1 - 5.7) \times 10^{-3}$.\cite{bib:tgc_sm}
In the minimal supersymmetric extensions of the Standard Model (MSSM), additional loop corrections  contribute
to the gauge couplings.\cite{bib:tgc_lep2} These are typically of the same size as the predictions from the Standard Model, 
but, for specially chosen parameters, the contributions can be enhanced and can 
get as large as $\Delta \kappa_\gamma \simeq 1.7 \cdot 10^{-2}$.\cite{bib:tgc_mssm}
Contributions to  gauge couplings from additional heavy gauge bosons (Z$^\prime$) are suppressed
by a factor $m_\W^2/\Lambda^2$, where  $\Lambda$ is the scale associated with the new gauge boson.
Only through delicate  construction is it  possible to invent  models with heavy gauge bosons
that are consistent with  \LEPI data, and have significant contributions to the 
gauge couplings.\cite{bib:tgc_lep2}

The gauge boson couplings influence quantities such as the partial width of 
the $\Z \rightarrow f\bar{f}$,
the branching ratio $B(b\rightarrow s\gamma)$, and, through
loop contributions,  the anomalous magnetic-moment of the muon.
The impact on the partial width   of the \Z  can be used to determine limits
on  anomalous gauge-boson couplings from fits to the precision data on the \Z. This yields:\cite{bib:tgc_lep1}
\[ \Delta \kappa_\gamma = 0.016 \pm 0.019 ~~~~ \Delta  g_1^Z = -0.017 \pm 0.018. \]
In the fit gauge couplings, all but the one being fitted are set to their Standard-Model predictions.
However, these indirect limits on  anomalous gauge-boson couplings depend on  assumptions
such as scales and form factors of the processes in question. They  therefore
provide tests of the gauge couplings  only in specific models.

\subsection{Neutral Triple-Gauge Couplings}
The triple gauge-boson couplings  $\Z\gamma \Z$, $\Z\gamma \gamma$ and $\Z \Z \Z$ 
vanish at lowest-order in 
 the Standard Model. Loop contributions to these vertices are of the order of 
$10^{-4}$.\cite{bib:tgc_neut}
Experimentally, the existence of such  gauge couplings can be inferred from
the total cross section and angular distributions of $\Z \Z$ and $\Z \gamma$ final states.
The $\Z \gamma$ final state is sensitive to the $\Z \gamma \Z$ and $\Z\gamma \gamma$ vertices. 
The most general description of such interactions that is compatible with
Lorentz and electromagnetic gauge invariance contains eight independent parameters 
($h_i^V$, $i=1..4$, and $V=\gamma,\Z$).   
The ZZ final state is sensitive to the $\Z \Z \Z$ and  $\Z \Z \gamma$ vertices.
  In this case, the couplings can be parametrized
using  four independent parameters  $f_i^V$, $i=4,5$, and $V=\gamma,\Z$.
The parameters $h_i^V$ and $f_i^V$ are in general independent of each other. 
Both t $f_i^\gamma$ and $h_i^Z$ are used to parametrize couplings at
 a $\Z \Z \gamma$ vertex, but in one case the vertex involves two real \Z bosons, and in the other
case it involves a real \Z boson and a real photon. 

\begin{table}
\tbl{Preliminary combined \LEP results on neutral triple gauge-boson couplings.
Shown are the   95 \% C.L. limits. In each case, only the listed parameter 
is varied, while the others are kept at their Standard-Model values.   
 \label{tab:tgc_lep_n}} {
\begin{tabular}{|l|c|} \hline
Parameter  &   95 \% C.L \\ \hline \hline
$h_1^\gamma$        & $[-0.056 , +0.055 ]$ \\  \hline
$h_2^\gamma$        & $[-0.045, +0.025]$ \\  \hline
$h_3^\gamma$        & $[-0.049, +0.008]$ \\  \hline
$h_4^\gamma$        & $[-0.002, +0.034]$ \\  \hline
$h_1^Z$             & $[-0.13 , +0.13 ]$ \\  \hline
$h_2^Z$             & $[-0.078, +0.071]$ \\  \hline
$h_3^Z$             & $[-0.20 , +0.07 ]$ \\  \hline
$h_4^Z$             & $[-0.05 , +0.12 ]$ \\  \hline
\end{tabular} \hspace*{.5cm}
\begin{tabular}{|l|c|} \hline
Parameter  &   95 \% C.L \\ \hline \hline
$f_4^\gamma$        & $[-0.17 , +0.19 ]$ \\  \hline
$f_4^Z$             & $[-0.31 , +0.28 ]$ \\  \hline
$f_5^\gamma$        & $[-0.36 , +0.40 ]$ \\  \hline
$f_5^Z$             & $[-0.36 , +0.39  ]$ \\  \hline
\end{tabular}
}
\end{table}
\begin{figure}
\begin{center}
\epsfig{file=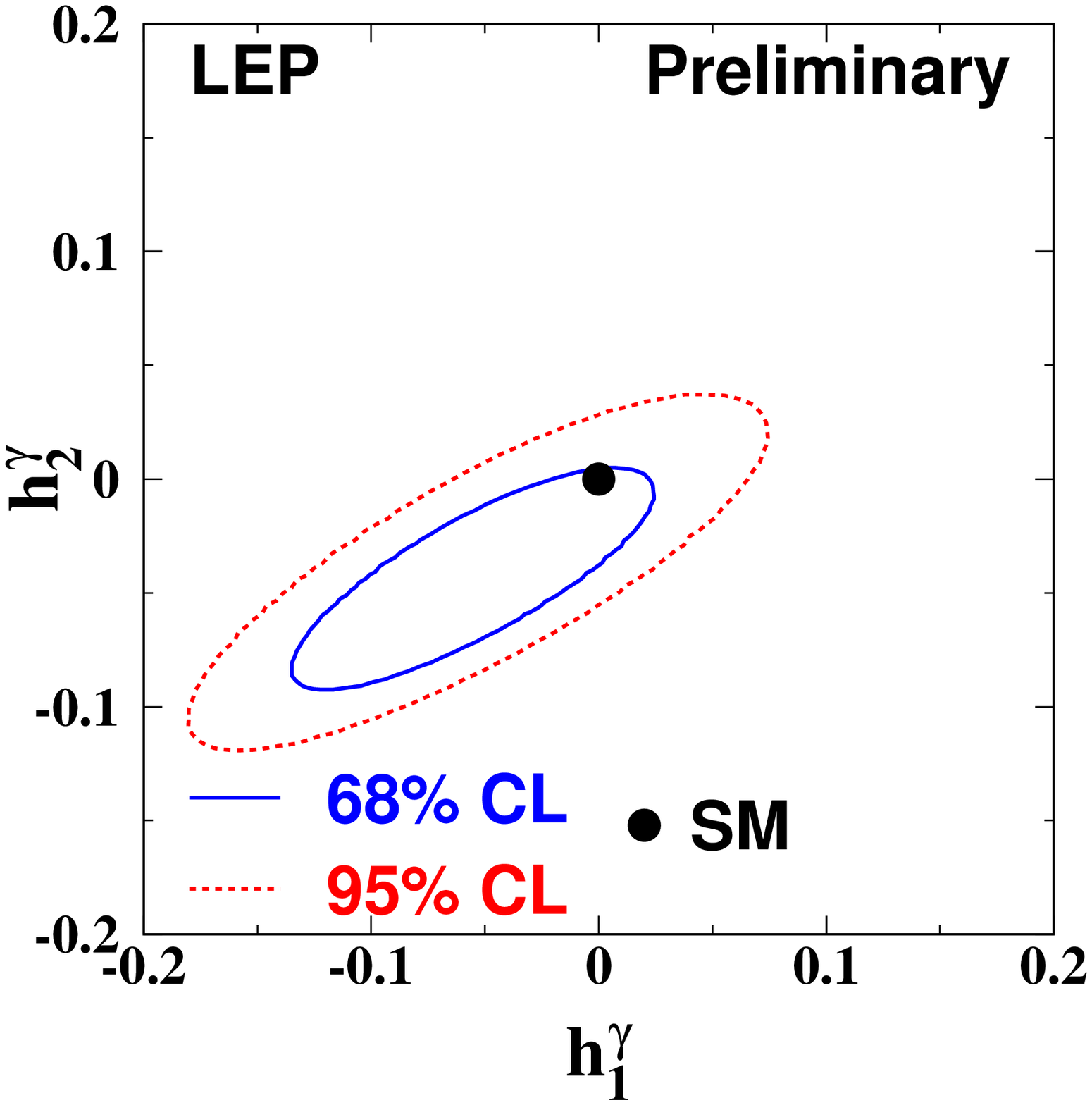,width=5.cm}
\epsfig{file=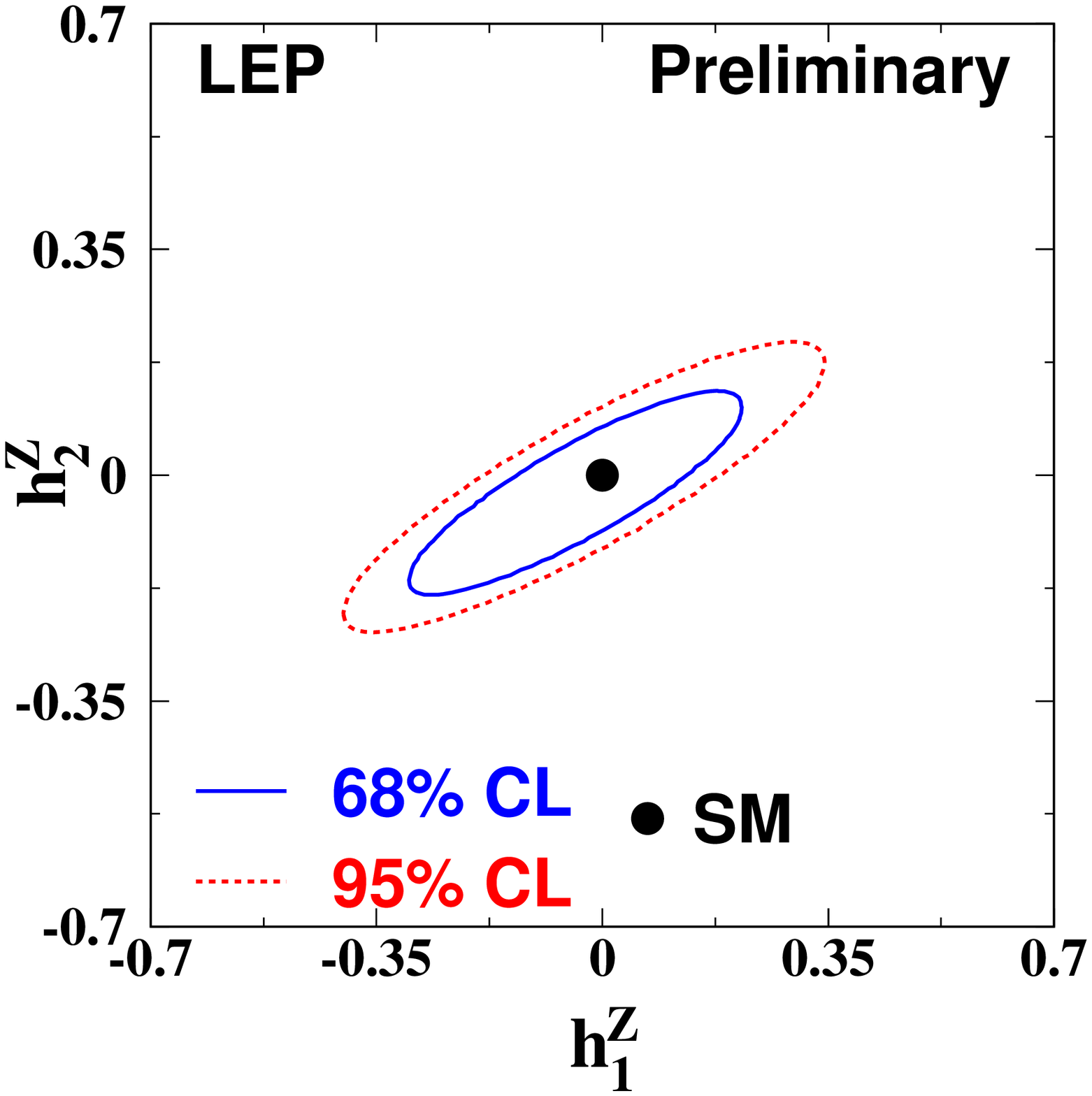,width=5.cm} \\
\vspace{-.3cm}
\epsfig{file=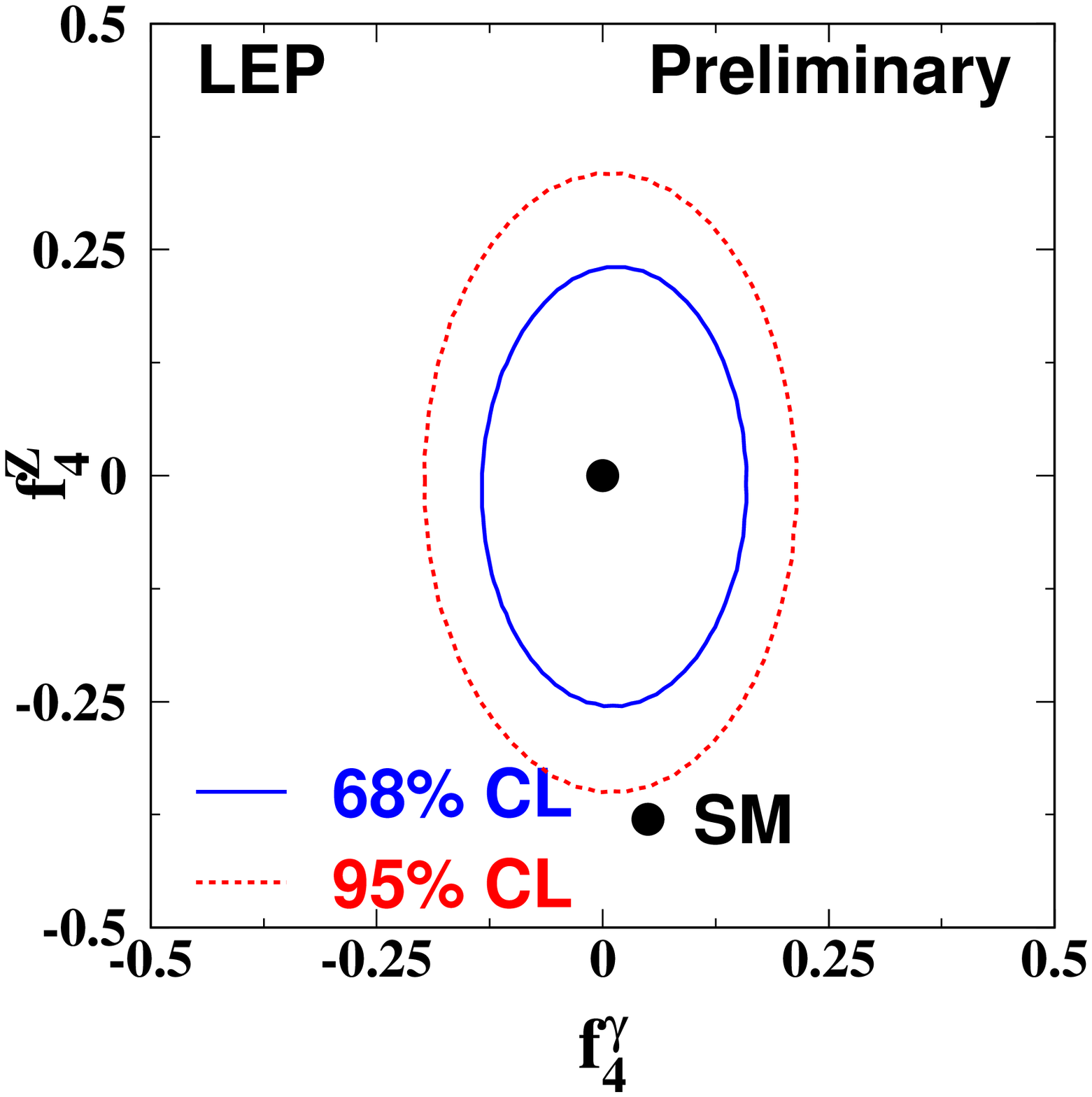,width=5.cm}
\epsfig{file=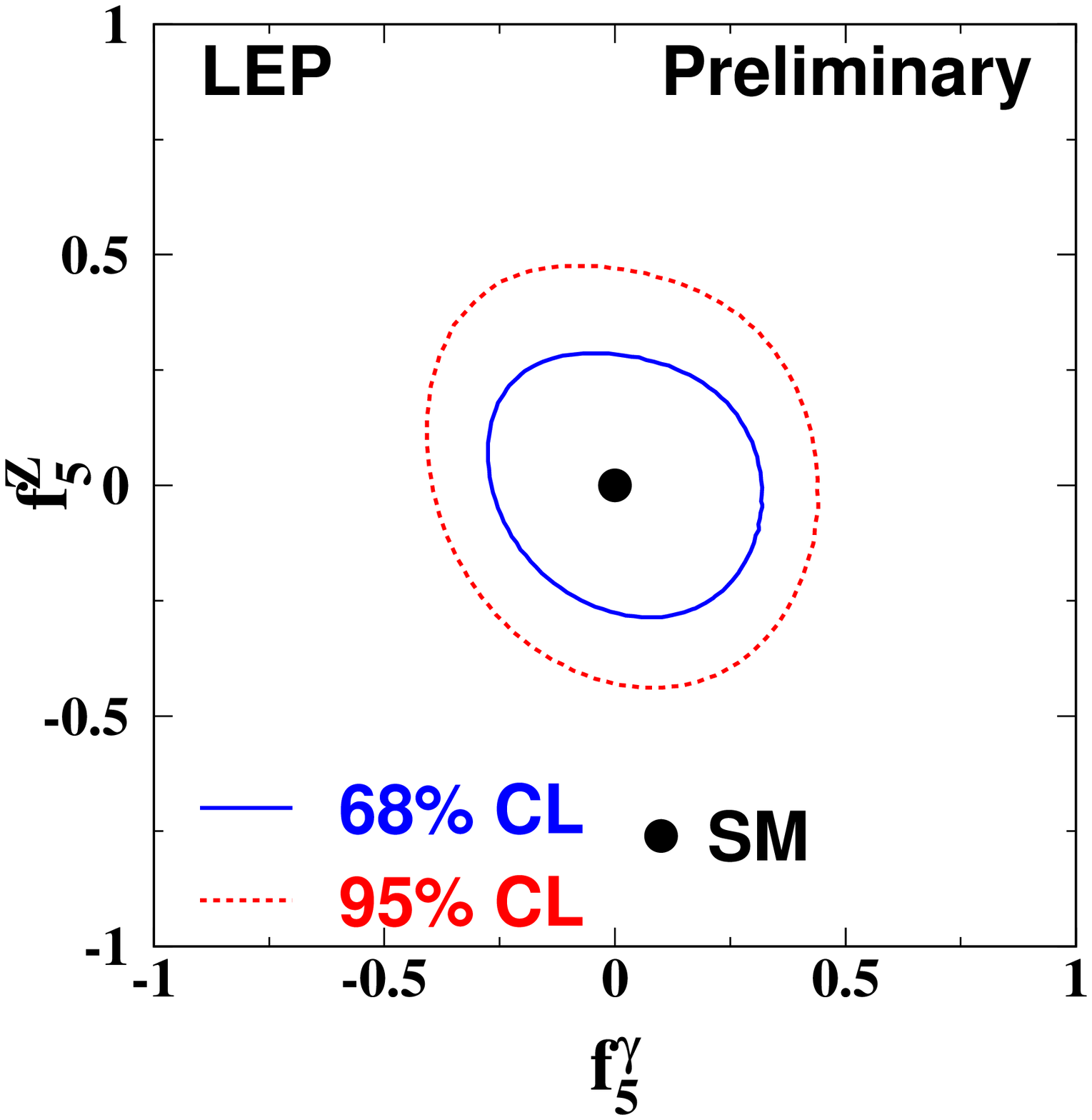,width=5.cm}
\end{center}
\vspace*{-.3cm}
\caption{Contours for 68 \% and 95 \% confidence regions for the simulaneous fit of the two neutral gauge
couplings, while the  others are kept at their  Standard-Model values.  \label{fig:tgc_n}}
\end{figure}
Preliminary combined results for the $h$ and $f$ parameters from  \LEP 
experiments~\cite{bib:tgc_lep_r} are shown in 
Table~\ref{tab:tgc_lep_n} for fits to single parameters, while keeping
all other parameters  fixed at their Standard-Model values.
Figure~\ref{fig:tgc_n} shows examples of fits where two parameters are fitted
simultaneously, while the other parameters are kept at their  Standard-Model predictions.
For  neutral gauge-boson couplings, the precision is  about an order of magnitude worse than
for  gauge couplings of the \W.

\subsection{Quartic Gauge Couplings}
\begin{figure}
\begin{center}
\epsfig{file=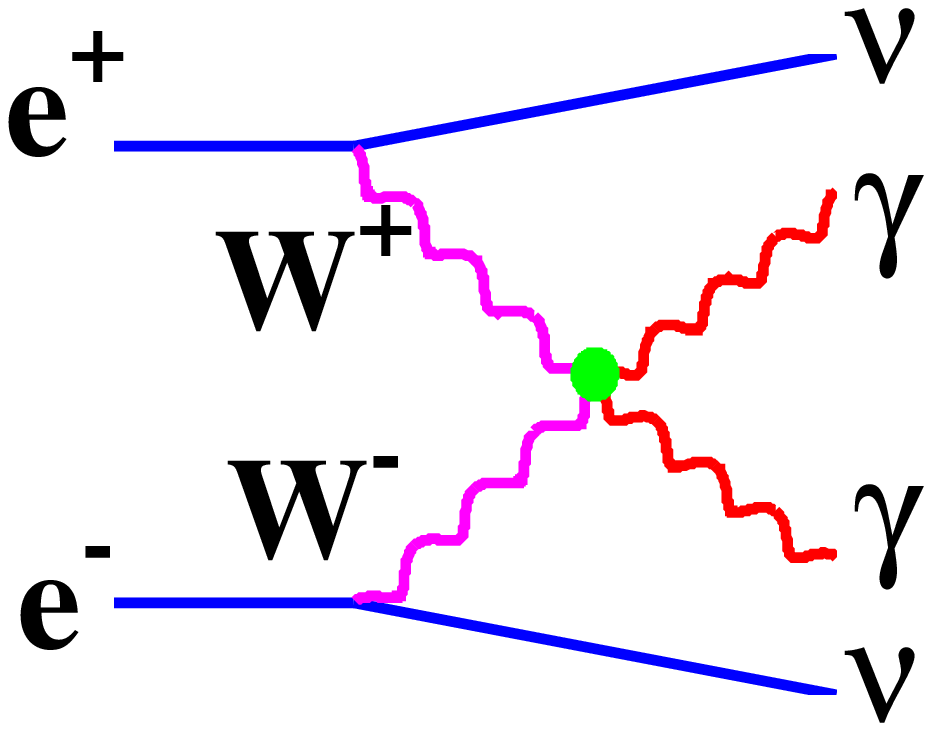,width=3.8cm}
\epsfig{file=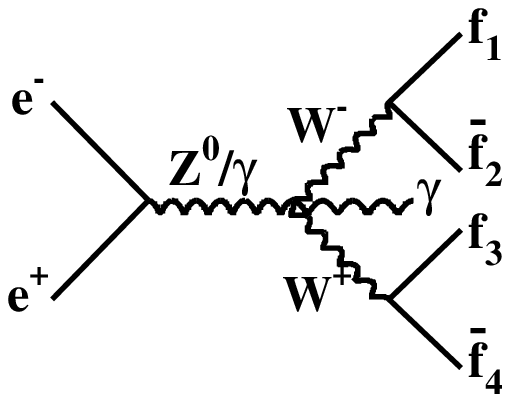,width=3.8cm}
\epsfig{file=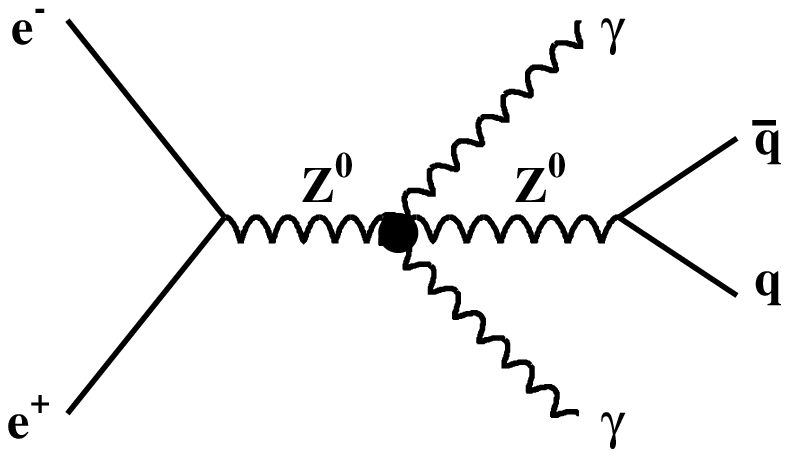,width=4.8cm}
\end{center}
\vspace*{-.3cm}
\caption{Feynman diagrams for quartic gauge couplings contributing to the  $\nu \nu \gamma \gamma$, 
$\W^+\W^- \gamma$ and, $\Z  \gamma \gamma$  final sates.  \label{fig:qgc_fey}}
\end{figure}

Due to the non-abelian nature of the electroweak interaction, the Standard Model 
predicts finite quartic gauge couplings $\W^+\W^-\W^+\W^-$, $\W^+\W^-\Z \Z $, $\W^+\W^-\Z \gamma$    
and $\W^+\W^- \gamma \gamma$. These do not play a significant role at \LEP energies,
but will be important at  the 
LHC\cite{bib:qgc_lhc} and at any future \TeV $e^+e^-$ linear collider.\cite{bib:qgc_tev} 
At \LEPII, the study of $\W^+\W^- \gamma$ and $\nu \nu \gamma \gamma$
final states can be used to infer the $\W^+\W^-\Z \gamma$  and  $\W^+\W^- \gamma \gamma$ couplings.
The coupling $\Z \Z \gamma \gamma$, which is not contained in the Standard-Model Lagrangian, can be studied
in $\Z \gamma \gamma$ events.  
Figure~\ref{fig:qgc_fey} shows the relevant Feynman diagrams.
The observation of $\W^+\W^- \gamma$  events has been  used to set the first direct limits
 on anomalous  quartic gauge  couplings.\cite{bib:qgc_opal,bib:qgc_l3} However, the indirect limits on anomalous quartic
couplings derived from the value of $\Delta r$ obtained from the precise \LEP/SLD \Z
data\cite{bib:qgc_lep_sld} are significantly better than the limits from the 
direct measurements at \LEPII.

\section{Kinematic Reconstruction of  \bW-Pair  Events }\label{sec:mw_evt}
The observables in 
\W-pair decays are the jets from hadronically decaying \W bosons and the 
charged leptons from leptonic \W-boson decays.
This section describes how the data can be used most effectively to extract information
on  the mass of the \W boson in each event.
To motivate the need for kinematic reconstruction of the events, we first discuss the
simpler case of semileptonic \W-pair decays.
In the ideal case, the mass of the hadronically decaying \W in a
\WWqqln event  can be calculated exactly
from the invariant mass of all its decay products.
Since the jets are formed through the addition of  the four momenta of the 
jet fragments (see Section \ref{sec:jet}), the
invariant mass of the two jets in the event, in principle,  equals the invariant 
mass of all the particles from the hadronic \W decay (the charged lepton and 
the unobserved neutrino  
are ignored in the  reconstruction of the jet).
For a real detector, however, the mass resolution is degraded,  largely because
of the energy resolution of jets. 

Part of the resolution can be regained
by  imposing  the constraints of energy and momentum conservation in each event.
This can be done in a kinematic fit to the final state. However,
the simplest approach is to scale the jet-pair mass by the ratio of the
beam energy to the sum of the energies of the two jets. This is equivalent to 
scaling   the jet energies and momenta of both jets by the same factor to make  the 
energy of the reconstructed \W  equal that of the beam. 
However, it should be recognized that the energies of the two \W bosons in an event do not exactly equal 
the beam energy, because of the natural  width of the \W, and the possibility of initial-state radiation.

\begin{figure}[t]
\begin{center}
\epsfig{file=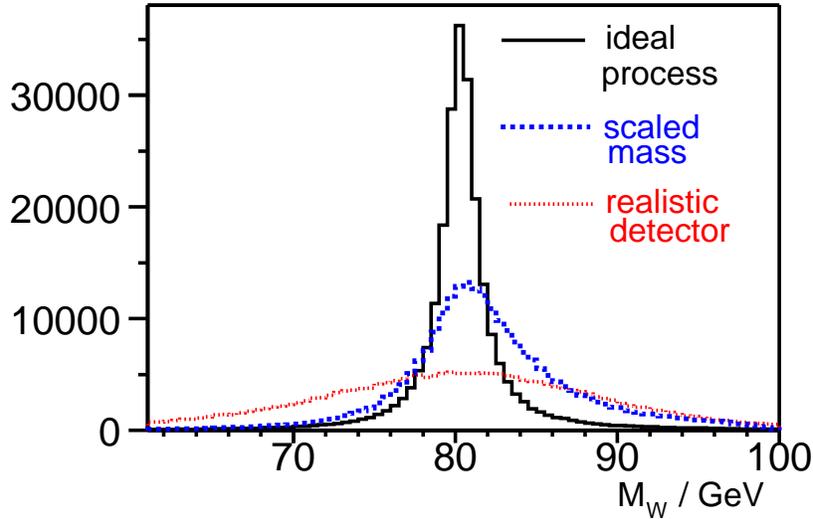,width=12.cm}
\end{center}
\caption{Mass of the hadronically-decaying \W in  semileptonic events, calculated
from the jets  for the ideal case (all particles are measured perfectly),
and for a realistic  detector,
both  with and without rescaling  
to the beam energy. The Monte Carlo events were generated 
at a center-of-mass energy of 200 \GeV. \label{fig:inv_mass}}
\end{figure}

Figure  \ref{fig:inv_mass} shows the invariant  mass of  the jet pair in 
semileptonic events for a ``perfect detector'' (ideal process), and
for the full simulation of the \opal detector, both  with and without  scaling
to  beam energy.
(For the perfect detector, it was assumed that all particles from  \W  decay 
are  measured correctly. The invariant mass of the two jets then equals  the mass
of the hadronically decaying \W boson, assuming energy and momentum  conservation in the hadronization process.)
The line shape  for the perfect detector is characterized  by the  natural width of the \W, and reflects 
the ultimate mass resolution.  Not surprisingly, 
the invariant mass determined from the two reconstructed jets has a very broad distribution, but 
part of the resolution is regained through a rescaling of the mass. More sophisticated approaches 
implement the constraints of energy and momentum conservation to improve mass
resolution, and will be discussed in the following section. The distribution of the scaled
mass  in Fig~\ref{fig:inv_mass} is observed to be biased towards larger masses. This is due to
the mass rescaling of events that contain  initial-state radiation, and  will
be discussed in Section~\ref{sec:kin_isr}.   

\subsection{Kinematic Fiting}\label{sec:kfit}
A kinematic fit can be used  to include energy-momentum constraints in any event,
and thereby improve 
the reconstruction of the \W mass. In such a fit, the  values of the momenta and
energies of the jets and leptons are 
modified within their experimental uncertainties, subject to
the desired constraints.
Technically, this is achieved through  minimizing  a $\chi^2 = \vx {\mathsf V }\vx^T$, where the vectors \vx 
contain the measured information for 
of the jet and lepton variables, and $\mathsf V $ is the covariant matrix for  their
uncertainties and correlations.\cite{bib:statistic} The constraints can be included either through Lagrange 
multipliers, or through ``penalty'' functions that  add  contributions to the $\chi^2$
if the constrains are not fulfilled (e.g., $(\sum E_{fit} - E_{cm})^2/(\delta_E)^2$,  where  $\delta_E$ is
the uncertainty in $ E_{fit}$, and $E_{cm}$ is the
value to which the fitted energy $E_{fit}$ should be constrained.)

Assuming a known lepton mass, the energies and momenta of the charged leptons can be described by three variables.
In the case of jets, this is more complicated since the jet mass
is not well defined, but has to be determined via the jet algorithm.
In principle, this can  be varied in the fit, but this leads to instability of the fits.
In practice, the jet mass is  either fixed in the fit or the $\beta=p/E$
of the jet is kept constant.
The  components of the vector \vx that  describe the energies and momenta of a lepton or jet
are typically  chosen in a way that makes it possible  both to treat the errors as Gaussian, and to
minimize their correlations. Reflecting the cylindrical symmetry of the detector two of the variables 
define the azimuthal and polar directions,
and the third gives the magnitude of the momentum.
The optimal choice for the third variable  depends on the
flavor of the particle. For electrons, for which the energy is measured in the calorimeter,
the energy is most appropriate. For muons, the momentum is  measured by the curvature of the trajectory in a magnetic
field, making $1/p$  a better  choice.
For jets, the situation is less clear, and  typically either $p$ or $\ln p$ are used in the fits.   

Constraints of total   energy and momentum conservation are applied in all fits. 
These  are referred to as 4C fits,
because of the presence of the four kinematic constraints.
In addition,  the constraint that both \W bosons have the same mass can be applied (5C fit).
This is equivalent to requiring that each  \W boson has half of the center-of-mass 
(or beam) energy.
The constraint does not reflect the exact underlying kinematics, but since the resolution
of the kinematic fit is comparable to the natural width of the \W boson,
it is a useful constraint to impose, if there is interest only in the average \W mass in the event.
It is also possible to constrain both \W masses to some fixed values (6C fit). This is useful
for determining the $\chi^2$ of the fit as function of the  \W mass, which in turn
can be used to calculate the probability of observing the given event as a function of 
the \W mass.
Combining such probability distributions  from many events,
can provide another measure of $M_\W$ and its uncertainty.

Semileptonic \W-pair events contain a neutrino that cannot be measured directly. The
three components  of the neutrino momentum can therefore be taken as free parameters, reducing the 
effective number of constraints of the fit to two (for the 5C fit) or one (for the 4C fit). 
Technically, the fits are
more stable when the momentum constraint is used directly to calculate the neutrino
momentum (from the missing energy), instead of leaving the neutrino momentum
as a free parameter, and then imposing the constraint via a Lagrange multiplier or a penalty
function. 
For  the case of  semileptonic \W-pair decays into a $\tau$, $\nu_{\tau}$, and two jets,
the analysis is more  complicated because the additional neutrino 
(or neutrinos if the $\tau$ decays leptonically) from the $\tau$ decay.
However because of the large large momentum of the $\tau$, the direction of the $\tau$  can be 
estimated simply from the
direction of its decay products, but its energy has to be treated as
a free parameter. It can be shown that, in this case, the 5C fit is equivalent
to fitting the two jets to  the constraint that the \W has the beam energy.
The measured direction of the $\tau$ does not affect  the value of the \W mass,
and  is used only to find the energy of the $\tau$ and the momentum of the $\nu_{\tau}$ that
fulfill all the constraints.

\subsection{Initial-State Radiation}\label{sec:kin_isr}
\begin{figure}
\begin{center}
\epsfig{file=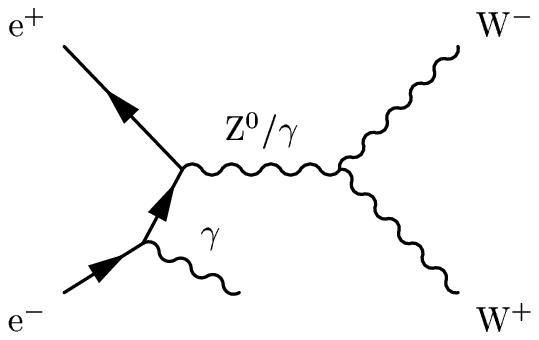,width=5.5cm} 
\epsfig{file=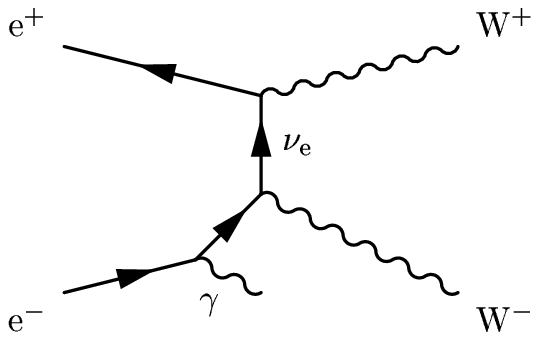,width=5.5cm}
\end{center}
\caption{Feynman diagram of   an event with initial-state radiation.
\label{fig:isr_fey}
}
\end{figure}
\begin{figure}
\parbox[t]{6.cm}{
\begin{center}
\epsfig{file=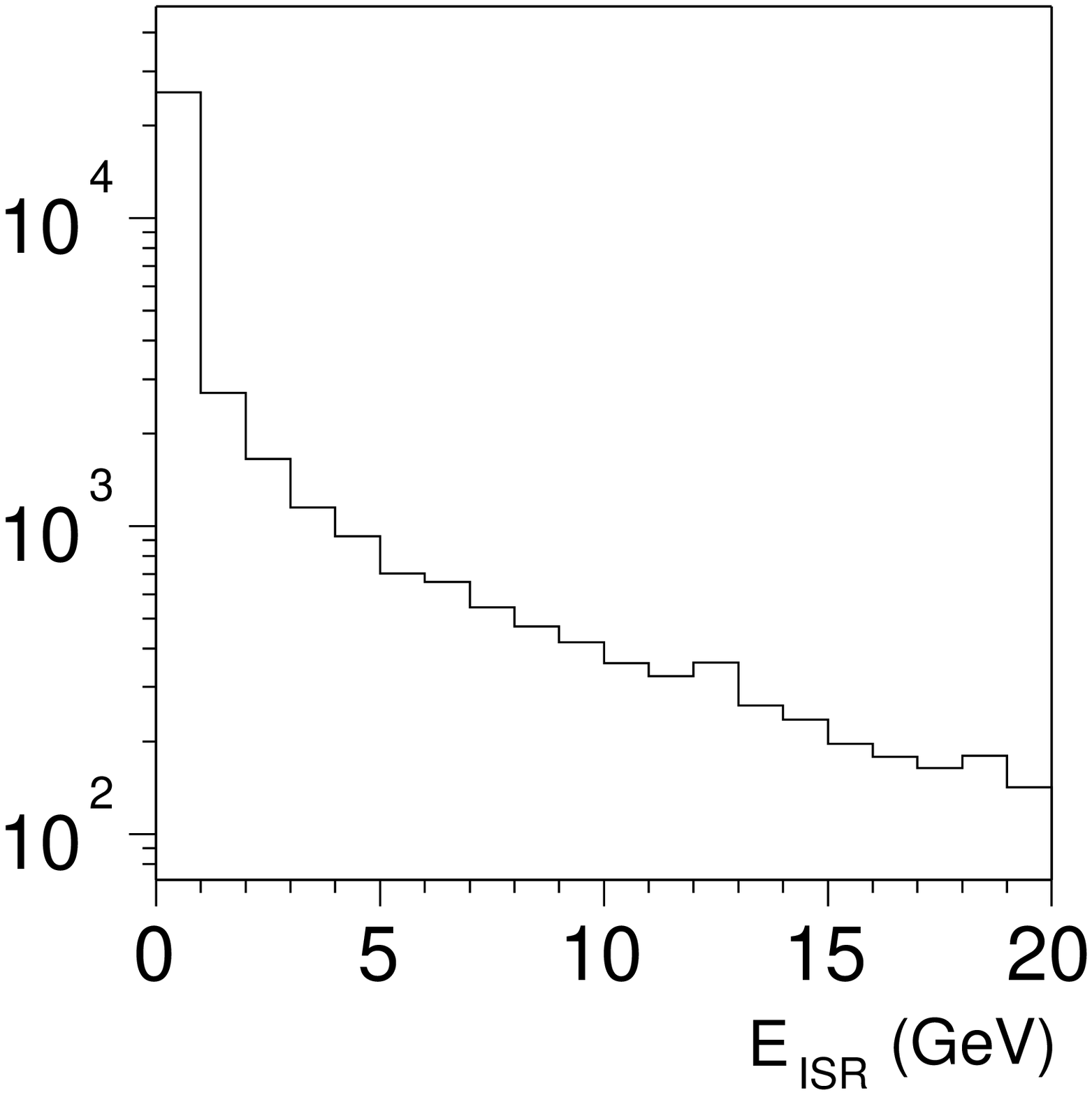,width=6.2cm}
\end{center}
\caption{Energy of the initial-state photons simulated with \koralw.
\label{fig:isr_spc}}
}
\hspace*{.3cm}
\parbox[t]{6.cm}{
\begin{center}
\epsfig{file=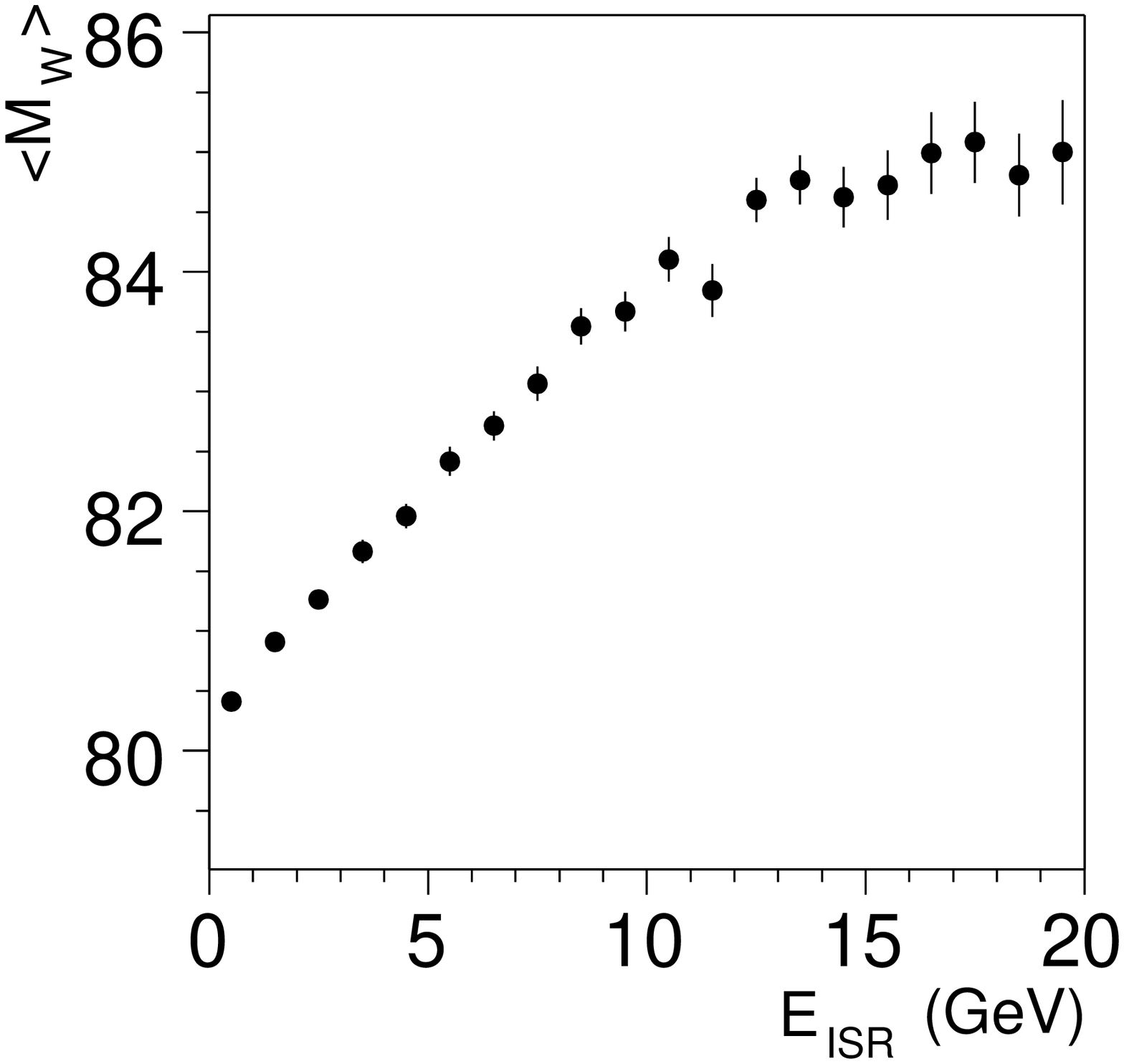,width=6.0cm}
\end{center}
\caption{Average reconstructed mass for
semileptonic \W-pair decays versus the energy of the initial-state
photon.
 \label{fig:kin_isr}}
}
\end{figure}
The kinematic fit assumes that the sum of  energies of the \W-pair 
decay products is equal to the center-of-mass energy. Of course, this is
not true when a photon is radiated by the electron or positron
prior to the creation of the \W pair (see Fig. \ref{fig:isr_fey}). 
The energy distribution of such initial-state photons ($E_{ISR}$), simulated  with the \koralw program,
is shown in Fig. \ref{fig:isr_spc}. In the fit, the center-of-mass energy is assumed to be the
same as the energy of the $\W^+\W^-$ system, which means that the reconstructed
energy of the jets (and of the lepton) will tend to  be biased  to
larger values.
As shown in Fig. \ref{fig:kin_isr}, this causes an overestimation of the \W mass
by  about 0.5 \GeV, and must be taken into proper account.
Because the initial-state photons are 
radiated mainly at small angles (along the 
beam pipe), and  are therefore not observed in the detector,
the effect of such ISR on the mass reconstruction
can be considered in a relatively straightforward statistical manner in 
the determination the \W mass, and will be discussed in Section~\ref{sec:mw_global}.

\subsection{Jet Pairing}\label{sec:jet_pair}
\begin{figure}[t]
\begin{center}
\epsfig{file=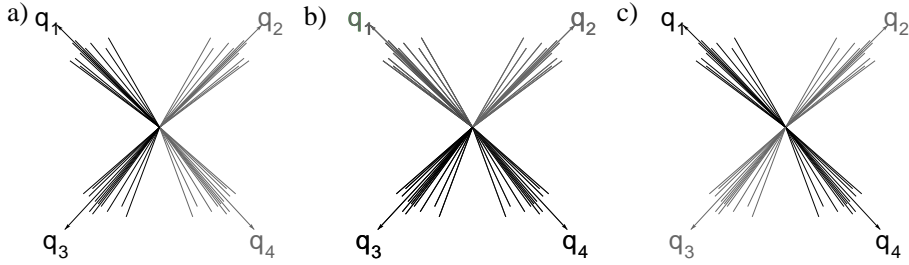,width=12.0cm}
\end{center}
\caption{Possible ways to combine four jets into two \W bosons.
         The two corresponding to each \W are marked by the 
         same shading. \label{fig:jet_comb}}
\end{figure}
In all-hadronic \W-pair decays, there are three possible ways to combine the four jets
into two \W bosons. Figure \ref{fig:jet_comb} illustrates the three combinations.
One way to decide which of  the combinations is best  for determining the mass 
is to convert the  5C kinematic $\chi^2$ fits into  fit probabilities.
The combination with the highest probability is most likely to be the correct
combination.
 But  the fit with the
next  highest fit probability often carries useful information about the \W mass.  
\begin{figure}
\begin{center}
\epsfig{file=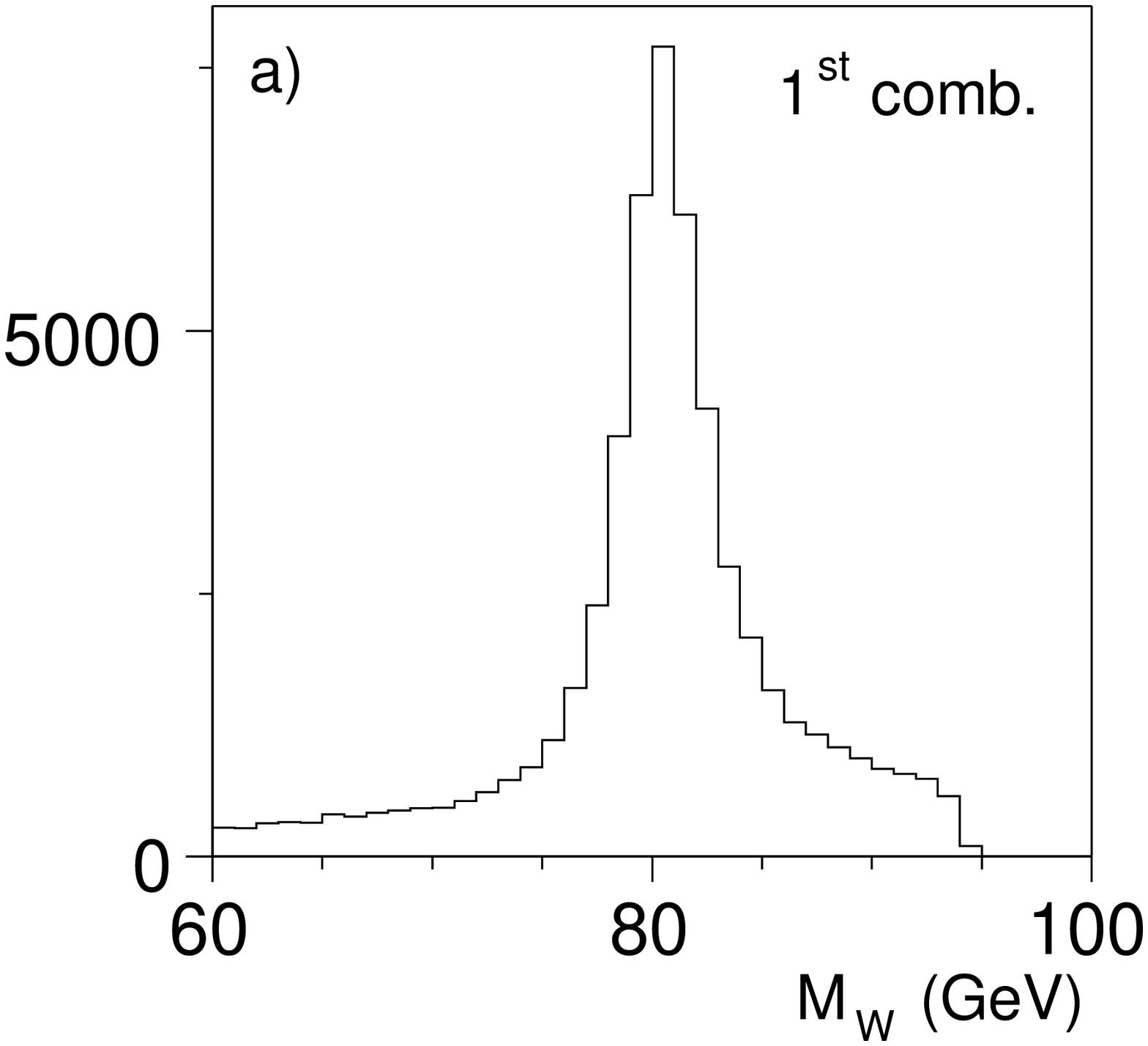,width=6.0cm}
\epsfig{file=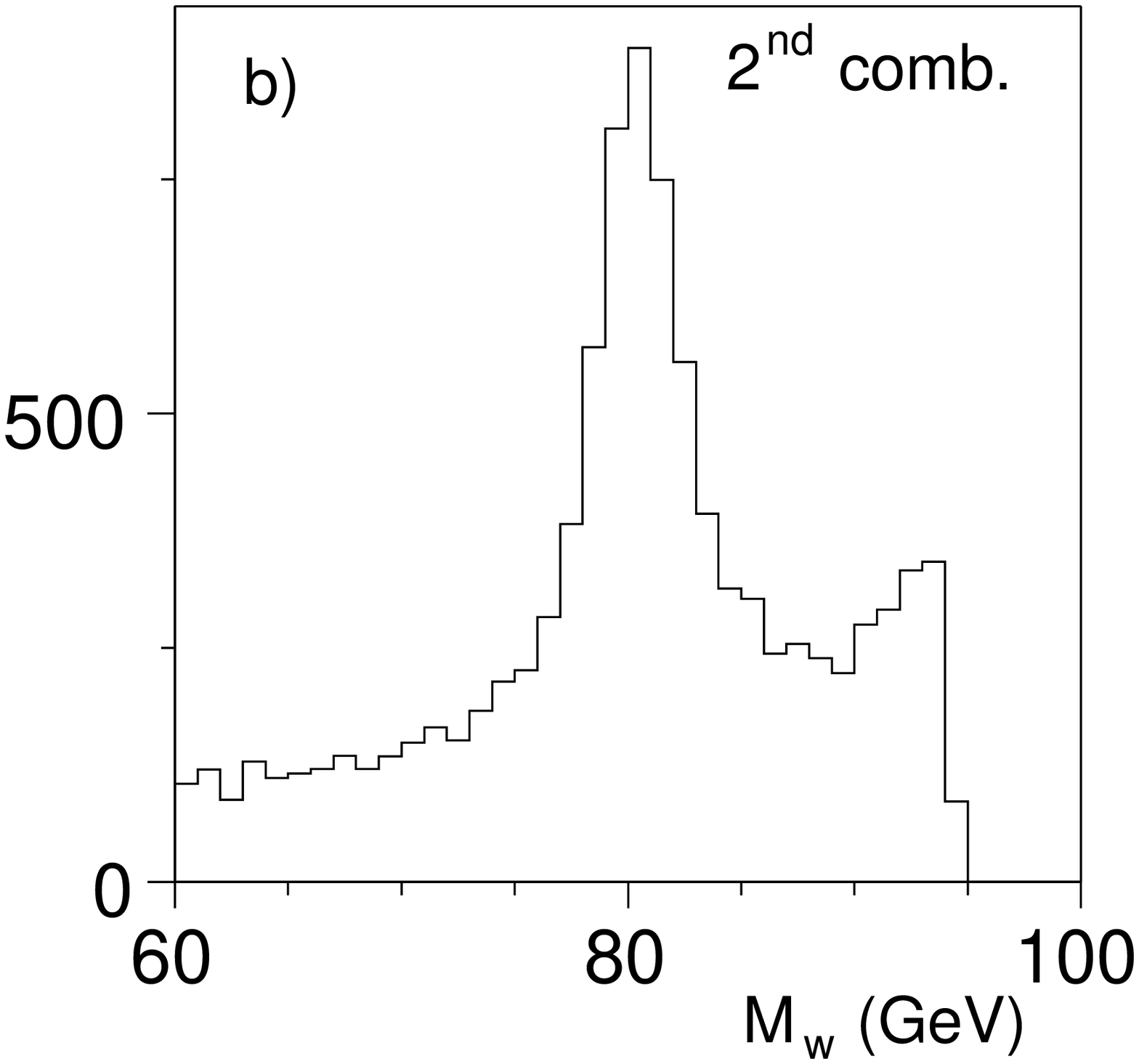,width=6.0cm}
\end{center}
\caption{Mass distribution for  the combination with the highest fit probability,
and  the next highest probability if $\prob_2 > 1/3 \times \prob_1$. \label{fig:jet_comb_bw}}
\end{figure}
Figure \ref{fig:jet_comb_bw} shows the mass distribution from the 5C fit 
for the combination with the highest fit probability, and for the combination with the
next highest fit probability, provided it has at least  $1/3$ of the probability  of the best fit. 
To ensure a reliable result, only fits with $\prob>0.01$ are
used for  reconstructing the mass. 
Monte Carlo  simulations indicate  that 90 \% of the events yield the correct combination  
 among those chosen in Fig~\ref{fig:jet_comb_bw}.

\begin{figure}
\begin{center}
\epsfig{file=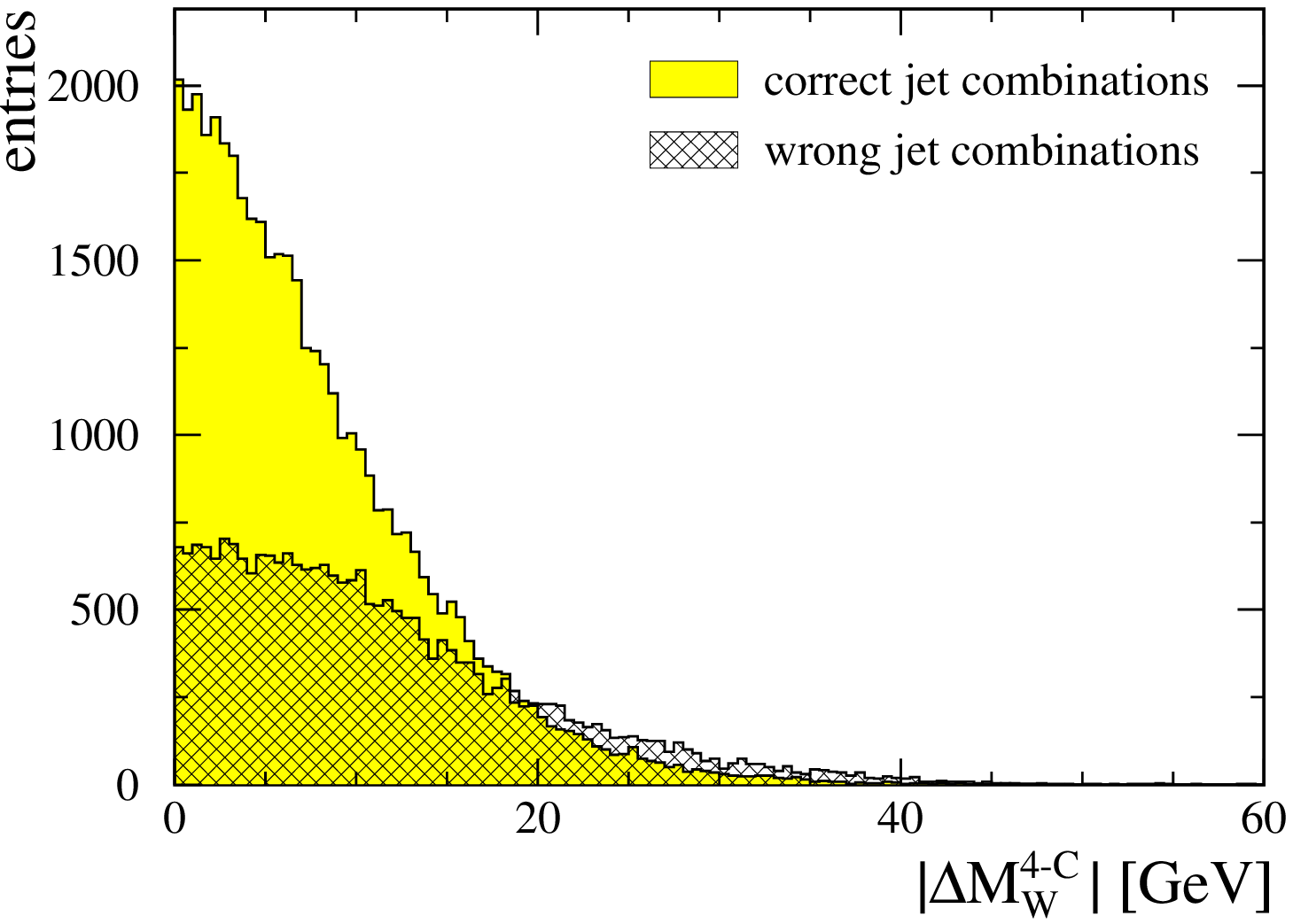,width=6.2cm}
\epsfig{file=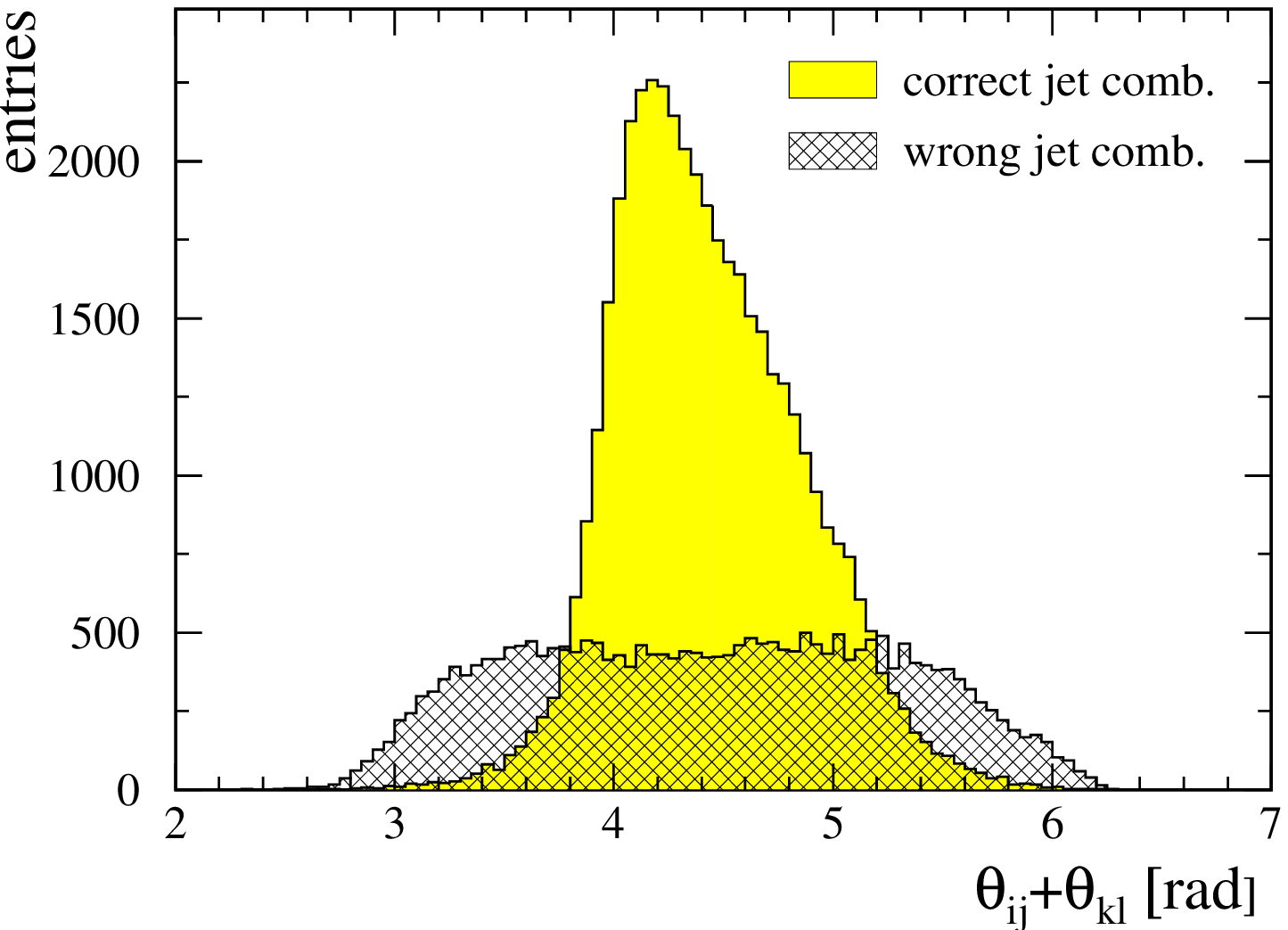,width=6.2cm}
\end{center}
\caption{Distributions used to distinguish between the right and wrong jet combination in
four-jet events. \label{fig:jet_lcor_4c}}
\end{figure}

\begin{figure}
\begin{center}
\epsfig{file=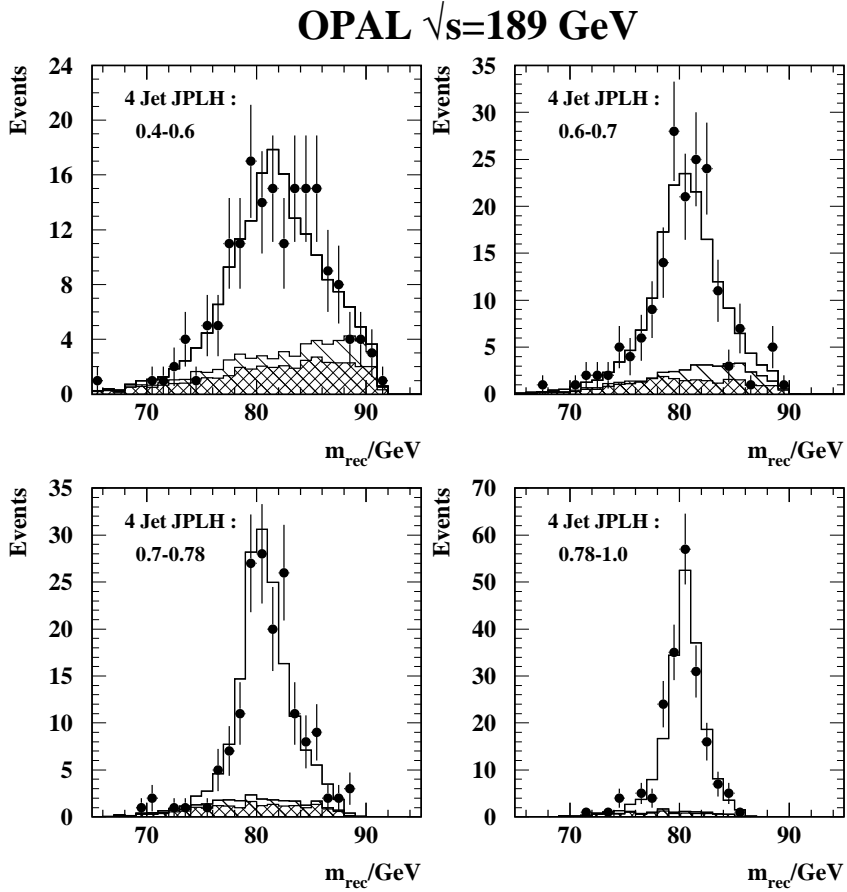,width=12.5cm}
\end{center}
\caption{Fitted \W mass in bins of the jet-pairing likelihood (JPLH).  The points are the
\opal results for a center-of-mass energy of 189 \GeV, and the solid lines show the 
 prediction from Monte Carlo. The contribution from background (other than  \W-pairs)  corresponds to 
the cross-hashed histograms, and the contribution from wrong jet pairing
 as singly-hashed histograms.
\label{fig:jet_lcor_mw}}
\end{figure}

Other  techniques can be  used to improve the probability of finding the correct jet pairing.
For the analysis of the data at $\sqrt{s}=189$ \GeV, 
the \opal collaboration, for example, used a likelihood discriminator based on:
\begin{itemize}
\item The difference  between the two \W masses (\dmwc) obtained in 4C fits
that used   only energy and momentum conservation but not the equal mass constraint.
\item  The sum of the two dijet opening angles $\theta_{ij}+\theta_{kl}$ in the laboratory frame.
\end{itemize}
Only combinations that had a 5C fit probability greater than 0.01, and
a fitted mass in the range $65~\GeV < M_\W^{5C}< 90~\GeV$, were considered in the analysis.
Figure \ref{fig:jet_lcor_4c} shows the value of \dmwc  and  $\theta_{ij}+\theta_{kl}$
for the right and the wrong combinations. The plot contains more
right than wrong combinations because 
most wrong combination do not fulfill 
the requirement  on  fit probability
and fitted mass ($M_\W^{5C}$).
The combination with the highest likelihood is correct  in 89 \% of the cases.
The output of the jet-pairing likelihood is correlated with the mass resolution. Figure \ref{fig:jet_lcor_mw}
shows the mass distribution for the published \opal data and the Monte Carlo prediction in bins
of jet-pairing likelihood.\cite{bib:opal_189} Clearly,  both the 
mass resolution and  background
fraction improve with  increasing values of the  jet-pairing likelihood. This information can also 
be used to strengthen  the statistical power of the final mass determination. 

\begin{figure}
\begin{center}
\epsfig{file=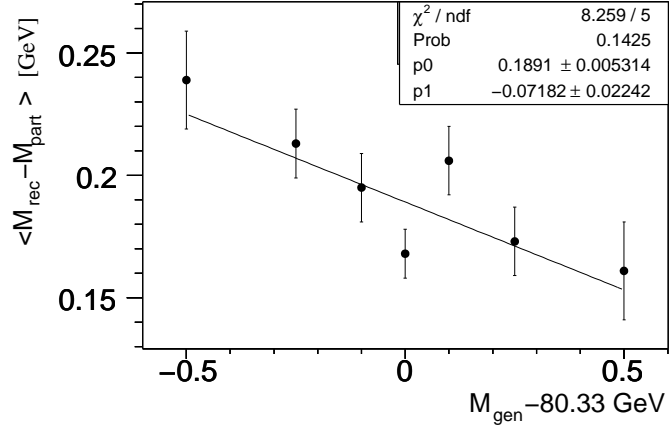,width=10cm}
\end{center}
\caption{Truncated mean of the difference between the reconstructed \W mass and the two-parton mass  for events 
with $\log{y_{45}} < -6$, and a jet pairing likelihood  greater than 0.6, as 
function of the \W  mass used to generate  the Monte Carlo sample.
The function  $<M_{rec}-M_{part}>=P_0 + P_1 (M_{gen}-80.33~\GeV)$ was fitted to the points.
A horizontal line ($P_1 = 0 $) would correspond to a lack of dependence on $M_{gen}$.
The overall shift is due to the bias from events  with initial-state radiation. 
\label{fig:jet_lcor_bias}}
\end{figure}

The  information from 
the angles between  jets is also correlated with the \W mass, and it can therefore bias
events with a high jet-pairing likelihood towards the reference mass used in the construction of
that likelihood. To illustrate this effect, Fig.~\ref{fig:jet_lcor_bias}  shows the truncated mean
of the difference of the reconstructed \W mass ($M_{rec}$) and the two-parton mass ($M_{part}$) for events 
with $\log{y_{45}} < -6$ and a jet-pairing likelihood  greater than 0.6, as a
function of the \W  mass used to generate  the Monte Carlo sample ($M_{gen}$). 
The truncated mean was calculated  only for events in which the parton and
reconstructed \W mass agreed within 5 \GeV. From the slope of the linear fit,
it can be seen that the mean reconstructed mass (for the selected
jet combination) is biased towards the mass used to construct 
the reference distributions in the jet-pairing
likelihood (about 70 \MeV per \GeV). 
Biases of this type have to be accounted for in the final extraction of the mass, and,  in general, reduce the
sensitivity of the analysis. In the extreme circumstance when the fit always returns the reference mass, 
the analysis would, of course, have no sensitivity.
 
\subsection{Five-Jet versus Four-Jet Events}
In the reconstruction of all-hadronic \W pairs, it is not possible to know  which 
hadron \ originates from which \W. 
Consequently, jets often contain hadrons from both \W bosons,
and, even for the ``correct'' jet pairing, the reconstructed \W will 
therefore contain energy from both \W bosons.
(This can be confirmed using  Monte Carlo events, where it is possible to match the 
reconstructed jets to the closest initial quark. The correct jet pairing is then 
defined as the pairing for which the matched quarks originated from the same \W.)
In events where a quark radiates an energetic gluon at large angle,  
the jet algorithm can sometimes assign  the hadrons from gluon fragmentation
to those from a quark originating from the other \W boson.
\begin{figure}
\begin{center}
\parbox[b]{.5cm}{a) \vspace*{5cm}} \parbox[t]{5.6cm}{\epsfig{file=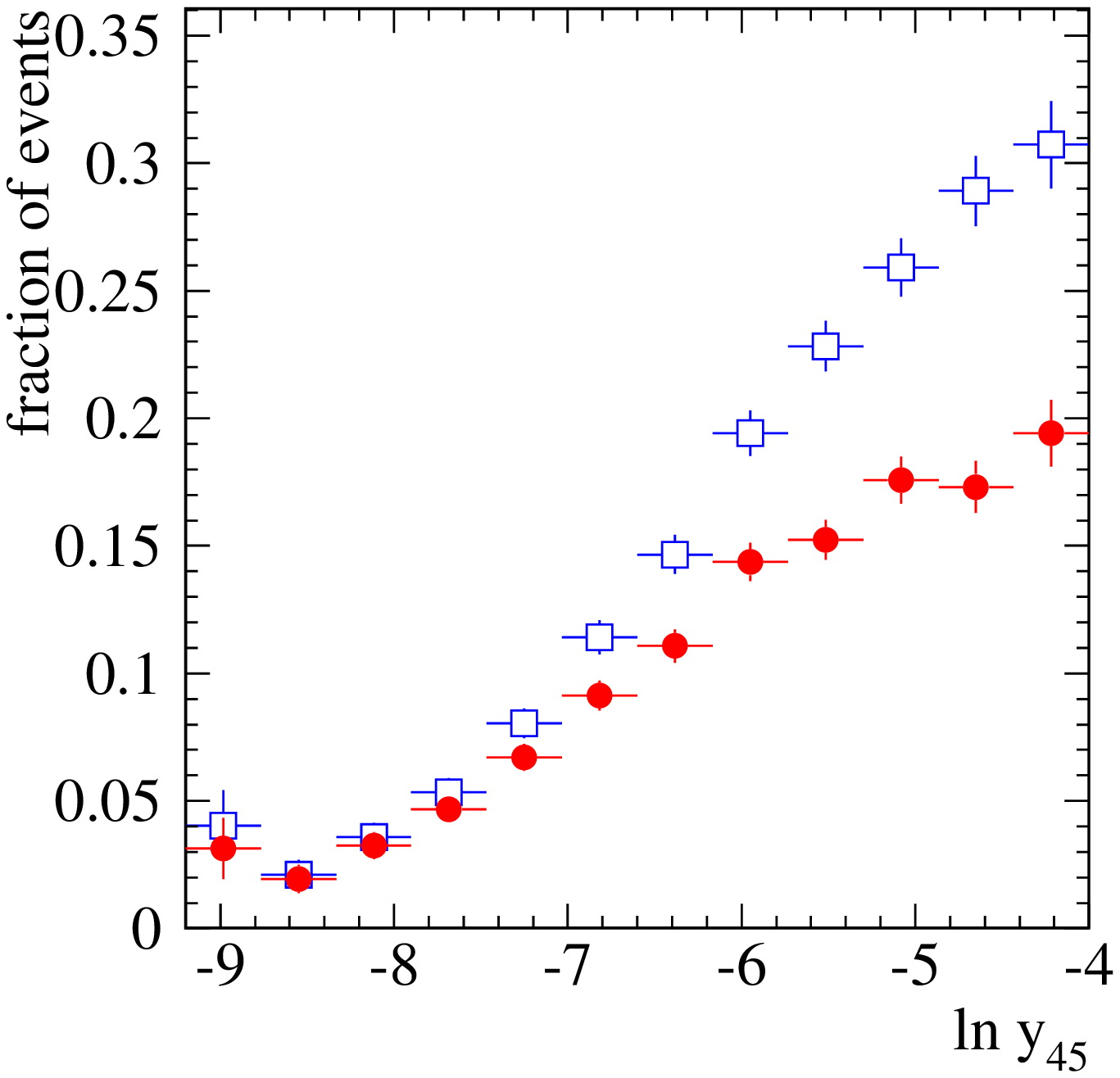,width=5.5cm}}
\parbox[b]{.5cm}{b) \vspace*{5cm}} \parbox[t]{5.6cm}{\epsfig{file=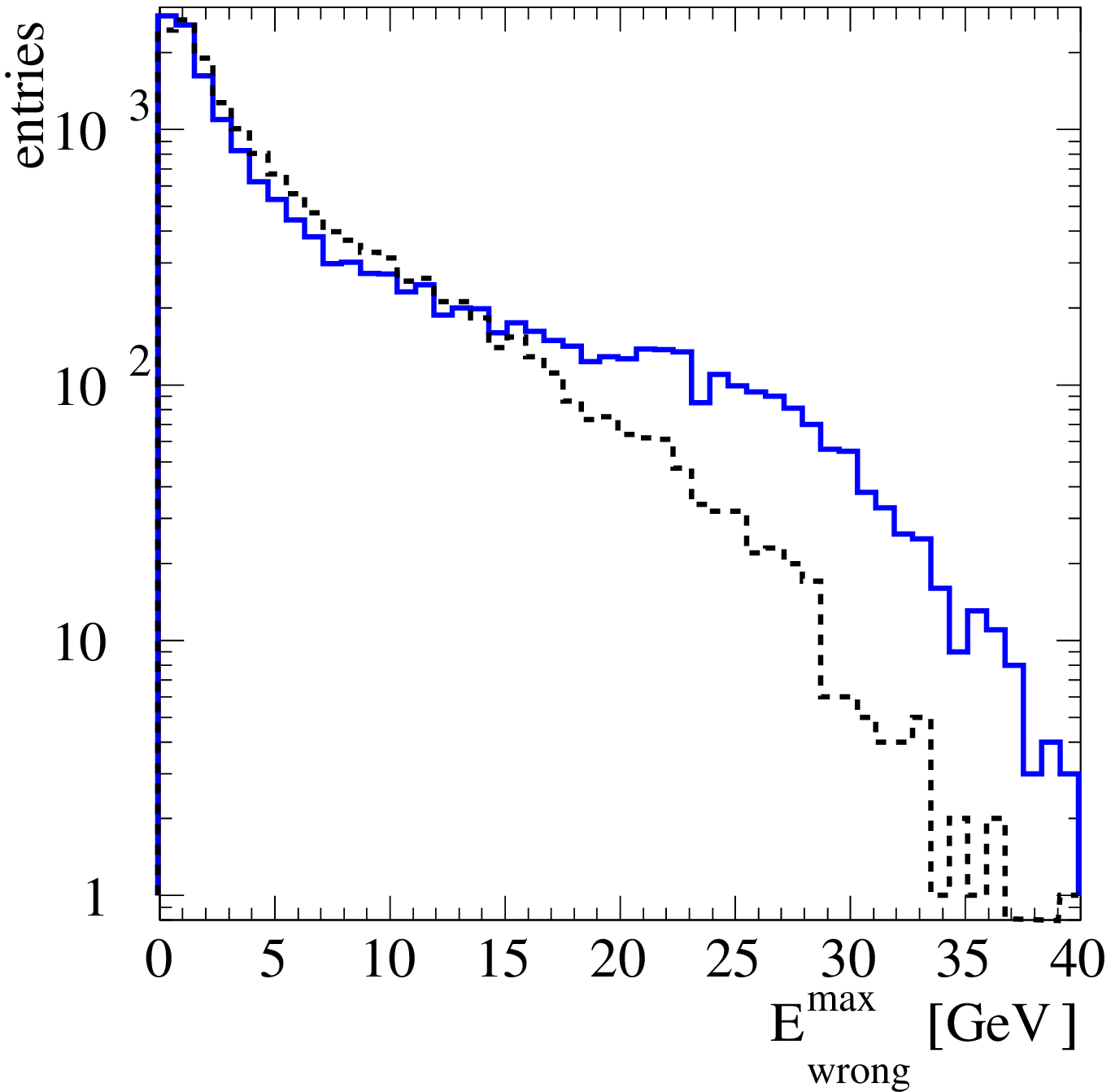,width=5.5cm}} 
\end{center}
\caption{a) Fraction of events with more than 10 \GeV energy assigned
to the wrong \W boson (for correct jet pairing) as function of $\ln y_{45}$ for four  jets (empty boxes) and 
5 jets (solid circles). b) Distribution of incorrectly associated energy
for events with $\ln y_{45}> -6.8$, for four jets (solid) and five jets (dashed) . \label{fig:wrong_e}}
\end{figure}
 In this kind of event, the reconstructed  \W mass is unreliable, and produces 
large uncertainties. Some of these events can be recovered if the jet algorithm
is used to reconstruct five instead of four jets. In this case, the hadrons from the
gluon can form a separate jet, leading to a clarification of the kinematics. 

When hadrons from the two  \W-decays are associated to the same jet, then even the correct
jet pairing will have a fraction of the measured energy assigned to the wrong W boson.
For the correct  jet combination,
Fig. \ref{fig:wrong_e} a) shows the fraction of events in which  more than 10 \GeV of 
energy is assigned to the wrong \W, as function of $\ln y_{45}$
(events with larger  $\ln y_{45}$ are more five-jet like, as discussed in Section \ref{sec:jet}).
This fraction is shown both for the case when the event is reconstructed as a four-jet and as a five-jet event.
Figure \ref{fig:wrong_e} b) shows the distribution of incorrectly  associated
energy for  events with  $\ln y_{45}> -6.8$ (five-jet like events), when these
 are  reconstructed as  four-jet and as  five-jet events.
Clearly, the fraction of events
with a large amount of wrongly associated energy is expected to be greater for  five-jet like events.
This effect is less pronounced if the event is reconstructed as a five-jet instead of
a four-jet event. 
Thus, as stated previously,the problem of wrongly associated energy can be reduced by reconstructing the event as 
having five jets.
There are ten ways to combine five jets into two \W bosons, in contrast to only
three possibilities for 4 jets.
Thus, in selecting the best  strategy,
the advantage of having less  wrongly associated energy in the correct jet pairing has 
to be weighed against the increased 
 difficulty of finding the right jet combinations.

For five-jet like events with $\ln y_{45}> -6.5$, 
the \opal analysis of the 189 \GeV data used a different likelihood discriminator to find the
right jet combination.
This  was based on the  mass in the 5-C fit, the difference  between the two masses in the 
4-C fits, the smallest opening
angle among two jets associated with the \W reconstructed from three jets, and the production angle of 
the \W reconstructed from three jets (relative to the beam axis). The probability of finding 
the correct jet pairing is $\approx 70$\%.
 \begin{figure}[h]
\begin{center}
\epsfig{file=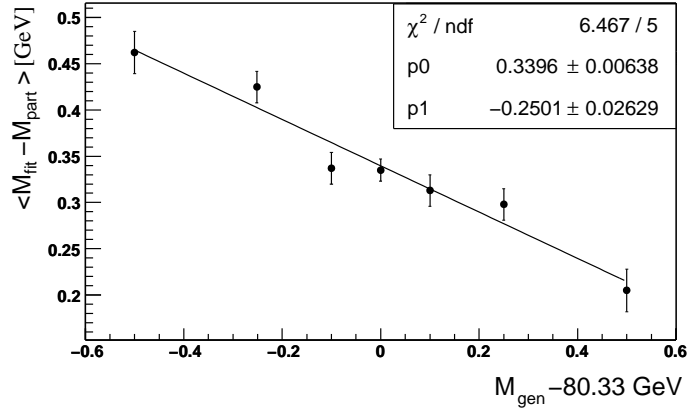,width=10cm}
\end{center}
\caption{Truncated mean of the difference between the reconstructed \W mass from five jets 
and the two-parton mass  for events   with $\log{y_{45}} > -6$, and a jet pairing likelihoods  greater than 0.4, as 
a function of the \W mass used to generate  the Monte Carlo sample. The function 
$<M_{fit}-M_{part}.=P_0 + P_1 (M_{gen}-80.33~\GeV)$ was fitted to the points.
A horizontal line ($P_1 = 0 $) would correspond to a lack of dependence on $M_{gen}$.
\label{fig:jet_lcor_mw_5}}
\end{figure}
Because the  mass from the 5-C fit is used in the jet-pairing likelihood, the mean reconstructed mass of the
selected jet pairings is more biased  to $M_{gen}$ than in the case of  four-jet events.
This is  shown in Fig. \ref{fig:jet_lcor_mw_5}. 

The question, for which values of $\ln y_{45}$ the event should be reconstructed as a four-jet
rather than a five-jet event, and how to pick the right combinations, depends critically  on the
technique used to calculate the \W  mass from the specific event information
(this will be discussed in Section \ref{sec:mw_global}).
If the analysis had, for example, only little sensitivity to wrong jet combinations, it
would then be  possible to reconstruct all events as five-jet events, and to use more than
just the ``best''  combination in any  event.

\section{Determination of the  Mass of the \bW Boson}\label{sec:mw_global}

This section will focus on how the  information described previously
can be used to determine the value of the mass of the \W boson. For the ideal case,
namely  a perfect detector and
exact  association of all emitted  particles to their respective \W bosons, the mass of the \W boson and its
width can be determined through  a fit of the distribution in reconstructed mass to a 
Breit Wigner function. 
However, as emphasized previously, quark fragmentation, 
 detector resolution, initial state radiation, 
event selection, and backgrounds have pronounced impact on $M_\W$ and $\Gamma_\W$. 
These effects can be estimated through Monte
Carlo calculations that include  full detector simulation.
The methods used to determine the mass of the \W boson  can be grouped into three 
classes, all of  which will be discussed in this section. In order to calibrate or 
to check any specific  procedure, 
the method is applied to many Monte Carlo samples that are of the same  size as
the data samples, and generated for different values of the \W-boson mass.

\subsection{Fitting with an Analytic Function} \label{sec:fit_ana}

\begin{figure}
\begin{center}
\epsfig{file=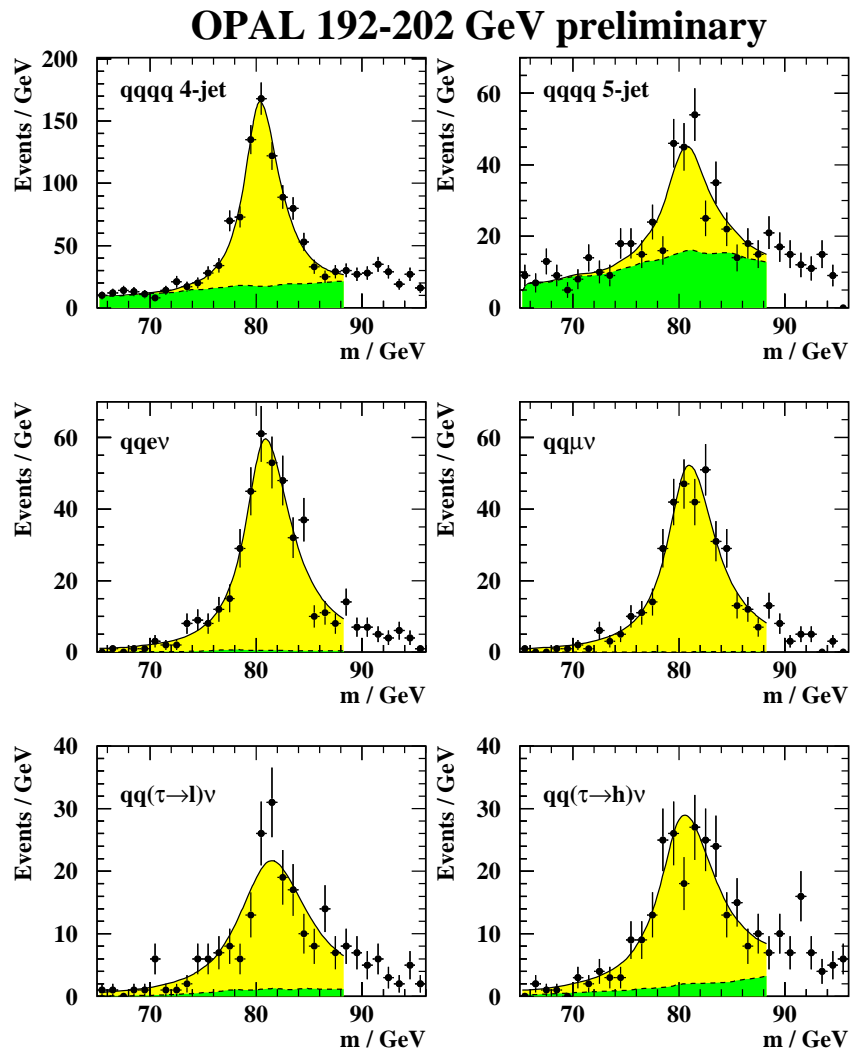,width=12.5cm}
\end{center}
\caption{Comparison of the reconstructed mass (points) with the result of the
fits. The dark area indicates the background contribution, which includes 
for the case of hadronic \W -pair decays  the impact of wrong jet combinations.
\label{fig:bw_dis}}
\end{figure}
On way to calculate the \W  mass from the reconstructed mass distribution is
to fit that distribution with some analytic function. The chosen  parametrization
must agree with the Monte Carlo predictions for different values of the \W mass,
if the extracted mass is to be reliable.
 The \opal collaboration,    
in a preliminary analysis of the 192-202 \GeV  data,\cite{bib:opal_ww192}  used a Breit Wigner function
for semileptonic \W-pair decays,  with a maximum at $m_0$ and,  different widths
 below ($\Gamma_{-}$) and above  the maximum ($\Gamma_{+}$), to fit the distribution in reconstructed mass $M_{rec}$:
\[ S(M_{rec}) = A \frac{M^2_{rec} \Gamma^2_{+(-)}}{(M_{rec}^2-m_0^2)^2+M_{rec}^2 \Gamma^2_{+(-)}}. \]
where A is a normalization constant. 
The reason for using  a different width above and below the maximum is that the asymmetry 
from  initial state radiation  biases the reconstructed mass 
towards larger values
(see Section~\ref{sec:kin_isr}).
The values of $\Gamma_{+(-)}$ depend on the resolution,  and are determined from Monte Carlo for
the different lepton channels. The data is fitted with the sum of this (pseudo) 
Breit-Wigner function and a contribution from background.
The background terms are obtained from 
Monte Carlo, and include the contribution from wrong jet combinations
in $W^+W^-$ events when $>10$\% of the hadronic energy is assigned incorrectly. 
Ignoring the  normalization, the only free parameter is the
maximum $M_0$ of the Breit-Wigner function. This parameter is closely related to
the \W mass. Monte Carlo samples generated with different values of the \W  mass are used
to determine a calibration that is  used to convert the fitted value of $M_0$
into a  true \W  mass ($M_\W$).

In the hadronic channel, the fitting  function is a product of $S(M_{rec})$) 
and a Gaussian 
$G(M_{rec})= \exp[-(M_0-M_{rec})^2/2\sigma^2]$. The width $\sigma$ of the Gaussian
is  determined from  Monte Carlo simulations that are used to optimize  agreement between the analytic function
and the Monte Carlo distribution in the reconstructed mass.
Figure~\ref{fig:bw_dis} displays  the bests fit to the data for different hadronic and semileptonic 
channels.

\subsection[Comparison with Monte Carlo Spectra]
{Comparison with Monte Carlo Spectra---Reweighting} \label{sec:fit_rew}

The \W mass can also be determined by searching for the value of  the \W mass 
that provides  best agreement between  data
and  Monte Carlo for any variables that are 
 sensitive  to the \W  mass.
A direct comparison of
data with  Monte Carlo  includes automatically all effects present in the mass
distribution, which are, after all,  part of the Monte Carlo simulation, e.g., detector resolution,
hadronization, initial-state radiation, background, biases due to  jet pairing,  etc.
It is not possible, however, to generate a set of fully simulated events for each
W-mass point used in such a comparison. It is not even  feasible to produce a
very fine grid (say, 0.02 \GeV) of mass points to provide templates as a function of
W mass.
However, with the help of a weighting procedure described below,
it is possible to produce  distributions  in any observables
for  arbitrary values of  \W  mass, from just a single sample of Monte Carlo events generated
at $M_\W^{gen}$. 
This method is particularly valuable for small excursions from $M_\W=M_\W^{gen}$.
The probability $p_i$ that some  observable falls into bin $i$  of a distribution  can be estimated
from $N$ Monte Carlo events generated at $M_\W^{gen}$ as: 
\[p_i = N_i/ \sum_j N_j. = N_i/N \]
where  $N_j$ is the number of generated Monte Carlo
events falling into bin $j$. The uncertainty on $p_i$ from the statistics of Monte Carlo is 
given by
\[ \sqrt{\frac{N_i((\sum_j N_j)-N_i)}{\sum_j N_j}}\frac{1}{\sum_j N_j} \approx  \frac{\sqrt{N_i}}{\sum_j N_j}=\frac{\sqrt{N_i}}{N}. \]
In  any simulation of \W-pair events, there is some  probability to produce 
two \W bosons that have masses $m_1$ and $m_2$. This  probability
will, of course, depend on the input value $M_\W$ in the simulation. 
In the following, we will call $p(\vx|M_\W)$ the probability density for generating   an event with a four-fermion 
configuration  \vx in a Monte Carlo simulation that uses a value of  $M_\W$ for the \W mass.
The  four-fermion  configuration \vx can be specified, for example, by the two masses $m_1$ and $m_2$ of 
the  \W  pair. In the most general case, however, \vx is given by the four-momenta of the four fermions. 
The reweighting technique is based on  the fact that the  difference between Monte Carlo simulations
that correspond to different \W  masses  is given by the probability  $p(\vx|M_\W)$.  
We can define  a weight function
\[  w(\vx,M_\W^{gen},M_\W^{rew}) =  p(\vx|M_\W^{rew}) / p(\vx|M_\W^{gen}) \]
which can be used to obtain 
a distribution that corresponds to a \W  mass  $M_\W^{rew}$, by taking events generated with $M_\W^{gen}$
and multiplying each by this weight.
Basically, this means  that the  probability to simulate a
configuration  \vx (i.e., \W masses $m_1$, $m_2$), for a  $M_\W^{rew} \ne M_\W^{gen}$, is
obtained through a weighting by $w$ of
Monte Carlo events generated at a mass  $M_\W^{gen}$, thereby generating
a distribution for a new mass $M_\W^{rew}$.
More formally, this can be seen from  the fraction of $N_i(M_\W^{gen})$ weighted entries in a bin $i$ of 
any distribution $B_i$:
\[ \frac{1}{N} \sum_{n=1}^{N_i(M_\W^{gen})} w(\vx_n,M_\W^{gen},M_\W^{rew}) =
 \int p(i,\vx,|M_\W^{gen}) w(\vx,M_\W^{gen},M_\W^{rew}) d\vx = \]
\[ \int p(i|\vx)p(\vx|M_\W^{gen})  \frac{p(\vx|M_\W^{rew})}{p(\vx|M_\W^{gen})} d\vx =
    \int p(i|\vx)p(\vx|M_\W^{rew}) d\vx = p(i|M_w^{rew}), \]
which is indeed the probability  $p(i|M_w^{rew})$ to find an event in bin $i$ for Monte Carlo
events  generated at $M_\W^{rew}$.
Here $p(i|\vx)$ is the probability for an event with a parton configuration \vx (generated at $M_\W^{gen}$)
to fall into bin $i$, and
$p(i,\vx|M_\W)$  is the probability to find  
an event with  a parton configuration \vx falling into bin $i$ for  Monte Carlo events generated at a mass $M_\W$.
Using the reweighting ansatz, the number of events in any bin $j$  is given by:  
\[ N_j(M_\W^{rew}) = \sum_{n=1}^{n=N_j(M_\W^{gen})} w(\vx_n,M_\W^{gen},M_\W^{rew})\]
and its error is given by:  
\[ \Delta N_j(M_\W^{rew}) = \sqrt{\sum_{n} w^2(\vx_n,M_\W^{gen},M_\W^{rew})}. \]
Using the above, we can define a  probability density $p_i(M_\W^{gen},M_\W^{rew})$
of finding a contribution in bin $i$ for mass $M_\W^{rew}$, based on a 
Monte Carlo sample generated at  a \W  mass $M_\W^{gen}$, as follows: 
\[ p_i(M_\W^{gen},M_\W^{rew}) = 
\frac{N_i(M_\W^{rew})}{\sum_j N_j(M_\W^{rew})} =
\frac{ \sum_{n=1}^{N_i(M_\W^{gen})} w(\vx_n,M_\W^{gen},M_\W^{rew})}
{\sum_j \sum_{n=1}^{N_j(M_\W^{gen})} w(\vx_n,M_\W^{gen},M_\W^{rew})} \]

The reweighting ansatz, can be extended to change both the generated W mass and its width.
The parton configuration \vx can be described by the two  \W masses $m_1$ and $m_2$.
The probability to have an event with \W masses $m_1$ and $m_2$ at a 
center-of-mass energy $\sqrt{s^\prime}$ of the
\W pair,    can be approximated  by:
\[ {\cal BW}(M_\W,\Gamma_\W,m_1) {\cal BW}(M_\W,\Gamma_\W,m_2) { PS}(s^\prime,m_1,m_2)
{ ISR}(s,s^\prime)  \]
where $M_\W$ and $\Gamma_\W$ are the  mass and width of the W boson, and 
 ${\cal BW}$ is the Breit-Wigner function:
\[ {\cal BW}(M_\W,\Gamma_\W,m_\W) = \frac{\Gamma_\W }{\pi \cdot M_\W}
\frac{m_\W^2}{(m_\W^2-M_\W^2)^2+(m_\W^2 \cdot \Gamma_\W/M_\W)^2} \]
$PS$ is a  factor leading to a decreased  probability when the sum  $m_1+m_2$ gets close
to the phase space limit of $\sqrt{s^\prime}$. $ISR(s,s^\prime) $ gives the probability for 
radiating an initial-state
photon,  which leads to the center-of-mass  energy  $\sqrt{s^\prime}$  of the \W pair for an $e^+e^-$
center-of-mass energy of $\sqrt{s}$.
The $ISR$ and $PS$ functions do not depend on $M_\W$, and will therefore cancel in the calculation of any weight.
Thus, the relative weight $w$ that we seek can be given by the ratio of the 
product of two Breit-Wigner functions:
{ \small
\[ w(m_1,m_2,M_\W^{gen},\Gamma_\W^{gen},M_\W^{ref},\Gamma_\W^{ref}) =
    \frac{  {\cal BW}(M_\W^{rew},\Gamma_\W^{rew},m_1) {\cal BW}(M_\W^{rew},\Gamma_\W^{rew},m_2)}
         {  {\cal BW}(M_\W^{gen},\Gamma_\W^{gen},m_1) {\cal BW}(M_\W^{gen},\Gamma_\W^{gen},m_2)} \]
}
The \ldrei collaboration uses the four-momenta of the partons to define the
parton configuration \vx to calculate  the following weights in their 
analysis of  data taken at $\sqrt{s}=183$ \GeV:\cite{bib:l3_ww183} 
{ \small
\[ w(p_1,p_2,p_3,p_4,M_\W^{rew},\Gamma_\W^{rew},M_\W^{gen},\Gamma_\W^{gen},s^\prime) =
\frac {{\cal M}^{4F}(p_1,p_2,p_3,p_4,\Gamma_\W^{rew},M_\W^{rew},\Gamma_\W^{gen},s^\prime)}
      {{\cal M}^{CC03}(p_1,p_2,p_3,p_4,\Gamma_\W^{gen},M_\W^{gen},\Gamma_\W^{gen},s^\prime)}. \]
}
where ${\cal M}^{4F}$ and ${\cal M}^{CC03}$ are the
the matrix element for  producing  four partons with four-momenta $p_i$ 
at a   center-of-mass energy  $\sqrt{s^\prime}$. 
${\cal M}^{4F}$ is calculated using the full set of four-fermion diagrams,
while ${\cal M}^{CC03}$  is calculated from the CC03 diagrams.
Here, the reweighting  also corrects for the fact that the reference Monte
Carlos are generated using a CC03 matrix element.

\begin{figure}[th]
\begin{center}
\epsfig{file=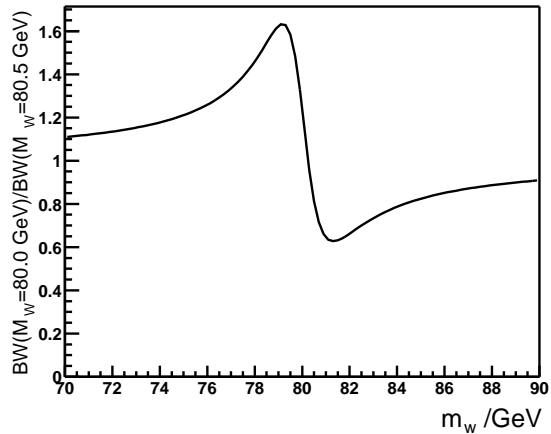,width=8cm}
\end{center}
\caption{Ratio of two Breit Wigner functions for $M_\W=80.0 $ \GeV and  $M_\W=80.5 $ \GeV, as a function of $m_\W$.
\label{fig:ratio_bw}}
\end{figure}
Figure  \ref{fig:ratio_bw} shows the ratio of two Breit Wigner functions for $M_\W=80.0$ \GeV and  $M_\W=80.5$ \GeV,
as a function of $m_\W$.
In the tails of the Breit Wigner functions, the weights stay close to unity, as long
as the difference between  $M_\W^{gen}$ and  $M_\W^{rew}$ is small compared to the width of the \W boson.
For larger differences between  $M_\W^{gen}$ and  $M_\W^{rew}$, the increase in uncertainty from 
large weights can be reduced by using Monte-Carlo samples  generated at
other  values of the \W  mass $M_{\W,k}^{gen}$. 
Under such circumstances, the error on the probability for an entry in  bin $i$ for  a \W  mass $M_\W^{rew}$
can be minimized by using  a weighted average of the probabilities calculated from different 
$M_{\W,k}^{gen}$.

\begin{figure}[th]
\begin{center}
\epsfig{file=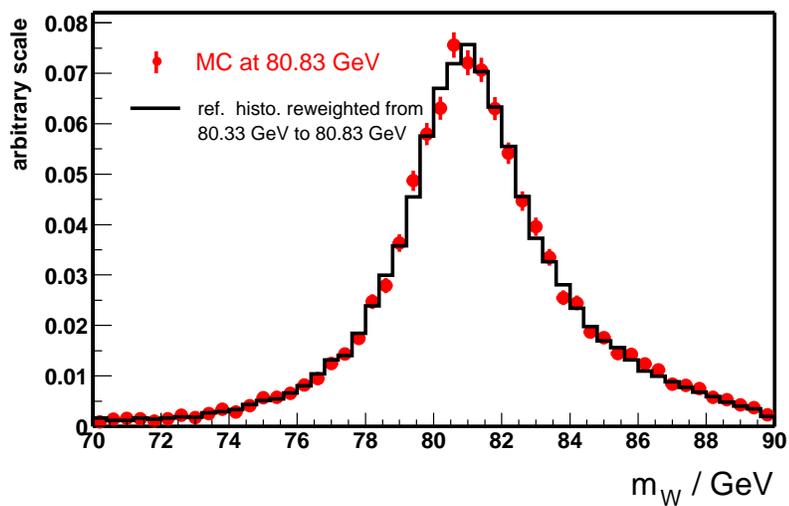,width=12cm}
\end{center}
\caption{Comparison of the reconstructed \W mass from a Monte Carlo generated at $M_\W=$80.83 \GeV (points with error
bars) with the corresponding distribution from a Monte Carlo generated 
at $M_\W^{gen}=$80.33 \GeV and reweighted to  $M_\W^{rew}=$80.83 \GeV (histogram).
\label{fig:rew_cal}}
\end{figure}
Figure  \ref{fig:rew_cal} shows a comparison of the reconstructed \W mass for a Monte Carlo sample  
generated at $M_\W=$80.83 \GeV  with the corresponding distribution generated from Monte Carlo 
at $M_\W^{gen}=$80.33 \GeV,  but then  reweighted to  $M_\W^{rew}=$80.83 \GeV. Clearly, the reweighted distribution
is in excellent agreement with the direct Monte Carlo.

The reweighted distribution can be used in a likelihood fit to extract the  mass and width
of  the \W boson. If there  is interest only in the mass, the width can be defined using
the value from the Standard Model, i.e.,  $\Gamma_\W = 3G_F M_\W^3/ (2\sqrt{2} \pi)(1+2 \alpha_s/(3 \pi))$.\cite{bib:w_br}
The simplest  approach for a fit is to use a bin size such  that the bin to bin fluctuations 
of the reweighted reference histograms are sufficiently small that they do not  affect the significance 
of the fit. However, this approach uses only the finite bin into which 
the observable (e.g., the reconstructed mass) falls,  and not its exact value,
and may thereby  limit the   
statistical power of the data.
In their analysis of the 183 \GeV data, the \ldrei collaboration compares two approaches that minimize this 
effect:~\cite{bib:l3_ww183}
In the  ``box method'', the likelihood of observing a given event is  calculated using 1000  Monte Carlo events 
that fall into  a box
 centered around the data point. The box size at the peak of the distribution is typically $\pm35$ \MeV.
In the tails, it is limited to a maximum value of $\pm250$ \MeV, even if the box contains less than 1000 events.
In a second approach, a fine binning is chosen, and the histograms are then smoothed with a cubic spline fit.
The results from these two  methods agree  within 15\% of their  statistical errors.
  
The \alephe collaboration uses multi-dimensional distributions in their analysis of $\sqrt{s}=189$ \GeV 
data.\cite{bib:aleph_ww189}
In the four-quark channel, they use the two  masses from the 4C fit
(requiring energy and momentum conservation, as discussed Section \ref{sec:kfit}).
In the $qqe\nu$ and  $qq\mu\nu$ channels, they  use the single mass from a 5C fit
(energy and momentum conservation and equal-mass constraint -- see Section \ref{sec:kfit}),   
the fitted error on that mass, and the mass of the di-jet system from a 4C fit. (There is a 43\% correlation
between the mass from 
the 4C and the 5C fit). Using this 3-dimensional distribution reduces the statistical uncertainty by $14\pm1 $\%
relative  to a one-dimensional mass analysis in the $qql\nu$ channel.

The \opal collaboration separates its 189 \GeV  data into subsets of different resolution or
background fractions.\cite{bib:opal_189} The semileptonic data is analyzed in bins that correspond to
 the fitted uncertainty.
In the four-quark channel, four-jet and five-jet events are treated separately, and
the four-jet data is analyzed in terms of bins of different jet-pairing likelihood (see Section \ref{sec:jet_pair}).
\begin{figure}
\begin{center}
\epsfig{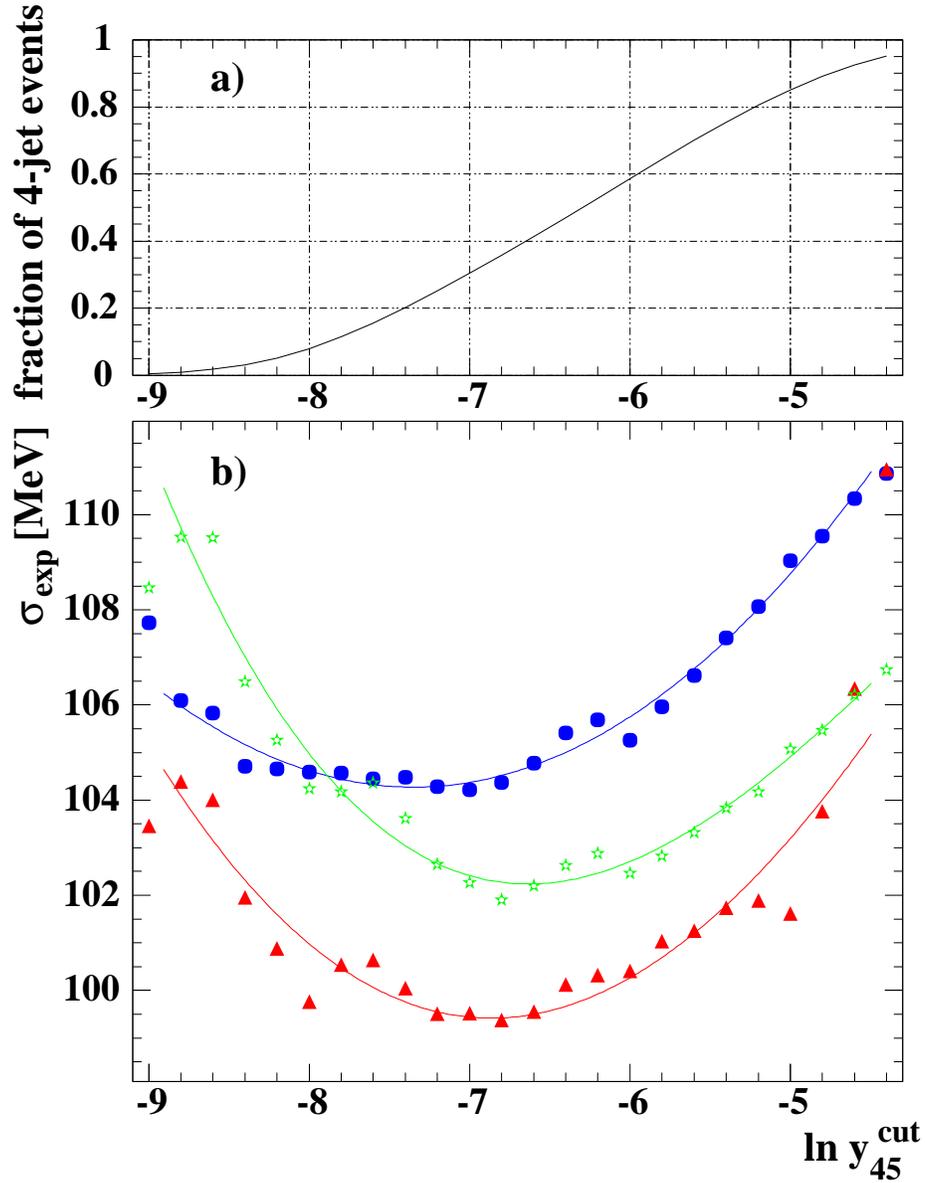}
\end{center}
\caption{The dependence of a) fraction of 4-jet events and, b)  the statistical uncertainty on the
\W mass, on $y_{45}^{cut}$, for the 4-quark analysis described in the text. 
The solid circles represent an analysis for which the reweighting is done for two classes of events, 
one for four-jet
events and one for five-jet events. The stars represent an analysis where the  fit
for reweighted four-jets is done in four bins of the jet-pairing likelihood. (Events in different bins 
have different resolutions and different
background fractions.) The triangles indicate the  improvement achieved if the jet
combination with the second-highest jet-pairing likelihood (Section~\ref{sec:jet_pair}) 
is used as an additional class of events
when the fit probability of this combination is greater than  $1/3$ of the  fit with highest
probability.
\label{fig:rew_5jet}}
\end{figure}
Figure  \ref{fig:rew_5jet} shows the dependence of the results on the value of
$y_{45}^{cut}$, which determines whether events are treated as four-jet or  as five-jet 
events.\cite{bib:tim} Figure \ref{fig:rew_5jet}~a) shows the remaining fraction of four-jet events when
all events with $y_{45}> y_{45}^{cut}$ are treated as four-jet events. The expected uncertainty on the
mass ($\delta M_\W$),
shown in  Fig. \ref{fig:rew_5jet}~b) is based on an integrated luminosity of 183 $\pb$
for a center-of-mass energy of 189 \GeV.
The solid circles represent an analysis where the reweighting is done in two classes, one for four-jet
events and one for only five-jet events. The stars represent an analysis where the fit for reweighted
four jets is done in four bins of the jet-pairing likelihood. As discussed in Section \ref{sec:jet_pair},
events in different bins 
have different resolutions and different
background fractions. The triangles indicate the additional improvement gained when the jet
combination with the second-highest jet-pairing likelihood (see Section~\ref{sec:jet_pair}) 
is used to define  an additional class of  events
for which the  fit probability of that combination is greater than  $1/3$ of the best-fit
probability. All distributions have a broad minimum in the range
$-8< \ln y_{45}^{cut} <-6$. The sharper increase at small four-jet fraction, especially for the analysis
using four bins for the four-jet events, is due to the fact that the reweighted reference histograms 
have poor statistics for small four-jet fractions.
One can clearly  see the improvement in the expected statistical error when the four-jet events are
split into more bins, and  also when the second-best jet combination is used.
 
The mass determination at \delphi is based on a convolution fit, which is described in the following
section.

\subsection[Determination of the Event Likelihood]
{Determination of the Event Likelihood---Convolution Analysis}\label{sec:conv}

The mass of the \W boson can also be determined 
from an overall minimization of a product of individual event likelihoods that are  calculated
from a convolution of a ``physics", or signal function, with a resolution function. 
Here, the total probability to observe a given event is split into two parts: One reflects 
the probability to observe
the  quantities derived from some kinematic fit (e.g.,~the two masses from  a 4C fit) 
when \W bosons of mass 
$m_1$ and  $m_2$ are produced in an  event. 
This part is given by
a resolution function. 
The other part of the probability, namely that the specified  masses  $m_1$ and  $m_2$ 
are produced in the $e^+e^-$ collision, depends on the  mass  and  width  of the \W boson (the dynamics).
This part of the probability is called the physics function.

The simplest ansatz is that the physics function is given by a Breit-Wigner
function in one or two variables, multiplied by a phase space factor, and 
the resolution function is given by a Gaussian  with a central value and width 
determined from the kinematic fit.
 
The importance of taking the background into proper account in the event likelihood is crucial,
 as can be seen by the  following example:
A single event with   a 
reconstructed mass of 60 \GeV can change the log-likelihood by 0.15, if the 
assumed \W boson mass is changed from 80 \GeV to 81 \GeV, and can therefore  affect significantly 
 the final fit. However, the most likely explanation for such an  event 
is that it is due to background, for which the likelihood should not depend at all, or, in the case
of a wrong jet pairing in a $\W^+\W^-$ event, only weakly, on the assumed mass of the \W boson. In the convolution
analysis,  this option can be implemented  by adding a term that does not depend on the \W mass, but 
describes the background.

As discussed in Section~\ref{sec:kin_isr}, initial-state radiation  shifts the reconstructed
mass  by a factor $\sqrt{s/s^\prime}$. Since $s^\prime$ is not
known on an event-by-event basis, any initial-state radiation can be taken into account 
through  a convolution of a shifted physics function with 
the  spectrum for initial-state radiation:
\[ f(m_w|M_\W,\Gamma_\W,s) = \int^s_0  f_{no ISR}(m_w \cdot \sqrt{s^\prime/s};M_\W,\Gamma_\W,s^\prime) 
                              { ISR}(s,s^\prime) ds^\prime .\]
where  $f_{no ISR}$ refers to the physics function before considering the initial-state radiation,
and   ${ ISR}(s,s^\prime)$ is the probability for emitting radiation in the initial state,  which reduces
the center-of-mass energy of the \W-boson pair from $\sqrt{s}$ to $\sqrt{s^\prime}$.
\begin{figure}[th]
\begin{center}
\epsfig{file=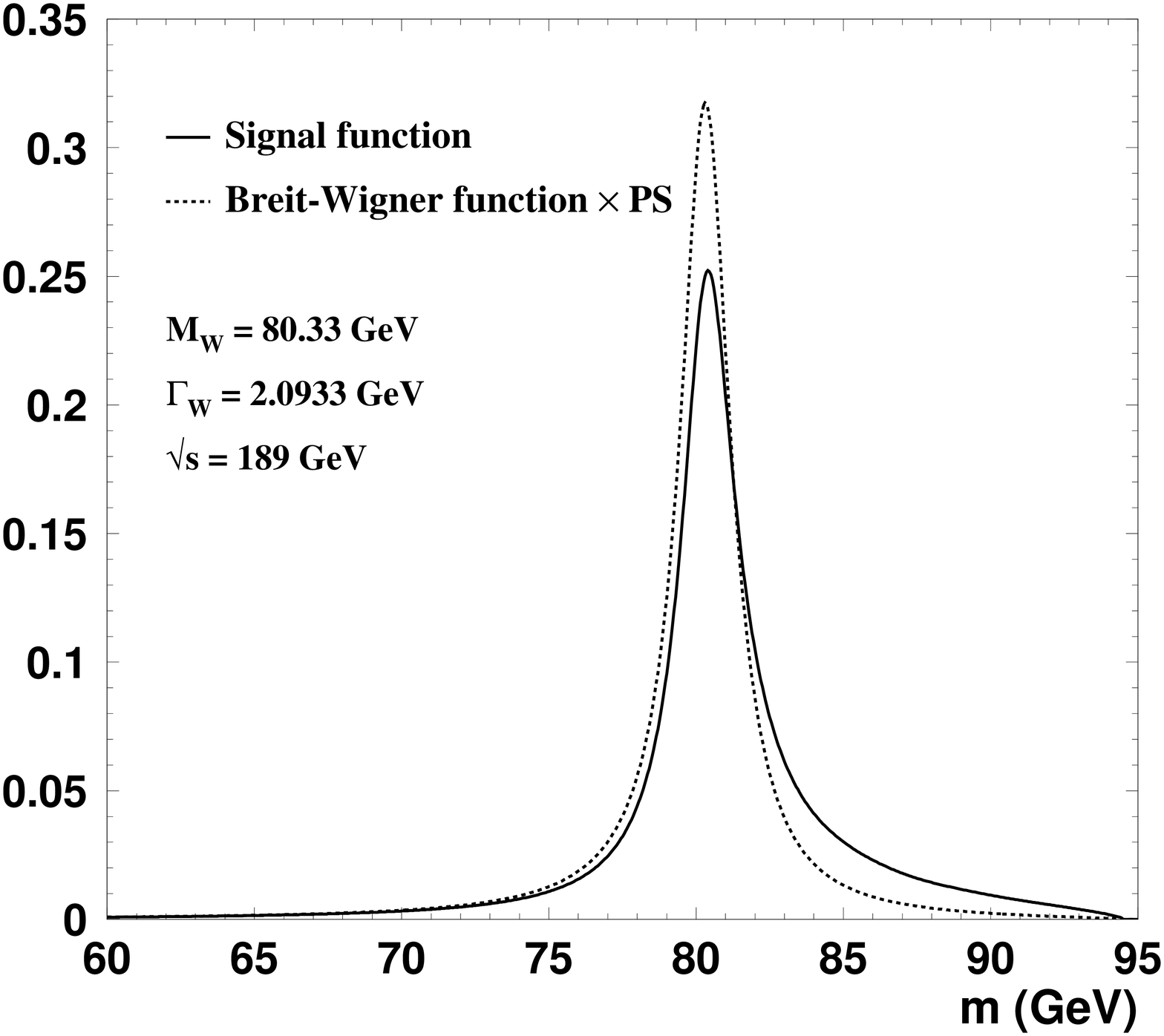,width=8cm}
\end{center}
\caption{Physics function before  (dotted) and after (continuous)  the convolution with initial-state
radiation.\protect\cite{bib:dub_phd} 
\label{fig:con_isr}}
\end{figure}
Figure  \ref{fig:con_isr} shows the dependence of the physics function on $M_\W$,
before and after the convolution with  initial-state
radiation. The shift to higher masses as a result of  initial-state radiation is visible in
Fig.~\ref{fig:con_isr}.
A  more natural way to include the effect of initial-state radiation is 
to convolute  the ISR spectrum with a scaled resolution function instead of the 
physics function, but this is very difficult for an analysis in which the resolution function
 is calculated  separately for each
event, basically because of the required computing time.

It is also possible to neglect  initial-state radiation at the first stage of analysis,
and to correct for the bias at a later step.
The \delphi collaboration followed this approach in their analysis
of  semileptonic data taken at  $\sqrt{s}=183$ \GeV.\cite{bib:delphi_ww183} 
For the data sample at $\sqrt{s}=189$ \GeV,\cite{bib:delphi_ww189} \delphi included an
ISR correction in their physics function, which resulted in a reduction of $\approx400$ \MeV  in the bias of
the fitted mass. 

\begin{figure}[th]
\begin{center}
\epsfig{file=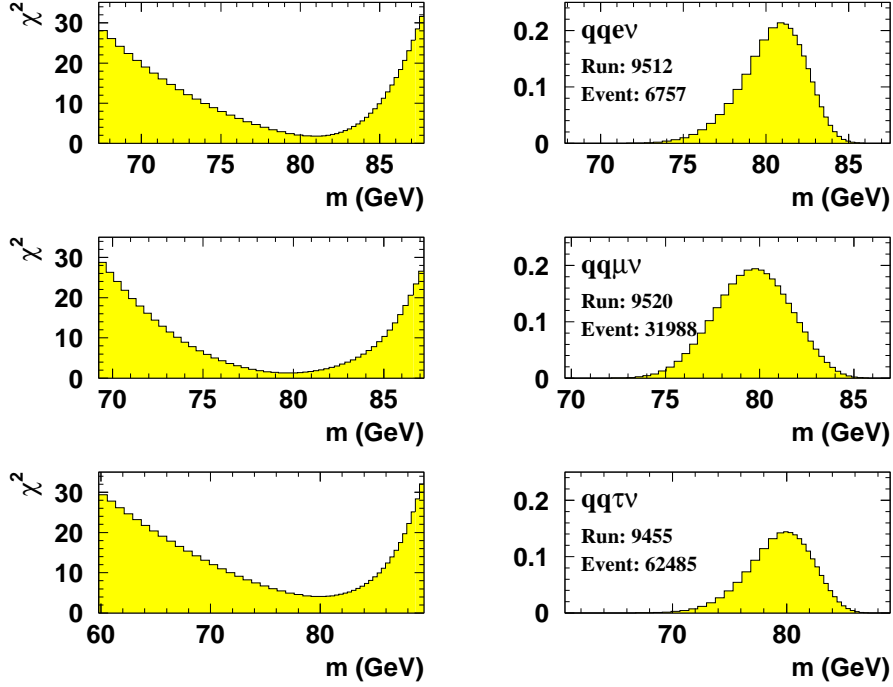,width=12.5cm}
\end{center}
\caption{$\chi^2$ distribution and resolution function as  functions of
 average \W mass for three  events from the \opal
detector.~\protect\cite{bib:dub_phd}
 \label{fig:conv_chi}}
\end{figure}
The uncertainties on  reconstructed mass in 5C fits
in the $qql\nu$ channel, 
can be  quite asymmetric. For such events, information would be lost if 
purely Gaussian errors  were   assumed for the resolution.
In order to treat these kinds of events more correctly, the resolution function
can be determined using the dependence of the 
$\chi^2$ of Sect.~\ref{sec:kfit} on $m_\W$. This $\chi^2$ dependence is
obtained from the 6C fit, in which, in addition to 
energy and momentum conservation, both \W masses are fixed to a specific value.
Figure  \ref{fig:conv_chi} shows examples of 
the $\chi^2$ dependence and the resolution functions for three $qql\nu$ events
in a one-dimensional  analysis,\cite{bib:dub_phd}
where both \W masses are fixed and sampled at  the same average mass.
For these events, one can  see the asymmetric nature of the $\chi^2$
dependence.

\begin{figure}[th]
\begin{center}
\epsfig{file=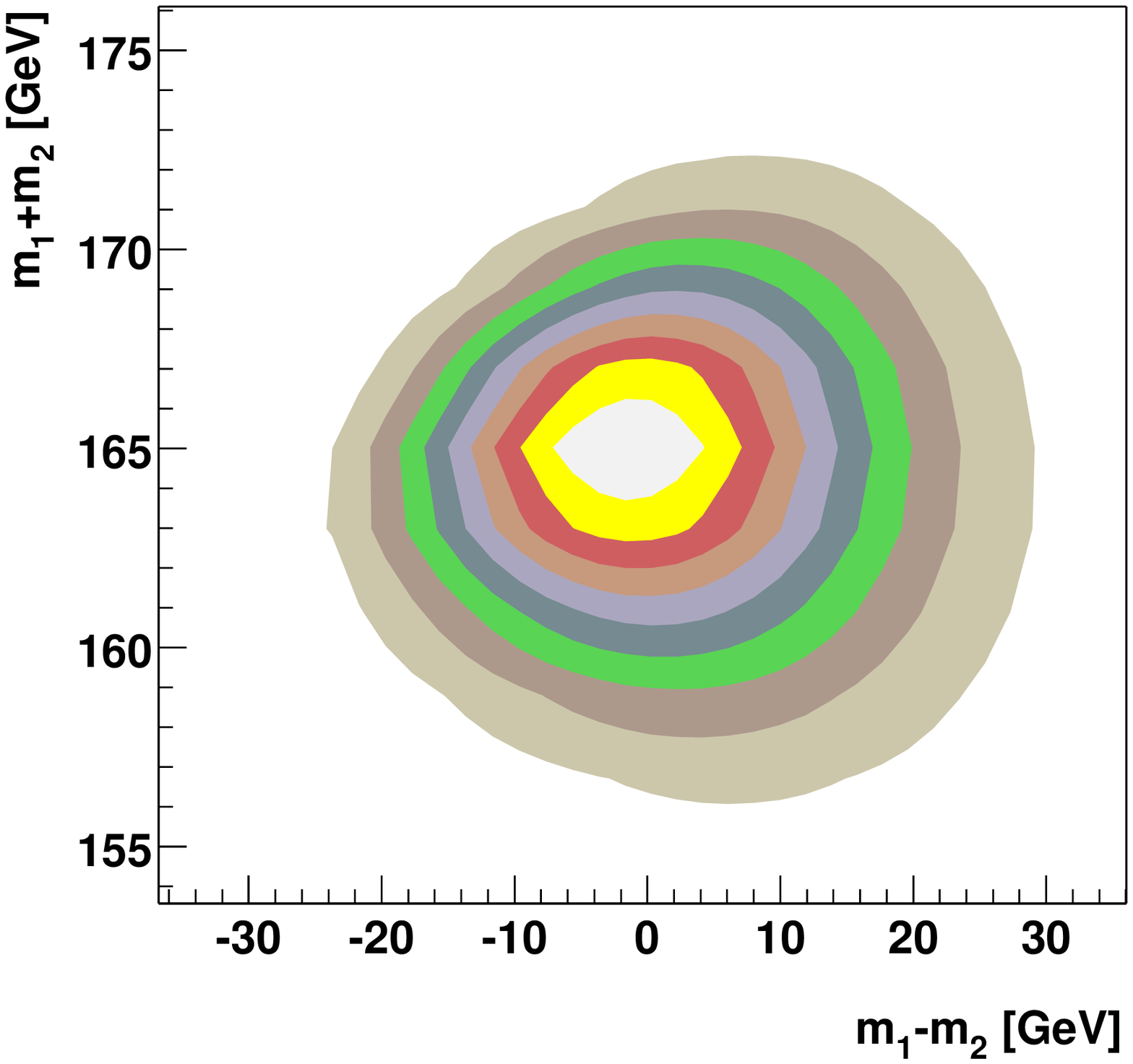,width=5.5cm}
\epsfig{file=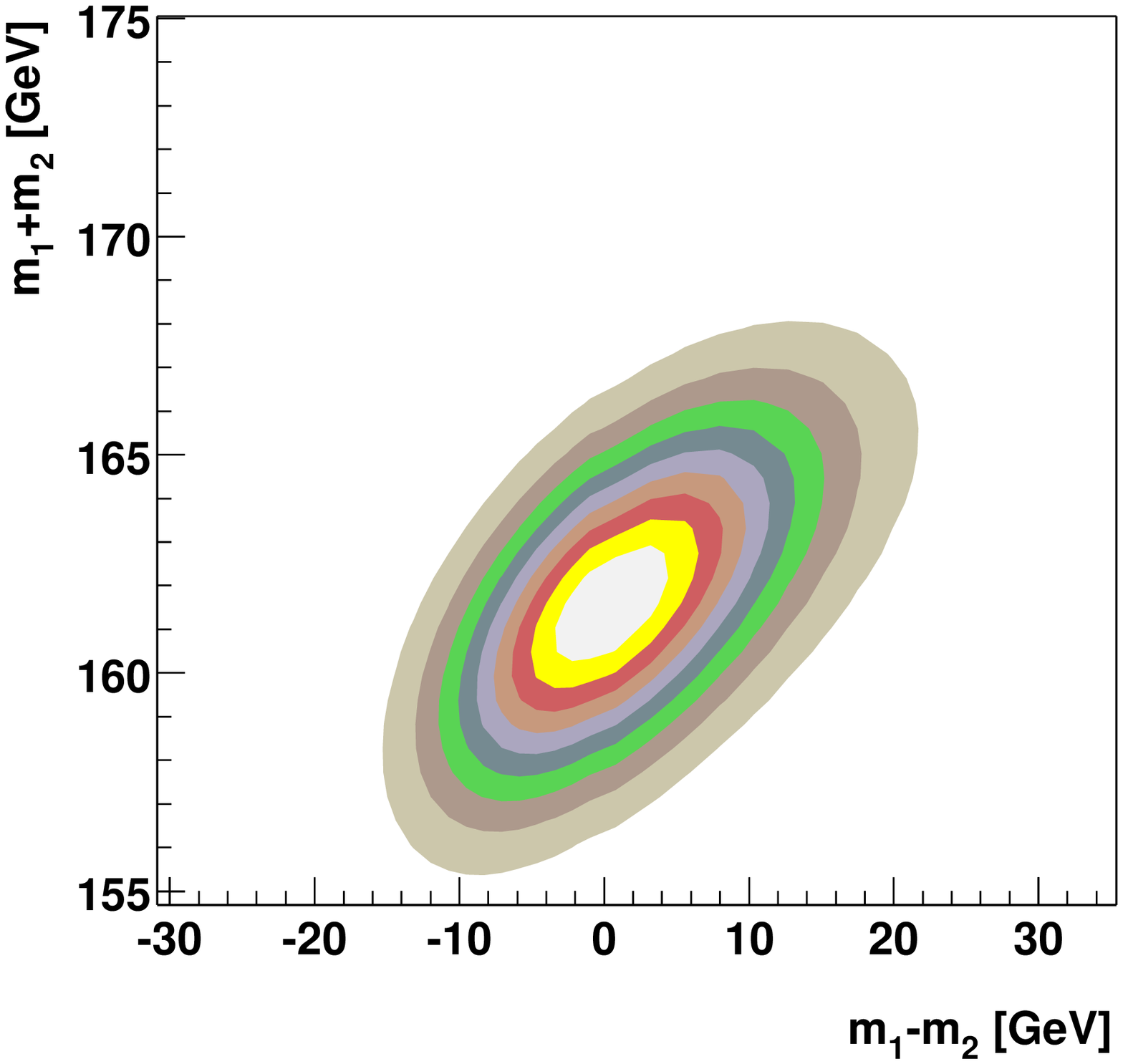,width=5.5cm}
\end{center}
\caption{Two examples of contours of equal  $\chi^2$  for the sum and difference of the two masses 
determined by a scan of 6C fits in semileptonic \W pair decays
 ($m_1$ is the mass of the hadronically decaying \W boson).  \label{fig:conv_chi_2d}}
\end{figure}

The  contours of equal  $\chi^2$  for two events, as a function of the \W masses assumed  in a scan of 6C fits,
 are shown in  Fig. \ref{fig:conv_chi_2d}. Clearly, a Gaussian error (determined by a 4C fit) would not be appropriate for describing
these events.

The problem of finding the correct jet pairing in the four-quark channel (see Sect. \ref{sec:jet_pair})
can be avoided in a two-dimensional convolution analysis. In this case, the 
resolution function, which yields the probability that the observed event originates
from two \W bosons with masses $m_1$ and $m_2$, can be calculated from  the sum of probabilities
for the different jet combinations.
From the three possible combinations in four-jet events (and the
ten  possible combinations in five-jet events), typically, only the right combination contributes significantly to the
sum in the region of $m_1 \approx m_2$. Since  the convolution of the resolution function with the physics
function (a two-dimensional Breit-Wigner function, with a maximum at $m_1=m_2=M_\W$) dominates
this region, and the contribution  to the probability from  wrong combinations is quite  small.

The \delphi collaboration uses  three different jet algorithms (\durham,
\diclus  and \camjet) in the analysis of their $\sqrt{s}=189$ \GeV all-jet data\cite{bib:delphi_ww189}
to select the jets  in a 4C fit.
Events for which the \durham algorithm gives a  value of  $y_{45}>0.002$ (see Section \ref{sec:jet}) 
are  reconstructed  as five-jet instead of four-jet events.    
The resolution function is then determined through a sum over all jet combinations and all jet algorithms,
with relative weights reflecting the probability of correct jet pairing. This probability is
determined from the jet charge and, in the case of five-jet events, from the transverse momentum of
the gluon candidate.
The possibility that an event contains initial-state photon radiation 
along
the beam pipe is considered by repeating the kinematic fit using a modified energy and momentum
constraint:\cite{bib:delphi_ww189}
\[ \sum_i (E,p_x,p_y,p_z)_i = (|p_z^{fit}|,0,0,p_z^{fit}). \]
For 16\% of the events, such a fit favors a solution with significant momentum emitted along the beam pipe
($|p_z^{fit}|/\sigma_{p_z^{fit}}>1.5$). For these events, an additional term,
using  jets from the modified  fit, is then included in the calculation of the resolution function. The relative
weight of this term is based on  the  probability for the ISR hypothesis.
The inclusion of this extra term improves the expected uncertainty on the mass  for these events by  15\%.
\begin{figure}[th]
\begin{center}
\epsfig{file=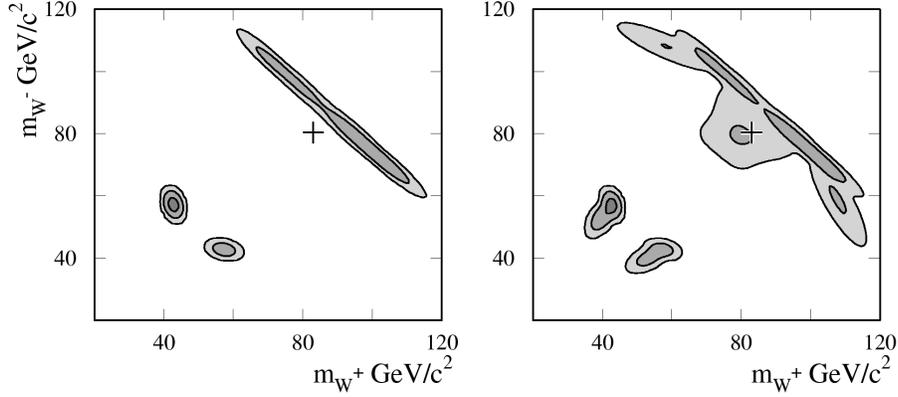,width=12cm}
\end{center}
\caption{Contours of the resolution  function for a four-jet event without (left) and with (right) the 
additional term from the fit of the ISR hypothesis . \label{fig:conv_delhi_2d}}
\end{figure}
Figure  \ref{fig:conv_delhi_2d} shows the contours of the resolution  function for a four-jet event 
with and without the additional ISR term in the fit. 
The diagrams  show the
contours from the two wrong jet combinations at low mass and away from
$m_{\W_+}=m_{\W_-}$, and the contour from the correct solution at $(m_{\W_+}+m_{\W_-})/2 \geq 80$ \GeV,
which has a significant contribution at $m_{\W_+}=m_{\W_-}$. 
Due to the initial-state radiation, the correct jet combination favors mass values above 80 \GeV, while
the inclusion of the additional ISR term gives an additional contribution at $m_{\W_+}=m_{\W_-} \approx 80$
to the physics function.

In a 2-dimensional convolution  analysis of  hadronic \W pairs  
at $\sqrt{s}=189$ \GeV,\cite{bib:opal_189} 
the \opal collaboration  uses a slightly  different approach: All events are reconstructed 
as five-jet events using the
\durham algorithm, independent of their value of $y_{45}$. 
As discussed above, jet combinations with large differences between the two reconstructed masses
do not have a large effect on the probability in a  two-dimensional convolution analysis. 
However, if an event is  reconstructed as a
five jet-event, the jet combination where only the gluon is associated wrongly
can have  contributions to
the resolution function close to the diagonal $m_1 = m_2$, when the gluon energy is sufficiently small.
The gluon jet is most likely to be combined with another jet in the reduction 
of five jets  to four.
All possible  jet pairings that  differ only in the association of these two jets to the \W bosons 
are grouped together. By construction, the group containing the correct jet combination also
contains the combination  where only the gluon is associated to the wrong
W boson.
A special jet-pairing selection was developed using the information from  all the jet combinations
within  a group of jets  to suppress the jet combination where only the gluon is associated incorrectly.\cite{bib:opal_189}
On average, three combinations are selected, which include the correct combination   92\% of
the time.
A good  feature of  treating all events as five-jet events is that it avoids the necessity
of choosing  a value of $y_{45}$  to decide whether an event is treated as a four or a five-jet
event.

All convolution analyses use approximations in their estimation of the physics and resolution functions:
The errors on the input variables are treated as uncorrelated Gaussian errors.
The jet masses are either fixed  in the fit to some initial input values, or they are
calculated  assuming a constant $\beta=p/E$ given by the initial  values of the jet energy and momenta.
In reality, both the 
jet masses and $\beta$ have uncertainties and  complicated correlations with the measured
jet energies. Because of this and other approximations (e.g.,
the complete neglect of ISR in the 183 \GeV \delphi 
analysis\cite{bib:delphi_ww183}), 
biases are expected in the extraction of the \W mass, and these
have to be corrected with the help of some calibration procedure based on
Monte-Carlo events.

In comparison with  other techniques, the convolution method has  the advantage 
of a more complete treatment of the fit errors on an event-by-event basis. The two-dimensional 
convolution analyses  in the four-quark channel also make it possible to include more jet pairings.

\subsection{Ensemble Tests} \label{sec:sample_test}
The different techniques to determine the \W mass 
are calibrated and tested using Monte-Carlo
event samples. The treatment includes  full detector simulation, 
and uses subsamples comparable in size to those in the data. 
The calibration response, which is defined by the generated (input) \W  mass as a function of
the extracted mass, is determined from the analysis of such  Monte Carlo samples generated 
at different \W-boson masses.
\begin{figure}[t]
\begin{center}
\epsfig{file=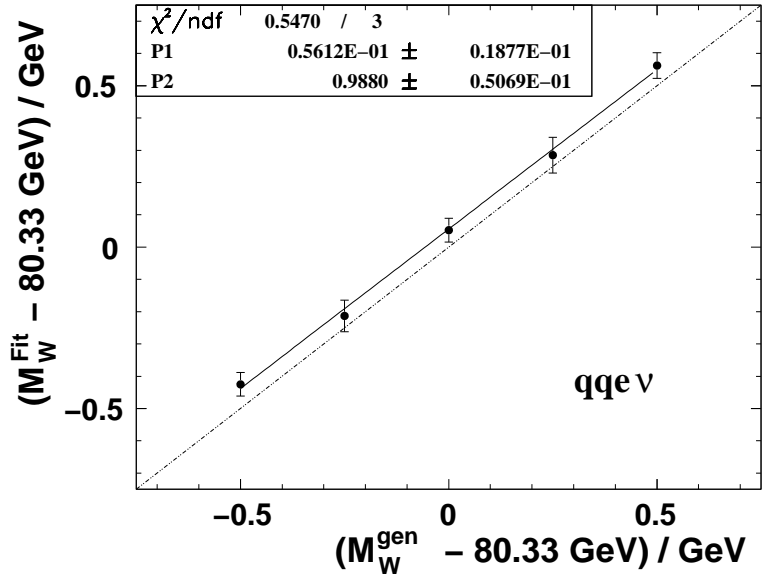,width=10.0cm}
\end{center}
\caption{The extracted mass ($M_\W^{Fit}$) as function of the input mass ($M_\W^{gen}$) 
for the convolution fit
to semileptonic \W-pair decays at $\sqrt{s}=189$
 \GeV from  \opal.\protect\cite{bib:opal_189,bib:dub_phd}
\label{fig:bias_en}}
\end{figure}
As  example, Fig.~\ref{fig:bias_en} shows  the mass correction for the
convolution analysis in the $\WWqqen$ channel ( in fact, this is  used only as a cross check in
the analysis of the \opal data at $\sqrt{s}=189$ \GeV~\cite{bib:opal_189,bib:dub_phd}).
The  masses were determined from  the mean value of the results of
statistically independent  data-sized subsamples generated with a \W-boson mass given
by $M_{\W}^{gen}$. The error bars indicate the uncertainty on the mean value.
Since the points are consistent with a linear dependence, they were fitted to
the function $M_{\W}^{Fit}-80.33~\GeV = P_1+ P_2 \cdot(M_{\W}^{gen}-80.33~\GeV)$. 
(The average value of $M_{\W}^{gen}$ was subtracted 
both from  $M_{\W}^{gen}$ and from  $M_{\W}^{Fit}$.) The result is used to correct the
extracted value ($M_{\W}^{Fit}$) to an unbiased measurement.
 The statistical
error on the extracted mass has  to be scaled  by the inverse slope of the response function
to account for the fact that a change in $\Delta M_{\W}^{Fit}$ 
 corresponds to a change of $ \Delta M_{\W}^{gen}= 1/P_1 \cdot \Delta M_{\W}^{Fit}$. 

In  mass extraction techniques that use a bias correction to
determine the final result, the uncertainty of the correction from limited
Monte Carlo statistics has to be taken  as an additional source of error. In the case of
the reweighting analyses, this kind of bias study  is used only  as a cross check, and if the result
is consistent with no bias, no further correction is required.
Nevertheless, in this case, an error of similar size, from statistical uncertainty
of the reference distributions used for the reweighting, has to be 
taken into account.

\begin{figure}[t]
\begin{center}
\epsfig{file=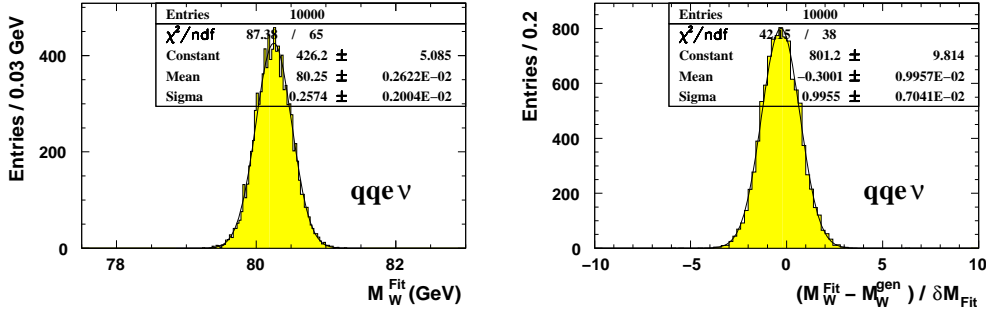,width=13.0cm}
\end{center}
\caption{Example of an ensemble  test for the convolution analysis
of semileptonic \W-pair decays at $\sqrt{s}=189$ 
\GeV.\protect\cite{bib:dub_phd}
\label{fig:pull_em}}
\end{figure}
Monte-Carlo ensembles  can also be used to estimate the expected statistical error, and
to check whether the error returned in  the analysis is a good estimator of statistical
uncertainty.  Figure~\ref{fig:pull_em} shows the distribution of the  result of $10,000$ data-size 
Monte Carlo samples and their  pull distribution $(M_{\W}^{Fit}-M_{\W}^{gen})/\delta M_\W^{Fit}$ from \opal.\cite{bib:dub_phd}
The \rms of the distribution of the extracted mass, or the width of a Gaussian fitted to
the distribution, can be used as an estimate of the expected statistical error 
of the measurement. For data samples with low statistics, the error returned
for different samples often shows large variations. The precision of
the fit fluctuates because different samples contain events with different sensitivity
to the measurement. However, the pull distribution can be used to check whether
the fitted uncertainty is reasonable. When the uncertainty   is  estimated correctly,
the  pull distribution is a  Gaussian centered at zero, with a width of unity. 
A different central value  indicates a bias,  while a larger (or smaller)
width indicates a under- (or over) estimation of  statistical uncertainty on the extracted mass.

The subsamples of events used in  Fig.~\ref{fig:pull_em} were sampled from a Monte-Carlo
pool of events only a factor of 90 larger than the data sample.
Any  given Monte Carlo event is therefore used multiple times
in different subsamples. 
The uncertainty on the \rms of a Gaussian distribution with $N$ uncorrelated entries is given 
by $\rms/\sqrt{2N}$. For 100 subsamples, this would lead to an uncertainty on
the expected error of about 7 \%. This is not sufficiently  precise  to compare the
statistical power of different methods used to determine the \W mass.
Since it is not technically feasible to produce Monte-Carlo samples with enough statistics
for such a comparison, alternative techniques, involving  multiple use of the
same event in different samples, have been tried. 

However, multiple use of the same events affects the statistical precision of the ensemble test, both
in the determination of any possible bias in the \W mass, as well as
in the determination of the expected uncertainty from the \rms of the results.
In addition, using the same events in different subsamples can decrease the
\rms of any result from such correlated  subsamples, and therefore lead to an underestimation of the expected error.
This issue was tested using  a simple ``toy'' Monte-Carlo simulation. 

In such a toy Monte Carlo, the   
parton-level \W masses are generated  using a random number  generator that provides 
a Breit-Wigner distribution. The usual result from some kinematic fit is replaced by the
average of the two \W masses,  and a random term generated with a Gaussian distribution
with  a width of 2 \GeV (a width that corresponds to the resolution of the kinematic fit).
The ensemble test of the toy Monte Carlo is based on subsamples  of 1000 events in which the individual masses 
are determined  from  reweighting analyses. The procedure has a statistical precision that is of the
same order as the mass determination of hadronic or semileptonic data
collected in a single \LEP experiment at a center-of-mass energy of 189 \GeV.
Ensemble tests are  performed using total Monte Carlo samples consisting of $N_{sam}=10,000$ to $100,000$ events, 
from which subsamples of an average of  1000 events are formed by selecting
any given event with a probability of $1000/N_{sam}$.
Since the probability to be selected in any sample is identical for all events, the multiple use of events should  not
introduce a systematic bias to the measurement.
The size of the total pool of  Monte-Carlo events, corresponding to  10 to 100 times the data sample, are typical for 
Monte Carlo studies  with  full detector simulation.
In order to check  the  veracity of the more detailed ensemble studies, the
tests were repeated, but using   500 statistically independent sets of $100,000$ toy Monte-Carlo events,
which were used in different ways to check for the presence of any bias and its expected uncertainty.

\begin{figure}[t]
\begin{center}
\epsfig{file=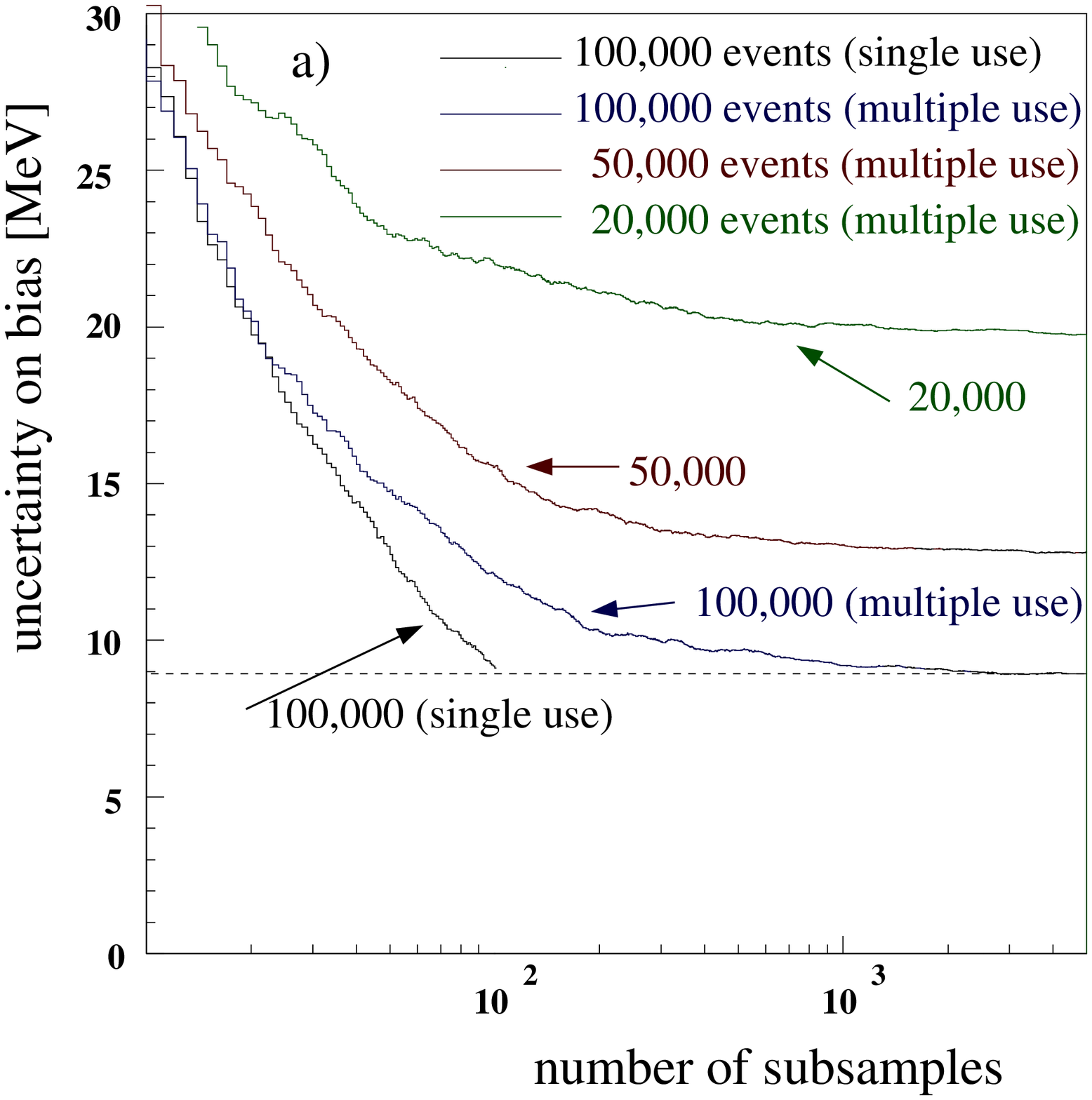,width=6.0cm}
\epsfig{file=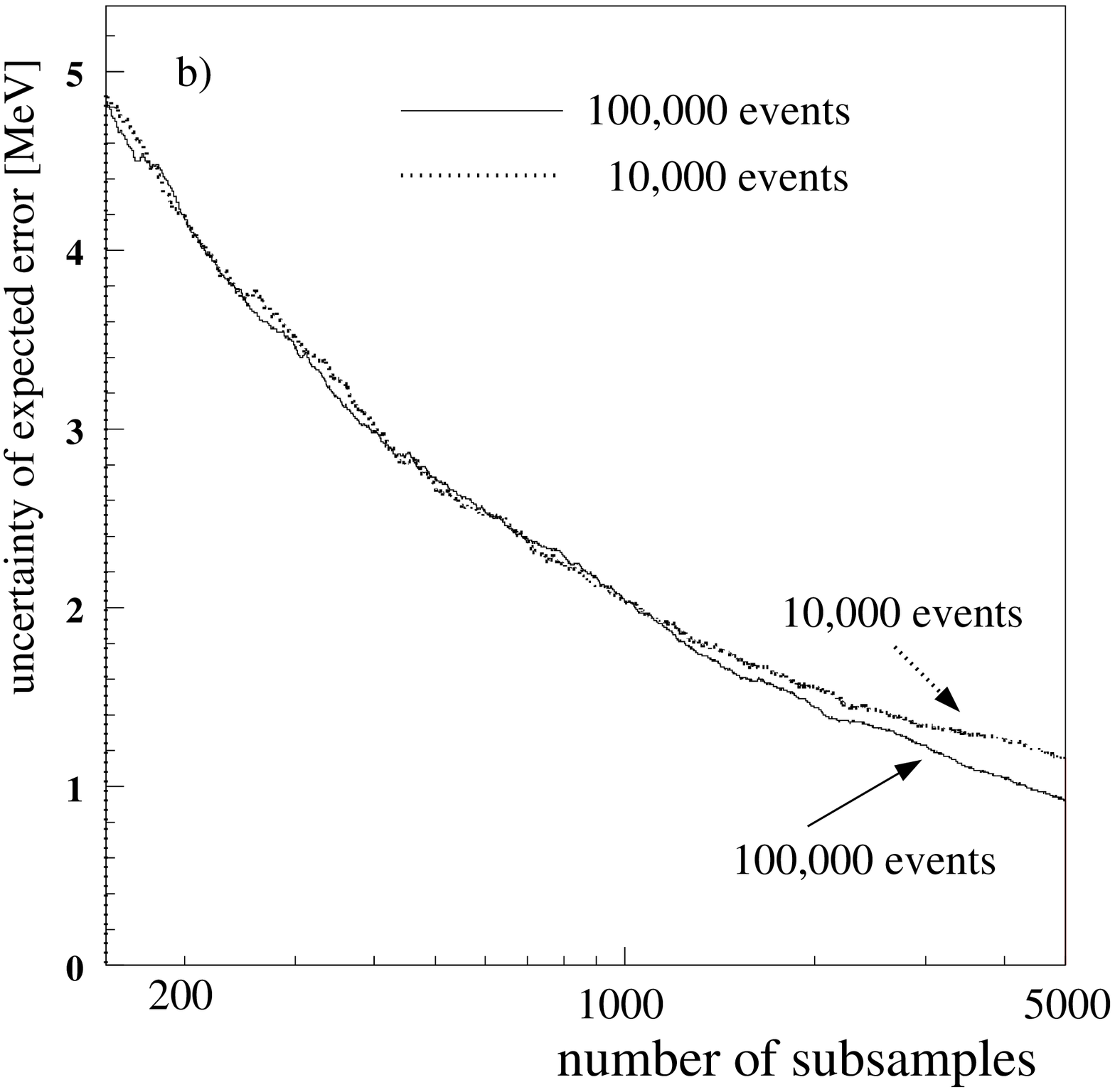,width=6.0cm}
\end{center}
\caption{Test of the multiple use of the same Monte Carlo events in subsamples
of 1000 events based on total Monte-Carlo samples of $10,000$ to $100,000$ events.
a) Statistical uncertainty on the bias determined in the subsample tests. b) Statistical uncertainty on the
expected error on the \W mass. \label{fig:sam_test}}
\end{figure}
Figure~\ref{fig:sam_test}a) shows the \rms of the bias, as determined from the toy ensemble tests,
as a function of the number of subsamples chosen from a fixed total of events.
It is clear  that the \rms decreases as the size of  the Monte Carlo sample increases.
This can be inferred from the drop in the uncertainty as the number of subsamples increases
(especially for the case when no events are ever used more than one time), and in the scaling
of the uncertainties with total number of events, when events are used a large multiple number of times.
But it also seems  that the multiple use of events does not improve 
the ultimate precision achieved on the bias,  compared to when  the events are used just once.
The dotted horizontal line indicates the result for the case when any of the  $100,000$ events
is used only once (this provides a maximum of $100$ independent samples). 
The additional uncertainty from using an event several times appears to decrease, and approach
the statistical power of the total sample, in the limit of  a 
 large  number of subsamples.

Figure~\ref{fig:sam_test}b) shows the reduction in the statistical uncertainty on the  expected
error on the \W mass  with  increasing number
of samples. For more than 1000 subsamples, the decrease is slower when
$10,000$ independent events are used (this is only 10 times the size of a subsample) rather than 
$100,000$ events, but, even for a total sample of $10,000$ events,
statistical uncertainties of about 1 \MeV can still be reached.

\begin{table}
   \tbl{Expected error on the \W mass (in \MeV), calculated 
using 5000 subsamples, for different total-sample and subsample sizes.
     \label{tab:sam_test}}{	
\begin{tabular}{|r|ccc|} \hline
                     & \multicolumn{3}{|c|}{Events per subsample} \\
Total Events in MC   & 1000 & 500 & 250 \\ \hline \hline
$ 50,000,000$  &  89 & 126 & 179 \\
$100,000$    &  88 & 125 & 178 \\
$50,000$      &  88 & 124 & 175 \\
$20,000$       &  87 & 123 & 172 \\
$10,000$       &  84 & 118 & 166  \\ \hline
\end{tabular}
}
\end{table} 

From Figure~\ref{fig:sam_test}b), one might be tempted to conclude that it is possible
to measure the expected error on the \W mass ($ \approx 50$ \MeV ) with a relative precision of a few percent,
even with total Monte-Carlo samples that are only  10 times larger than data.
But, in addition to the statistical uncertainty on the expected 
error, corrections for systematic underestimation of the expected
error from the multiple use of  the same events must also be considered.
This is because the analysis  of the subsamples are not independent, 
and the \rms of any results therefore tends to be smaller than for independent subsamples.
In particular, Table~\ref{tab:sam_test} gives the expected error on the \W mass,
calculated from  $5000$ subsamples, based on the total 
$50,000,000$ independent Monte Carlo events, with  prediction 
for smaller batches of Monte Carlo events and for different number of events per 
subsample.
The impact on the uncertainty becomes  more pronounced as the number of events per subsample
is reduced.
Nevertheless, as seen from Table~\ref{tab:sam_test} and Fig.~\ref{fig:sam_test}b), it is still  possible to 
estimate the expected error on the \W mass with a precision of the order of 1\% using Monte-Carlo
samples of about 100 times the data sample of $\approx 1000$ events.

\section{Systematic Uncertainties}\label{sec:syst}
As described in Section \ref{sec:mw_global}, the determination of the mass of the \W boson 
is based  on a comparison of data with  a  prediction from Monte Carlo. This can be done either
by using any reference distributions (templates)  from the Monte Carlo simulations, or through direct
comparison of the mass  determined from data with that  from Monte-Carlo samples (with bias correction).

As discussed previously, 
the modeling  of  \W-pair production has uncertainties 
due to  contributions from the four-fermion diagrams, initial-state radiation, and general
QED corrections. 
Simulation of the hadronization and of  detector response contributes  major
sources of systematic uncertainties, but
the reliability of many features of the  Monte-Carlo simulation can be checked using the large amount of
data that is available  at center-of-mass energies close to the \Z mass (\Z data),
where the
cross section for $e^+e^- \rightarrow f\bar{f}$ is very large due to the \Z resonance.
Nevertheless, when both \W bosons in $\W^+\W^-$ production decay hadronically, final-state interactions between their
decay products provide  an additional source of systematic uncertainty  that cannot be studied
with  \Z data. These final-state interactions  can be classified
into two groups: Bose-Einstein correlations between identical particles from  different
W bosons, and   color-reconnecting effects  caused by interactions between the decay products
of the two \W bosons that can lead to an the exchange of color.
In addition, from the  assumption of energy conservation, any uncertainty on the beam energy 
causes a corresponding uncertainty in the
reconstructed \W mass. All these issues are discussed below in greater detail.

\subsection{ Uncertainties from QED and Four-Fermion Processes}
The largest QED correction to the reconstructed \W mass is due to initial state radiation (ISR).
As mentioned before, neglecting this effect would result in
a shift of the reconstructed mass of about 0.5 \GeV.
The Monte-Carlo generator \koralw\cite{bib:koralw} describes the ISR up to $\ord (\alpha^3)$. Multiple low-energy or collinear photon
radiation is taken into account through  exponentiation in the leading-log or next-to-leading-log approximation.
The uncertainty due to ISR can be estimated by comparing the \koralw predictions with  predictions
of \excalibur,\cite{bib:excalibur} which uses a different scheme  for implementing the ISR.
The disadvantage of comparing two independent Monte-Carlo predictions is that it requires large 
statistics. To reduce the statistical  uncertainty 
of the comparison to 10\%
of the statistical uncertainty of the data, requires  Monte-Carlo samples with 100 times the size of the data.
A more powerful method involves the comparison of distributions  derived
from the same Monte-Carlo events, but use different event weights. As the distributions are highly 
correlated, these weights are close to unity, and the statistical uncertainty in the difference
between the distributions is therefore greatly reduced. 
The event simulation with \koralw includes weights for each event, and  
it is therefore possible to reweight the distributions so that they correspond effectively to lower-order calculations.
The comparison of the standard \koralw predictions with predictions corresponding to lower
orders in $\alpha$,  in principle,  overestimates  the uncertainty, but the resulting
error is still significantly smaller than the statistical uncertainty from  the comparison
of independent samples. 

In  the complete $\ord (\alpha)$ QED calculation, 
the  two  pairs of fermions from the \W bosons  are not
completely independent because of loop corrections.
 This correction can therefore influence the invariant mass spectrum
of the fermion pairs.\cite{bib:raconww}
The programs \yfsww\cite{bib:yfs} and \raconww\cite{bib:raconww} include these corrections
in the double-pole approximation\cite{bib:double_pole} for the CC03 diagrams.
Distributions for \yfsww can be compared with those from \koralw with the
help of event weights. 
Since the complete $\ord (\alpha)$ QED corrections are only known for the CC03, and not for
the full four-fermion matrix element,
one can either use the difference between the two  estimates from \koralw and
\yfsww  for the CC03 diagrams as a systematic error,
or  as an estimate of the $\ord (\alpha)$ QED corrections,
and  correct the results obtained with the  four-fermion
matrix element by this amount. 
In the latter case, the systematic error has to be estimated either from the difference 
between \yfsww and \raconww, or by using
\yfsww with different options, and  switching  on and off some parts of the higher orders.
Since the double-pole approximation is only applicable for the CC03 diagrams, it is 
not sufficient to use just \yfsww for  generating  the Monte-Carlo samples.

In fact, inital $\W^+\W^-$ analyses were based on  Monte-Carlo samples calculated using 
only the CC03 diagrams.
The systematic uncertainty from the missing interference with other
four-fermion diagrams  was taken
as the full difference between the CC03 and the four-fermion prediction. This 
difference was again estimated from \koralw events, using event weights 
to reweight the four-fermion matrix element to the CC03 matrix element.
Since the effect of completely neglecting the interference with
other four-fermion diagrams is typically $< 30$ \MeV, no systematic error 
from higher-order uncertainties on this interference is assigned in
the current analysis, which is based on the full four-fermion matrix element.

\subsection{Detector Effects}
\begin{figure}[h]
\parbox[t]{6.cm}{
\begin{center}
\epsfig{file=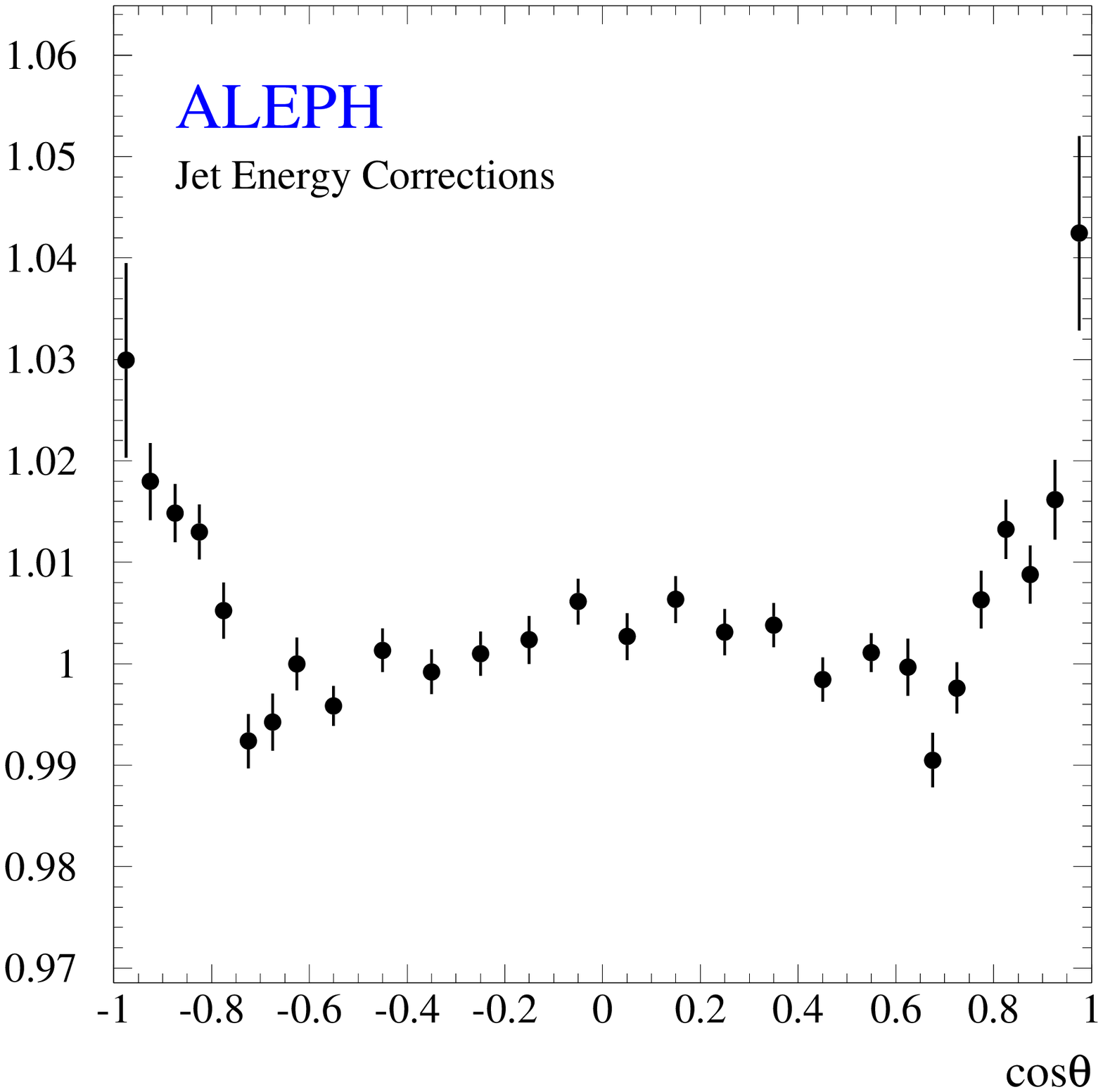,width=5.8cm}
\end{center}
\caption{ 
The double ratio of  \hfill \newline
$(E_{jet}/E_{beam})_{data}/(E_{jet}/E_{beam})_{MC}$ 
derived from a comparison of  1998 \Z data 
with Monte Carlo simulations.\protect\cite{bib:aleph_ww189}
\label{fig:sys_det1}}
}
\hspace*{.5cm}
\parbox[t]{6.cm}{
\begin{center}
\epsfig{file=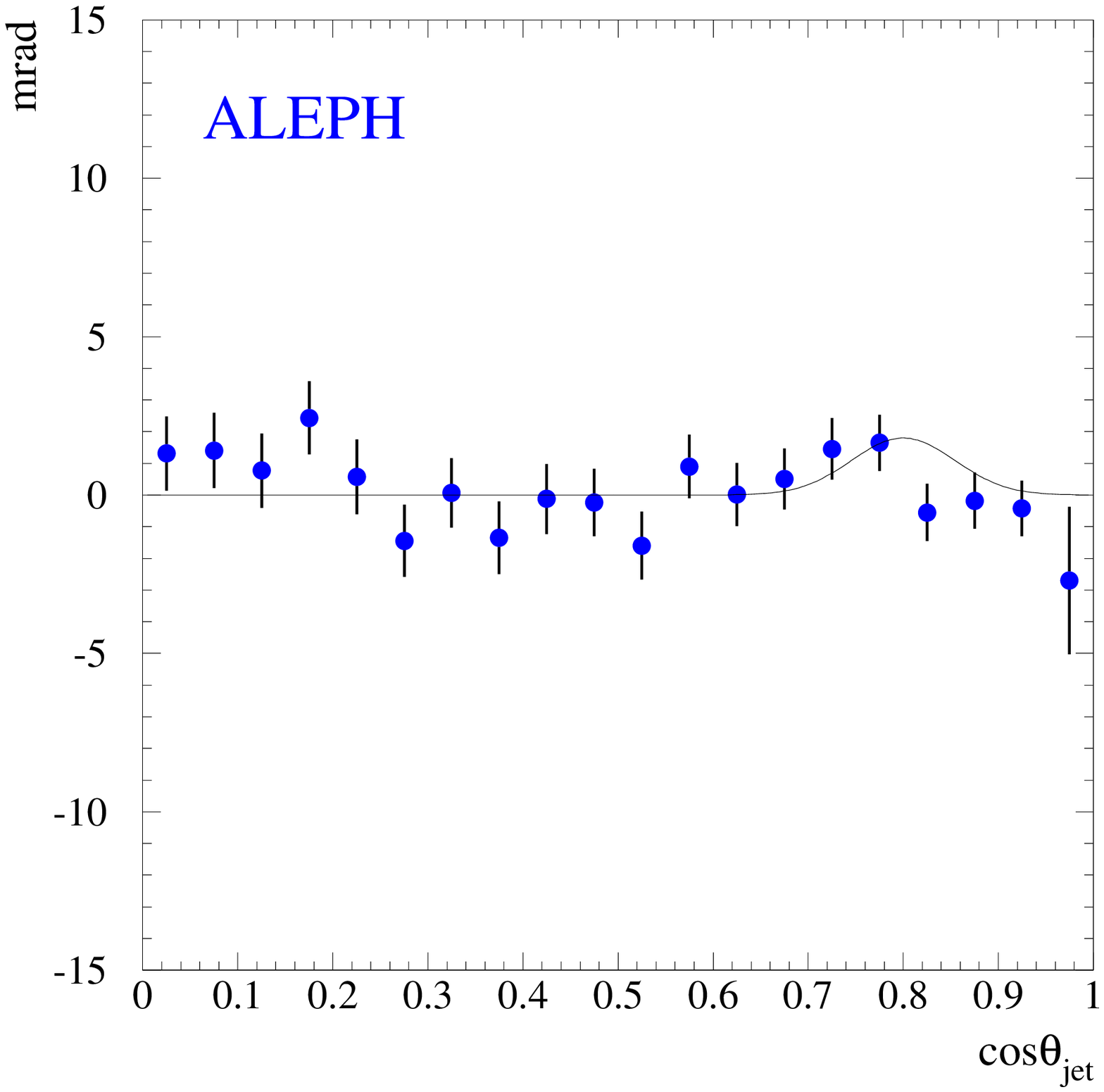,width=5.8cm}
\end{center}
\caption{Difference in the polar direction of 45 \GeV jets between the
measurement with charged tracks and with clusters in the electromagnetic calorimeter.
The points show the results from the 1998 \Z calibration data, and the line shows the
fit to the high statistics 1994 data sample.\protect\cite{bib:aleph_ww189} 
 \label{fig:sys_det2}}
}
\end{figure}
Monte-Carlo simulations used for determining the mass of the \W include a full
\geant-based detector simulation.\cite{bib:geant} Well understood data sets are used to check how well the
jet and lepton energies and directions are simulated in the Monte Carlo. Differences between data and the simulation
are used to calculate correction factors. The uncertainties on these corrections
are used to estimate the remaining systematic error from  detector effects.

During  each year of \LEPII operation, in addition to the high energy data,
the \LEP experiments also  collected   data
at the \Z resonance. Since these data were collected with  exactly the same detector configuration,
it has been used to calibrate the detector response year by year.
At the \Z resonance, decays into two jets  produce events where 
both jets originate from primary partons, which have energies equal to the 
beam energy and  are back to back. Comparing how well the measured and simulated jets fulfill 
this requirement, checks how well the jet energy scale and the  resolutions on jet energy and direction 
are simulated for jets of $45$ \GeV in energy. The statistics of the data are sufficient to perform this study as 
a function of the 
polar angle of the jets. Figure~\ref{fig:sys_det1} shows an example from the \alephe collaboration.\cite{bib:aleph_ww189}
The energy scale for jets  below $45$ \GeV can be checked when \Z decays produce three jets. Applying energy and
momentum conservation, the energies of the three jets can be calculated from their directions
 and the measured jet masses, and compared with the direct measurement
of  jet energy. The precision on  jet energy scale is compromised by  uncertainties 
from hadronization that enter into   the
determination of the jet directions and masses. 
Nevertheless, these comparisons can  be used to estimate a possible energy dependence of the jet-energy scale. 

The jet-energy scale and its dependence on  polar angle 
for energies  energy above $45$ \GeV can be checked using 
two-jet events from  higher-energy  data collected  above the 
\W pair threshold.
The energy scale for leptons, its resolution, and the angular resolution, can be determined in
the same way as for jets, but using two-lepton events. The dependence on  energy can be
checked from events with a lepton pair and a photon. Two-lepton events in the highest-energy data, 
both with and without an additional photon, can be used to determine the
energy scale for leptons for energies above $45$ \GeV. 

The understanding  of  jet and lepton properties can also be checked by comparing 
measurements using different  components of the detector. The jet direction can  be determined either
from charged tracks in the central tracking chamber, or from clusters in the electromagnetic
calorimeter (mainly created by photons from $\pi^0$ decays -- see Fig.~\ref{fig:sys_det2}). 
Electrons can be measured 
in the tracking chamber and in the electromagnetic calorimeter.   
The muon direction can be measured in the tracking device and the outer muon chambers. In addition,
muons produce signals in the calorimeters expected for minimum ionizing particles (MIP) and these which can be used for
further verification.

A systematic uncertainty of $\approx 10$ \MeV can be attributed to detector effects 
for the combined  \W-mass measurement at \LEP.\cite{bib:lep_ww_results}

\subsection{Hadronization}\label{sec:had}
Hadronization is the process in which partons transform into hadrons that form
massive jets observed in the detector. Since energy and momentum are conserved in  hadronization, the invariant
mass of all hadrons from \W decay must equal  the mass of the \W boson,  
independent of the nature of the hadronization process. However, effects from associating particles to the
wrong \W boson and, more importantly, the detector response, depend on details of the hadronization.
The systematic uncertainty on the measurement of the \W  mass is estimated by comparing
different Monte-Carlo simulations and different sets of parameters used in these simulations.
Each Monte Carlo contains phenomenological parameters that are adjusted, such  that the
high statistics data at \LEPI are described as well as possible. 
Unfortunately, none of the
current models provide a set of parameters that describe all aspects of the data simultaneously.

To determine which aspects of  hadronization have the largest influence on the
measurement of the \W mass, it is informative to study the scaled hadronic mass for semileptonic \W pair decays;
  \begin{align} 
 M &  =  \sqrt{(E_1+E_2)^2-(\vec{\vp}_1+\vec{\vp}_2)^2} \cdot E_{beam}/(E_1+E_2)   \nonumber \\ 
   &    = \sqrt{M_1^2+M_2^2 + 2 E_1 E_2 +2 |\vec{\vp}_1| |\vec{\vp}_2| \cos{\theta_{12}}} \cdot E_{beam}/(E_1+E_2). \nonumber 
\end{align}
$E_i$, $\vec{\vp}_i$  and $M_i$ are the energies,  momenta and masses of the individual jets,
$\theta_{12}$ denotes the angle between them, and $E_{beam}$ is the beam  energy. 
The scaling factor $E_{beam}/(E_1+E_2)$ 
approximates the effect of  energy conservation and  the equal-mass constraint used in the kinematic fit. 
It effectively rescales the 4-momenta of the  jets such that they have an energy corresponding to the beam energy.
The effects of hadronization on the scaled hadronic mass are similar to those on the mass from
a kinematic fit, but it provides an easier way to understand the impact
 of  hadronization on the determination of the \W mass. 
From the  above equation, one can see that the scaled hadronic mass depends basically on the three 
quantities $(M_1^2+M_2^2)/(E_1+E_2)^2$, $E_1 E_2/(E_1+E_2)^2$ and $\cos{\theta_{12}}$.
All three  depend on the nature of hadronization, but, as implied above, the effects will
cancel exactly if the scaled hadronic mass is calculated directly from the hadrons simulated
by the Monte-Carlo programs, before considering any detector effects. The effects of 
hadronization on the reconstructed mass can therefore be estimated   better by studying  the
difference of these quantities  before and after implementing  detector simulation. 
\begin{figure}
   \centerline{
    \epsfig{file=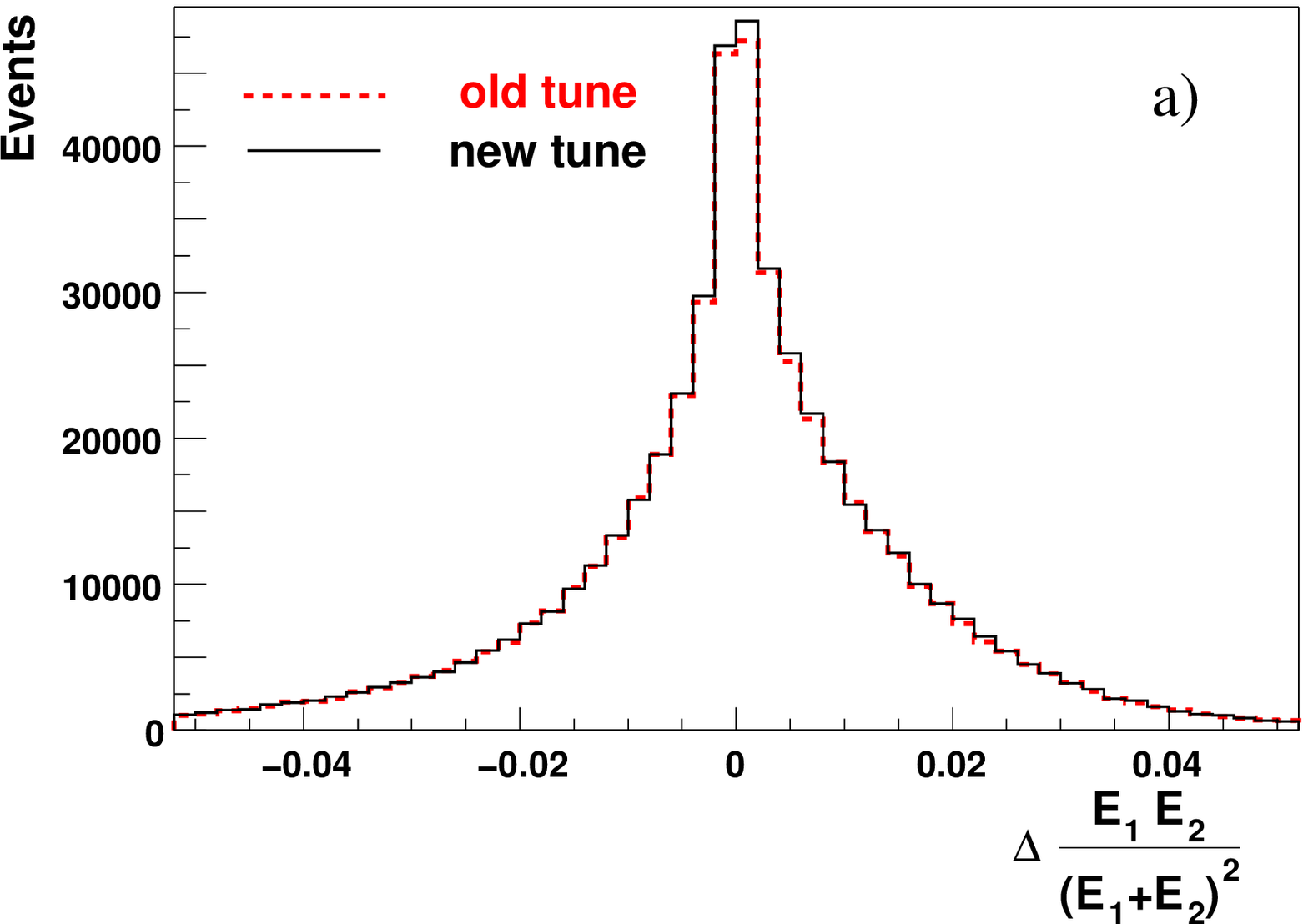,width=6.35cm}
    \epsfig{file=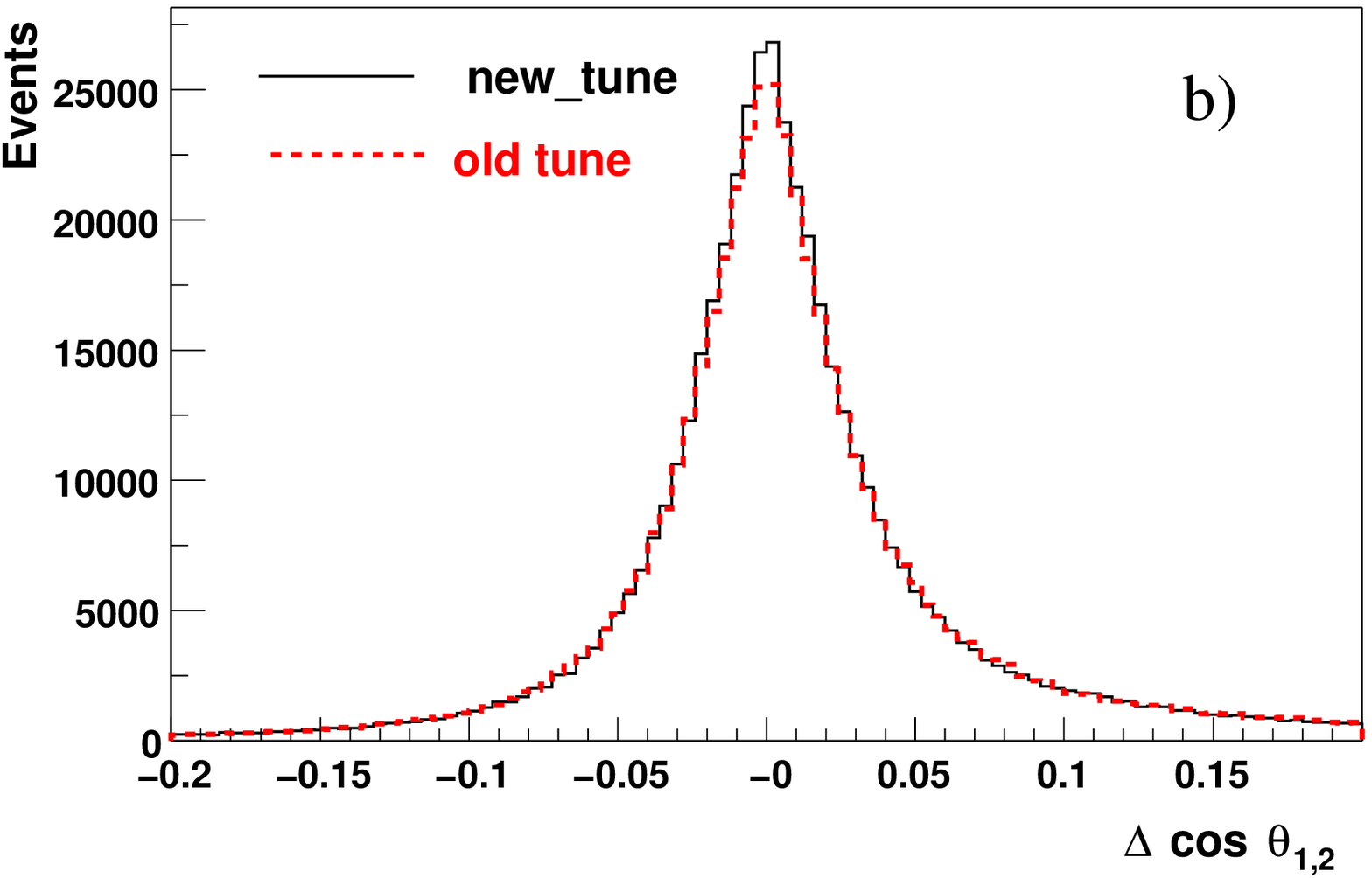,width=6.35cm}}
   \centerline{
    \epsfig{file=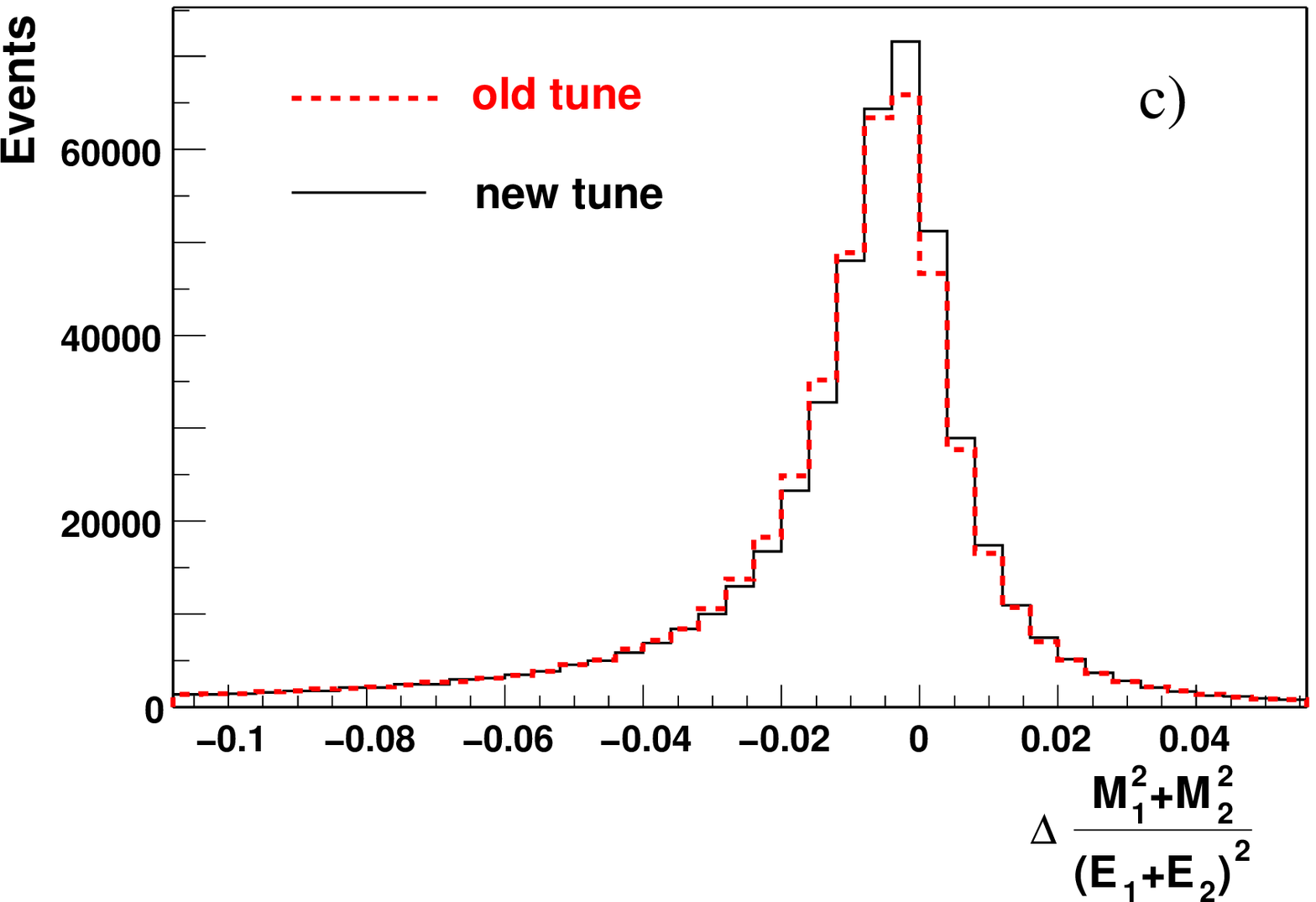,width=6.35cm}
    \epsfig{file=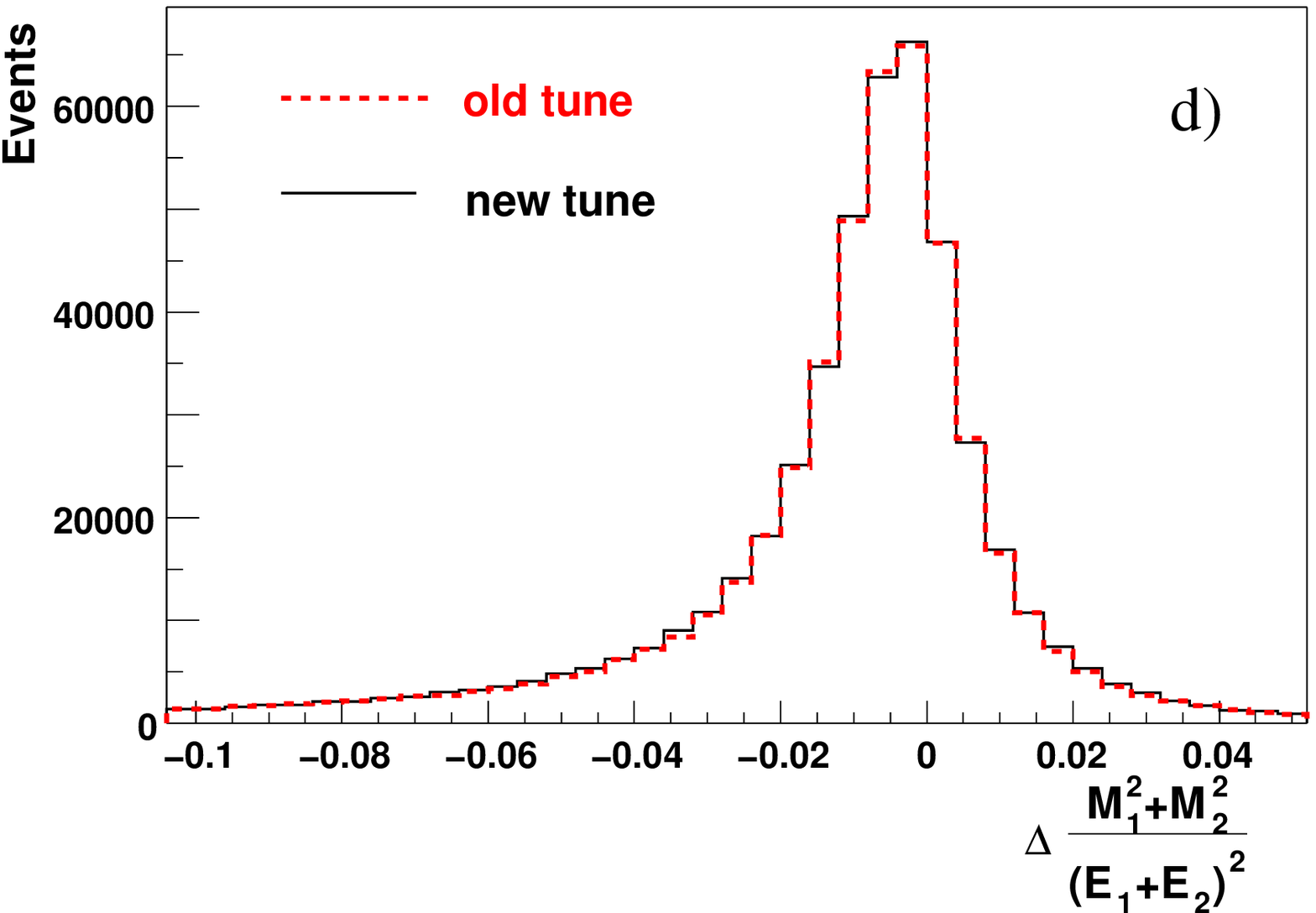,width=6.35cm}}
   \caption{Difference in the indicated variables, measured both without and with  detector simulation.
           In a)-c), the prediction of the new \jetset tune is compared with that of the
           old one. In d), the prediction for the old \jetset tune is reweighted to the
           same baryon and kaon fraction as in the new tune.    \label{fig:had_com}}	
\end{figure}
Figure~\ref{fig:had_com}a)-c) shows these differences for two sets of Monte-Carlo parameters 
in \jetset,\cite{bib:phytia} as tuned by the \opal collaboration. In the initial tuning,\cite{bib:opal_jtold}
the Monte-Carlo parameters  were  optimized primarily to provide a good description of inclusive quantities described by 
event shape variables such as thrust, differential jet rates, and jet masses. 
The more recent tuning  includes  exclusive quantities such as 
particle fractions and their energy spectra.\cite{bib:opal_jtnew} 
The differences between these two versions of \jetset provide a  difference in
the reconstructed \W mass of 40-50 \MeV, depending on the exact method of analysis.    
The greatest  difference is in $(M_1^2+M_2^2)/(E_1+E_2)^2$,
with a shift of 0.0006 in the mean value, corresponding to a mass shift of 40 \MeV. The shifts
observed in the two other variables correspond to mass shifts of $\leq 10$ \MeV in the \W mass.

The two  Monte-Carlo tunes predict  significantly different detector corrections to jet masses.
This can be attributed to their  different baryon and kaon fractions. 
Events generated with the old tune contain on average 30\% more baryons and 10\% more charged kaons
and $K_L^0$-mesons.
As described in Section~\ref{sec:jet}, jets are formed from charged tracks and from clusters in the
electromagnetic and hadronic calorimeter. A pion mass is assumed for tracks, while a mass of
zero is assumed for the pseudo-particles formed from  calorimeter clusters. Neglecting
the true particle masses, influences both the jet mass and energy. The effects on  jet energy are
 corrected on average  by the jet-energy calibration based on \Z  data, 
and the impact on the reconstructed \W mass is  small due to the
energy constraint in the kinematic fit.  
However, the effect on  jet mass cannot be obtained from \Z data. This is because jets from the 
\Z decay are back to back, and it is  possible to determine the imbalance in energy and  momentum  of
the two jets assuming only energy and momentum conservation, but  not  whether
the momentum scale is wrong. The latter can be determined for  systems with known energy and momentum,
but only when the total momentum does not vanish (e.g., the \Z in $Z\gamma$ event with photons measured  
in the detector).

\begin{figure}
   \centerline{
    \epsfig{file=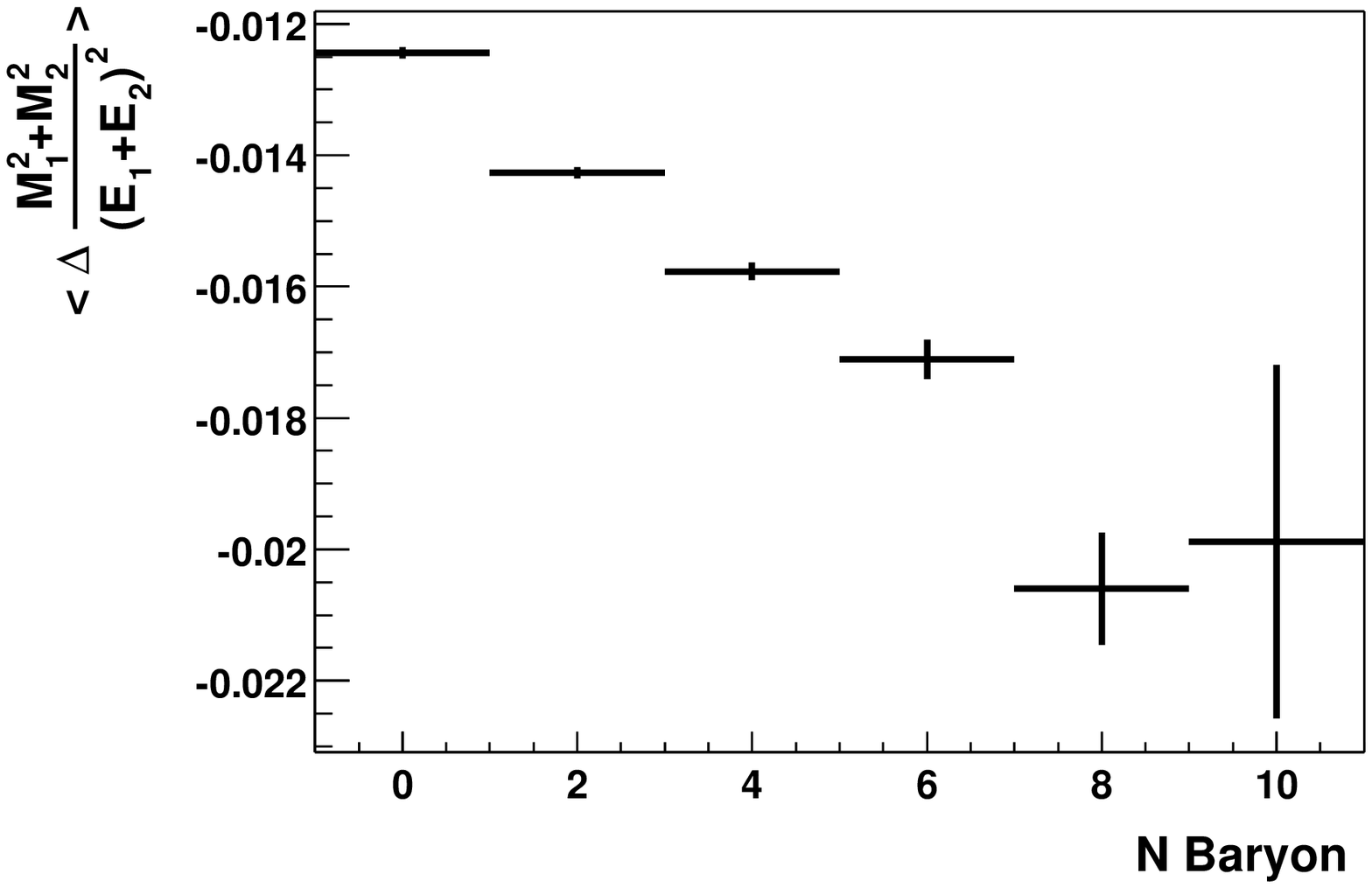,width=6.8cm}
    \epsfig{file=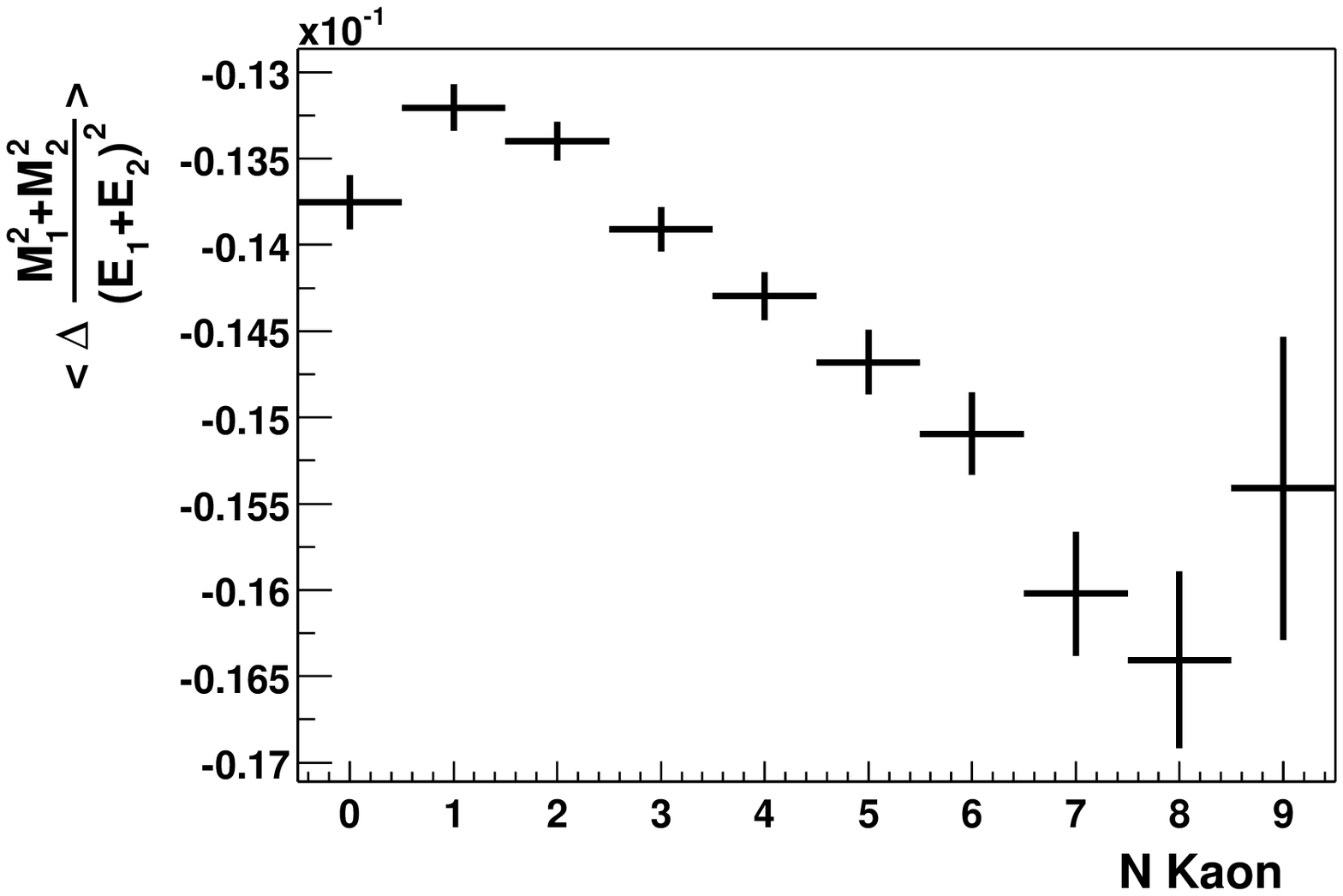,width=6.8cm}}
   \caption{Average difference between $(M_1^2+M_2^2)/(E_1+E_2)^2$ without and with  detector simulation
            for semileptonic \W decays, as function of the number of baryons (left), or charged kaons and
            $K_L^0$-mesons (right), in the event.    \label{fig:had_bias}}	
\end{figure}
Figure~\ref{fig:had_bias} shows the  difference between mean value of$(M_1^2+M_2^2)/(E_1+E_2)^2$, without and 
with detector simulation, for semileptonic \W decays, as a function of the number of baryons or 
charged kaons and $K_L^0$ in the event. 
Clearly, one must conclude that
the detector bias on  jet mass depends
on the number of baryons and kaons in an event.
In order to check  whether the entire effect can be attributed to this difference, the  distribution in
 $(M_1^2+M_2^2)/(E_1+E_2)^2$ 
for the old \jetset tune was re-calculated using event weights, creating the same baryon and kaon 
multiplicity as in the new tune. This reweighted distribution is compared to 
the distribution of the new tune in Fig~\ref{fig:had_com}d), and indicates that
both Monte Carlos now show the same shift in the jet mass
when they  have the same baryon and kaon fractions.

When  the reweighting technique is used to determine the \W  mass from a Monte Carlo sample
with adjusted baryon and kaon fractions, 
the correlation of  these fractions with quantities that should not be affected by the reweighting
(e.g., properties of the \W at parton level) have to be properly taken into account.
\begin{figure}
   \centerline{
    \epsfig{file=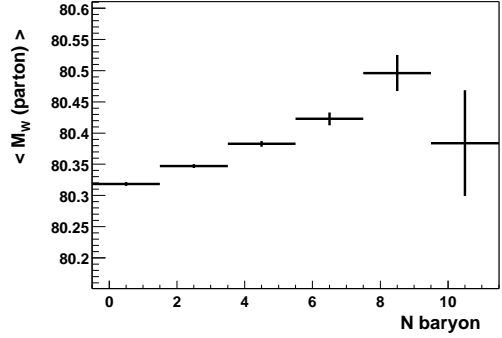,width=8.2cm}}
   \caption{Average \W mass at parton level,
            for semileptonic \W decays, as a function of the number of baryons  in the event.    \label{fig:had_bias_mw}}	
\end{figure}
Figure \ref{fig:had_bias_mw}, which shows the average mass of the \W boson at the parton level as function of
the number of baryons per event, indicates the correlation between the two quantities. In events with a higher \W mass,
the production of baryons and kaons is more likely due to the larger phase space in the decay.
To avoid a bias from such correlations, the weight factors affecting  the average number of baryons and kaons
in an event are normalized separately in bins of  \W mass and energy at the parton level. This normalization
 guarantees that the average
weight for events in  each bin of  \W  mass and energy  is the same, and that the parton mass
and energy distributions are therefore not changed by  kaon and baryon reweighting. 

\begin{table}
   \tbl{Shifts in \W mass for  different Monte Carlo models and tunes, without and with
    correction for  different baryon, charged kaon, and  $K_L^0$ fractions. \label{tab:had_fit}}{	
\begin{tabular}{|c|cc|cc|} \hline
 & w/o    &  with  & Change in      &  Change in  \\
 &  rew.             &   rew.       &  Baryon frac.   & Kaon frac.   \\ \hline
JT(old) - JT(new)                 &  $-50\pm10$  & $+3\pm10$ &  1.32 &  1.10 \\
\herwig - JT(new)                  &  $-15\pm10$  & $-6\pm10$ &  0.92 &  1.08 \\
\ariadne 4.09 - JT(new)            &  $-27\pm10$  & $+3\pm10$ &  1.23 &  1.03 \\
\ariadne 4.11 - JT(new)            &  $-17\pm10$  & $-4\pm10$ &  1.09 &  1.01 \\
reweight  JT(new)                 &         & $-8\pm10$ &  1.10 &  1.00 \\
reweight  JT(new)                 &         & $-10\pm10$&  1.00 &  1.05 \\ \hline
\end{tabular}
}
\end{table} 
The  uncertainty on hadronization can be estimated using data-sized subsamples in 
ensembles of the kind described in Section~\ref{sec:sample_test}, 
and different Monte-Carlo generators. 
The results shown in Table~\ref{tab:had_fit} are based on Monte-Carlo samples  from $10^6$ $W^+W^-$ pairs.
The mass is extracted for semileptonic \W-pair decays using a reweighting analysis. The reference distributions
are  determined from the new tune of \jetset by \opal,\cite{bib:opal_jtnew}  while the  data-sized 
subsamples are taken for the
old  tune, a \herwig tune, and two different \ariadne tunes. 
The Table shows   the results both with
and without correcting the Monte-Carlo models for  baryon and kaon multiplicity, and  the ratio of 
these multiplicities  to the reference values  in the new \jetset tune. 
The two last  lines show the effect  of changing either the baryon or kaon fraction in the new  tune.
It can be seen from the Table that the differences between the Monte-Carlo simulations can be accommodated
by their different baryon and kaon fractions.  
The uncertainty from   multiplicity is estimated  
 to a precision of $\approx 10$ \% and  $\approx 5$ \% for baryons and kaons, respectively, and 
the remaining uncertainty due to the hadronization is smaller than 10 \MeV.

\subsection{Mixed Lorentz-Boosted \Z Method}
In the estimation of systematic errors  from detector effects and
hadronization, many issues, such as the  uncertainty on
energy and direction of  jets and leptons, and the difference between 
hadronization models, are considered separately. Since the uncertainty on most of these contributions is 
limited by Monte Carlo statistics, adding the uncertainties in quadrature leads to a total error that
could be reduced
if such  effects could be considered simultaneously. This would also have the advantage 
of gaining a better  treatment of  the correlation between  errors. The ansatz of the
mixed Lorentz-boosted \Z (MLBZ) method\cite{bib:mblz} is based on a treatment of \Z events
in a way that would provide a  direct  estimate
of the  uncertainty in the measured \W  mass from detector-resolution 
and hadronization  effects, purely through a comparison of  data with simulated events.

Events with  features similar to hadronic \W-pair decays are constructed from two \Z events.
First, the 4-momenta of all measured particles in a \Z event are Lorentz boosted in one
direction by the Lorentz factor $\gamma = E_{b,W}/m_\W$, where  $\gamma$ is the average boost 
expected for a \W boson with a mass of $m_\W=80.35$ \GeV/c$^2$ at a beam energy  $E_{b,W}$.
The boosted event is then combined with a second event treated in exactly the same way, except that the
Lorentz boost is performed in the opposite direction.
In order to optimize the available statistics, each \Z event is  combined with several
other \Z events. The correlations introduced between such  events due to  multiple
use of the same  \Z event are taken into account in the estimate of statistical uncertainty
of the method. The newly created events are then passed through a standard mass analysis, assuming a 
center-of-mass energy $\sqrt{s}= 2 \gamma E_{b,Z}$ in the constrained fit, where $E_{b,Z}$ is the
beam energy for the \Z data  ($2 E_{b,Z} \approx M_{\Z}$).   
Since the technique used to determine the mass is largely invariant under a 
rescaling of  all  masses and energies, 
any shift in the mass of MLBZ-events  determined from simulated \Z events  ($M_{MLBZ}^{sim}$)
compared to the mass determined from \Z-data  ($M_{MLBZ}^{data}$) provides a measure of
the shift expected in the  measurement of the mass of the \W boson:
\[ \Delta M_\W^{MLBZ} = ( M_{MLBZ}^{data} - M_{MLBZ}^{sim}) \cdot \frac{M_\W}{M_\Z} \]
$\Delta M_\W^{MLBZ}$  therefore  estimates  the systematic errors on the \W mass for an imperfect 
 detector-simulation   and effercts from an imperfect simulation of the hadronization. 
The two most important biases in the reconstruction of the mass in MLBZ events are:
\begin{itemize}
\item A positive shift of the mass due to ISR photons that  are lost along the beam pipe and 
not taken into account in the kinematic fit. 
\item A negative bias due to the smearing of the  direction of the momentum vectors of the jets from 
\Z decay. The true direction of the 
momenta are always back to back, and any smearing of the reconstructed direction will always lead to a 
reduced opening angle between the two jets and therefore to a reduced invariant mass.
\end{itemize}
These biases can be canceled out  by comparing the reconstructed mass from MLBZ-events calculated form 
\Z-data and  from simulated \Z events.

Since the events are correlated through multiple use in  combinations of two \Z events
in one MLBZ event, care has to be taken in the estimation of the statistical accuracy.
Both the \rms of statistically independent samples and the ``Jackknife'' method,\cite{bib:jackknife}
have been used to estimate  the statistical precision of the method as  
about $300/\sqrt{n_{\Z}}$ \MeV for a \Z sample with $n_{\Z}$ events. 
For example, this corresponds to an uncertainty of about 3 \MeV for the 1997 \delphi data.\cite{bib:mblz}

There are several limitations to the MLBZ method, mainly due to 
the difference between W and \Z decays,
the different event topologies of \Z and \W-pair decays, 
and the fact that  two seperate events are mixed. In particular,
\begin{itemize} 
\item MLBZ events cannot be used to study any bias from initial-state radiation, since there is far less ISR 
 at the \Z pole.
\item Effects of final state interactions (FSI) between decay products of the two
W bosons cannot be examined.
\item The results from MLBZ events might be influenced by different flavor composition of
 \W and \Z decays. The influence of b-quark decays can be estimated by studying the mass bias
as a function of a b-tag probability in \Z events (due to the relatively long life time of
B hadrons, the fraction of b quarks in a sample can be changed by changing criteria  on  tracks
to originate from the primary vertex).
\item 
The MLBZ method is sensitive  mainly to a jet energy scale of 45 \GeV.
\item 
Effects from detector response on  particle density, such as  double-track resolution, etc., 
can only be examined  if the particles originate from the same \W or \Z. 
\item The back-to-back geometry often leads often to an over estimation  of  uncertainties
 that are sensitive to this symmetry, such as  acceptance near the beam pipe, or at the  edges of the central
and forward detectors.
\item The back-to-back topology of  \Z decays also reduces sharply the sensitivity 
to single-jet mass effects that are important for understanding the  uncertainty in hadronization.
Qualitatively, momentum and energy conservation in \Z events can
be used  only to determine the energy scale, because rescaling of the momentum would not influence the
constraint that the total momentum should be zero.
This  can be examined more quantitatively if one approximates the mass from the
constrained fit by  the scaled hadronic mass: 
\[ m_{scale} = m_{inv} \cdot \frac{E_{cm}/2}{E_1+E_2} = \sqrt{(E_1+E_2) - (\vec{p_1}+\vec{p_2})} \cdot \frac{E_{cm}/2}{E_1+E_2}. \]
where $E_i$ and $\vec{p_i}$ are the energies and momenta of the two jets forming the \W or \Z boson, and $E_{cm}$ is
the center-of-mass energy in the system used for the constrained fit. (In the approximation that both
\W bosons from a \W-pair  have the same mass, $E_{cm}/2$  equals the  \W  energy.)
When the change in the center-of-mass energy is taken into account, 
this quantity is nearly invariant under a Lorentz boost, because  both $E_{cm}$ and  $(E_1+E_2)$ 
are increased by the same $\gamma$ factor
 (neglecting the fact that  $|\vec{p_1}+\vec{p_2}| \ne 0$, because of detector resolution effects),
and $m_{inv}$ is therefore invariant under  Lorentz transformations.
We can consequently estimate the bias on $m_{scale}$ prior to implementing  the Lorentz boost. 
We can parametrize the difference of the measured jet momenta relative to those  in a perfect detector as
$\vec{p_1}+\vec{p_2}-(\vec{p_1}^{true}+\vec{p_2}^{true})=\Delta \vec{p}_{res} + \Delta \vec{p}_{M}$, where 
$\Delta \vec{p}_{res}$ is due to the resolution of the momentum measurement, and $\Delta \vec{p}_{M}$ is due
to a bias in the reconstruction of the  mass of single jets, and the biases in $|p/E|$  of the jets.
Using this parametrization, the average bias from the momentum measurement  is:
\begin{align}
 < m_{scale}^2-m_{true}^2 >  = & < (\vec{p_1}^{true}+\vec{p_2}^{true})^2-(\vec{p_1}+\vec{p_2})^2 >  \nonumber \\
                             = &  -<(\vec{p_1}^{true}+\vec{p_2}^{true}) \cdot  \Delta \vec{p}_{M}>-  \nonumber \\ 
                               &      <(\vec{p_1}^{true}+\vec{p_2}^{true}+  
                                  \vec{p}_{M}) \cdot \Delta \vec{p}_{res}>- \nonumber  \\
                               &
                                <( |\Delta \vec{p}_{M}|^2 >  -  <( |\Delta \vec{p}_{res}|^2 >.  \nonumber
\end{align}
The last term is the   bias of the scaled mass towards lower values due to momentum resolution.
In the case of the \Z boson, the first term vanishes because of momentum conservation 
($\vec{p_1^{true}}+\vec{p_2}^{true}=0$). For \W pair events, $\vec{p_1^{true}}+\vec{p_2}^{true}$ is different from 0.
Furthermore, $ \Delta \vec{p}_{M}$ is due to a mismeasurement of the size of the momenta, and its direction
is therefore strongly correlated with the directions of $\vec{p_1}^{true}$ and $\vec{p_2}^{true}$, leading to a 
non vanishing contribution to the bias in the \W  mass from a bias in  individual jet masses. 
\end{itemize}
The MLBZ method provides  an interesting cross check for the estimation of uncertainty from detector
effects  and hadronization.
But, in order to fully exploit it in the evaluation of these errors, all above-mentioned points have
to be  taken into full account. 

\subsection{Beam Energy}
The kinematic fit constrains the total energy of the \W pair to a center-of-mass energy  assumed to be
twice the beam energy. If the  beam energy assumed in the
kinematic fit is larger then the true beam energy  by a factor 
$(E_{beam}+\Delta E_{beam})/E_{beam}$, all jet and lepton energies
reconstructed by the kinematic fit will be larger  by this factor relative to 
a kinematic fit using the correct beam energy. The reconstructed \W mass is therefore also
larger by this factor, leading to a shift on the \W mass of:
\[ \Delta M_\W = \frac{M_\W}{E_{beam}} \Delta E_{beam} \]

The beam energies were  determined  by the \LEP Energy Working Group with
a precision of 20 to 25 \MeV,\cite{bib:lep_ebeam} as described below.
  
\subsubsection{Determination of the \LEP Beam Energy}
By far, the  most precise method for determining the beam energy at \LEP is based on resonant 
depolarization.\cite{bib:res_dip} 
At \LEPI energies, the precision in the beam energy using this
method is $< 1$ \MeV. 
This method is based on determining the 
spin tune, $\nu$, which is proportional to the beam energy $E_{beam}$:\cite{bib:lep_ebeam}
\[ \nu = \frac{(g_e-2)}{2} \frac{E_{beam}}{m_e c^2} \]
where $(g_e-2)$ is the anomalous magnetic moment of the electron, and $m_e$ is the electron mass. 
Depolarization effects increase sharply with  beam energy, leading to an insufficient
build up of transverse polarization at center-of-mass energies used in runs at
\LEPII. The measurements of beam energy  at \LEPII are therefore based on an estimate of the total integrated
magnetic field ($B$) along the beam trajectory, which is proportional to the beam energy:\cite{bib:lep_ebeam}
\[ E_{beam} = \frac{e}{2\pi c} \int_{\LEP} B dl \]
The integrated field is estimated from the continuous measurement of 
16 NMR probes situated
in several  of  the 3200 \LEP  bending dipoles. Each of the NMR probes  samples the 
field in only a small region of the magnet.

Energy measurements based on NMR probes can be calibrated in the range of 40-55 \GeV
through the  precise  measurement of beam energy using resonant depolarization. 
With this calibration, the beam energy can be determined  with each of the probes.
These measurements, taken  over all runs in a given year, 
yield  spreads among  the different NMR probes with an \rms of $\sigma_{NMR} =$ 30 to 50 \MeV, 
leading to a contribution to the systematic uncertainty in absolute energy  of 
$\sigma_{NMR}/\sqrt{16}=$ 8 to 13 \MeV.

The extrapolation of the calibrated energy from the  the 40 \GeV to 55 \GeV range
to center-of-mass energies used for physics runs was checked using
flux-loop measurements, as follows. 
Each dipole magnet is instrumented with a flux loop, which can be used to measure the change 
in magnetic field as the magnet is ramped up. The measurements are performed  during dedicated magnet cycles,
 outside of the normal runs.
 The flux loops sample 98\% of
the field in each of the main bending dipoles, excluding fringe fields at the ends of
the magnets. This corresponds to 96\% of the total bending field, because certain special
magnets are not instrumented with such flux loops.
The changes in  magnetic field are calculated and compared with
the NMR measurements, but are not used to correct the
results from NMR. The difference between the two measurements is used only to estimate
a systematic uncertainty  from the extrapolation of the NMR measurements to the energies of interest.
This uncertainty was 20 \MeV for the year 1997, and 15 \MeV for the following years.

For the year 2000, another important source of systematic uncertainty had 
to be investigated  because of the impementation of  
bending-field spreading (BFS). For any fixed RF acceleration voltage, the
 power loss from synchrotron radiation
depends on  the structure of the bending field. During t 2000,  approximately 100 
previously unconnected corrector magnets were powered in order to exploit this dependence
for boosting the beam energy by 0.18 \GeV.
The increase in energy from the BFS was calibrated in test runs with the help of a
beam spectrometer,\cite{bib:beam_sectro} by testing how much the RF frequency had to be adjusted in order to achieve
the same  bending  without BFS as with the BFS.    
The systematic uncertainty of this calibration was estimated to be 13 \MeV.
The beam spectrometer was used to obtain the bending angle in a laminated steel dipole magnet
using six beam-position monitors, each of which measured the charge induced by the 
beam on four electrodes. The beam position was determined from the
relative sizes of the induced charges, which were  measured with a precision of
$2 \cdot 10^{-5}$ to achieve a required  precision in position of  $1\mu m$. 

The total uncertainty  on the  beam energy estimated by the \LEP Energy
Working Group\cite{bib:lep_ebeam} for  individual years is between 20 and 25 \MeV, with a year-to-year 
correlation of 82\%.

\subsubsection{Beam-Energy Measurement using Radiative  Fermion-Pairs}
The cross section for  radiative fermion-pair events $e^+e^- \rightarrow f\bar{f} \gamma$
has a maximum when  the invariant masses of the fermion pair equals the
\Z mass. 
When the di-fermion mass is reconstructed using the constraint  
of  energy conservation,  its value depends on the ratio of  assumed
to  true beam energy. Using the known \Z mass, the reconstructed mass  can therefore also be used to determine
the beam energy. 
This kind of  analysis uses  techniques similar to those used in the determination of the mass of the \W, and can be inverted to 
provide  an additional cross check on the systematics from  hadronization
and detector uncertainties.
The masses of leptonically-decaying \Z bosons are related to the directions
of the leptons relative to the photon, as follows:\cite{bib:ebeam-opal}
\[ \frac{s^\prime}{s} = \frac{ \sin \theta_1+ \sin \theta_2 - |\sin( \theta_1+ \theta_2)|}
                             { \sin \theta_1+ \sin \theta_2 + |\sin( \theta_1+ \theta_2)|}, \]
where  $\sqrt{s}$ is the center-of-mass-energy of the event, $\sqrt{s^\prime}$ the invariant 
mass of the lepton pair, and $\theta_1$ and $\theta_2$ the angles between the
leptons and the photon.
When  the event contains an observed energetic photon, its direction can be used in the analysis, otherwise,
an unobserved photon is assumed to be emitted along the beam pipe.
Muon pairs comprise the most sensitive lepton channel. In the case of electron pairs, 
additional background from t-channel processes must be taken into account, and for $\tau$ pairs
there is a substantial loss in  efficiency.

The dominant systematic error in this energy measurement 
arises from the uncertainty in the  polar angles of the leptons.
Using a  wrong   ratio  of the length to  width of the detector for
determining  the polar angles leads to biases in  beam energy. This
``aspect ratio'' cannot be checked with  \Z-calibration data because a wrong value  still
yields back-to-back leptons. The uncertainty is therefore estimated by comparing results from different
detector components (the tracking system, the calorimeter, and the muon detector) with each other.

In hadronic $f\bar{f}\gamma$ events, the two fermions are reconstructed as jets by forcing all observed particles,
except isolated photons, into  two jets. The beam energy can be determined in a manner similar to that used in the
lepton analysis (from the directions of the jets relative to an isolated photon or the beam axis).
The mass of the jet mass, however, cannot be neglected. In this case $s^\prime/s$ is given by:
\[ \frac{s^\prime}{s} = \frac{ \sin \theta_1+ \sin \theta_2 + 
                              |\sin( \theta_1+ \theta_2)| \left(1-\frac{2(|p_1|+|p_2|+|p_{\gamma}|)}{\sqrt{s}}\right)}
                             { \sin \theta_1+ \sin \theta_2 + |\sin( \theta_1+ \theta_2)|}. \]
The sum of the momenta can be approximated as:\cite{bib:ebeam-l3}
\[ |p_1|+|p_2|+|p_{\gamma}| = \sqrt{s}- \sum_{jet} \frac{1}{2} \frac{m_j^2}{E_j} + {\cal O}(\frac{m^4_j}{E^3_j}). \]
In this approach, the jet direction can  be determined  either by all particles, or from  charged tracks,
or just from the calorimeter information.
\begin{figure}
   \centerline{
    \parbox[b]{.3cm}{a) \vspace*{4.4cm}} \parbox[t]{5.8cm}{ \epsfig{file=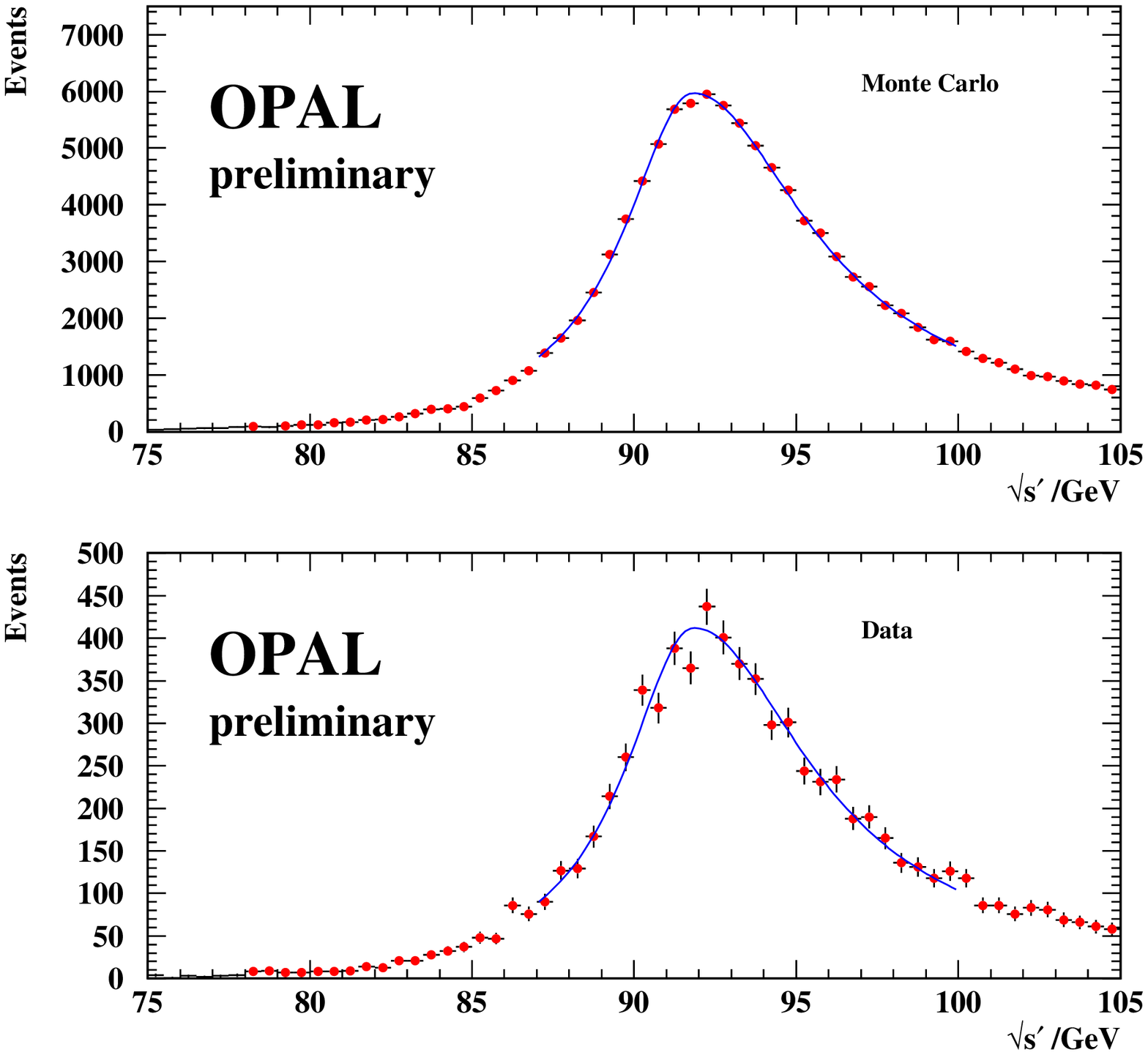,width=5.9cm}}
    \parbox[b]{.3cm}{b) \vspace*{4.4cm}} \parbox[t]{6.5cm}{ \epsfig{file=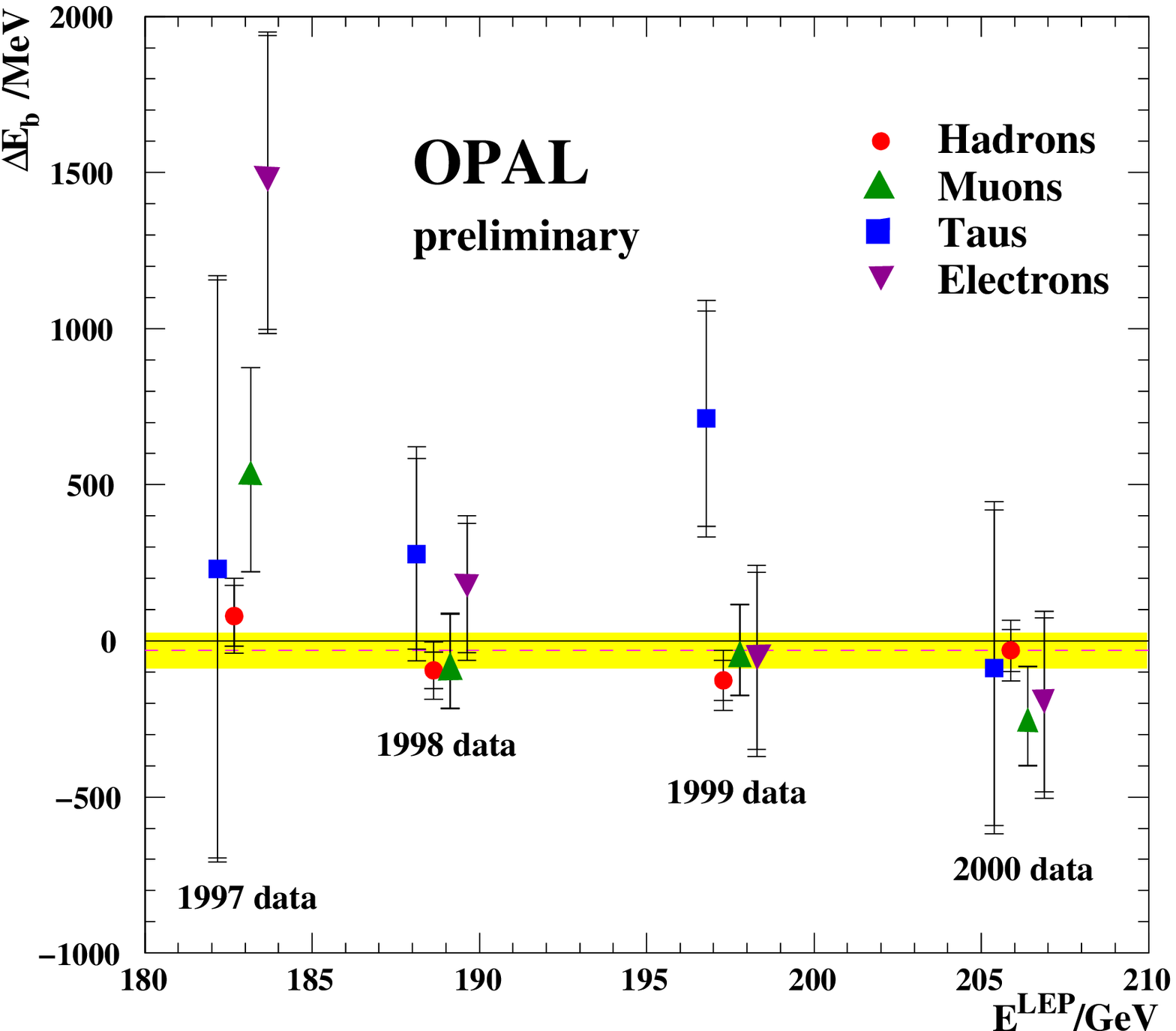,width=6.6cm}}
   }
   \caption{a) Invariant mass of the $q\bar{q}$ system in $e^+e^- \rightarrow q\bar{q} \gamma$ Monte-Carlo
         events and data
         at $\sqrt{s}=206$ \GeV. The curves are  the fitted functions used to determine
         the beam energy. b) Difference  between the measured beam energy and the one predicted
         by the \LEP Energy Working Group ($\Delta E_b$). For clarity, the muon points are displaced rightwards by
         0.5 \GeV, the $\tau$ points leftwards by 0.5 \GeV, and the electron points rightwards by 1 \GeV. The
         dashed line indicates the mean $\Delta E_b$, with its $\pm 1$ standard derivation band given by the
         shaded region.
  \label{fig:ebeam_h}}	
\end{figure}
Alternatively, the $q\bar{q}$ invariant mass  can be determined from  
a kinematic fit, using energy and momentum conservation, similar to that implemented in the \W-mass measurement. 
The beam energy can then be determined by 
fitting an analytic function to the mass distribution. Figure~\ref{fig:ebeam_h}~a) shows the mass distributions
for Monte-Carlo events and data at $\sqrt{s}=206$  \GeV from the \opal collaboration.\cite{bib:ebeam-opal}
The data are well  described by the fitted function. 
For the preliminary \opal analyses, the largest systematic uncertainty in this channel arises from 
differences in hadronization models. These yield different baryon and kaon fractions,
which can bias  mass reconstruction, as discussed in Section~\ref{sec:had}. The effect on the
$q\bar{q}$ invariant  mass in  radiative  events is even larger then for \W pairs,  because of the larger
Lorentz factors. Other major sources of systematic error are from uncertainty  on the measurement of the jet
energy and direction.  Figure~\ref{fig:ebeam_h} b) shows the difference  between the \opal measurement
of the beam energy and the prediction from the \LEP Energy Working Group. The average of all channels
over the years is\cite{bib:ebeam-opal}
\[ \Delta E_{beam} = 31 \pm 41(stat.) \pm 36(syst.) ~\MeV \] 
The \ldrei collaboration expresses their results for the hadronic channel in terms of a measurement of the \Z 
mass:\cite{bib:ebeam-l3}
\[ m_{Z} =  91.226 \pm 0.034 (stat.) \pm 0.072 (syst.)~\GeV \]
Both results show no significant deviation from expectation.

\subsection{Color Reconnection}
The lifetime of the \W boson is smaller than typical time scales for parton hadronization. The decay products
of two hadronically decaying \W bosons can therefore have significant space-time overlap. In principle,
the two \W bosons do not hadronize independently, and can exchange gluons during
the hadronization process. 
Consequently, any particles produced during  fragmentation
cannot be associated unambiguously  with any one of the \W bosons.
The exchange of colored gluons that affects the color flow between
the partons is called color reconnection.
In order to estimate possible effects of this color reconnection,
one requires a modification in standard models of hadronization.

The models of Sj\"ostrand and  Khoze (SK models)\cite{bib:sk} are based on the 
\pythia Monte-Carlo generator (see also Section~\ref{sec:mc}).
Calculations indicate that the probability of gluon exchange during the perturbatively-described
parton-shower process is small. The color reconnection in the SK models is therefore  based on
a modification of  string fragmentation.
As described in Section~\ref{sec:mc}, a string connects all partons forming 
a color singlet: it starts at a quark, ends at an anti-quark, and  has kinks at  gluons vertices. 
In the case of color reconnection, the string configurations are
changed such that partons from both \W bosons can belong to the same string.
In the SK-I model, strings are assumed to have  finite width, and this affects  the phase space overlap
between any two strings. If this is large enough, the strings are reconnected 
at the point of largest space-time overlap. 
The width of the string is a parameter of the model that can be used to produce an arbitrary 
overall reconnection probability. 
In the SK-II models, strings  are assumed to be infinitesimally narrow, and color reconnection 
occurs the first time  two strings cross.
This affects primarily the middle of the string. Particles  produced at this point
tend to have low energy, and are far from the main jets, which are dominated by the high energy
hadrons formed at the ends of the strings, and which reflect the direction of the original quarks
that initiate the parton shower.

In the \herwig Monte Carlo, color reconnection is implemented in the cluster fragmentation.\cite{bib:herw_cr}
If color reconnection is enabled, gluons from different \W bosons are allowed to form a cluster.
New associations of partons into a cluster are considered if they lead to a smaller 
space-time extent of the cluster. When such an association exists, it is realized with a 
probability of $1/9$, reflecting the probability that the two partons form a color singlet. 

In the \ariadne (AR) model, the implementation of color reconnection takes place  in 
the inherent simulation of the QCD shower.\cite{bib:ar2}
The perturbative shower is described by radiation of gluons from a color dipole.
Higher-order effects are implemented through the requirement that a gluon be radiated only with
smaller transverse momentum  than the previous radiation.
When the option of color reconnection is switched on, the dipole that radiates a gluon can  be formed
by partons from the same or different \W bosons. 
The decision as to which partons form dipoles is
 based on the minimization of the string length $\lambda$, as defined  by:\cite{bib:ar_gh}
\[ \lambda = \sum_{1}^{n-1} \ln(p_i+p_{i+1})^2/m_0^2. \]
where the string consists of $n$ partons with 4-momenta $p_i$
(with neighboring partons in the string forming dipoles), and 
$m_0$ is a hadronic mass scale  of $\approx 1$ \GeV. In this model, partons close 
in momentum space are more likely to be color connected. 
The minimization of the string length
$\lambda$ introduces  not only color reconnection between the two \W bosons, but it also changes
the assignment of partons to color dipoles within a color singlet (a \W or \Z). In models
without  color reconnection, this assignment is based  purely on the order of gluon emission
in the dipole cascade.
The AR3 model,  which considers all gluons in the simulation of color reconnection,
is theoretically disfavored because gluons radiated with  energies greater than $\Gamma_\W$ 
are perturbative, and therefore  radiated incoherently by the two initial
color dipoles from separate \W bosons.\cite{bib:sk}
In the AR2 model,\cite{bib:ar2} only gluons with
energy less then $\Gamma_\W \approx 2$ \GeV are allowed to influence the opposite W.
As mentioned previously, 
the modeling of color reconnection in \ariadne 
also contains contributions to 
events that
contain only one hadronically decaying \W or \Z. 
It is certainly inconsistent to apply 
fragmentation parameters to a color-reconnection model when these  were derived from 
models without any color reconnection applied fitted to  high statistics \Z data. 
The cleanest way to minimize  confusion between color reconnection and hadronization
effects (for different tunes of Monte Carlos) is to compare the AR2 model to
a model where color reconnection is allowed  for partons that originate from the
same \W or \Z, but not allowed for partons from different \W-bosons (as in, e.g., the AR1 model). 
By construction, both Ariadne models are identical for \Z decays.

The rearranging of  strings and clusters due to color reconnection, affects mainly the low energy
particles that are emitted far from the jet axis. The  analyses that attempt to   
constrain   color-reconnection models, or try to reduce the impact  of color
reconnection on the determination of the \W mass, concentrate therefor on such particles.

\subsection[Reducing the Influence of Color Reconnection on the  Determination of the Mass of the \W]
{Reducing the Influence of Color Reconnection on the  Determination of the Mass of the \bW}
The scaled di-jet mass  can also be used to examine the
influences of color reconnection on the extracted \W mass. 
As discussed in  Section~\ref{sec:had}, the measurement of the  \W mass is sensitive to   
the opening angle between the individual jets and to the jet masses.
\begin{figure}
   \centerline{
\epsfig{file=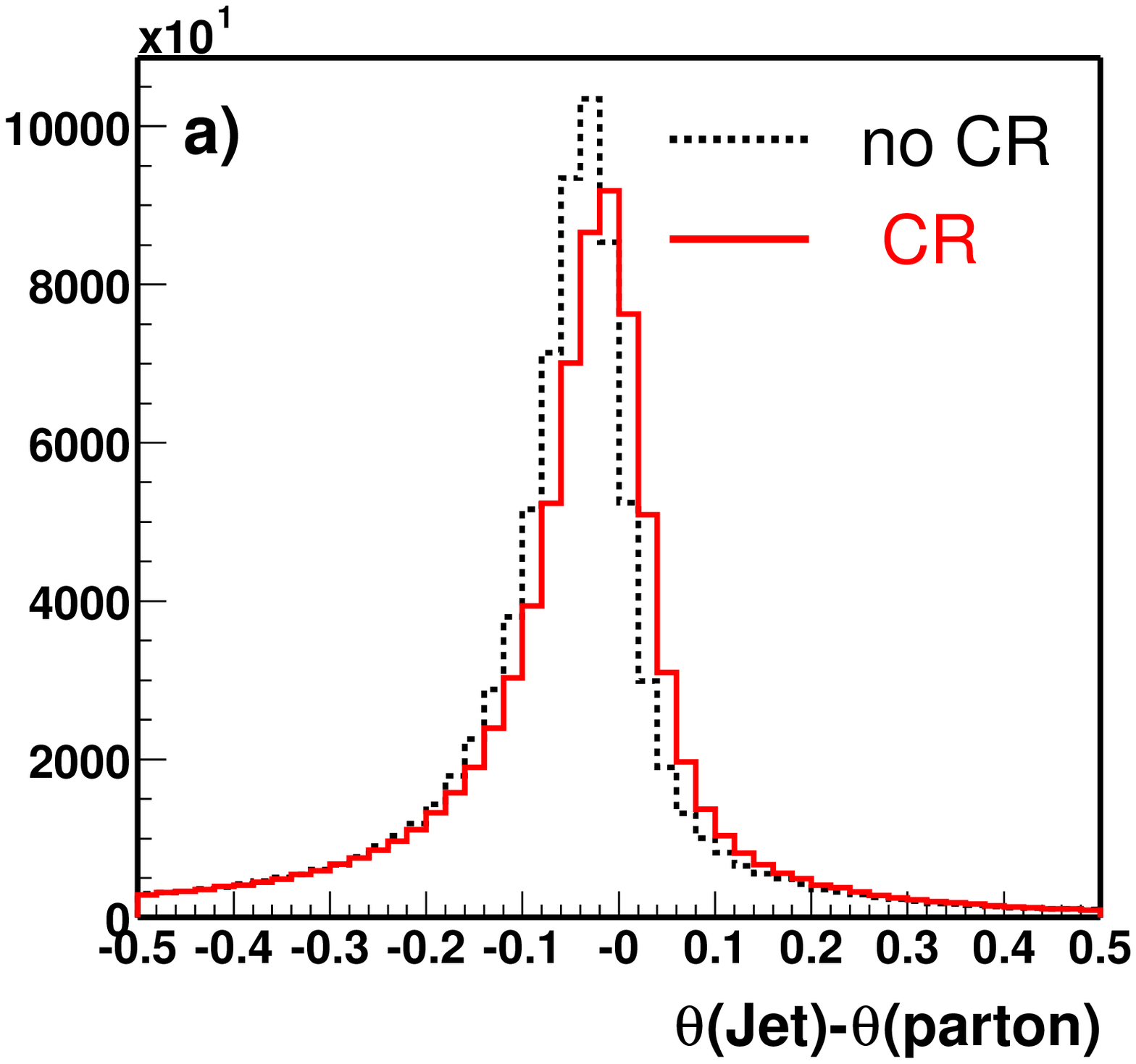,width=6.3cm}
\epsfig{file=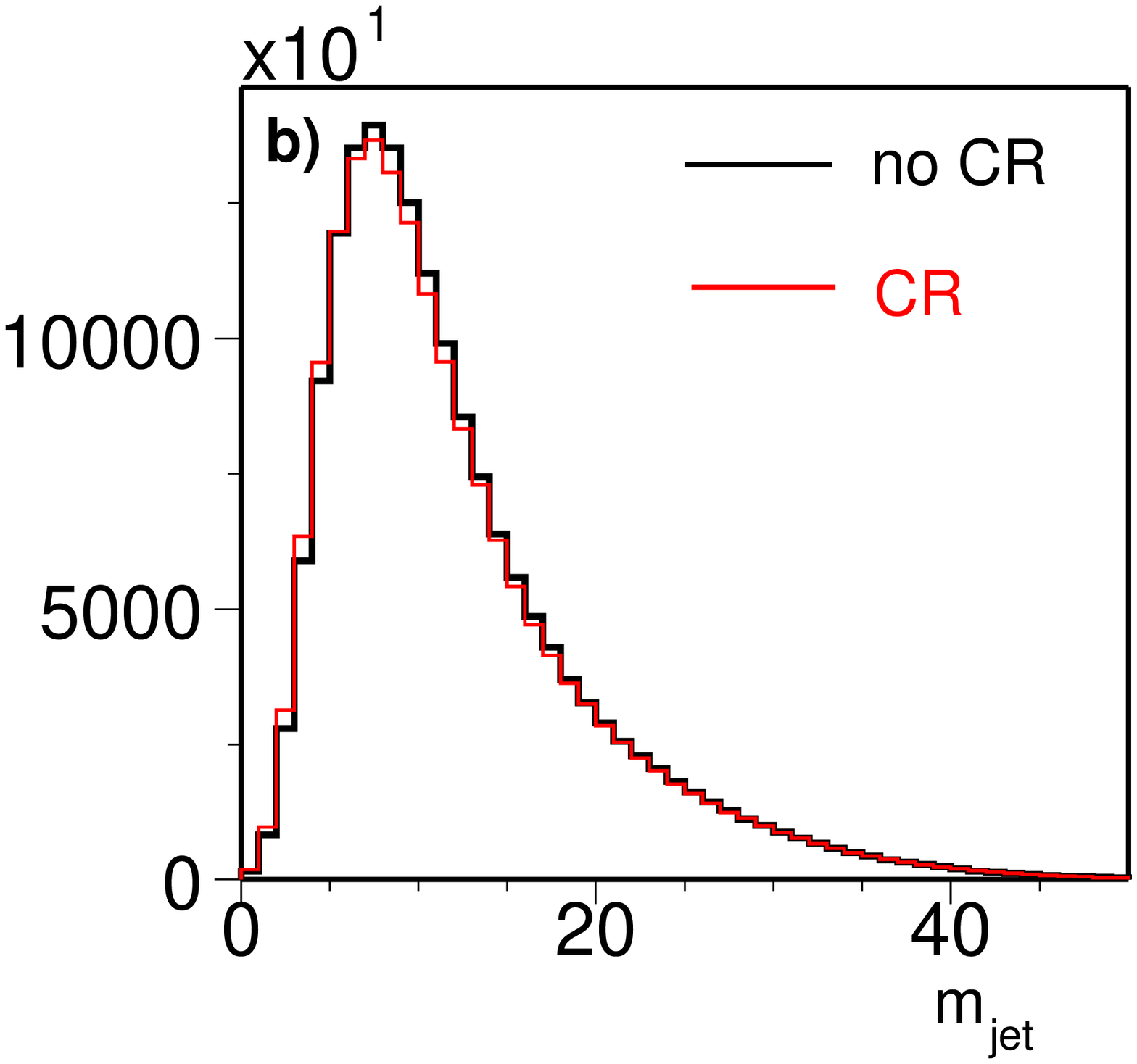,width=6.3cm}
}
   \caption{Comparison between the SK-I model with full color reconnection and \pythia 
without color reconnection for: a) the difference in the opening angle in the 
$\W \rightarrow q\bar{q}$ decay, and
b) the jet mass'  when calculated using jets and the original partons. 
 \label{fig:cr_ang}}	
\end{figure}
Figure~\ref{fig:cr_ang} a) shows the difference in the opening angle of the two jets belonging
to the same \W boson, calculated  either from the  information in the detector or from the initial quark-antiquark
pair.  
The jets acquire mass in the process of hadronization,  but, assuming no cross-talk between
the two \W-bosons, and because of energy and momentum conservation, the
hadronization process should not affect the \W mass. The contribution of the individual jet masses 
to the invariant mass of the pair must therefore  be compensated by a smaller angle between
the two jets relative  to the angle between the original quark-antiquark pair.
In addition to this effect, the figure shows the difference in the angle between the SK-I model with full
color reconnection, and  the one without  color reconnection. 
These differences in the opening angles lead to a change of 350 \MeV in the scaled
di-jet mass.  The effect of color reconnection on the individual jet mass, shown in
Fig.~\ref{fig:cr_ang} b), is relatively small, and would contribute  to a change of only about 25 \MeV.
The main effect of the color reconnection relevant to the  measurement of the \W-boson mass is therefore a
change in jet direction.

\begin{figure}
\parbox[t]{6.1cm}{
   \centerline{\epsfig{file=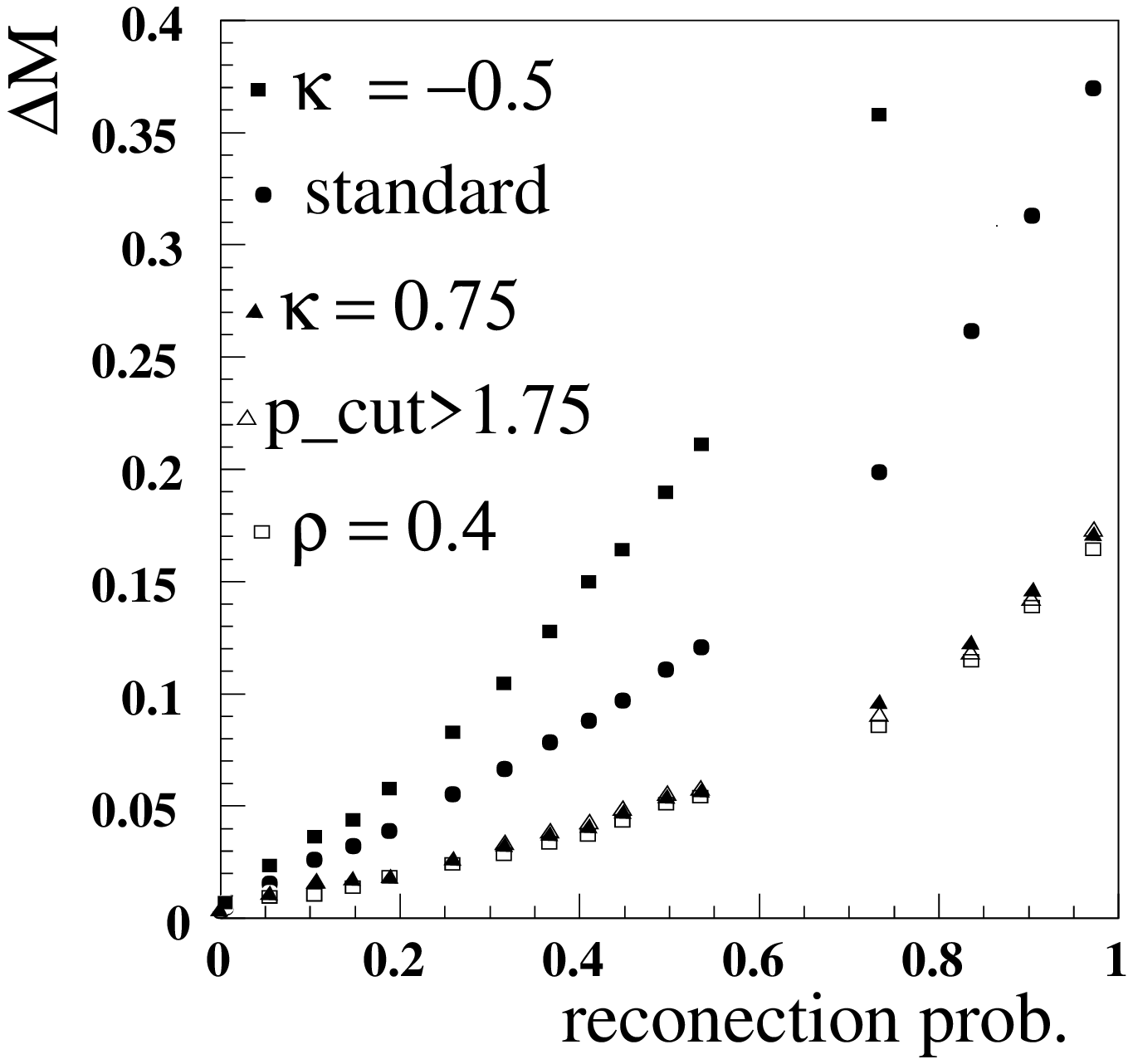,width=5.8cm}}
   \caption{Mass bias as a function of color reconnection probability for the SK-I model. \label{fig:ski}}	
}
\hspace*{.3cm}
\parbox[t]{6.1cm}{
  \centerline{\epsfig{file=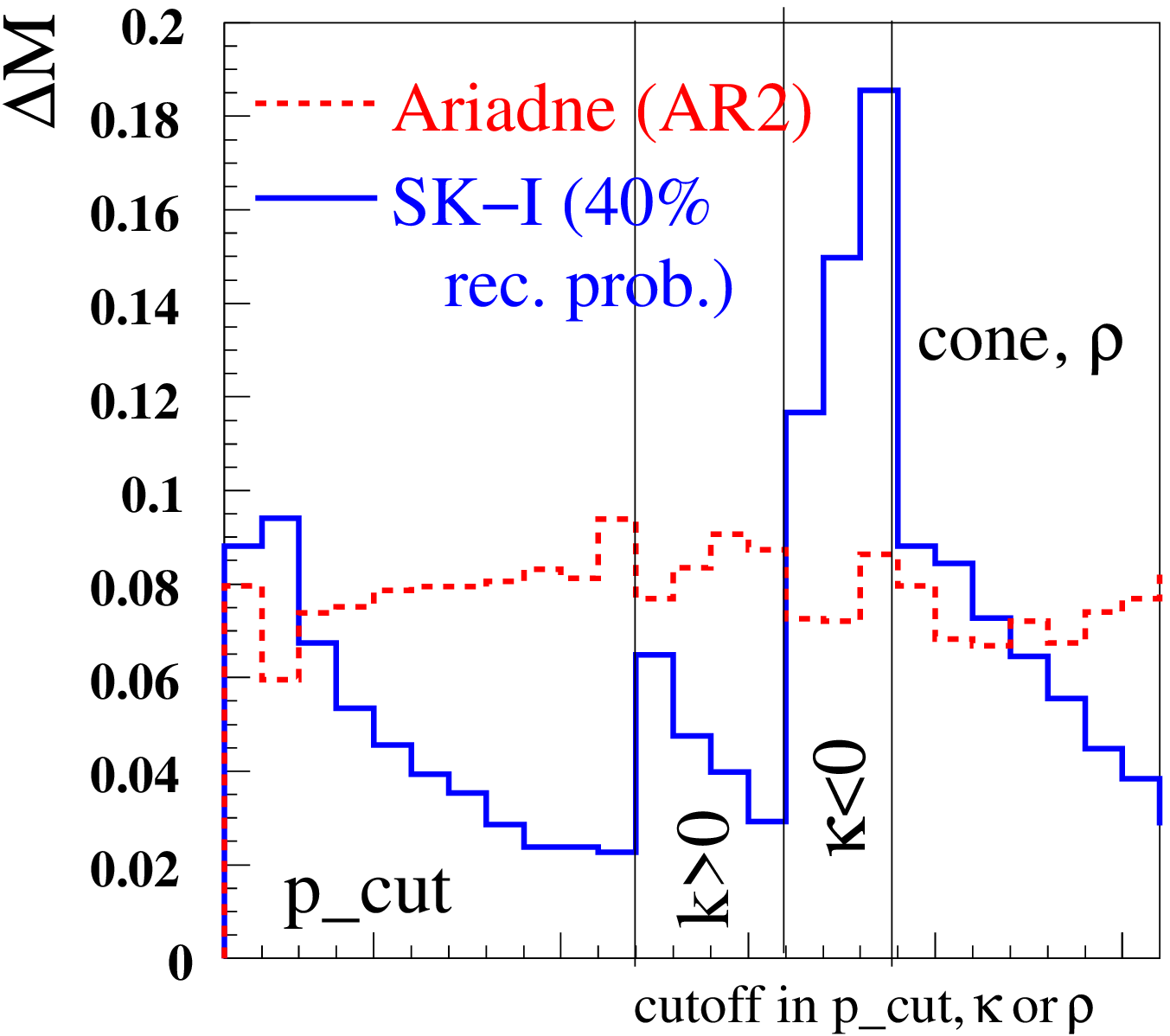,width=5.8cm}}
   \caption{Comparison of mass bias for the \ariadne (AR2) and \pythia (SK-I) model. \label{fig:ari}}	
}
\end{figure}
Three strategies were tried to reduce the impact of color reconnection on jet direction.
First, either particles with
a momentum larger than $p_{cut}$ were used to calculate the jet directions, or  those  
 within a cone with a half opening angle of $\rho$ around the jet axis. In a third  approach,
the jet direction was determined from the vector sum $\sum \vp \dot |\vp|^\kappa$ 
over all particles in a jet, where
each particle momentum is weighted by an extra factor $|\vp|^\kappa$. Positive $\kappa$ emphasizes
high-momentum particles, and  negative $\kappa$ low-momentum particles.
The three approaches are compared with the standard analysis in Fig.~\ref{fig:ski}. The
values of $p_{cut}$, $\rho$ and $\kappa$ were chosen such that  the statistical 
uncertainty from the  the determination of the jet direction  was of similar in the
three analyses. The figure  shows the mass difference between a Monte Carlo without color
reconnection and  the SK-I model, as a function of the color-reconnection probability.
All three approaches  reduce substantially the mass bias from color reconnection,
and,  as expected, the bias  increases for negative values of $\kappa$.
Figure~\ref{fig:ari} shows the mass bias for the SK-I and AR2 models, for different values of 
$p_{cut}$, $\rho$ and $\kappa$.
In the SK-I model,  the mass bias from color reconnection decreases with  
larger cutoffs. The figure also shows that the mass bias can be increased relative
to the standard analysis by increasing the influence of low-momentum particles on 
jet direction (i.e., by using  negative $\kappa$ values). In the \ariadne model, the 
mass bias is not affected through modifications in  the calculation of the jet direction.
This difference in behavior might be due to the fact that in the AR2 model, 
the color reconnection  affects the gluon radiation in the QCD shower. 
(In the SK models, only  string fragmentation is affected by color reconnection)
Because of  the 2 \GeV requirement on  gluon energy, it was expected that only low-momentum 
particles would be influenced in AR2 by  color reconnection, which is clearly not the case.

\subsubsection{Limits on Color Reconnection Models}
The \W  mass is not the only observable that is influenced by effects from color reconnection.
Without color reconnection,  the particle multiplicity of hadronically-decaying \W-pairs
should, naively, be exactly twice
that of  hadronically-decaying \W bosons in semileptonic decays. However, this does not follow when
color reconnection is present. Measuring the difference between the multiplicity in hadronic
\W-pair decays and twice that in semileptonic decays, can provide a measure of color reconnection.
Most of the systematic uncertainty in the  measurement of  multiplicity can be avoided in the 
difference  measurement,
with the main contribution being   due
to different event selections  and different backgrounds in the two channels.

Because of color reconnection, more particles are expected to be emitted 
between jets from different W bosons.
Therefore, models can be tested by comparing the particle flow in the region between jets from the same
W with  the  particle flow between jets from different \W bosons. 
Because the four jets in a hadronic \W-pair decay are in
general not in one plane, and since the average angles between the  jets from the same and from different
W bosons are different, there is much flexibility for  defining the regions between  jets
to  be used in the analysis.

The \ldrei collaboration applies quite strict requirements for their events, so as 
to guarantee well-defined separated event topologies.\cite{bib:l3_col}  
The efficiency for complete event selection 
is on average 12\%, and the probability to pick the correct jet pairing is 91\%.
\begin{figure}
   \centerline{\epsfig{file=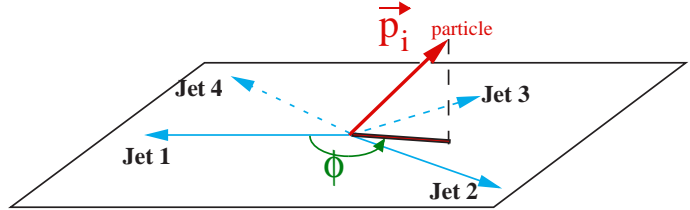,width=9cm}}
   \caption{Illustration of the method used to construct distributions of particle flow. \label{fig:pflow_def}}	
\end{figure}
\begin{figure}
   \centerline{\epsfig{file=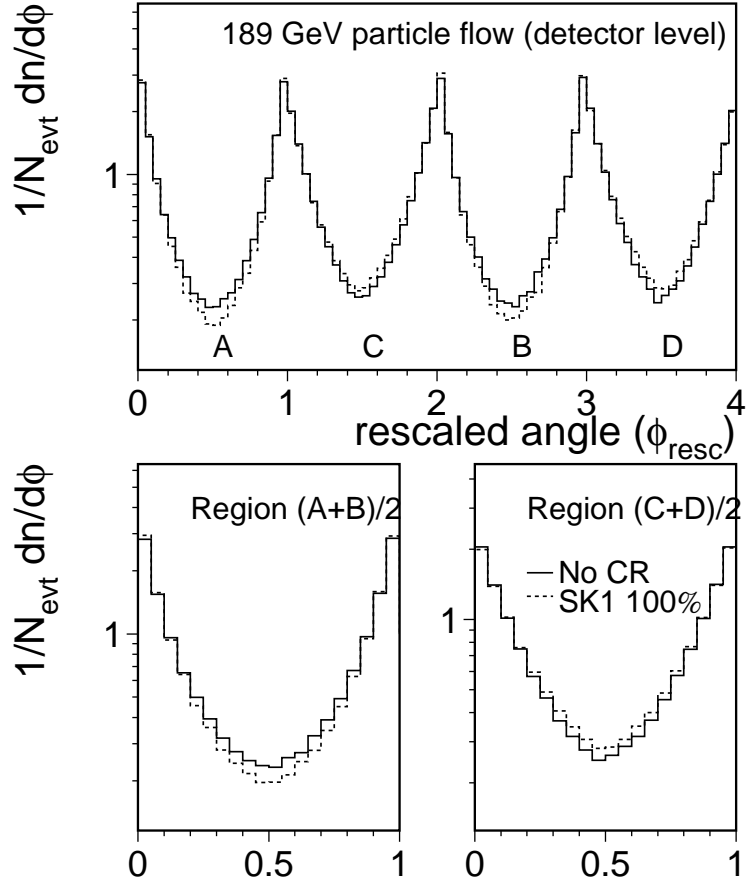,width=10cm}}
   \caption{Distribution in  particle flow predicted for $\sqrt{s}=189$ \GeV. \label{fig:pflow_mc}}	
\end{figure}
The  definition of particle flow is illustrated in Fig.~\ref{fig:pflow_def}. 
A plane is defined by the most energetic jet (jet-1) and the
jet belonging to the same \W (jet-2). The azimuthal angle between jet 1 and the projection 
of all particle  and jet momenta onto that
plane, is defined by the sense of rotation from  jet-1 (towards  jet-2).  
The fact that the angles between  jets  differ from one event to another,  
and that the events are not 
planar, is parametrized by a rescaled angle, as follows. 
A particle $i$  for which the above-defined projection
falls between jet-j and jet-k is  reprojected onto the plane  spanned by those two jets.
The angle $\Phi_i$ between jet-j  and  this projection is  divided by
the space angle $\phi_{jk}$ between  jet-j and jet-k. 
This is termed the rescaled angle $\Phi_{resc}^i=\Phi_i/\phi_{jk}$. 

In order to characterize  the particle flow, each particle enters the $\Phi_{resc}$ 
distribution with unit weight. For  energy flow,
the weight equals the energy of the particle.
\begin{figure}
   \centerline{\epsfig{file=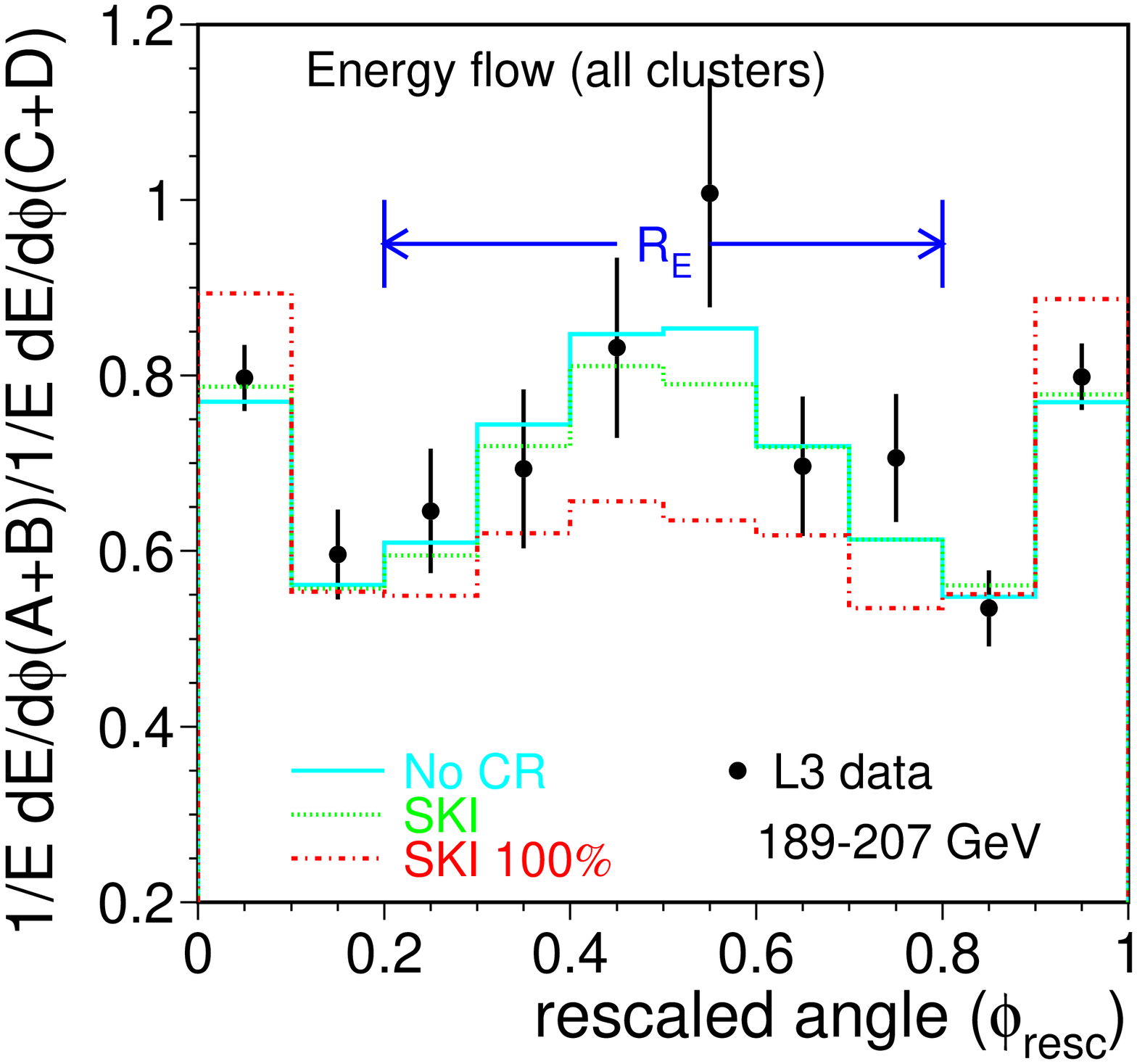,width=6.cm}
               \epsfig{file=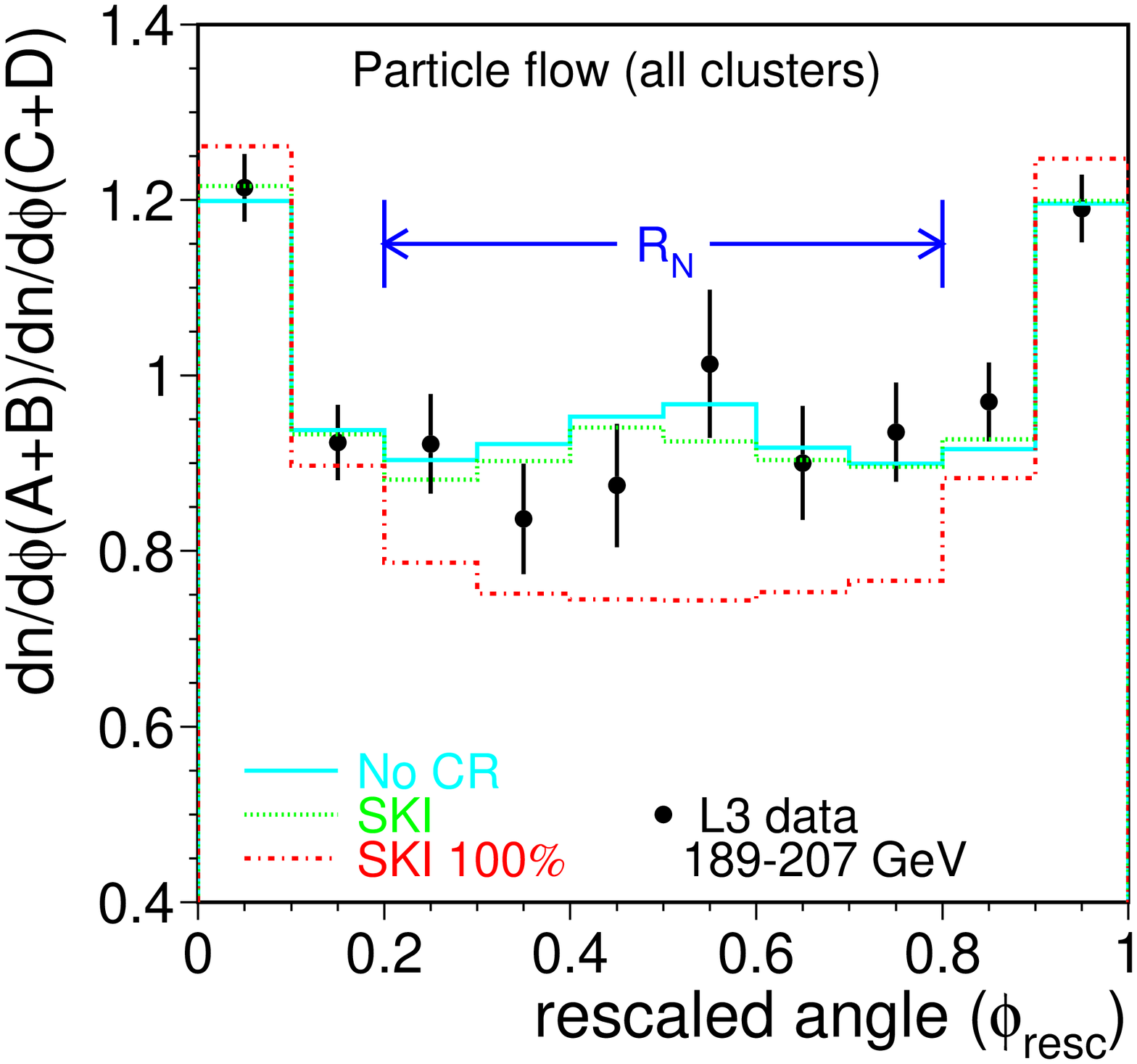,width=6.cm}}
   \caption{Ratio of flow in the regions between jets for the same \W (A and B), to the flow in the 
    region between jets from different \W bosons (C and D)  \label{fig:pflow_dat}}	
\end{figure}
Figure ~\ref{fig:pflow_mc} compares the particle flow for  \pythia without color reconnection and the SK-I
model with full color reconnection. 
In order to show the four regions of flow in one diagram, 
the rescaled angle for particles in the region between jet-2
and jet-3 (region C) is increased by 1, for particles between jet-3 and 4  (region B) 
by 2, and for particles 
between jet-4 and 1 (region D)  by 3.
We see that the Monte-Carlo simulation with  color reconnection predicts fewer  particles
in regions A and B (between jets from the same \W), and more particles in the regions C and  D (between jets
from different \W bosons). By measuring the ratio of particle flow between jets from the same and different
W bosons, the dependence on absolute flow in the inter-jet region can be minimized, while maintaining  sensitivity to
color reconnection.  
Figure ~\ref{fig:pflow_dat} compares the \ldrei measurement for  particle and energy flow
with different Monte-Carlo predictions.\cite{bib:l3_col}

The particle flow analysis of the \opal Collaboration is based on less stringent 
criteria for event topology,\cite{bib:opal_pflow}  which provide an  overall efficiency for \W-pair all-jet events 
of 42\%. Comparing this analysis to a more restricted one similar to the analysis of L3,\cite{bib:l3_col} 
shows an improvement in the sensitivity to color reconnection models of about 15\%-30\%.
It should be pointed out that the more restrictive \opal analysis  is  distinct
from  the \ldrei analysis, as it uses the \opal jet pairing, jet ordering, and  the result of a
kinematic fit for  defining jet directions.
Weighting has been tried for the particle flow using a factor $\ln (1/x_p)$, where 
$x_p$ is the  particle momentum divided by that of the beam.
Although  this enhances the contribution from low-momentum
particles, which should in principle make the analysis  more sensitive to color reconnection effects,
it dose not provide any  significant improvement in sensitivity. 

\subsection{Bose-Einstein Correlations}
The production of identical bosons (e.g., pions) close together
in phase space can be affected by Bose-Einstein correlations, even when the two bosons originate from 
different \W decays. Any observed enhancement would suggest 
that the hadronization of the two \W bosons from hadronic \W-pair
decays is not independent. This, in turn, could influence
the reconstructed \W mass.

\subsubsection{Bose-Einstein Correlations within \Z and \W Bosons}
\begin{figure}[h]
   \centerline{\epsfig{file=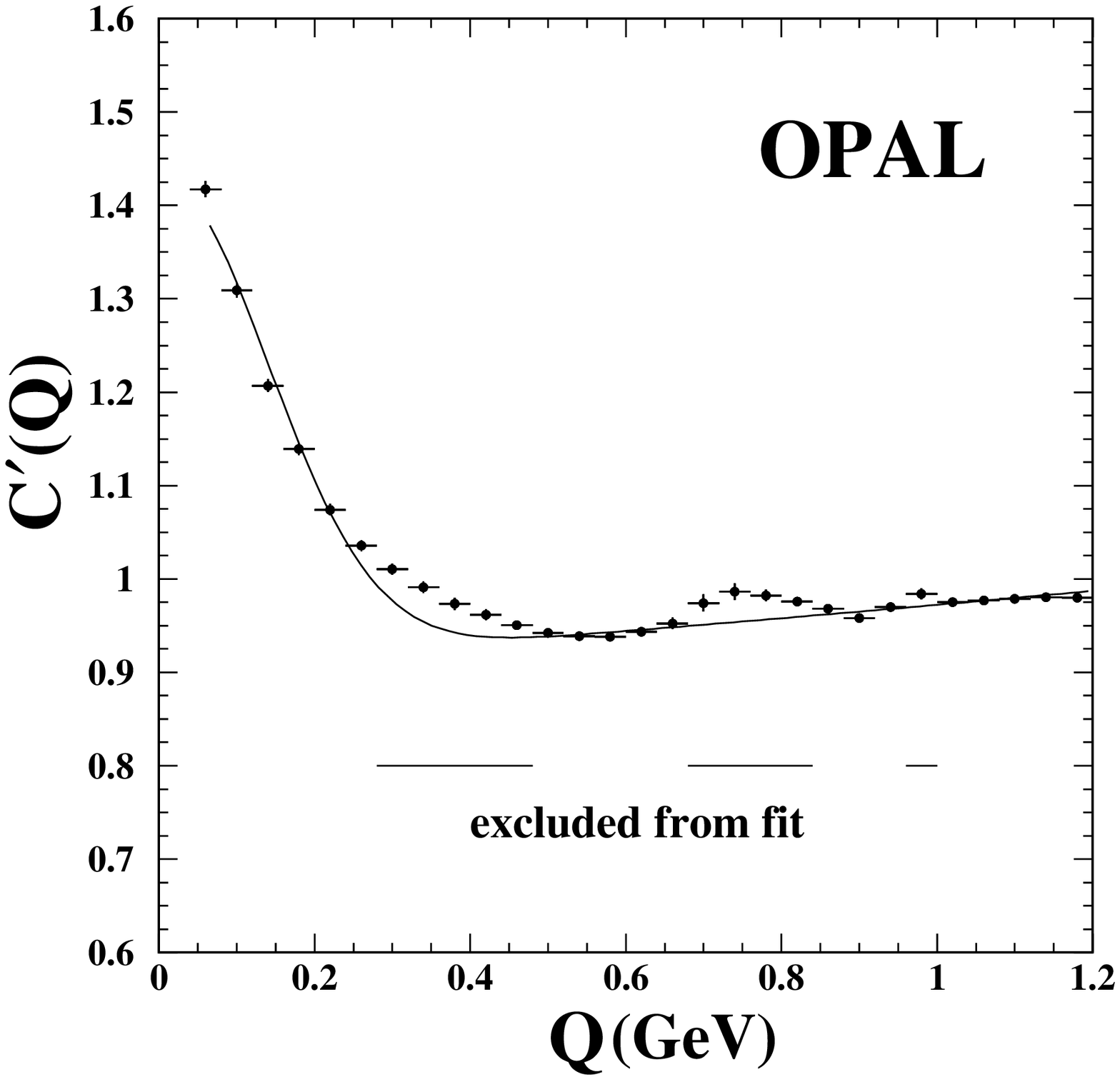,width=6.5cm}
               \epsfig{file=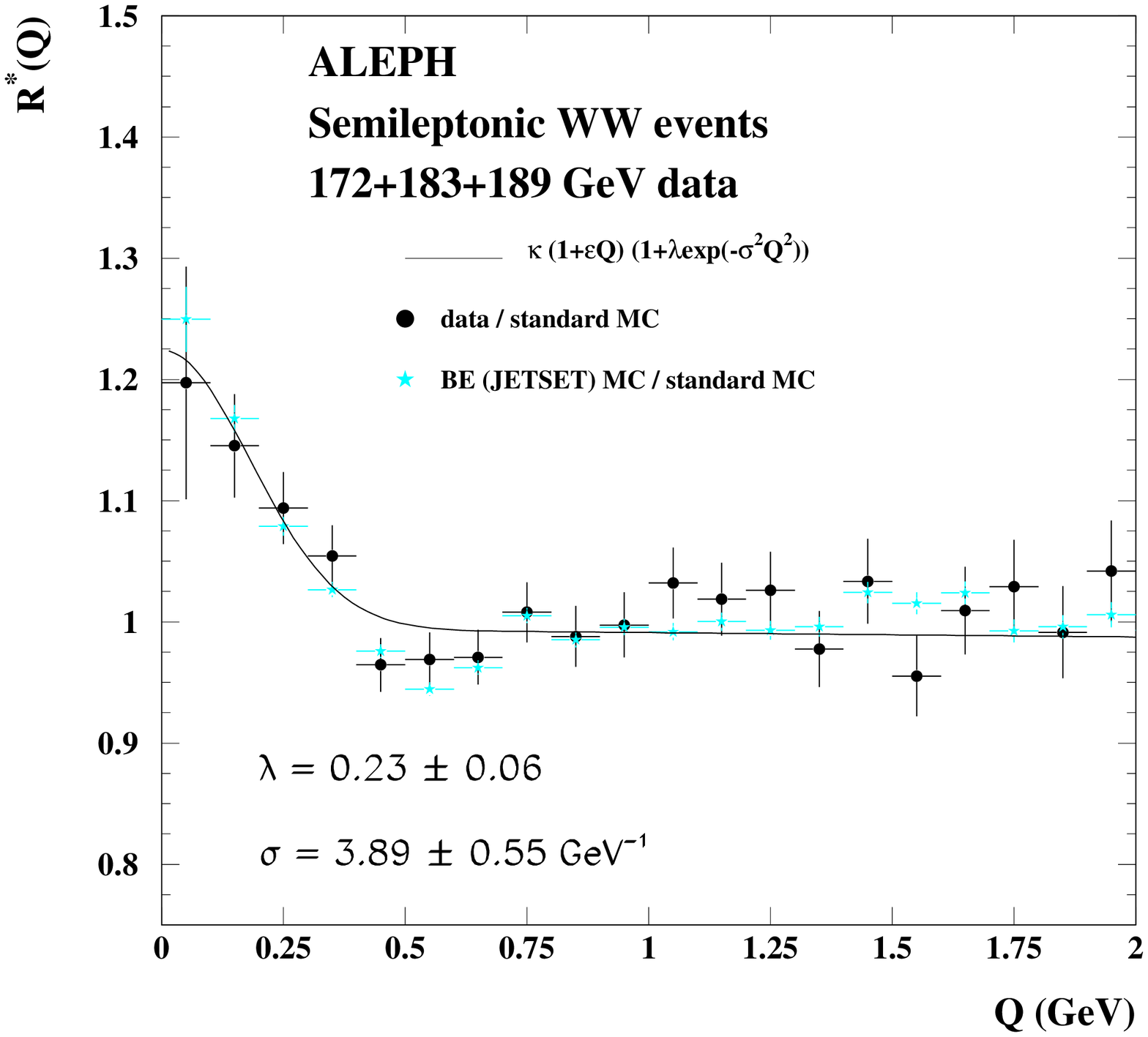,width=6.5cm}}
   \caption{Observation of Bose-Einstein correlations in  
\Z events,\protect\cite{bib:bos_z_o} 
and  \W-pair events containing only one hadronic \W-decay.\protect\cite{bib:bos_w_a} 
The distribution was obtained by comparing like-sign boson
pairs with opposite-sign pairs.  The correlation from resonance  decays to particles of opposite charge has
not been fully corrected,
and such regions have therefore been  excluded from 
the fit. \label{fig:be_wz}}	
\end{figure}
Figure~\ref{fig:be_wz} shows the well-established Bose-Einstein correlations
observed within both \Z and \W boson decays. They can be
observed through  two-particle correlations that can be parametrized as:\cite{bib:bose_bas} 
\[ f_2(Q)= \frac{\rho(\ip_1,\ip_2)}{\rho(\ip_1)\rho(\ip_2)}=(1+f_{\pi}(Q)\lambda e^{-Q^2/R^2})(1+\delta Q +
 \epsilon Q^2). \]
where $\ip_1$ and $\ip_2$ are the 4 momenta of the two particles, $Q^2= -(\ip_1-\ip_2)^2$, $\rho(\ip_i)$ and $\rho(\ip_1,\ip_2)$
are the single and two-particle density functions, and  $f_{\pi}(Q)$ is the probability that the two
observed charged tracks are pions. The parameter $\lambda$ describes the strength of the correlation, and
$R$ can be interpreted as the  size of the source for the correlated particles. The factor $1+\delta Q + \epsilon Q^2$,
with free parameters $\delta$ and $\epsilon$, 
reflects long-range correlations (e.g., correlation due to charge and momentum conservation, and constraints
from  phase space). 

In a full quantum-mechanical description of  hadronization, 
Bose-Einstein correlation can be incorporated by symmetrizing the overall wave function.
Since such a description is not available, models are used to accommodate any Bose-Einstein 
correlations. 
There are basically two approaches used in Monte-Carlo simulations:
In a global approach, events are weighted
in a way that yields the desired correlation functions after the reweighting.
The weight for each  event can be defined as the product of the $f_2(Q)$ values for all pairs 
of identical particles. 
In the local approach, the desired two-particle correlation function is generated by modifying
the momentum of the particles and thereby pulling identical particles closer together in phase space.

At a first glance, the global event weighting is more appealing, because changing the probability
that an event  is produced with a given particle correlation is what would be expected from
symmetrizing the wave function. This ansatz, however, has two major problems. Technically, because of the high multiplicities,
some of the event weights can get very large, and produce large fluctuations.
Moreover, global event weights  in general
violate the factorization hypothesis, which requires that properties determined by the
perturbative  parts of the interaction, such as the width of the \Z, the b fraction, and
the three-jet rate, are not affected by  hadronization. This hypothes has been  well tested,
in that the distributions in  the  above observables agree with theoretical predictions.

The problem of factorization can be circumvented  by implementing a veto algorithm applied only 
to the hadronization phase of the simulation. 
This ansatz\cite{bib:be_nov}
is based on the  \pythia program (see Section~\ref{sec:mc}),
in which quarks and gluons are simulated   first  by a 
perturbative QCD shower, and then hadronized via a string model.
The selection procedure takes each  
simulated event with a probability proportional to the
product of the $f_2(Q)$ values for all pairs of identical particles
in an event, however, if the event is rejected by 
the veto algorithm, then without changing the  quark and gluon configuration from its perturbative simulation,
the simulation of the hadronization is repeated. In this way, observables that
depend  only on the  perturbative process are not affected by the rejection algorithm. 
Only the particles produced in the primary interactions are used
in the calculation of weights
for the rejection algorithm.
This keeps the number of identical particles small and the weights of manageable size.

\begin{figure}[h]
   \centerline{\epsfig{file=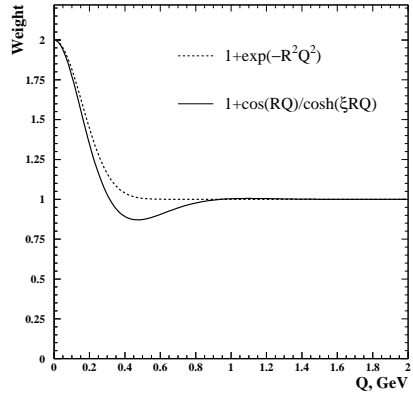,width=5.5cm}
               }
   \caption{Weight function ($w$) for global reweighting. \label{fig:be_vato}}	
\end{figure}

In  the model of Kartvelishvili and Kvatadze 
(K\&K model),\cite{bib:be_kk} motivated by a string model, the event weights are 
based on the following ansatz  for the matrix element:\cite{bib:be_bo}
\[ M_{12} = exp[(i\kappa-b/2)A_{12}] \]
where  $\kappa$ is the string tension, $b$ a breaking probability,  and $A_{12}$  a space-time area of the
string containing the particle 1 and 2.
Symmetrizing this matrix element, ($A_{12} \ne A_{21}$) leads to a weight:
\[ w = 1 + \sum_{perm} \frac{\cos(\kappa (A_{12}-A_{21}))}{\cosh(\frac{b}{2} (A_{12}-A_{21}))}
     = 1 +  \sum_{perm} \frac{\cos(RQ)}{\cosh(\xi RQ)} \]
where we have substituted $\kappa (A_{12}-A_{21}) \rightarrow RQ$ and  $\frac{b}{2} (A_{12}-A_{21}) \rightarrow \xi RQ$.
This functional form produces weights that  are smaller than unity for some values of $Q$ (see Fig~\ref{fig:be_vato}). 
The parameter $\xi$ can be tuned such that the average weight equals unity, and this 
keeps the  width of the \Z from changing, and keeps the weights  reasonably small.
An advantage  of this model is that it can be applied a posteriori to fully simulated events.

Factorization  in the local approach is preserved  by construction,
because the underlying parton structure of an event is not influenced
by the local disturbance of the momenta.
The small value of $\approx 1$ fm for R  indicates furthermore
that the effect of Bose-Einstein correlation is localized
to only parts of a string, but that ends of the strings fragment
nearly independently of each other.
In the local approach of L\"onnblad and Sj\"ostrand,\cite{bib:pyboye}
the value of $Q$ for a pair of  identical  bosons with masses $m$  is  changed by $\delta Q$, as defined by:
\[ \int^Q_0 \frac{q^2 dq}{\sqrt{q^2+4m^2}} = 
    \int^{Q+\delta Q}_0 f_2(Q) \frac{q^2 dq}{\sqrt{q^2+4m^2}}. \]

This change cannot be achieved in a way that satisfies  both energy
and momentum conservation. In order to conserve  momentum, and to 
change the configuration as little as possible, the momenta of the
two particles are altered by equal amounts in the direction defined
by their momentum difference:
$\delta \vp_1 = -\delta \vp_2 = c(\vp_1 - \vp_2)$;
the factor c is chosen to obtain a change of $Q$ by amount $\delta Q$.
This procedure leads to typical shifts of several hundred \MeV  in total energy, for
\Z events.
Different schemes can be used  to ensure energy conservation.
In the original model ($BE_0$), the momenta of all particles are
rescaled by a common factor to yield energy conservation.
This rescaling has only minor impact  on most other observables, but
it introduces a negative bias in the reconstructed \W mass in hadronic \W-pair decays.

To avoid this  bias, other schemes  have been developed, in which  energy 
conservation is  achieved without a global rescaling of all particle momenta.
In the $BE_3$ and $BE_{32}$ schemes, the functional form of $f_2(Q)$ is modified
in a way  that $f_2(Q)$  gets smaller than 1 for certain values of Q. 
In these schemes, 
$f_2(Q)$ contains a free parameter, which  is determined in each event
in such a way that the energy shift from pairs with $f_2(Q)>1$ is compensated
by those from pairs with $f_2(Q)<1$.
In the $BE_m$ and $BE_\lambda$ schemes, for each pair of identical
particles, a pair of non-identical particles with configuration 
close to the original  pair is also selected. The opening angle of this pair
is changed such that the resulting energy shift compensates  exactly  the
energy shift from the pair of identical particles.
The two schemes differ in their definition of how a pair of 
non-identical particles is selected.
The implementation of Bose-Einstein correlations as a momentum
transfer between two identical particles with small $Q^2$
looks technically as a final-state interaction. Nevertheless, this implementation
does not imply that the underlying  process has 
its origin in such an interaction.

\subsubsection{Bose-Einstein  Correlations between Particles from Different \W Bosons.}
As we mentined, it is not possible to determine a priori whether any  two particles in hadronically decaying \W-pair events
originated from the same  or from
different \W bosons.
A model must therefore be formulated 
to test whether the observed correlation in hadronic \W-pair
decays is due only to  Bose-Einstein correlations between particles from the same \W, or whether 
there is  an additional contribution from a correlation between different \W bosons.
This can be done by comparing the two-particle density function in all-hadronic and single-hadronic \W-pair
decays, by checking the validity of the  relation:
\[ \rho^{WW}(1,2) = 2 \rho^W(1,2)+\rho^{WW}_{mix} \]
 where $\rho^{WW}(1,2)$, $\rho^W(1,2)$, and $\rho^{WW}_{mix}$ represent, respectively, the two-particle densities 
from hadronic \W-pair decays, the hadronic part of ``semileptonic'' \W-pair decays, and  
two  ``semileptonic'' events that are combined to represent all-hadronic \W-pairs. 
The last term is needed to describe any contributions to
two-particle density for particles arising from different \W bosons. 

The mixed events are constructed by adding the hadronic parts of two ``semileptonic'' events, after applying
appropriate rotations and boosts in order to reflect the desired kinematics. Since not all detector effects are 
invariant under boosts and rotation, it is important to select pairs for which the 
required transformations cause the smallest  possible bias. 

The observed particle correlations must also be corrected for
background. This is important because of the  significant contributions from 
$e^+e^- \rightarrow \Z/\gamma \rightarrow q\bar{q}$ events.
The hadronic $\Z/\gamma$ decays that pass the \W-pair 
selection have a two-particle density distribution that is sufficiently different  from that of hadronic
W-pairs, and that must be taken into account.
The validity of the   above  relation can be checked  by examining the difference:
\[ \Delta \rho(Q) =  \rho^{WW}(1,2) - 2 \rho^W(1,2)-\rho^{WW}_{mix} \]
or checking the ratio
\[D(Q) = \frac{  \rho^{WW}(1,2) }{2 \rho^W(1,2)+\rho^{WW}_{mix}}. \]

To parametrize any discrepancy between data and Monte-Carlo models, we define the integral:
\[ J \equiv \int_0^{Q_{max}}\Delta \rho dQ \]
In the \ldrei measurement,\cite{bib:be_l3} $Q_{max} = 0.68$ is chosen  as the point where 
predictions for the Monte-Carlo with and without Bose-Einstein correlations differ by less 
than one standard deviation.
\begin{figure}[h]
   \centerline{\epsfig{file=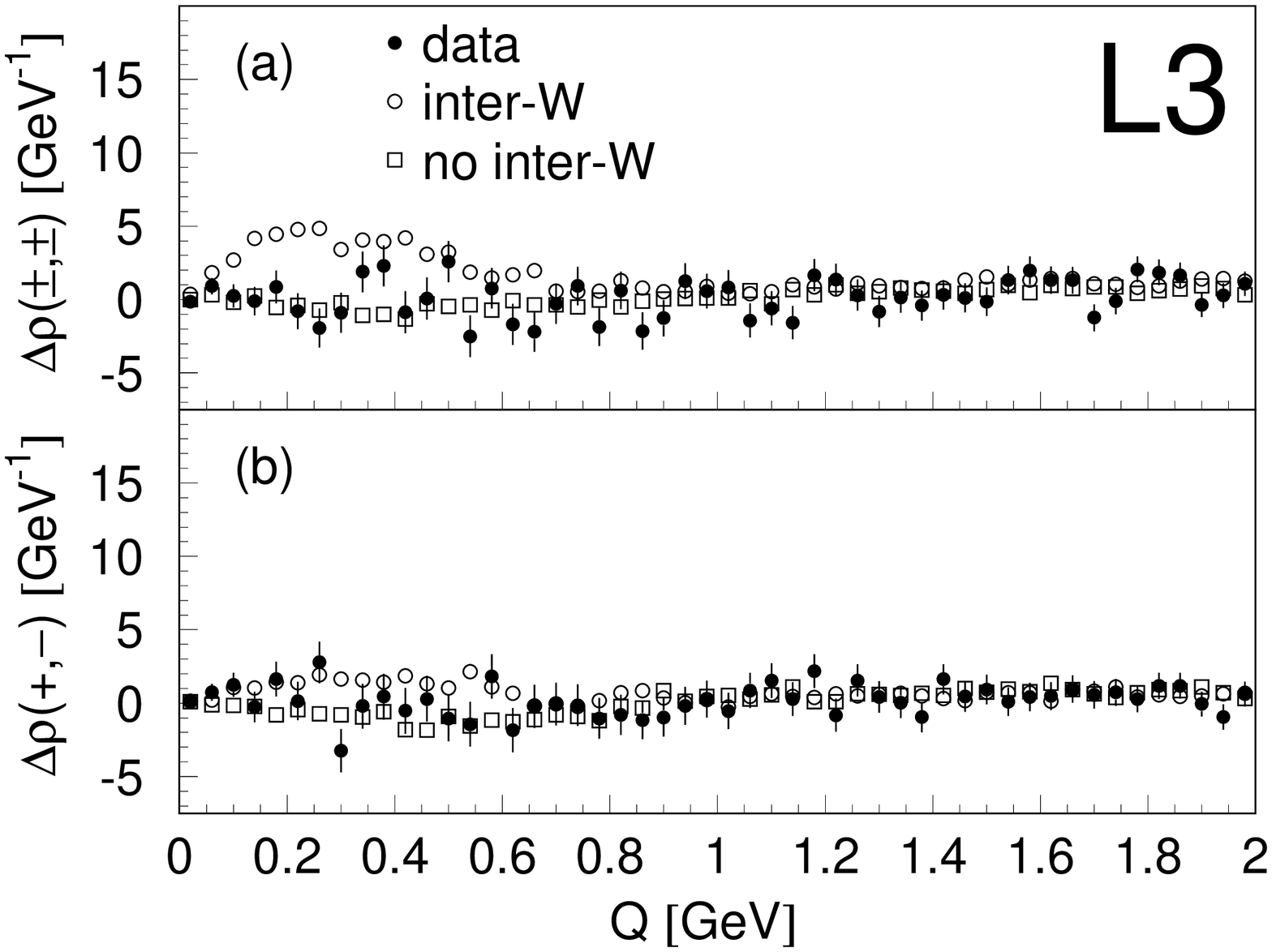,width=7.2cm}
                 \epsfig{file=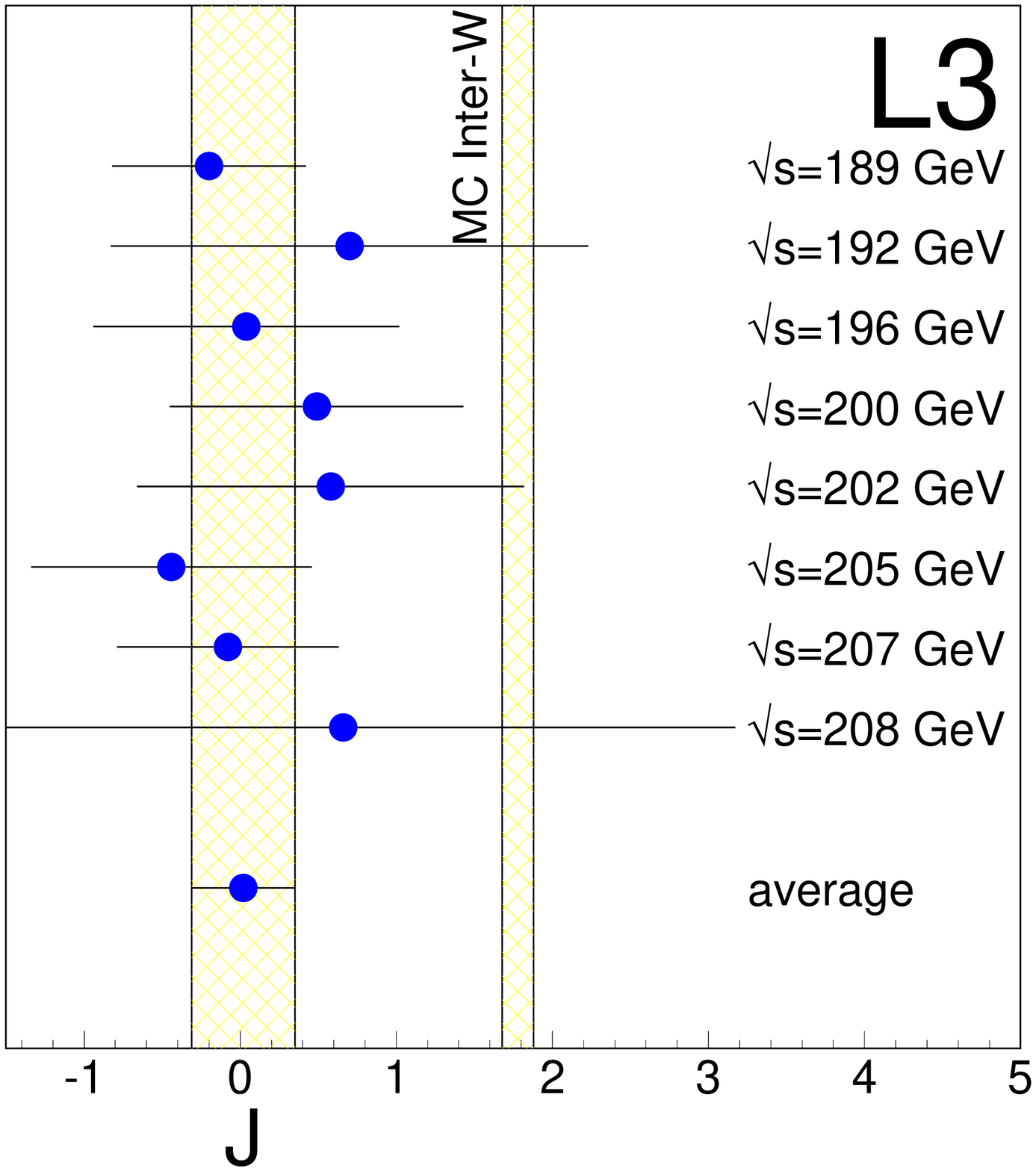,width=5.7cm}
               }
   \caption{Comparison of $\Delta \rho$ and $J$
for \ldrei data with Monte-Carlo models  both with and without Bose-Einstein correlations between particles
from different \W bosons.\protect\cite{bib:be_l3} 
\label{fig:be_l3}}	
\end{figure}
Figure~\ref{fig:be_l3} shows a comparison of the \ldrei data with the LUBOYE ($BE_{32}$) Monte-Carlo\cite{bib:pyboye} both
with and without Bose-Einstein correlations. The data clearly disfavor Bose-Einstein correlations
of  same strength between particles from different \W bosons  as for paritcles within
the same W.

\subsection{Combining of \LEP Results}
In order to combine the separate mass measurements into one \LEP result, the correlation in
the systematic uncertainties  between  different channels, years of data taking, 
and between experiments, must 
be taken into account. The detector systematics and the uncertainties from limited 
Monte-Carlo statistics are the only errors that are assumed to be uncorrelated between 
experiments. The four \LEP experiments claim significantly different 
systematic uncertainties from final state interactions.\cite{bib:al_mw,bib:del_mw,bib:l3_mw,bib:opal_mw}
Nevertheless, studies of Monte Carlo
samples with identical events passed through all the  detector simulations suggest
that the experiments have the similar sensitivity to final-state effects, and that
the differences are due primarily to different choices of  model parameters used to estimate
the uncertainty. In combining the results, the same systematic errors from FSI are used
for all four experiments. Similar studies are in
preparation to gauge  the hadronization uncertainties. The  uncertainty from hadronization and detector effects 
on the measurement of  jet properties are considered to be fully correlated between the
hadronic and semileptonic \W-pair decays. The uncertainty from beam energy
is fully correlated between channels and experiments. The year-by-year correlation
is taken from the estimates of the \LEP Beam Energy Group.\cite{bib:lep_ebeam}
Table~\ref{tab:sys_com} lists the systematic errors on the preliminary determination 
of the combined mass of the \W boson.\cite{bib:lep_ww_results}
\begin{table}
 \tbl{Decomposition in uncertainties for the combined \LEP \W-mass results. 
          Detector systematics include uncertainties
          in jet and lepton energy scales and resolution. The `Other'
          category refers to errors, all of which are uncorrelated
          between experiments, arising from: simulation statistics,
          background estimation, four-fermion treatment, fitting, method 
          and event selection. The error decomposition 
          in the $qql\nu$ and $qqqq$ channels refers to separate fits to 
          the two channels. The last line indicates the statistical error for
          a combination that minimizes the statistical error and not the total
          error.
 \label{tab:sys_com}}{
  \begin{tabular}{|l|r|r||r|}\hline
       Source  &  \multicolumn{3}{|c|}{Systematic Error on $M_\W$ ($\MeV$)}  \\  
                             &  $qql\nu$ & $qqqq$  & Combined  \\ \hline   
 ISR/FSR                     &  8 &  8 &  8 \\
 Detector Systematics        & 12 &  8 & 11 \\
 Hadronization               & 19 & 18 & 18 \\
 \LEP Beam Energy             & 17 & 17 & 17 \\
 Color Reconnection         & $-$& 90 & 9 \\
 Bose-Einstein Correlations  & $-$& 35 &  3 \\
 Other                       &  4 &  5 &  4 \\ \hline
 Total Systematic            & 29 & 101 & 30 \\ \hline
 Statistical                 & 33 &  36 & 30 \\ \hline\hline
 Total                       & 44 & 107 & 42 \\ \hline
 & & & \\
 Statistical in absence of Systematics  & 32 & 29 & 22 \\ \hline

\end{tabular}
}
\end{table}

\section[Mass from Leptonic \bW-Pair Decays]
{Determination of the \W Mass  from Purely Leptonic \bW-Pair Decays}\label{sec:mw_lvlv}
An event, in which  both \W bosons decay leptonically contains two unobserved neutrinos.  It is therefore
not possible to  completely reconstruct the kinematics of such an event.
However, on a statistical basis, the observed charged leptons carry information 
about the \W-boson mass, that is, their  momenta
depend on the mass of the \W-boson. In fact, positions of edges of kinematic distributions,  
reflecting limits of phase space,  can be used to measure the \W mass.
The simplest variable is the energy of a single charged lepton, which can be
expressed as the following  function of the \W  mass:
\[ E_l = \frac{\sqrt{s}}{4} + \cos \theta^*_l \sqrt{\frac{s}{16}- \frac{M_\W^2}{4}}. \]
where $s$ is the square of the center-of-mass energy, and $\theta^*_l$ is the angle between
the lepton direction measured in the \W rest frame and the direction of the \W in the laboratory
frame (``helicity'' frame). The maximum and minimum in the energy spectrum occurs at $| \cos \theta^*_l| =1 $, 
where the lepton
is emitted along or opposite the  direction direction of the \W.
These values clearly depend on the \W mass.

The \opal collaboration uses another ``edge'' variable sensitive to the
W mass.\cite{bib:opal-lvlv}   
Energy and momentum  conservation and the assumption that both \W bosons have the same mass
yield only five constraints. 
The kinematics of the reaction is therefore underconstraind.
But the assumption that the neutrino momentum vectors lie  in the plane 
defined by the momenta of the charged leptons, defines   two solutions for neutrino momenta,
and the event can therefore be  reconstructed. Using this kind of reconstruction, a ``mass'' of the \W can
be expressed as a function of the momenta of the charged leptons, as follows:
\[ M_{\pm}^2 = \frac{2}{|\vp_l+\vp_{\bar{l}} |^2}\bigg( (P\vp_l-N\vp_{\bar{l}}) \cdot
(\vp_l+\vp_{\bar{l}}) \pm \]
\[  \sqrt{ |\vp_l \times \vp_{\bar{l}}|^2 
     \left[ |(\vp_l+\vp_{\bar{l}})|^2 (E_{beam}-E_l)^2-(P+N)^2\right]} \ \bigg) 
\]
where $P$ and $N$ are given by:
\[ P = E_{beam} E_l - E_l^2+ \frac{1}{2}m_l^2 \ \ \ \ 
   N = -E_{beam} E_{\bar{l}} - \vp_l \cdot \vp_{\bar{l}} + \frac{1}{2}m_{\bar{l}}^2. \]
$E_{beam}$ denotes the beam energy, and $\vp_l$, $\vp_{\bar{l}}$, $E_l$, $E_{{\bar{l}}}$,
 $m_l$, and $m_{\bar{l}}$ are, respectively, the momenta, energies, and masses of the charged lepton and 
antilepton. 
Of course, in general, the two  neutrinos  are  not emitted in the same plane as the two charged leptons,
and  the reconstructed mass values are therefore not always close to the \W mass.
Nevertheless, the larger
of the two solutions has a distribution that is in general beyond the 
W mass, and has a lower limit (from events where the two neutrinos are nearly in the same 
plane as the charged leptons) at the mass of the \W boson. In the following, this solution 
will be referred to as  pseudo-mass.

Measurements of the \W mass that rely on lepton energy or on the pseudo-mass 
have only  small correlations because
these distributions correspond to information from different regions of the phase space. 
For the case of  lepton energy, 
mainly events in which a lepton is emitted along the
\W direction affect the mass determination. For the pseudo-mass, events in which the neutrinos 
and charged leptons are nearly in the same plane contribute sharp information to the mass measurement.
Both the pseudo-mass and the energy spectrum rely on a precise measurement of lepton
energy, which is not possible for $\tau$-leptons because of the additional neutrino in each $\tau$
decay. In events, where one of the \W bosons decays into a $\tau$, only the energy spectrum
of the other lepton is used.

The statistical uncertainty of these dilepton  mass determinations  is larger than for
hadronic or semileptonic channels,  because only $\approx 10$\% of the \W-pairs decay leptonically
($\approx 5$ \% decay into electrons and muons), and because only events near the edges of phase space
contribute useful information to the mass determination.
The \alephe collaboration uses the lepton energy spectrum in their analysis,\cite{bib:aleph_ww189}
while the \opal collaboration  uses both the lepton energy
spectrum and the pseudo-mass.\cite{bib:opal-lvlv}

\subsection{Mass Determination}
Both the reweighting technique (see Section \ref{sec:fit_rew}) and the comparison with an 
analytic function  (see section \ref{sec:fit_ana}) can be used
to determine the \W  mass from the lepton energy and the pseudo-mass.
Compared to the hadronic and semileptonic channel, the reweighting technique
has two disadvantages: First, the distributions do not have clear maxima  at specific values, 
but are  broad spectra  with edge cutoffs. Consequently, only a small fraction of the events
contribute to the region that is important for determining the mass. In addition, the larger
statistical error requires a reweighting over a range of the order of the
width of the \W boson. This increases the statistical uncertainties on the reweighted distributions,
and necessitates the use of very large Monte-Carlo samples for different input \W masses
to get stable results.
The need for large Monte-Carlo samples is reduced if the \W mass is determined by comparing
the lepton energy and pseudo-mass distributions with analytic functions. The lepton 
energy $E_l$ can be parametrized as:
\[ f = \frac{1}{e^{\frac{-(E_l-P_1)}{P_2}+1}} \frac{1}{e^{\frac{E_l-P_3}{P_4}+1}}(P_5+P_6 E_l), \] 
where $P_1$ and $P_3$ are given by the position of the rising and falling edges of the energy 
spectrum, $P_2$ and $P_4$ reflect the steepnesses of  the edges, which depend on the lepton energy
resolution, $P_6$ describes the rise of the spectrum with energy, and parameter $P_5$ is determined 
for each set of values of the other parameters by normalizing the function $f$ to unit area or
to the number of events in the data.
\begin{figure}[t]
\parbox[t]{6.2cm}{
   \centerline{\epsfig{file=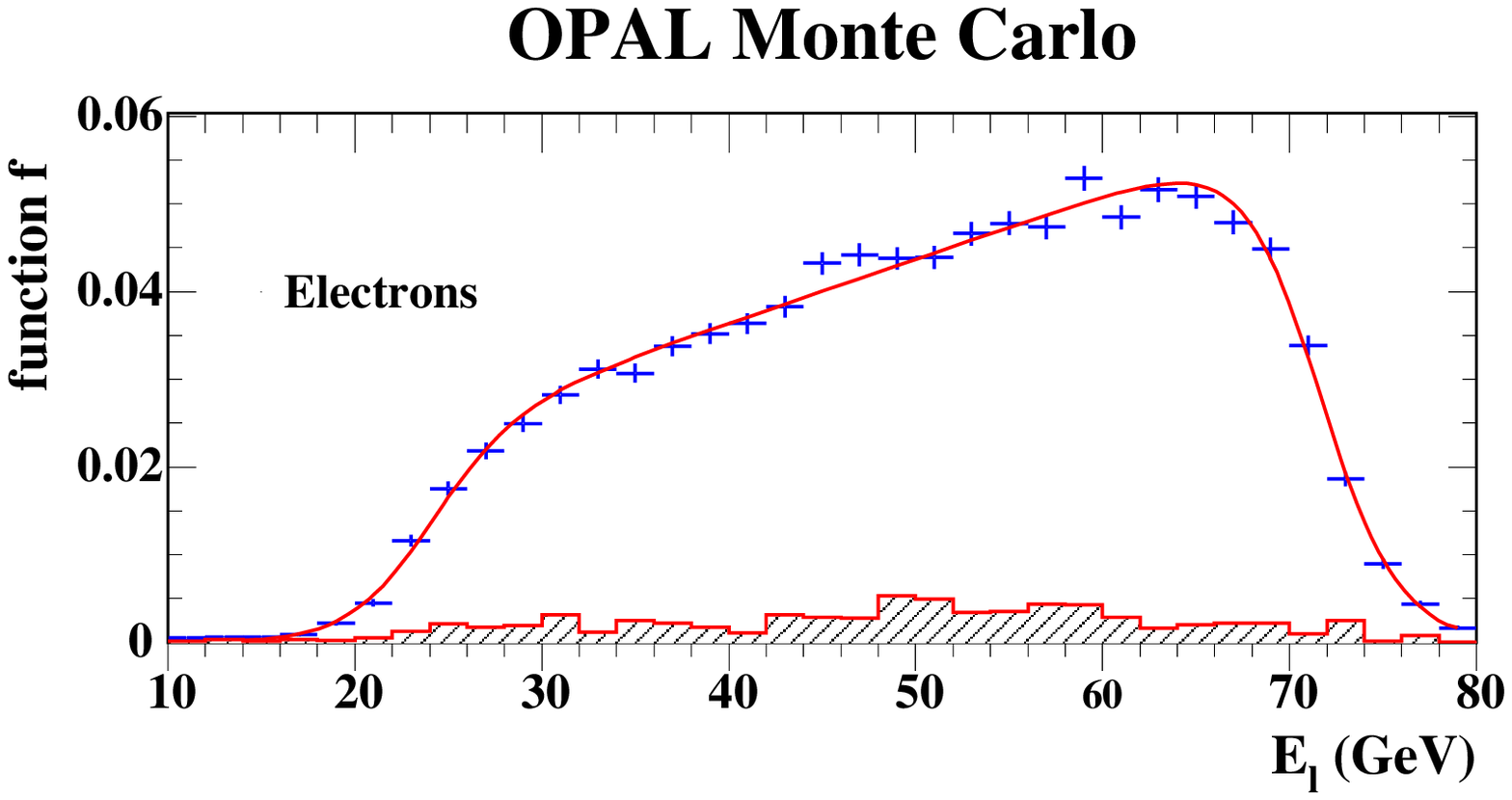,width=6.9cm}}
   \caption{Fit of the electron energy spectrum for Monte Carlo events at $\sqrt{s}=189$ \GeV and
               $M_\W= 80.33$ \GeV. The hashed region indicates the background.\label{fig:lvlv_e}}
} 
\hspace*{.3cm}
\parbox[t]{6.2cm}{
   \centerline{ \epsfig{file=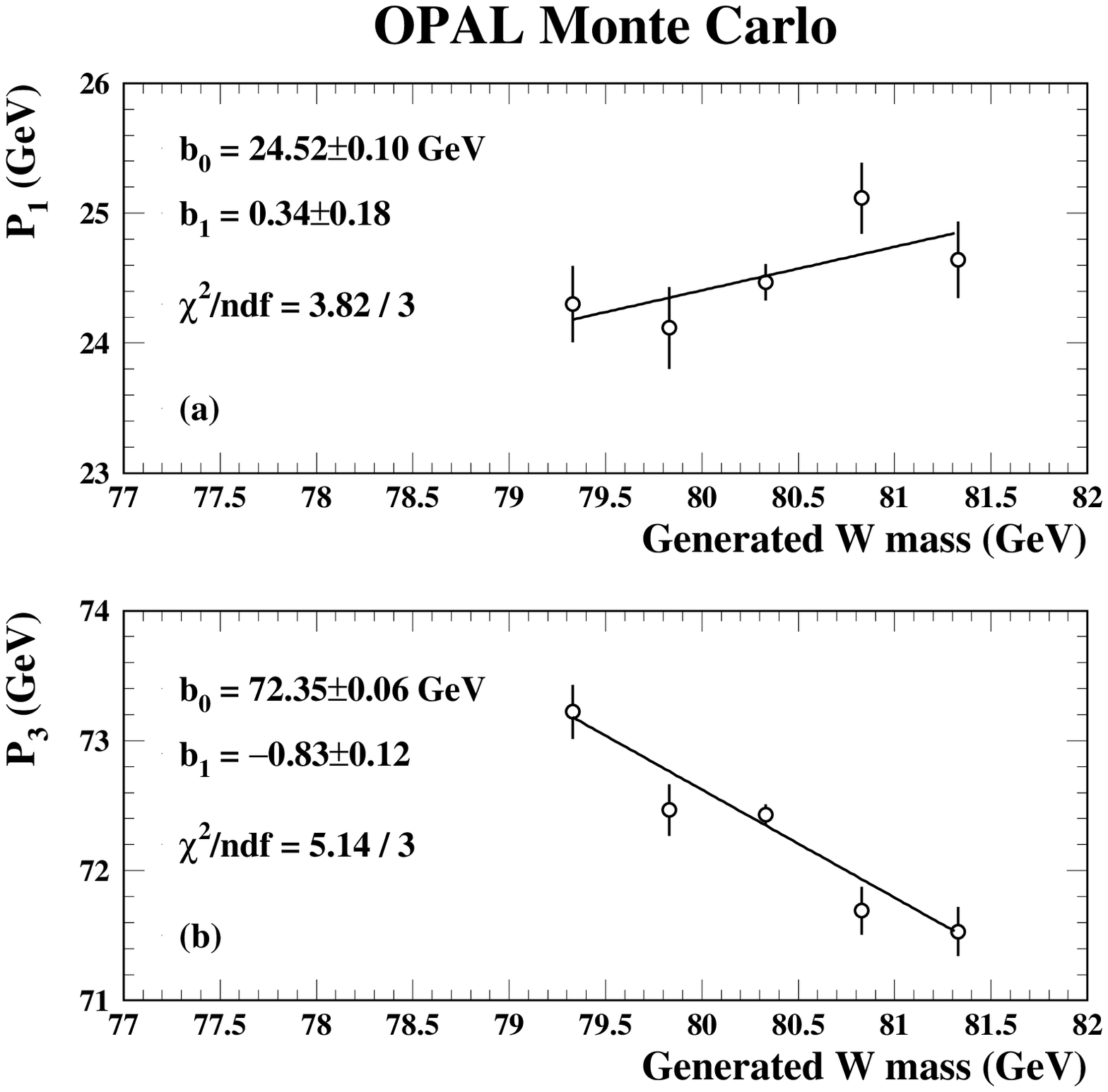,width=6.5cm}}
               \caption{Linear fit [$R=b_0+b_1 \times (M_\W - 80.33 \GeV)$] of the coefficients
               $P_1$ in (a) and  \newline $P_3$ in (b).  \label{fig:lvlv_e2}}
}	
\end{figure}
Figure~\ref{fig:lvlv_e} shows a fit of the electron energy spectrum to a Monte-Carlo sample
at  $\sqrt{s}=189$ \GeV and $M_\W= 80.33$ \GeV/c.
Figure~\ref{fig:lvlv_e2}  shows that in the
relevant mass range  the parameters $P_1$ and $P_3$ depend linearly on the \W-boson mass. 
The other parameters $P_2$, $P_4$ and $P_6$ show no significant mass dependence.

The pseudo-mass distribution can be parametrized with the function
\[ f^\prime = \frac{P_1}{e^{\frac{-(x-P_2)}{P_3}}+1}+P_4, \]
where $P_2$ is given by the position of the edge in the pseudo-mass distribution, while 
$P_3$ reflects the steepness of the edge. The constant term $P_4$ is needed to describe 
events with low pseudo-mass that have mismeasured  lepton momenta.
The parameter $P_1$ can be  determined from the normalization of the  function $f^\prime$ to unit area or
to the number of events in the data.
\begin{figure}[h]
\parbox[t]{6.2cm}{
   \centerline{\epsfig{file=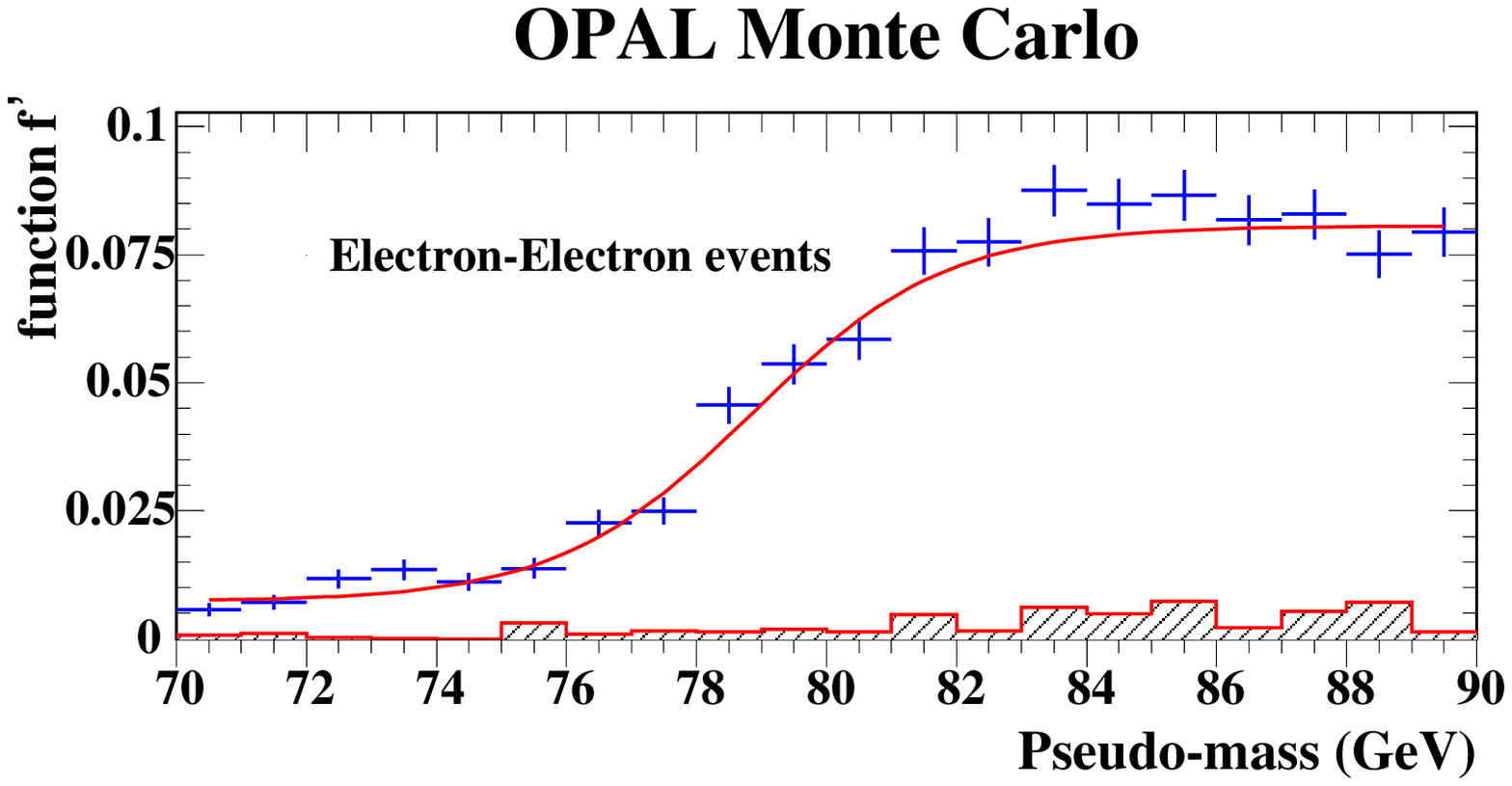,width=6.8cm}
}
\caption{Fit of the pseudo-mass distribution from electron-electron  
                Monte Carlo events  at $\sqrt{s}=189$ \GeV and
               $M_\W= 80.33$ \GeV. The hashed region indicates the background.
                \label{fig:lvlv_p}}	
}
\hspace*{.3cm}
\parbox[t]{6.2cm}{
    \centerline{ \epsfig{file=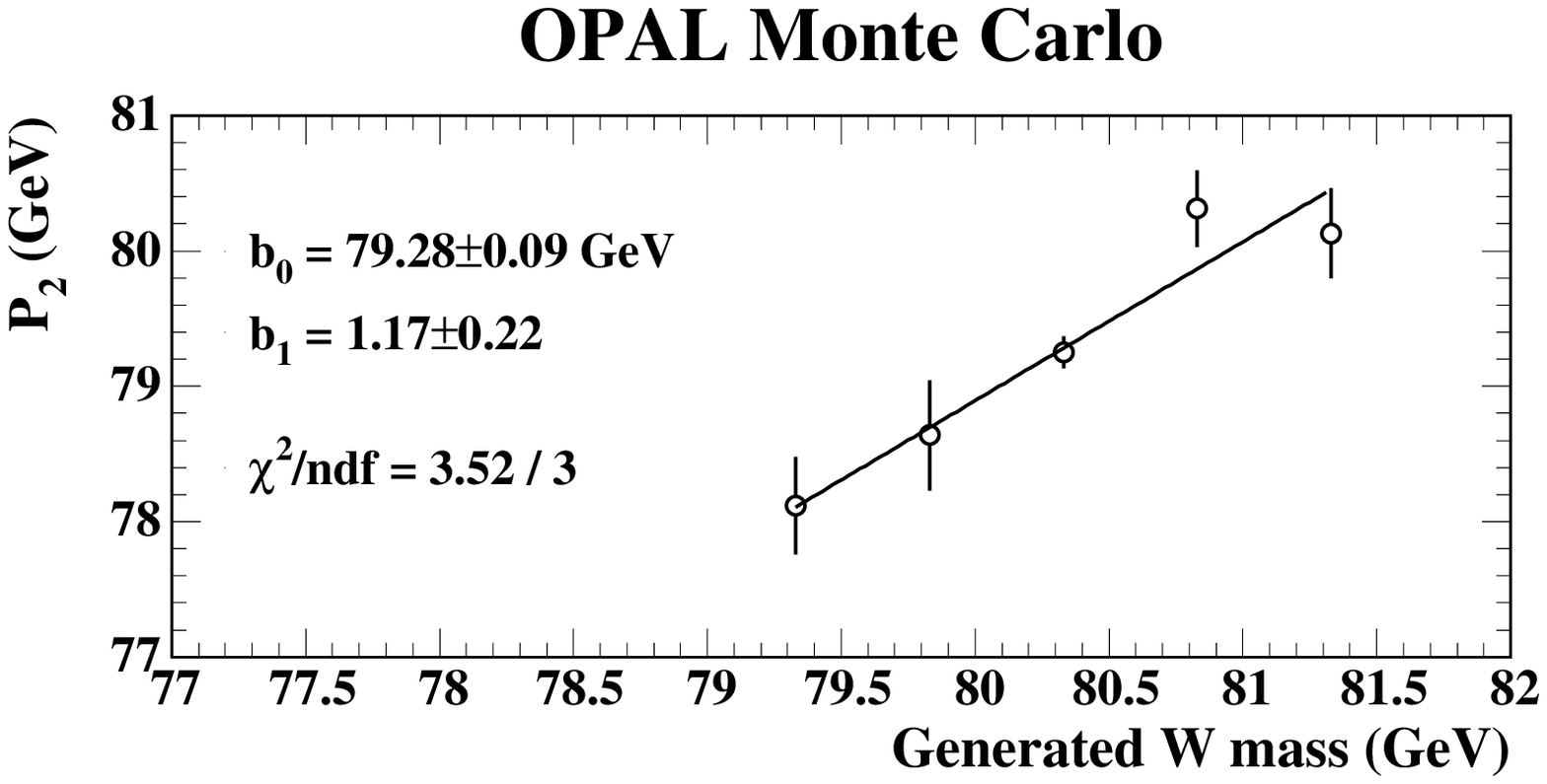,width=6.8cm} }
\caption{Linear fit of the coefficient $P_2$ to
 $b_0+b_1 \times (M_\W - 80.33 \GeV)$
.  \label{fig:lvlv_p2}}	
}
\end{figure}
Figure~\ref{fig:lvlv_p} shows a fit of the pseudo-mass distribution 
for electron-electron events ($\W^+W^- \rightarrow e^+\nu e^- \bar{\nu}$)   for a Monte-Carlo 
sample at  $\sqrt{s}=189$ \GeV and
$M_\W= 80.33$ \GeV.
Figure~\ref{fig:lvlv_p2} shows that the parameter $P_2$ depends  linearly on the \W mass. 
The other parameters show no significant mass dependence.

Since the energy resolution of the \opal detector is significantly better for electrons than
for muons, the lepton energies are  fitted  separately.
For the case of the pseudo-mass, $ee$, $e\mu$ and $\mu\mu$ final states are all treated separately.
\begin{figure}[h]
   \centerline{\epsfig{file=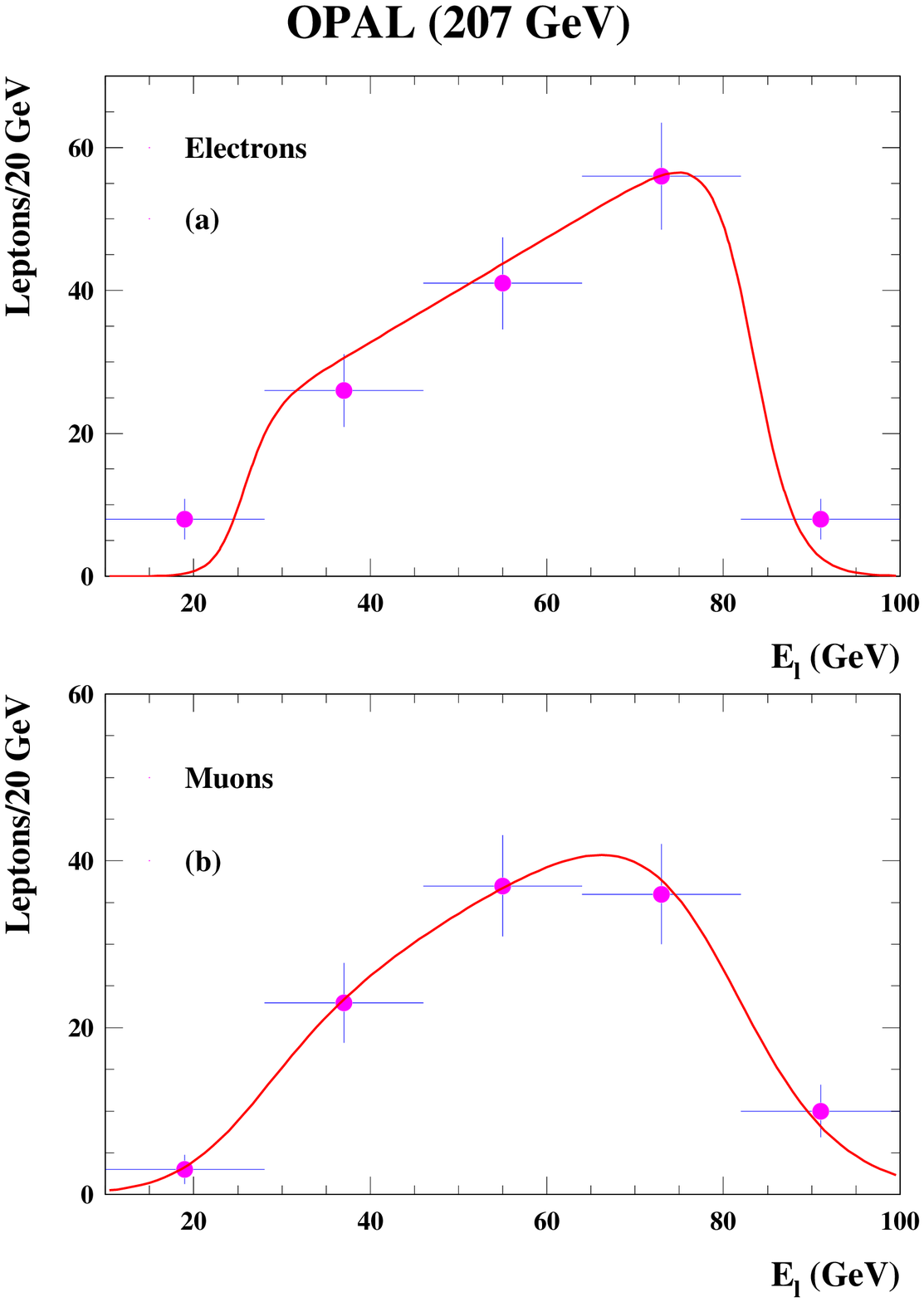,width=6.8cm}
                 \epsfig{file=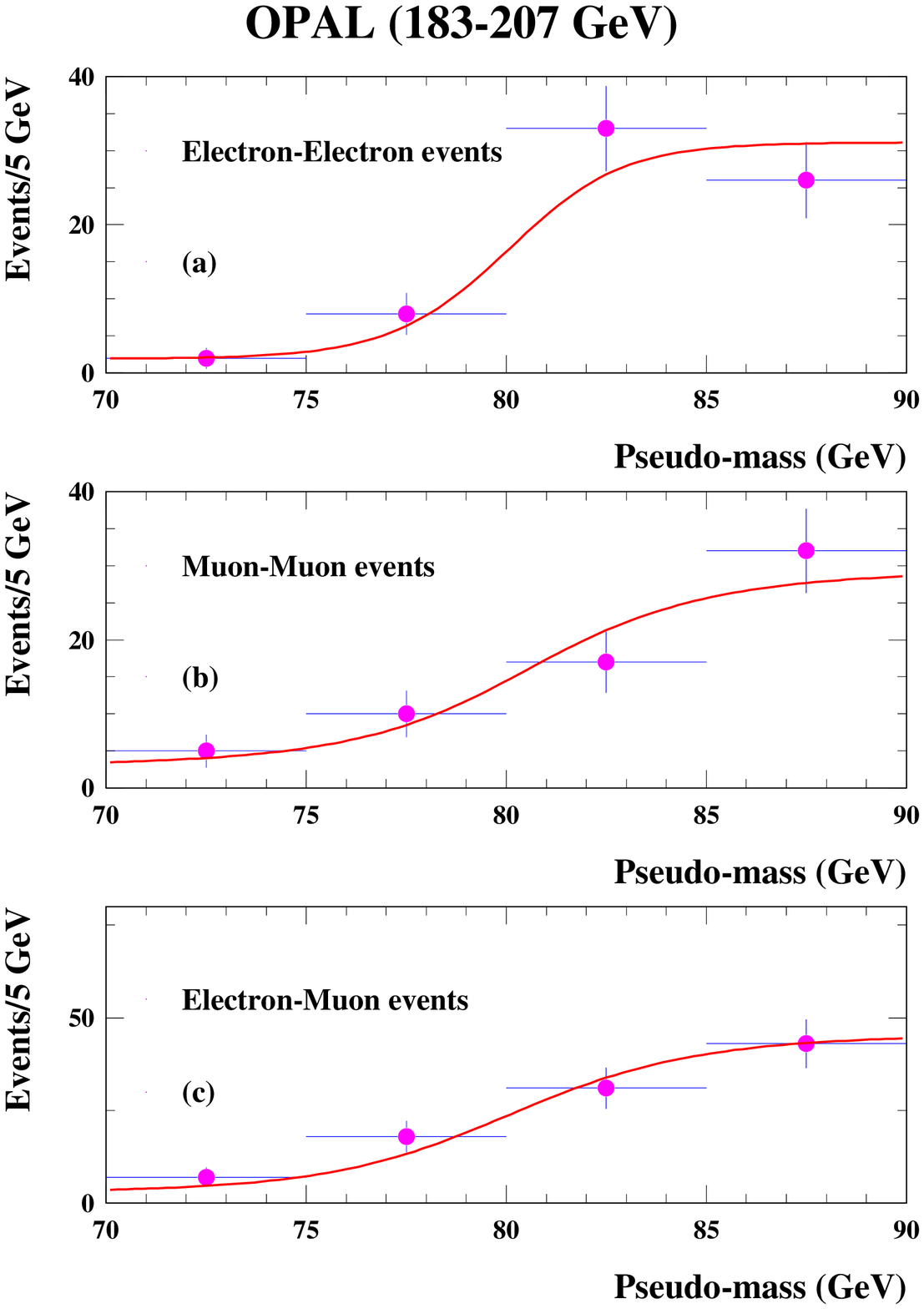,width=6.8cm}
               }
\caption{Fit to the \opal data for the lepton energy spectrum and the pseudo-mass. \label{fig:lvlv_data}}	
\end{figure}
The \W  mass is then determined from a simultaneous fit to the two lepton-energy spectra
and the three  pseudo-mass distributions  at each  center-of-mass energy. 
The parameters that do not depend on the \W  mass are fixed to the values determined by  Monte-Carlo. 
The parameters
sensitive to \W mass are parametrized  as linear functions
of $M_\W$ using Monte-Carlo samples at different \W input masses,
as illustrated in Figures~\ref{fig:lvlv_e}b and c, and~\ref{fig:lvlv_p}b. 
Figure~\ref{fig:lvlv_data} shows the fit to data for the lepton energy and  pseudo-mass at $\sqrt{s}=207$ \GeV.
For illustration purposes, the pseudo-mass distribution is for data
combined from  all center-of-mass energies, and  compared with the corresponding sum of the fitted functions at
different center-of-mass energies. The figure suggests  sharper edges in  distributions
for electrons relative to those for muons, reflecting the better energy resolution for electrons.

To determine the degree of correlation between the mass measurements using the pseudo-mass and
the lepton energy,  separate fits were performed to the energy spectrum and to the pseudo-mass for 
a large set of data-sized Monte-Carlo samples. 
A correlation coefficient of $11\pm  1\%$ was determined
from the distribution  of the  result  for the
lepton energy versus  results  for the pseudo-mass.  
This was confirmed by the pull of the common fit for lepton energy and pseudo-mass.
After correcting  the statistical error by the factor of 1.11 from this the correlation, 
the width of the pull distribution became $1.00 \pm 0.02 $, thereby confirming the internal
consistency of the procedure.

\subsection{Results}
The systematic errors for  the \W mass in the fully leptonic channels
are estimated in a way similar to that discussed in Section~\ref{sec:syst}. 
The result from the \opal collaboration, analyzing the complete \LEPII  data set, is:\cite{bib:opal-lvlv} 
\[ M_\W(l\nu l \nu) =  80.41 \pm 0.41 \pm 0.13 ~\GeV. \]
The first error is statistical and the second systematic. The dominant source for the systematic
errors are the uncertainties in the lepton energy scale and in the resolution. These are limited by
the available statistics from the \Z calibration runs.
The measurement in the fully leptonic channel 
has no contribution from  uncertainties from hadronization; also the error due to the  uncertainty in beam energy
is smaller than for semileptonic events because of the different dependence of the upper and lower edges of the
energy spectrum,  and the edge of the pseudo-mass, on 
beam energy, which for dileptons provide  partial cancellations.   
Future high-luminosity experiments  should therefore be able to reduce the systematic 
uncertainty to a level below that in semileptonic \W pair decays. 

The uncertainty on the  measurement of the \W mass in the fully leptonic channel is large compared to the
other channels. Nevertheless, its inclusion improves the overall precision by about 1\% corresponding to
an additional luminosity of 2\%. More importantly, the analysis is  complementary to those
used in the other channels, and therefore serves as a cross check. Improvements to the \W-mass measurement
will likely come from measurements at hadron colliders (\tevatron and LHC). These are more sensitive
to leptonic \W decays, and the interesting  comparison of  \W mass in  different decay channels can 
probably be carried out only  at electron-positron colliders.

\section{The Mass and Width of the \bW boson}~\label{sec:mw_results}

\subsection{Results}
\begin{figure}[h]
   \centerline{\epsfig{file=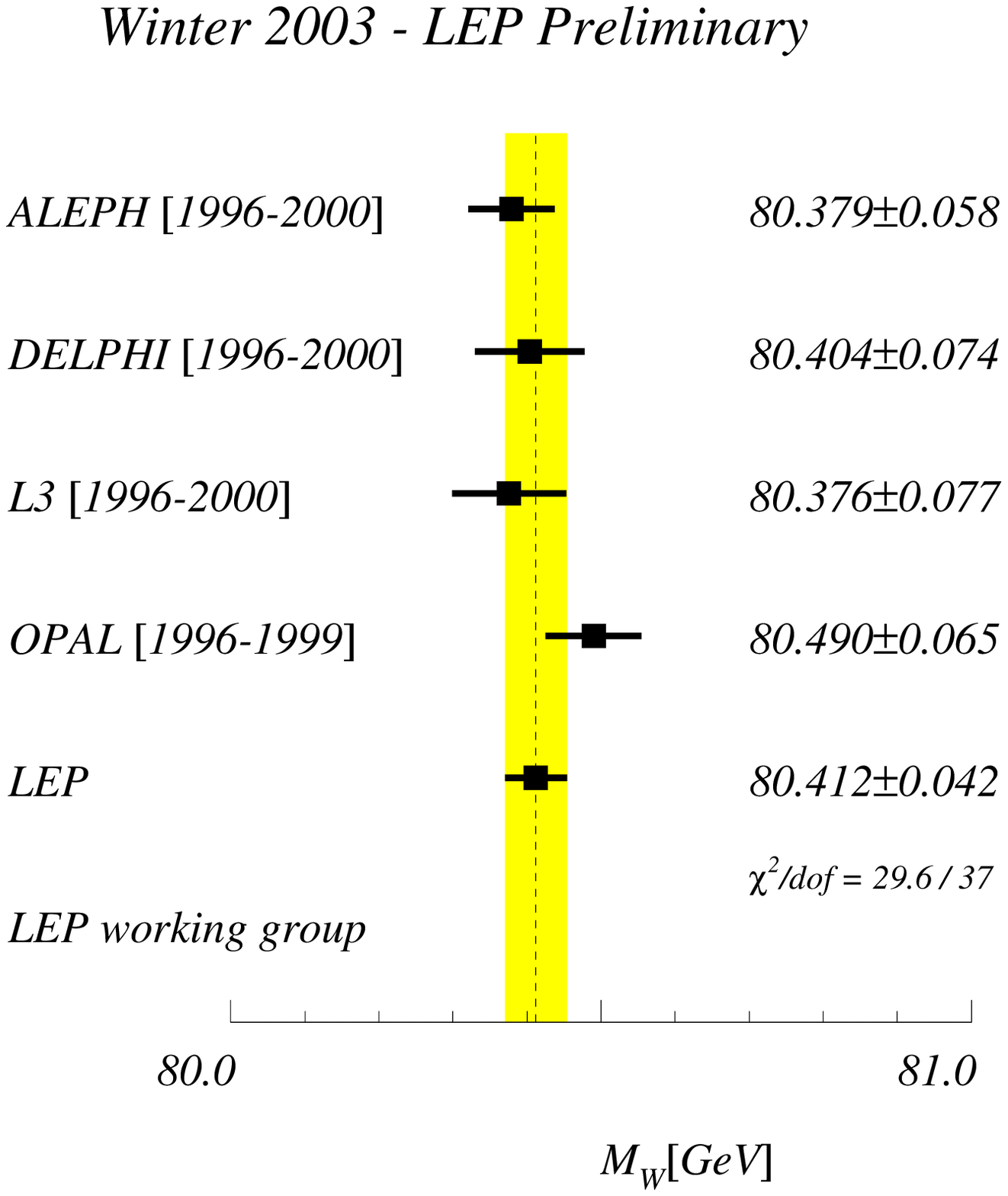,width=6.3cm}
                 \epsfig{file=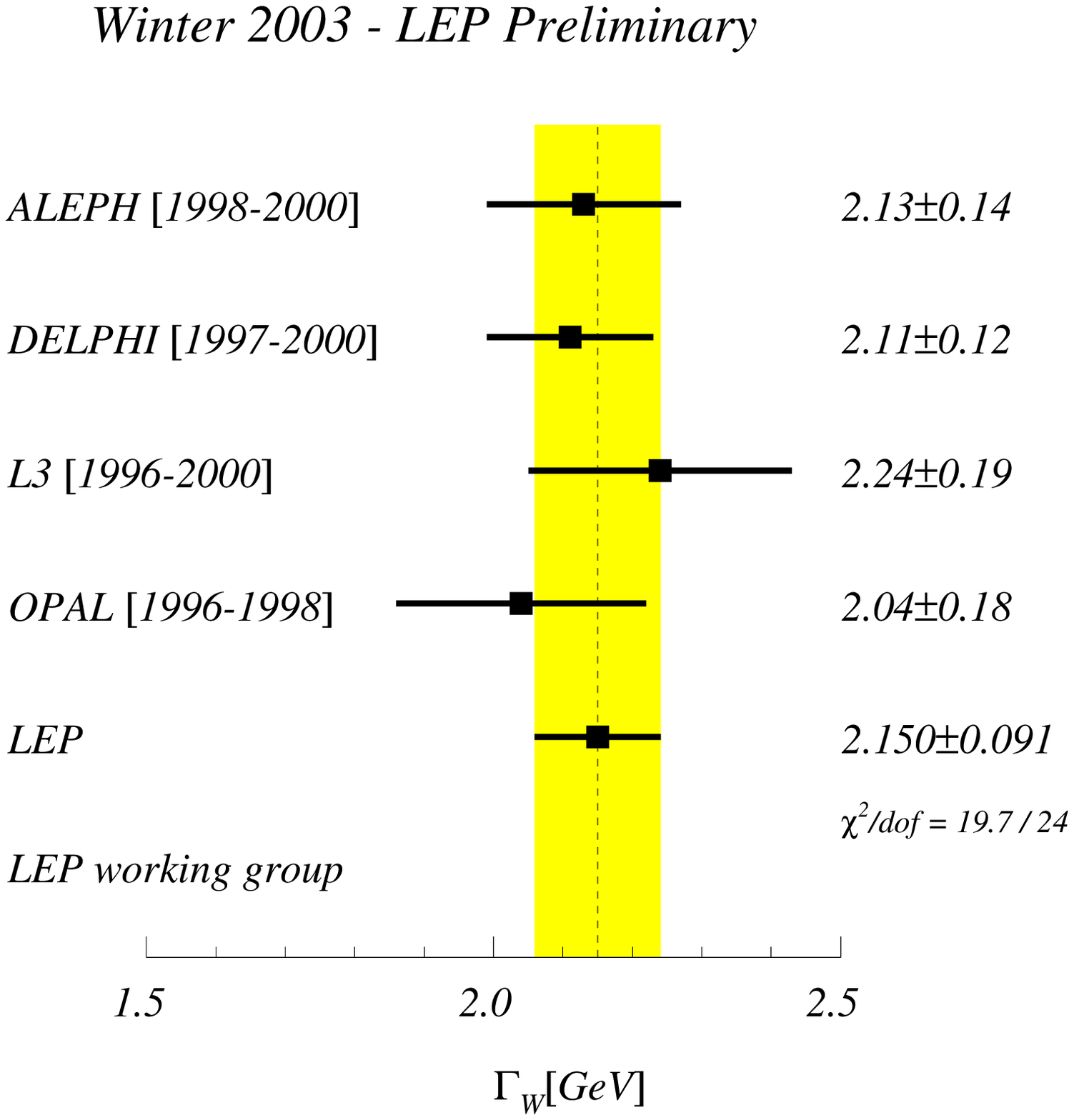,width=6.3cm}
               }
\caption{\label{fig:mw_res_com} 
          The combined preliminary results for the measurements of the
          \W mass and \W width   
          obtained by the four \LEP collaborations.\protect\cite{bib:ww2003-lep} The combined
          values take into account correlations between experiments
          and years of running, and hence, in general, do not give the same central 
          value as a simple average. In the \LEP combination of 
          the $\WWqqqq$ results, common values of errors are used for the color reconnection  and
          Bose-Einstein correlation. The individual and combined $M_\W$ values 
          include the results from
          the measurements of the $\W^+\W^-$  cross sections at threshold.}
\end{figure}

\begin{figure}[t]
   \centerline{\epsfig{file=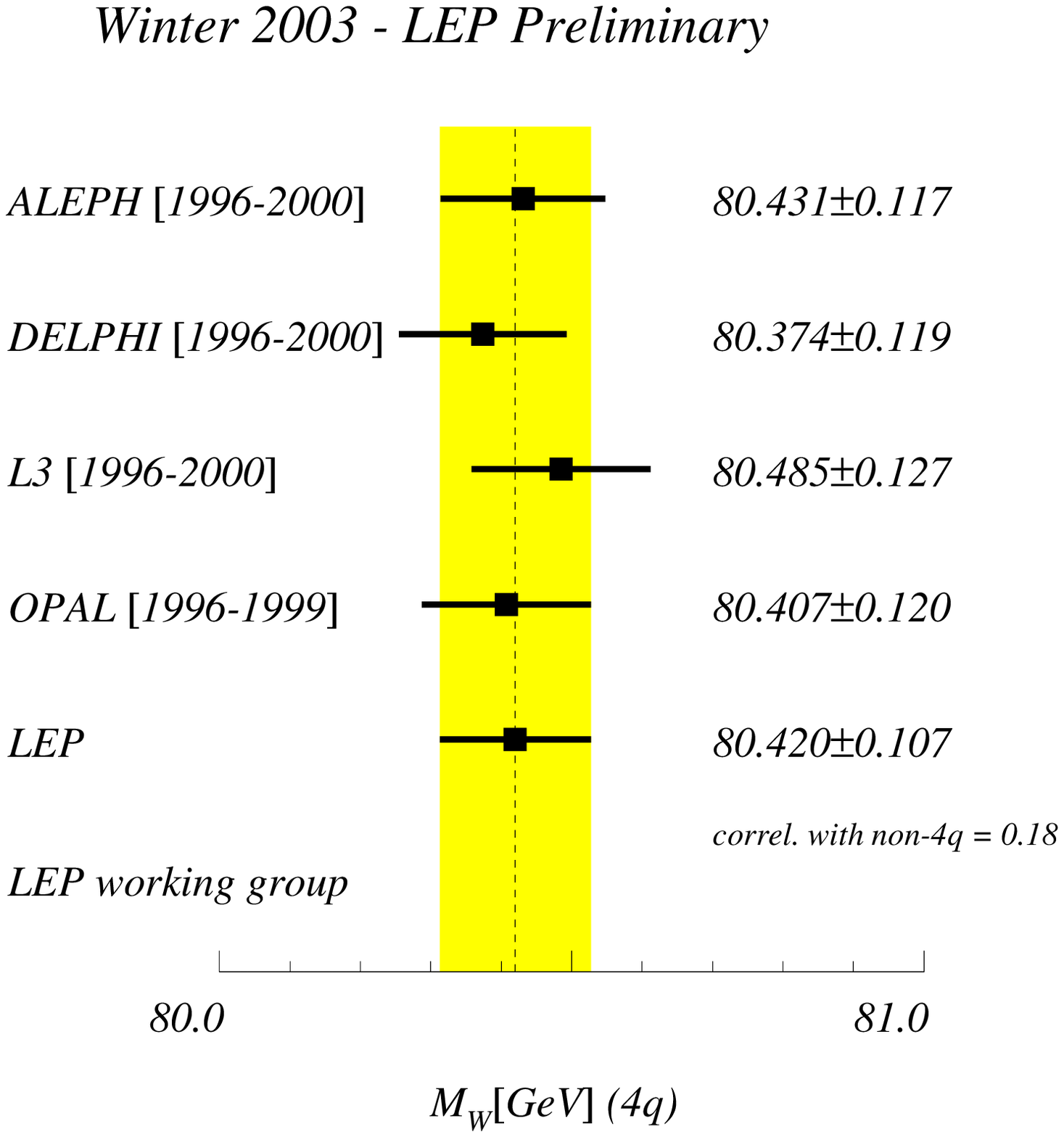,width=6.cm}
                 \epsfig{file=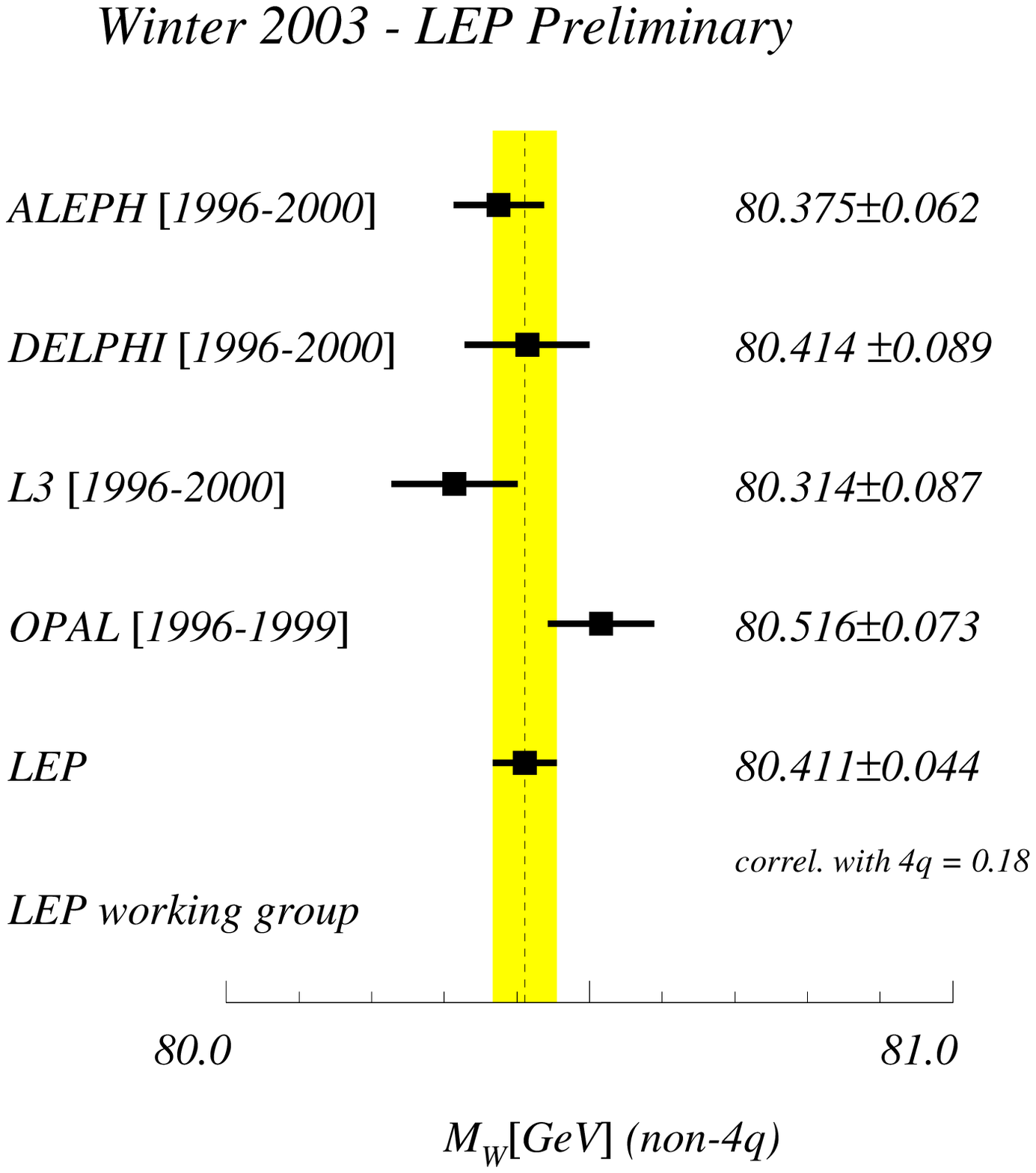,width=6.cm}
               }
\caption{\label{fig:mw_res_sep} 
          Preliminary  measurements of \W-mass
          in  semileptonic  and hadronic  channels 
          obtained by the four \LEP collaborations.\protect\cite{bib:ww2003-lep}
          The combined values take into account 
          correlations between experiments, years of running, and correlations between the two channels.
          In the \LEP combination of 
          the $\WWqqqq$ results, common values  of errors are used for color reconnection  and 
          Bose-Einstein correlation. 
          The \alephe and \ldrei $\WWqqln$ and $\WWqqqq$ results are
          correlated since they are obtained from a fit to both channels 
          that take  account of inter-channel correlations. }
\end{figure}

The preliminary results of the direct  reconstruction of the \W mass in all four \LEP experiments
are given  in Table~\ref{tab:wmass_experiments}.\cite{bib:al_mw,bib:del_mw,bib:l3_mw,bib:opal_mw,bib:ww2003-lep}
 Figure~\ref{fig:mw_res_com} shows the values obtained for the \W  mass and its width.
Figure~\ref{fig:mw_res_sep} shows  results separately for the hadronic 
and semileptonic \W-pair channels.
The preliminary combined result for the four \LEP experiments and all
channels, including the threshold measurements of Ref.\cite{bib:thr}, is:\cite{bib:lep_ww_results,bib:ww2003-lep}
\begin{eqnarray*}
           M_\W = 80.412\pm0.029(\mathrm{stat.})\pm0.031(\mathrm{syst.})  \GeV.
\end{eqnarray*}

\begin{table}
\tbl{Preliminary \W mass measurements from direct reconstruction
         ($\sqrt{s}=172-209$~\GeV).\protect\cite{bib:al_mw,bib:del_mw,bib:l3_mw,bib:opal_mw,bib:ww2003-lep}
          The first error is statistical and the second systematic. 
          Results are given for the
         semileptonic, the fully-hadronic channels, and their combination.
           The $\WWqqln$ results from the 
         \alephe and \opal collaborations include mass information from 
         the $\WWlnln$ channel.  \label{tab:wmass_experiments}
       }{
\footnotesize
  \begin{tabular}{|r|c|c||c|}\hline
    \multicolumn{1}{|c|}{ }
           & \WWqqln         & \WWqqqq         & Combined        \\   
Experiment & $M_\W$  \GeV        & $M_\W$ \GeV        & $M_\W$ \GeV        \\ \hline
     \alephe 
           & $80.377 \pm 0.062 $ & $80.431 \pm 0.117 $
& $80.385 \pm 0.059$ \\ 
     \delphi 
           & $80.414 \pm 0.089$ & $80.374 \pm 0.119$
& $80.402 \pm 0.075$ \\ 
     \ldrei 
           & $80.314 \pm 0.087$ & $80.485 \pm 0.127$
& $80.367 \pm 0.078$ \\ 
     \opal 
           & $80.516 \pm 0.073$ & $80.407 \pm 0.120$
& $80.495 \pm 0.067$ \\ \hline
\end{tabular}
}
\end{table}


As a cross check,  the difference in the \W  mass determined from
hadronic and semileptonic \W-pair decays is:\cite{bib:lep_ww_results,bib:ww2003-lep}
\begin{eqnarray*}
 \Delta M_\W(\qqqq-\qqln) =  +22\pm43~\MeV.    
\end{eqnarray*}

A significant value for $\Delta M_\W$ could indicate that 
final-state interaction effects are biasing the measurement  of $M_\W$ determined from $\WWqqqq$ events.
Since $\Delta M_\W$ is primarily of interest as a check on the possible 
effects of final-state interactions, the errors from color reconnection and Bose-Einstein correlation
 are ignored  in this determination.

\subsection{Implications for the Standard Model}
\begin{figure}
\parbox[t]{6.2cm}{
\centerline{\epsfig{file=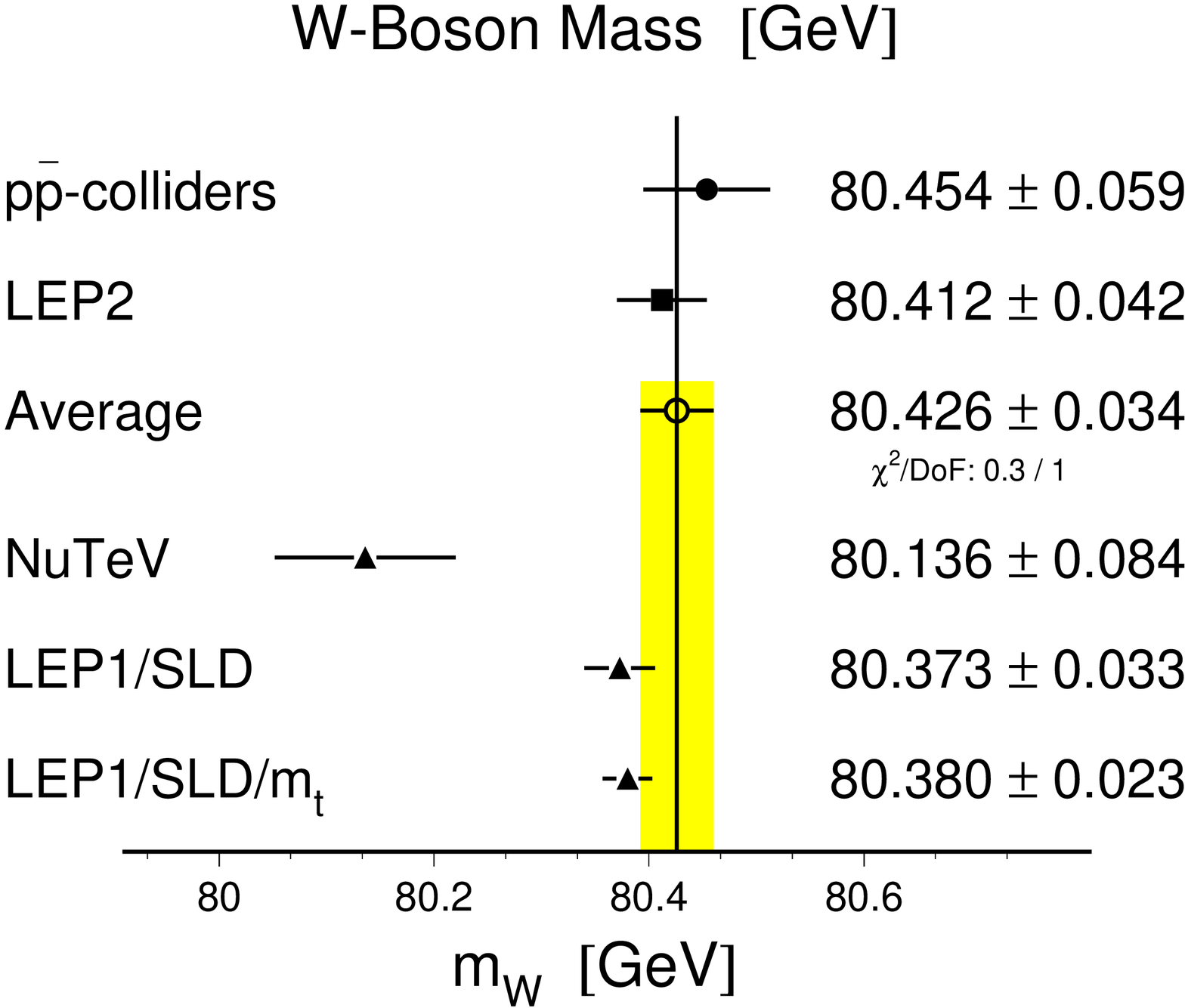,width=6.1cm}}
\caption{\label{fig:mw_comp} 
   Comparison of direct  measurements of $M_\W$ at \LEPII and
the \tevatron with indirect measurements.\protect\cite{bib:ww2003-lep}}
} \hspace*{.3cm}
\parbox[t]{6.2cm}{
\centerline{\epsfig{file=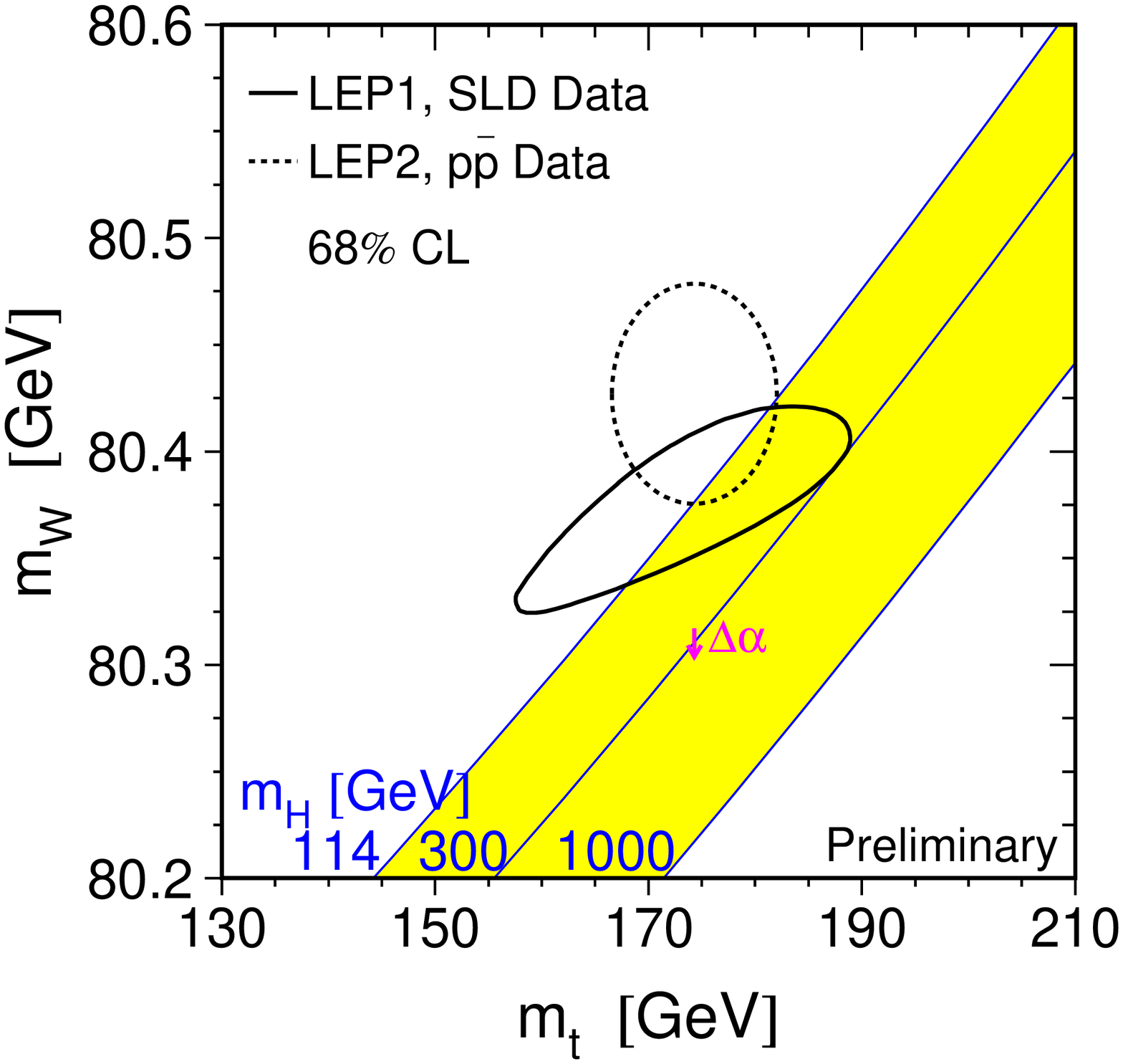,width=6.1cm}}
\caption{\label{fig:res_mw_mt} 
Comparison of direct measurements of $M_\W$ and $m_{t}$ with the results
of a Standard Model fit to electroweak data.\protect\cite{bib:ww2003-lep}
The gray bands show the prediction  for different Higgs masses.}
}
\end{figure}
The   properties of the \Z boson measured at \LEPI and SLD can be used
to predict the \W  mass. 
Figure~\ref{fig:mw_comp} compares the direct measurements 
from \LEPII and  the \tevatron\cite{bib:mw_tev} with indirect predictions from
the ratio of neutral-current to charged-current reactions in
neutrino-nucleon scattering from \nutev,\cite{bib:mw_nut} and with predictions
from fits to precision electroweak data.\cite{bib:elw_fits} 
The direct measurements of $M_\W$ now have similar uncertainty  as the 
predictions from fits to the Standard Model. Figure~\ref{fig:res_mw_mt} compares the
direct measurements of the mass of the top quark\cite{bib:mt_tev} and of the \W boson  
with the prediction of the Standard Model,
and with a fit to electroweak precision measurements. 
The gray  bands show the dependence  of the \W  mass 
on the top and Higgs masses in the Standard Model.
(As a result of loop corrections in the \W propagator, $M_\W$ depends 
quadratically on the top mass and logarithmically on the Higgs mass.)
The dependence is shown for three  values of Higgs mass,
with the  value of $m_H=114$ \GeV corresponding to the 95 \% lower bound on $m_H$ 
obtained from the direct
search for the Higgs boson.\cite{bib:higgs}
Similarly, the properties of the \Z boson depend 
on the Higgs and top-quark masses, and this can be used 
 to determine the \W  and the top-quark mass without reference to the direct
mass measurements.  
The fact that
the two  measurements agree within their uncertainties
 constitutes a particulary stringent test of the Standard Model at the 
level of quantum-loop corrections. 

\begin{figure}
\centerline{\epsfig{file=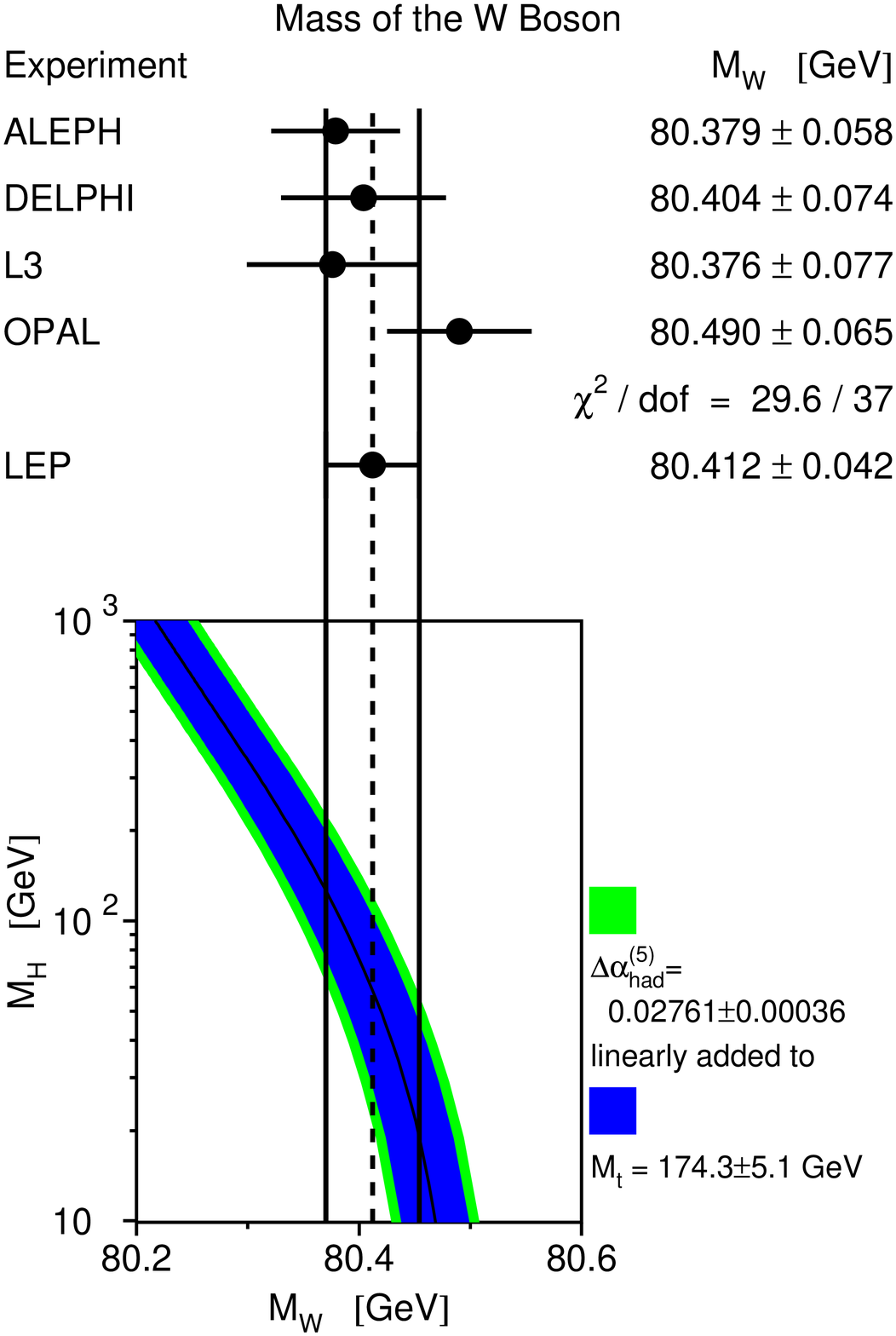,width=5.8cm}
            \epsfig{file=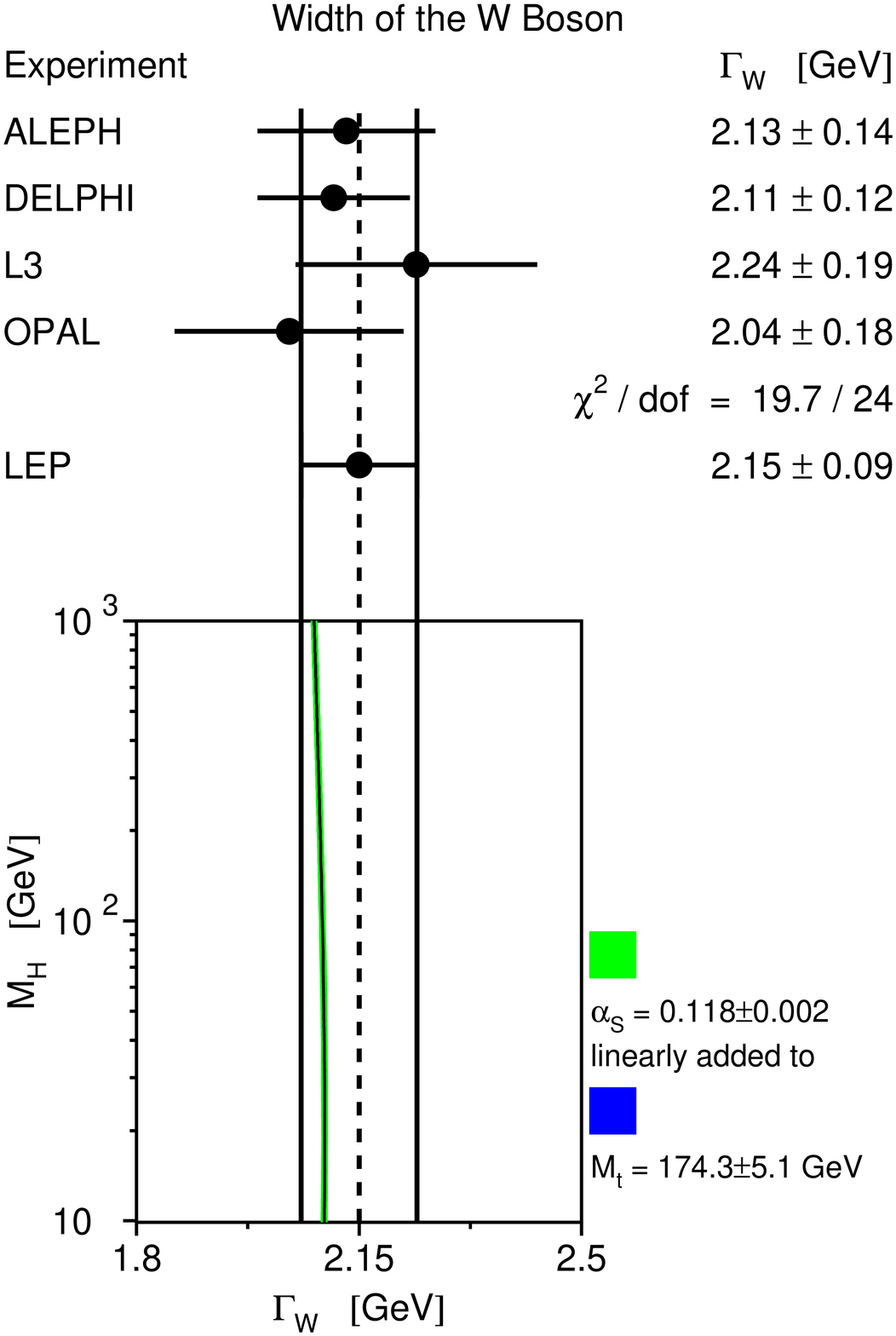,width=5.8cm}}
\caption{\label{fig:res_higgs} 
Dependence of the \W mass and width on Higgs mass.}
\end{figure}
The impact of the mass and width of the  \W boson on  the Higgs mass
is illustrated in Fig.~\ref{fig:res_higgs}, where the \LEP  measurements are
compared to the  predictions of the Standard Model as a function of  Higgs
mass. The value of the \W  mass suggests a light Higgs,
while the width  is not sufficiently precise to  influence fits to
the Standard Model.

\section{Summary and Outlook}
\subsection{Summary}
From 1996 to 2000, the \LEP accelerator operated at  center-of-mass energies above the
production threshold for W-boson pairs. One of the main goals was the precision determination
of the \W  mass. 
The successful operation of the \LEP accelerator and detectors, together with the great efforts
to understand the source of systematic uncertainties in the
measurement of the mass of the \W boson,
made it possible to reach this goal.
The preliminary result from the
combination of all \LEP experiments and all channels is:\cite{bib:lep_ww_results,bib:ww2003-lep}
\begin{eqnarray*}
           M_\W = 80.412\pm0.029(\mathrm{stat.})\pm0.031(\mathrm{syst.})  \GeV.
\end{eqnarray*}
he precision of this measurement is comparable with the precision of its indirect prediction
from electroweak fits to data measured at the \Z 
resonance by the experiments at \LEP and SLD.
The fact that the direct measurements of the \W  and top-quark masses agree 
with the Standard Model is 
an important test of  electroweak theory at the level of quantum-loop corrections.

The kinematic reconstruction of four-fermion events in the region of  phase space
dominated by W-pair production provides the key ingredient to  the precise determination of the 
mass of the \W boson. The mass resolution is improved greatly by 
forcing  energy and momentum conservation
in the kinematic reconstruction of the final state. When  both W bosons decay hadronically,
the mass extraction  can be improved by treating events with hard-gluon radiation 
as having five instead of four jets.

\LEP experiments  use different approaches for determining the \W 
mass from the information content of each event. One
approach is to fit the reconstructed mass spectrum with an analytic function, and 
then extract  the \W mass  from the  parameters of the fit, with any
additional bias corrections
estimated from Monte-Carlo simulations. In a second approach, the results from the
kinematic fits for each event are compared to Monte-Carlo-generated reference distributions,
which can be recalculated for arbitrary
\W masses through  a reweighting procedure. In a third approach,
event likelihoods are determined from results of a kinematic fit through the convolution
of a matrix element for the process  with a resolution function, and 
the \W  mass is extracted from a maximum likelihood fit that is  
based on the mass-dependence of the event likelihoods.

To bring  the systematic uncertainties of the mass
measurements to levels comparable to the statistical errors,
required consideration of many detailed issues, among which are that:
The results of  kinematic fits to final states have explicit  dependence on  center-of-mass energy,
with the consequence that the uncertainty on beam energy 
contributes to the systematic error on the \W-boson mass;
the detector response to jets and leptons be  studied using control samples
of \Z events  collected in parallel with the $\W\W$ events, but at $e^+e^-$ energies near the \Z resonance;
the systematic uncertainties from fragmentation of partons
into hadrons be estimated from comparisons of different Monte-Carlo models,
and these models  tuned to describe previous high-statistics \Z data.
(One important aspect of the hadronization uncertainty is the sensitivity of the \W mass  to the 
fraction of baryons and kaons  produced in parton fragmentation). 

For events where both \W bosons decay hadronically, the effects of interactions and 
interference between the two hadronic systems  (Bose-Einstein correlations
and color reconnection) are not as yet fully understood,  but can affect  significantly 
the systematic uncertainty on the \W mass in these final states.
As a consequence, these events contribute less than 10\%
to the final combination of mass results, despite the fact that their statistical precision is
comparable to that of the semileptonic \W-pair decays. 

The measurement of the total and differential four-fermion cross sections clearly demonstrates
 the presence  of the $WWZ$ and $WW\gamma$  gauge-boson couplings. 
From the agreement of these measurements with the Standard Model, it 
is possible to set limits on various anomalous contributions to the couplings,
which are in general in the range of a few percent.
Neutral gauge-boson couplings are studied using final states
dominated by \Z pair production (at center-of-mass energies above the 
threshold for \Z pairs).
Final states involving photons are used to study quartic gauge boson couplings.

\subsection{Outlook}
In the future, both hadron-collider experiments, and any next generation linear electron-positron
collider, will be able to contribute to the improvement of the precision on the \W  mass.

It is not possible, for various reasons,  to fully
reconstruct  \W events in hadronic collisions, but primarily this is because 
the interactions are between quarks and gluons contained within
the colliding hadrons. Consequently, neither the center-of-mass energy nor the momentum
parallel to the beam direction of the hard interaction are known a priori.    
Also, future hadron colliders will be operated at luminosities at which many overlapping 
and unrelated  primary interactions will be  recorded at the same time. \
In addition, the reconstruction of  hadronic
\W-boson decays is exceptionally challenging because of the presence of
huge background from strong-interaction processes at
hadron colliders. The \W  mass at hadron colliders must therefore be determined 
from the momenta of charged leptons in leptonic \W decays and from the imbalance in momentum in
the plane perpendicular to the beam (missing transverse momentum). 

The current value of the  mass of the \W boson from  the \tevatron proton-antiproton 
collider is $M_\W = 80.456 \pm 0.059~\GeV$, and  is limited both by the statistical and
systematic uncertainties in the determination of the lepton energy and the missing transverse
momentum.\cite{bib:mw_tev} Both the lepton energy and the missing transverse momentum can be calibrated 
using leptonic \Z decays. The uncertainty of these calibrations is currently dominated 
by the  statistics of \Z events.

After a major upgrade of the \tevatron accelerator and detectors, a new data-taking period
started in Spring 2001. The integrated luminosity is expected to increase by a factor of $\approx 100$ in
the next $\approx 6$ years.  This increase will  reduce both  the statistical  and
the systematic uncertainty currently limited by the statistical uncertainty of the calibration samples.
Uncertainties from  parton density functions (PDF), higher-order QCD corrections,
and non-uniformities of the electromagnetic calorimeter, however, will not scale with
integrated luminosity. 
The total uncertainty expected to be achieved  on the \W  mass  is $\approx 30$ \MeV.\cite{bib:mw_tev2}  

The proton-proton collider LHC is expected to start  taking data in $\approx 2007$.
At the LHC, the  rates for \W and \Z bosons will be so high that the statistical
uncertainties of the measurement (or from the calibration samples) will be negligible.
Even with these large samples of \Z bosons, 
the dominant errors on the measurement of the \W  mass  will be from the energy and
momentum  of electrons and muons. It will be quite challenging to calibrate the
detectors to the level needed to reach a
total error of 20 \MeV.\cite{bib:atlas_tdr}

At any future linear electron-positron collider, the highest precision for the 
\W-boson mass should be reached by a scan of the \W-pair production cross section 
near threshold.\cite{bib:mw_tesla} 
With an integrated luminosity of $100~\fb^{-1}$, the \W mass can be determined with an experimental
precision of $\approx 6$ \MeV, assuming the same efficiency and purity 
reached at \LEP, and assuming a total error of 0.25\% on the luminosity.
A full four-fermion calculation of the cross section as a function of the
\W  mass (with radiative corrections)  will be required to reach
this ultimate level of accuracy.


\section{Acknowledgments}
This kind of review is clearly not possible to undertake without the support of many colleagues. 
I am grateful to the members of the \LEP \W Working Group and the members of the four
\LEP collaborations working on \W physics for their inputs.
I would like to thank Tim Christiansen, J\"org Dubbert, 
G\"unter Duckeck and Tom Ferbel for helpful discussion and advice in preparing this
document. 
I also want to express my gratitude to Dorothee Schaile for leading our group at LMU
in a way that created an atmosphere that stimulated creative contributions and a free exchange
of ideas, and I  thank her for encouragement and for her support of my 
scientific and academic development, which made my
work in the field of \W physics at \LEP and this review possible.
\thispagestyle{empty}
\clearpage{\pagestyle{empty}\cleardoublepage}

\thispagestyle{empty}
\clearpage{\pagestyle{empty}\cleardoublepage}

\end{document}